\documentclass[apj,twocolumn,twocolappendix,numberedappendix]{openjournal}
\usepackage{newtxtext,newtxmath,enumitem,multirow,ctable,verbatim}
\usepackage[T1]{fontenc}
\usepackage[dvipsnames]{xcolor}
\usepackage[breaklinks,colorlinks,citecolor=blue,urlcolor=blue]{hyperref}

\newcommand{\Alf}{{Alfv\'en}}

\newcommand{\paperone}{Paper {\small I}}

\newcommand{\orcidauthor}[3]{\author{\href{http://orcid.org/#1}{#2$^{#3}$}}}
\newcommand{\tblexp}[1]{& {#1}\\ }
\newcommand{\tblspace}{& \vspace{-0.1cm} \\ }
\newcommand{\tbltopic}[1]{{#1}}
\newcommand{\tblpred}[1]{& \textcolor{blue}{ \ \ \ \ \ \ \ \ \ \ -- predicts: #1}\\ }
\newcommand{\tbltrad}[1]{& \textcolor{orange}{ \ \ \ \ \ \ \ \ \ \ -- traditional: #1}\\ \hline}
\newcommand{\tblsec}[1]{{#1}}

\shorttitle{CRs As CCs: Observations}
\shortauthors{Hopkins et al.}

\begin{document}

\title{\vspace{-0.8cm}Observational Implications of Cosmic Ray-Inverse Compton ``Boosted'' Cool Cores in Clusters\vspace{-1.5cm}}

\orcidauthor{0000-0003-3729-1684}{Philip F. Hopkins}{1*}
\orcidauthor{0000-0002-1616-5649}{Emily M. Silich}{1}
\orcidauthor{0000-0002-8213-3784}{Jack Sayers}{1}
\orcidauthor{0000-0002-7484-2695}{Sam B. Ponnada}{2}
\orcidauthor{0000-0002-1159-4882}{Isabel S. Sands}{1}
\affiliation{$^{1}$Division of Physics, Mathematics, and Astronomy, California Institute of Technology, Pasadena, CA 91125, USA}
\affiliation{$^{2}$Department of Space, Earth and Environment, Chalmers University of Technology, Gothenburg, Sweden}

\thanks{$^*$E-mail: \href{mailto:phopkins@caltech.edu}{phopkins@caltech.edu}},

\begin{abstract}
X-ray luminous, cool-core (CC) galaxy clusters contain powerful sources accelerating cosmic rays (CRs). 
High-energy CRs responsible for $\sim$\,GHz synchrotron lose energy rapidly, but there remains a long-lived ($\gtrsim\,$Gyr-old) population of $\sim 0.1-1\,$GeV CRs, propagating to $\sim 100\,$kpc distances and radiating primarily via inverse-Compton (IC) scattering of microwave background photons. 
We explore the observable consequences of such CR-IC emission. 
This produces remarkably thermal X-ray spectra, which could contribute significantly to the emission in CC centers. 
These naturally connect to ultra-steep radio sources and radio mini-halos at younger ages, but become undetectable in most radio, traditional hard-X-ray, and $\gamma$-ray searches (though future MHz imaging may detect them), while reproducing typical apparent density, temperature, entropy, and mass deposition rates of strong CCs. 
This would provide an alternative and qualitatively different resolution of the ``cooling flow problem'': clusters may appear as strong CCs because of strong CR-IC, while not actually cooling so rapidly. 
We show this predicts many observed correlations between AGN/jet properties, radio galaxy and central minihalo properties, cooling radii, X-ray cavity radii and apparent X-ray ``cooling luminosity'' $L_{\rm X,\,cool}$. 
Since $L_{\rm X,\, cool}$ is actually from CR-IC, the observed radio-X-ray ($L_{\rm radio}-L_{\rm X,\,cool}$), apparent ``cavity power'' ($P_{\rm cav}-L_{\rm X,\,cool}-L_{\rm radio}$), and strong CC-AGN correlations are predicted almost trivially and without free parameters.
Since CR-IC leads to an X-ray overestimate of the thermal pressure, the ratio of SZ to X-ray-inferred pressures should drop in CC centers: we show this is a robust tracer of CR-IC. 
CR-IC also suppresses abundances inferred from X-ray single-temperature fits relative to optical/UV measurements in CC centers ($\lesssim 10-50\,$kpc; with weak effects on X-ray temperatures). 
Both of these appear to be seen in sufficiently-resolved CC clusters. 
We show effects on cluster cosmology, hydrostatic mass estimation, and standard ``non-thermal pressure''/turbulence estimators are small.
We discuss redshift evolution, which could provide strong constraints on the models, noting observed surface-brightness profiles at high-$z$ may also imply CR-IC contributions.
\end{abstract}


\maketitle

\section{Introduction}
\label{sec:intro}

X-ray luminous or ``strong cool core'' clusters contain powerful radio (and sometimes $\gamma$-ray) sources at their center, which unambiguously demonstrate that a significant population of relativistic electrons is being accelerated in the cores \citep[e.g.][for a review]{mcnamara:2007.agn.cooling.flow.review.cavity.jet.power.vs.xray.luminosity.scalings.emph.compilation}. The high-energy leptons -- energies $\gg 10-100\,$GeV -- produce most of the observed radio and $\gamma$-ray emission, with the radio emission primarily in close proximity to (in both space \textit{and} time) the acceleration/injection sites owing to their short lifetimes $\lesssim 10^{7}\,$yr (implying CR propagation distances $\lesssim$\,a few kpc, for reasonable diffusivities/streaming speeds). But \citet{hopkins:2025.cr.ic.clusters.ideapaper} (hereafter \paperone) noted that given this population of high-energy leptons, any reasonable model predicts that there should be a corresponding population of low-energy ($\sim 0.1-1\,$GeV) leptons, containing most (if acceleration spectra are anything like those constrained by direct CR observations in the local interstellar medium [LISM]; \citealt{cummings:2016.voyager.1.cr.spectra,bisschoff:2019.lism.cr.spectra}) of the total CR lepton energy. These leptons have much larger lifetimes, $\sim 10^{9}$\,yr, implying that they should diffuse or stream out to radii $\sim 100\,$kpc before losing most of their energy. This is much longer than the typical lifetime of strong AGN, so the duty cycle of such ancient diffuse, low-energy CR halos (ACRHs) would be much larger than their sources, implying they should be present in a significant fraction of clusters.

As we show below, these ACRHs may have some loose physical relationship to certain types of radio halos observed (e.g.\ giant or mini-halos, phoenixes, radio relics/gischt, ghosts, etc.), but they are not the same. The emission from the diffuse, low-energy CRs filling the $\sim 100\,$kpc volume is largely undetectable in radio (primarily at $\sim$\,MHz frequencies), ACRH lifetimes/duty cycles are long, and ACRHs do not necessarily have any relationship with strong shocks or cluster mergers. On the other hand radio halos are often detected much more strongly at $\gtrsim$\,GHz, requiring (as is well-known in the literature) some strong diffusive shock acceleration or other re-acceleration mechanism to explain the existence of much higher-energy CRs at these larger scales (often associated with mergers). In a sense, some ACRHs could be the ``ghost halos'' of radio mini-halos or radio galaxies, \textit{before} being ``revived'' by any shocks/compression. Radio observations imply such ACRHs must exist with a much larger duty cycle than the observed radio halos themselves, but (to our knowledge) some of their potential consequences in the X-ray have been largely unexplored. 

In this paper, we expand upon \paperone\ and ask the simple question: ``What would such ACRHs look like?'' 
As argued in \paperone, we show that the dominant emission/loss mechanism for ACRHs should be inverse Compton (IC) scattering of cosmic microwave background (CMB) photons, to soft X-ray ($\sim$\,keV) energies. 
While we are certainly not the first to argue that IC could contribute to X-ray emission in some clusters \citep[see e.g.][]{raphaeli:1979.cluster.xray.emission.cr.ic.vs.thermal,sarazin:1988.book.cluster.xray.emission,sarazin:1999.cr.electron.emission.cluster.centers.xrays,hwang:1997.cluster.center.emission.in.euv.from.cr.ic,gitti:2002.perseus.minihalo.xray.inverse.compton,gitti:2004.minihalo.abell.2626.reaccel,bonamente:2007.abell.3112.clear.xray.inverse.compton.required.luminosity.fits.models.gamma.rays.too,murgia:2010.ophiuchus.cluster.minihalo.xray.inverse.compton,bartels:2015.radio.inverse.compton.cluster.minihalo.prospects,gitti:2016.radio.minihalos.coolcore.clusters.candidates.review}, ACRHs are distinct from the younger, more radio-bright CRs usually modeled. 
We show that because the highest-energy CRs are lost much closer in space and time to their sources, the resulting ACRH X-ray IC spectrum closely resembles thermal (free-free) emission, while the volume-filling ACRHs are largely invisible in more ``traditional'' harder X-ray searches, radio, and $\gamma$-ray searches for non-thermal particles. Recently \citet{hopkins:2025.crs.inverse.compton.cgm.explain.erosita.soft.xray.halos,lu:2025.cr.transport.models.vs.uv.xray.obs.w.cric} showed that around lower-mass (Milky Way and Andromeda-mass) halos, such ACRHs from the more modest population of more well-known SNe-accelerated CRs generate soft, thermal-like keV X-ray spectra with an extended surface brightness profile out to $\sim 100\,$kpc that appear to be observed in very deep soft X-ray stacks from eROSITA \citep{zhang:2024.hot.cgm.around.lstar.galaxies.xray.surface.brightness.profiles}. 
And \citet{ponnada:2025.intrinsic.tension.xray.tsz.halos.lstar.galaxies.without.crs} showed that the combination of X-ray and Sunyaev-Zeldovich observations, if correct, strongly rules-out any thermal model for that emission. 
\paperone\ argued that at the cluster scale, if such an ACRH contributes a significant fraction of the soft X-ray emission in the central $\lesssim 100\,$kpc, the \textit{apparent} X-ray spectra and profiles of quantities like gas density, entropy, and metallicity closely resemble many observed strong cool-core ([S]CC) clusters, and we show this is true for a wide variety of cluster masses and ACRH strengths, as well as a number of additional cluster properties (Table~\ref{tbl:properties}). If true, this would imply a radical revision to our understanding of the physics in these systems. 

We show that a simple model of these ACRHs -- assuming they dominate the apparent cooling luminosity in soft X-rays in cooling cores -- naturally explains a wide variety of apparent correlations and otherwise unusual properties observed in CC clusters without any free parameters, because quantities like the apparent cooling flow luminosity would actually arise from the relativistic electrons. 
We will show that while X-rays are a essential and important constraint here, there is fundamentally no measurement from X-rays \textit{alone} which can rule out CR-IC as an important contributor to the continuum luminosity, so it is critical to examine multi-wavelength predictions from radio-through optical/UV emission lines through $\gamma$-rays. 
We show that this scenario predicts a distinct, clearly-testable signatures in Sunyaev-Zeldovich observations, metallicity profiles (contrasting X-ray and optical/UV lines), and ultra-low-frequency radio halos, while other unique signatures (for e.g.\ neutral/molecular emission-line ratios in IR/optical/UV) are discussed in future work. Remarkably, these signatures appear ubiquitous in the best-studied SCCs, suggesting that this may be a common phenomena in apparent CC clusters. The signatures discussed in this paper are summarized in Table~\ref{tbl:properties}. 

We stress that we discuss these using simple toy models for CR injection and propagation and cluster gas properties, meant to illustrate the key ``typical'' behavior expected for comparison to cluster populations in an ensemble sense. Quantitatively reproducing any specific cluster and its detailed multi-wavelength properties would require more detailed models fit to those systems, which we will consider in future work.

In \S~\ref{sec:cr.spectrum}, we model the ACRH CR spectrum, then in \S~\ref{sec:softxray} calculate the corresponding soft X-ray spectra  (\S~\ref{sec:spectra}) and radial scalings (\S~\ref{sec:radial}). In \S~\ref{sec:multi.wavelength} we calculate the pan-wavelength spectrum and compare to hard X-ray IC searches (\S~\ref{sec:hardxray}), optical/UV/IR (\S~\ref{sec:uv.oir}), $\gamma$-ray (\S~\ref{sec:gamma}), and radio (\S~\ref{sec:radio}) constraints. We then explore the predictions, if CR-IC from ACRHs powers apparent CCs, in \S~\ref{sec:coolcores}, making predictions for the surface brightness/temperature/density/entropy/pressure profiles (\S~\ref{sec:cc.profiles}), classical/spectral cooling-flow problem (\S~\ref{sec:cf.problem}), Sunyaev-Zeldovich ``deficits'' (\S~\ref{sec:sz}), central metallicity suppression (\S~\ref{sec:z.drops}), X-ray spectral fitting and temperature inference (\S~\ref{sec:temp.fitting}), apparent ``cavity/jet power-cooling luminosity'' correlations (\S~\ref{sec:pjet}), cooling radii and CC sizes (\S~\ref{sec:rcool}), and various radio-X-ray and cooling rate/AGN correlations (\S~\ref{sec:Lradio.LX}). In \S~\ref{sec:discussion} we further discuss the implications of this interpretation for the true strength of cool cores (\S~\ref{sec:cfs}), AGN feedback/cooling flow models (\S~\ref{sec:feedback}), energetics of AGN feedback and demographics of CCs in simulations and observations (\S~\ref{sec:energetics}), implied CR pressures and dynamical effects and relation to cavity sizes/extents (\S~\ref{sec:pressure}), and implications for cosmology and cluster mass modeling constraints (\S~\ref{sec:cosmology}). We discuss the connection to radio mini-halos and other radio relics, and speculate on possible evolutionary scenarios, in \S~\ref{sec:evolution}. In \S~\ref{sec:redshift} we discuss redshift effects in CR-IC scenarios (\S~\ref{sec:redshift:cr}) noting their effects on apparent thermodynamic (\S~\ref{sec:redshift:cr:weak}) and surface brightness (\S~\ref{sec:redshift:cr:sb}) profiles, and degeneracies with AGN (\S~\ref{sec:redshift:cr:agn}), compared to pure-thermal scenarios (\S~\ref{sec:redshift:thermal}). We conclude in \S~\ref{sec:conclusions}.

\begin{footnotesize}
\ctable[caption={{\normalsize Cluster Properties Predicted/Explained by ACRHs \&\ Associated CR-IC}\label{tbl:properties}},center,star]{| c | l |}{\tnote{Table lists cluster property/observation, with the physical origin in the CR-IC scenario, plus quantitative CR-IC predictions shown in this paper (\textcolor{blue}{blue}), and 
for comparison the explanation/prediction in the ``traditional'' (no ACRH/CR-IC, i.e.\ thermal-emission-only) scenario/interpretation (\textcolor{orange}{orange}).}
}{
\hline
\tblspace
\tbltopic{Classical CF}  \tblexp{``Apparent'' cooling rate, boosted by CR-IC, mimics CF, with much weaker actual cooling present}
\tbltopic{Problem} \tblpred{$\dot{M}(r)$ rises, lack of cold gas, star formation, $\dot{M}_{\rm cool} \gg \dot{M}_{\rm spectral}$}
\tblsec{\S~\ref{sec:softxray},\,\ref{sec:cf.problem},\,\ref{sec:cfs}} \tbltrad{some input (e.g.\ AGN) finely-balances $L_{\rm cool}$ at each radius}
\tblspace
\tbltopic{Sunyaev-Zeldovich} \tblexp{X-ray pressure $P_{X}$ over-estimated from CR-IC contribution, SZ still traces (smaller) true thermal pressure $P_{\rm SZ} \approx P_{\rm true}$}
\tbltopic{Deficits} \tblpred{$P_{\rm SZ}/P_{X} \sim P_{\rm true}/P_{\rm apparent,\,X-ray} < 1$ in CC centers}
\tblsec{\S~\ref{sec:sz}} \tbltrad{not allowed (observations must be incorrect)}
\tblspace
\tbltopic{Abundance Suppression/} \tblexp{CR-IC contributes thermal continuum but weakly influences lines, dilutes apparent $Z$ in X-ray observations}
\tbltopic{Sub-Solar Central $Z$} \tblpred{fitted $Z$ from 1-$T$ X-ray spectra falls below optical/UV-observed $Z > Z_{\odot}$ at $R\lesssim 10\,$kpc}
\tblsec{\S~\ref{sec:z.drops}} \tbltrad{no explanation/opposite AGN model predictions}
\tblspace
\tbltopic{CC Surface Brightness,} \tblexp{CR-IC generically gives steeply-rising $I_{X}$ and apparent $n$ at $\sim$\,keV: emissivity $\propto e_{\rm cr,\,\ell} \propto r^{-2}$}
\tbltopic{Density Profiles} \tblpred{$I_{X} \propto R^{-1}$, $n_{\rm apparent} \propto r^{-1}$ in CC centers}
\tblsec{\S~\ref{sec:cc.profiles}} \tbltrad{CF-like density profile but no actual CF, so universal profile is ``coincidence''}
\tblspace
\tbltopic{CC Temperature,} \tblexp{ACRH CR-IC spectrum indistinguishable from thermal: weakly influences ``true'' central temperature}
\tbltopic{Entropy Profiles} \tblpred{$T_{\rm apparent}$ of IC-emission weakly-dropping, $K_{\rm apparent}\propto T/n^{2/3} \propto r^{+2/3}$ in CC centers}
\tblsec{\S~\ref{sec:cc.profiles}} \tbltrad{some but not runaway ``real'' cooling, so coincidental/fine-tuned $T$, $K$ profiles (generally not predicted)}
\tblspace
\tbltopic{Radio -- X-ray Cooling} \tblexp{CR-IC powered by $\sim$\,Gyr-averaged lepton injection rate: radio powered by recent $\sim 10^{7}-10^{8}$\,yr-old injection}
\tbltopic{Rate Correlation} \tblpred{$L_{\rm radio} \propto L_{X,\,\rm cool}$, ubiquity of strong radio sources+AGN in SCCs and vice versa, local $I_{\rm radio}-I_{X}$ correlation}
\tblsec{\S~\ref{sec:Lradio.LX}} \tbltrad{cooling $\rightarrow$ cold gas (not seen) $\rightarrow$ SF $\rightarrow$ SMBH fueling $\rightarrow$ radio; slope, normalization just coincidence}
\tblspace
\tbltopic{Cavity/Jet Power --} \tblexp{Apparent cavity/jet power $P_{\rm cav} \propto pV/t_{\rm bouyancy}$ scales with apparent X-ray $P_{X}$ from CR-IC, cooling $L_{X,\,\rm cool}$ from same}
\tbltopic{Cooling-Rate Correlation} \tblpred{$P_{\rm cav} \propto L_{X,\,\rm cool}^{2/3}$, without freely-adjusted parameters}
\tblsec{\S~\ref{sec:pjet}} \tbltrad{$L_{\rm X}\rightarrow L_{\rm radio}$ chain, plus nonlinear radio-to-cavity power to balance $L_{\rm cool}$; nonlinear slope not predicted}
\tblspace
\tbltopic{Jet Power -- Radio} \tblexp{Trivially follows from above for ACRHs: same lepton populations power both}
\tbltopic{Luminosity Correlation} \tblpred{sub-linear $P_{\rm cav} \propto L_{\rm radio}^{0.5-0.7}$, again without fitted parameters}
\tblsec{\S~\ref{sec:Lradio.LX}} \tbltrad{coincidence of acceleration efficiency+cavity size+expansion time+external halo+BH properties}
\tblspace
\tbltopic{Cooling Radii/} \tblexp{Scaling of apparent CR thermal energy and emissivity with $e_{\rm cr}$ fixes $e_{\rm cr}$ at a given ``apparent'' cooling time $t_{\rm cool}$}
\tbltopic{CC Sizes} \tblpred{$R_{\rm cool}(t_{\rm cool} < {\rm few\,Gyr}) \propto L_{X,\,\rm cool}^{1/3}$, without free parameters}
\tblsec{\S~\ref{sec:rcool}} \tbltrad{no explanation/coincidence}
\tblspace
\tbltopic{Radial $\dot{M}_{\rm cool}$} \tblexp{Quasi-universal $\dot{M}_{\rm cool} \propto L_{\rm X,\,cool}/T$ follows from CR-IC scalings above}
\tbltopic{Scaling} \tblpred{$\dot{M}_{\rm cool} \propto R$ in CC centers}
\tblsec{\S~\ref{sec:cf.problem}} \tbltrad{no explanation (standard cooling flow models predict $\dot{M}_{\rm cool} \propto $\,constant)}
\tblspace
\tbltopic{Cavity/Bubble} \tblexp{Roughly trace radii where ACRH CR pressure becomes comparable/larger than virial pressure in bubbles}
\tbltopic{Extent/Distances} \tblpred{$D_{\rm cavity} \propto (L_{\rm X,\,\rm cool}\, a^{4})^{1/3}$ or $(L_{X,\,\rm cool}\, a^{4}/T_{X})^{1/2}$}
\tblsec{\S~\ref{sec:pressure}} \tbltrad{not predicted, coincidence of bouyant rise time and particular times of previous AGN episodes}
\tblspace
\tbltopic{Hard-IC,} \tblexp{Extended ACRH undetectable: CRs primarily leptonic, older ($\gtrsim$\,Gyr), CR spectrum truncated above $\sim 0.1-1\,$GeV}
\tbltopic{Synchrotron, $\gamma$-rays} \tblpred{soft, thermal-like X-ray (no hard IC), synchrotron all at $\lesssim$\,MHz, very weak $\sim$\,MeV $\gamma$-rays}
\tblsec{\S~\ref{sec:multi.wavelength}} \tbltrad{no explanation for what happens to softer CR leptons from central synchrotron/$\gamma$-ray sources}
\tblspace
\tbltopic{Apparent Lack of} \tblexp{CR-IC causes pressure over-estimate (above), offsetting deficit from real CR pressure. CRs do not appear in linewidths}
\tbltopic{Non-Thermal Pressure} \tblpred{mass/kinematic/potential reconstruction accurate to within $<10\%$ in CC centers, better at larger $R$}
\tblsec{\S~\ref{sec:pressure},\,\ref{sec:cosmology}} \tbltrad{halos must be in actual near-hydrostatic equilibrium}
\tblspace
\tbltopic{Cosmology/Lensing} \tblexp{CR-IC cuts off exponentially from losses at $\gg 100\,$kpc, no effects on properties at $\sim R_{500}$}
\tbltopic{Large-Scale SZ} \tblpred{cluster-averaged or core-excised properties (e.g.\ $L_{X}-M$, $L_{X}-T$) same as ``traditional'' model}
\tblsec{\S~\ref{sec:cosmology}} \tbltrad{same as ACRH model}
\tblspace
\tbltopic{``Theoretical'' Cooling} \tblexp{AGN feedback needed, but models tend to blow out or overheat central gas so thermal X-rays do not resemble real CCs}
\tbltopic{Flow Problem} \tblpred{CR-IC from ACRH mimics CC profiles for few Gyrs, even if feedback rapidly blows out/heats central gas}
\tblsec{\S~\ref{sec:feedback},\,\ref{sec:energetics}} \tbltrad{models must be incorrect, requires fine-tuning to make realistic CC population}
\tblspace
\tbltopic{Redshift ($z$)} \tblexp{CR-IC surface brightness $S_{X} \propto e_{\rm cr}$ at fixed $R$ and wavelength (but spatial extent shrinks), at higher $z$}
\tbltopic{Evolution} \tblpred{central $S_{X}$, $K$ evolve weakly ($S_{X} \propto \dot{E}_{\rm cr,\,\ell}/v_{\rm st}$), compact CR-IC becomes degenerate with AGN at high-$z$}
\tblsec{\S~\ref{sec:redshift}} \tbltrad{SCC central thermodynamic properties must evolve strongly with $z$ to explain observed $S_{X}$}
}
\end{footnotesize}

\section{The CR Spectrum and Radial Profile}
\label{sec:cr.spectrum}

\begin{figure}
	\centering\includegraphics[width=1.02\columnwidth]{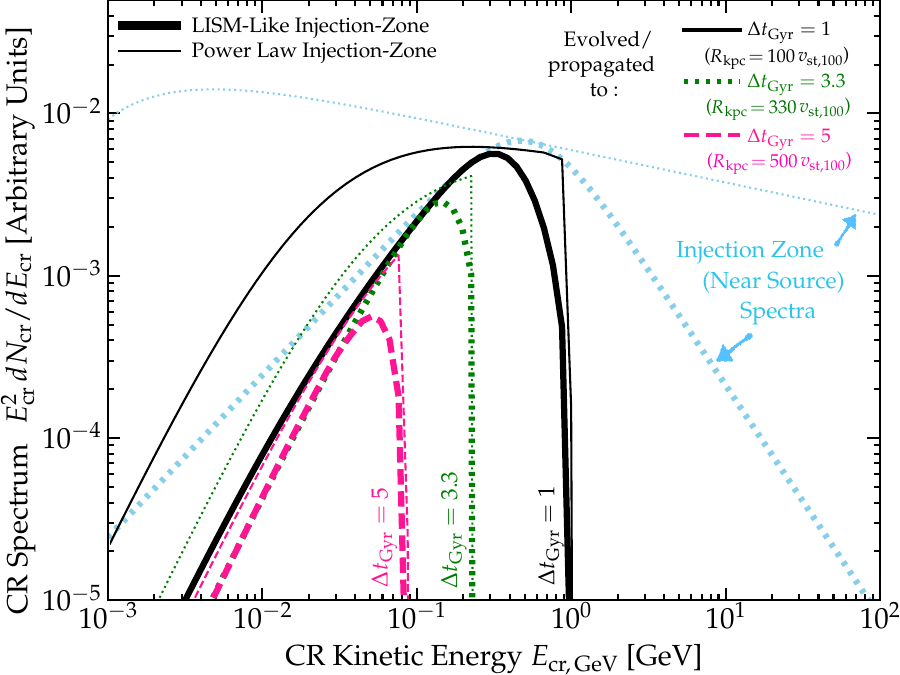} 
	\caption{Predicted mono-age CR lepton spectra for ACRHs at different radii and/or ages. First, the ``injection zone'': the region very near CR acceleration in space and time (CR age $\Delta t_{\rm Gyr} \equiv \Delta t / {\rm Gyr} \rightarrow 0$, and spatial extent/radius $R \rightarrow 0$), where the injection can be thought of as volume-filling. A plausible quasi-steady-state injection-region spectrum (injection balancing losses in the region) is the local ISM (LISM) spectrum, corresponding to synchrotron index $\alpha \sim 1$. For reference, we also compare a very shallow ``pure power-law'' (PPL) injection spectrum (1D $dN_{\rm cr}/dp_{\rm cr} \propto p_{\rm cr}^{-\delta}$ with $\delta=-2.2$, or $\alpha \sim 0.5$) with no escape or losses. We show the spectra evolved to age $\Delta t_{\rm Gyr}\gtrsim 1$, or equivalently for the CRs that propagate to radii $R \sim 100\,{\rm kpc}\,\Delta t_{\rm Gyr}\,v_{100}$, including Coulomb, bremsstrahlung, and IC+synchrotron losses. At $\gtrsim 10^{8}$\,yr (or $R \gg$\,kpc, for constant streaming speeds in a homogeneous medium), the CR spectra become broadly similar independent of injection spectrum (slightly \textit{softer} if we assume the PPL spectrum in the inner region), dominated by $\sim 0.1-1$\,GeV leptons. 
	\label{fig:cr.spectra}}
\end{figure}

\subsection{Outline}
\label{sec:cr.spectrum:outline}

Consider a CR injection ``episode,'' during which leptons are ``injected'' (by which we mean ``escape the injection region'' as defined below) through some process in or around a cluster core (not necessarily within the BCG), with total integrated energy injection rate 
\begin{align}
\dot{E}_{\rm cr,\,\ell} \sim  10^{43}\,{\rm erg,\,s^{-1}}\,\dot{E}_{43}\ . 
\end{align}
A variety of processes could contribute to CR acceleration (pulsar wind nebulae, colliding winds, accretion shocks, supernovae, AGN jet acceleration or termination shocks, wind cavity shocks, etc.) -- we are for now agnostic to these details, and ignore hadrons, though we return to them below. We are also agnostic to whether the injection formally comes from primary/secondary acceleration, or re-acceleration. As the CRs propagate through the cluster to radii $R$, they will form some spherically-averaged profile of leptonic energy density $e_{\rm cr,\,\ell}(R)$, with 
\begin{align}
\label{eq:ecr.r} e_{\rm cr,\,\ell} \sim \frac{f_{\rm loss} \dot{E}_{\rm cr,\ell}}{4\pi\,v_{\rm st,\,eff} R^{2}} \sim 0.5 f_{\rm loss} {\rm eV\,cm^{-3}}\,\frac{\dot{E}_{43}}{R_{100}^{2}\,v_{100}}
\end{align} 
in terms of some effective ($R$-dependent, in principle) isotropically-averaged effective streaming speed $v_{\rm st,\,eff} \equiv v_{100}\,100\,{\rm km\,s^{-1}}$, radius $R \equiv R_{100}\,100\,{\rm kpc}$, and a loss correction $f_{\rm loss}$.\footnote{Eq.~\ref{eq:ecr.r} follows immediately from energy conservation for appropriate $f_{\rm loss}$ and $v_{\rm st,\,eff}$ \citep[e.g.][]{quataert:2021.cr.outflows.diffusion.staircase,quataert:2022.isothermal.streaming.wind.analytic.cr.wind.models}, but for examples demonstrating its applicability and accuracy in CR-MHD simulations following detailed CR transport even with locally-variable losses and CR scattering rates, see \citet{su:turb.crs.quench,hopkins:cr.mhd.fire2,hopkins:cr.transport.constraints.from.galaxies,hopkins:2020.cr.transport.model.fx.galform,hopkins:2021.sc.et.models.incompatible.obs,hopkins:2022.cr.subgrid.model,ji:fire.cr.cgm,ji:20.virial.shocks.suppressed.cr.dominated.halos,butsky:2022.cr.kappa.lower.limits.cgm,ponnada:2023.fire.synchrotron.profiles,ponnada:2023.synch.signatures.of.cr.transport.models.fire}.}
Here $v_{\rm st,\,eff}$ represents the mean bulk propagation speed of CRs, and so includes contributions from ``true'' \Alf{ic} and super-\Alf{ic} CR streaming, advection, and diffusion (with effective $\kappa_{\rm eff} \sim v_{\rm st,\,eff}\,R$) 
Note that we will show most of our results are not particularly sensitive to $v_{100}$, but the value we scale to here is modest -- this is sub-sonic, comparable to \Alf\ and advective and/or buoyancy speeds in cluster centers.

Importantly, high-energy leptons ($E_{\rm cr,\,GeV} \gg 1$, where $E_{\rm cr}\equiv E_{\rm cr,\,GeV}$\,GeV is the CR kinetic energy) will rapidly lose energy -- even if synchrotron losses are negligible ($|{\bf B}| \lesssim B_{\rm cmb} \equiv 3.2\,(1+z)^{2}\,{\rm \mu G}$) and there is no strong local radiation field, the IC loss timescale from just the CMB is 
\begin{align}
\label{eqn:tloss} \Delta t_{\rm loss}^{\rm IC} \approx E_{\rm cr}/\dot{E}_{\rm cr}^{\rm IC} \sim 1.2\,(1+z)^{-4}\,E_{\rm cr,\,GeV}^{-1}\,{\rm Gyr}\ .
\end{align} 
For e.g.\ $\gtrsim 30-100\,$GeV CRs this becomes $\lesssim 10^{7}\,$yr -- so even if there is continuous injection of CRs, these high-energy leptons lose their energy within a distance $\sim (v_{\rm st,\,eff}/100\,{\rm km\,s^{-1}})\,{\rm kpc}$ from their source. 

Fig.~\ref{fig:cr.spectra} shows more detailed calculations of the CR spectrum as CRs propagate away from the ``injection zone,'' including the full expressions for losses from IC+synchrotron, Coulomb+ionization, and bremsstrahlung, compiled in \citet{hopkins:cr.spectra.accurate.integration} from \citet{blumenthal:1970.cr.loss.processes.leptons.dilute.gases,1972Phy....60..145G,Mann94}. If we consider a single mono-age CR population (with a fixed injection spectrum) in a homogeneous medium a time $\Delta t = \Delta t_{\rm Gyr} \,$Gyr since injection, then the effect of IC losses (where the CMB dominates) is entirely determined by $\Delta t$, with the dimensionless loss parameter $\tau_{\rm IC} = \Delta t / t_{\rm loss,\,IC}^{0} \sim 0.8\,(1+z)^{4}\,E_{\rm cr,\,GeV}^{0}\,\Delta t_{\rm Gyr}$. We can add synchrotron losses in principle by taking $t_{\rm loss,\,IC}^{0} \rightarrow t_{\rm loss,\,IC}^{0}/[1+(B/B_{\rm cmb})^{2}]$, but we will assume $B<B_{\rm cmb}$ to begin. In this limit the CR energy evolves simply as $E_{\rm cr}(t) = E_{\rm cr,\,0}/(1+\tau_{\rm IC})$, and the spectral correction is straightforward.\footnote{Note, this is distinct from the sometimes-quoted argument that losses simply steepen the power-law slope to a new power-law. That is only true in the injection zone assuming steady-state injection balancing IC losses. But away from the injection zone, we can think of each concentric spherical shell having an ``injection spectrum'' steepened by the shell interior, plus its losses, leading to a super-exponential spectral cutoff, as the equations above correctly model.} Moreover, for CRs injected at a source near the cluster center and streaming/diffusing outwards, we can relate $\Delta t$ to observed radius simply by 
\begin{align}
\label{eqn:age.radius.relation} \Delta t \approx \frac{R}{v_{\rm st,\,eff}} \sim {\rm Gyr}\,R_{100}\,v_{100}^{-1} \ .
\end{align} 
We also include Coulomb, ionization, and bremsstrahlung losses, relevant at lower CR energies, but these depend on the background gas density, for which we assume (for illustrative purposes here) $n_{\rm gas} \sim 10^{-3}\,{\rm cm^{-3}}$ in this calculation (similar to a NCC, for reasons below). Coulomb and ionization losses scale similarly and Coulomb dominates for the well-ionized conditions of interest here, with $t_{\rm loss,\,Coulomb} \sim 50\,{\rm Gyr}\,E_{\rm cr,\,GeV}\,(10^{-3}\,{\rm cm^{-3}}/n_{\rm gas})$ (up to logarithmic corrections), so is generally small for all but $\lesssim 100\,$MeV CRs, and $t_{\rm loss,\,brems} \sim 30\,{\rm Gyr}\,(10^{-3}\,{\rm cm^{-3}}/n_{\rm gas})$ (also up to logarithmic terms) so is generally small. The loss timescale is generically maximized at CR energies $\sim 0.1-1\,$GeV, with values $\sim 10^{9}$\,yr, so the majority of the time and volume will feature CR spectra dominated by these energies -- our ACRHs.

Note that our modeling here in Fig.~\ref{fig:cr.spectra} differs from \paperone\ in that we model (for now) a mono-age CR population at some $\Delta t$, while in \paperone\ (and in \S~\ref{sec:cc.profiles} below) we assumed steady-state CR injection in the center of a cluster with a given diffusion coefficient+streaming speed used to propagate the CRs to different distances. For purposes of modeling spectra, we focus here on $\Delta t$, following the convention in radio synchrotron studies, because it allows us to be agnostic to the actual (uncertain) CR transport parameters and injection rates/luminosities/time-histories from sources. Further, because the equations modeled here are linear, one can generate mixed-age CR emission spectra from the sum of various mono-age populations (up to non-linear effects of e.g.\ time+space-varying loss rates, which we discuss below). 
But for the range of simple toy models we consider in this paper (including those where we do explicitly model CR propagation with diffusion+streaming+advection), one can reasonably approximate the CR and emission spectra treating the CR ``age'' $\Delta t$ and ``distance from source'' $R$ as roughly equivalent, related by Eq.~\ref{eqn:age.radius.relation}. 

\subsection{Meaning of the ``Injection Zone'' and its (Small) Contribution to Emission}
\label{sec:cr.spectrum:injection}

For the injection spectrum, it is useful to consider not strictly the injected spectrum by one specific acceleration mechanism in an infinitesimally small volume (e.g.\ a shock thickness), but the mean spectrum of the ``injection zone,'' by which we simply mean the region over which one has either some volume-filling injection or injection balanced by losses. This is typically $\sim 10\,$kpc (the size of the galaxy) for e.g.\ injection by radio galaxies, or the size of strong radio arcs/lobes, or the X-ray ``cavities'' associated with the termination of the active jets (all similar sizes). Exiting this region, a plausible guess is an LISM-like lepton spectrum (since this is set too by some quasi-universal acceleration mechanism balanced by losses) -- for which we take the fit in \citet{hopkins:cr.multibin.mw.comparison} to the combination of Voyager, Pamela, AMS-02, and other data sets as in \citet{cummings:2016.voyager.1.cr.spectra,2018PhRvL.120z1102A,bisschoff:2019.lism.cr.spectra}.\footnote{As noted in \citet{hopkins:2025.crs.inverse.compton.cgm.explain.erosita.soft.xray.halos}, there is some factor $\sim 2$ uncertainty in the exact LISM spectral shape at $\sim1\,$GeV owing to large corrections for Solar modulation, but this is effectively implicit in our definition of $e_{\rm cr,\,\ell}$ and does not change our conclusions.} This is theoretically plausible (indeed, for our definition of the ``injection region,'' it would trivially be correct if the central radio galaxy were Milky Way-like), but also motivated by observations: the high-energy slope of the LISM spectrum (equivalent to synchrotron $\alpha \sim 1$) is much closer to what is typically seen in synchrotron observations of radio galaxies and lobes and other injection regions in clusters, compared to standard pure-power law injection spectra ($dN_{\rm cr,\,inj}/dE_{\rm cr,\,inj} \propto E_{\rm cr,\,inj}^{-\delta_{\rm inj}}$, with $\delta_{\rm inj} \sim -2$, or $\alpha \sim 1/2$). But we stress we adopt this only because it provides a simple, motivated, and plausible ``guess,'' for illustrative purposes in the toy models here. If one actually wishes to model the specific radio (and potentially X-ray CR-IC, hard X-ray, $\gamma$-ray) spectra of a real cluster, then the effective injection-zone spectrum would of course be one of the modeling parameters fit to the known non-thermal multi-wavelength emission in the radio galaxy. But even if we were to assume a shallow, pure power-law injection spectrum, we show in Fig.~\ref{fig:cr.spectra} that this only makes a large difference in or near the injection region, because outside of this region the ACRH CR spectrum is quickly reshaped by losses to be peaked at lower energies.\footnote{Note, we could also allow for different loss rates: i.e.\ stronger magnetic fields and/or radiation or higher densities in the injection zone, or a continuous radial profile of these quantities. This also makes no difference to our conclusions, because they become degenerate outside this region. For example, if in the injection zone we assume a higher density and $B$, such that the CR loss timescale in that zone is shorter by a factor $f$, then if the modified loss time is still longer than the CR escape timescale from the region $\sim R_{\rm inj}/v_{\rm st,\,eff}$, it will be a negligible correction, while if it is shorter, it will be equivalent to the spectra in Fig.~\ref{fig:cr.spectra} with a larger value of $\Delta t$. Since the low/high-energy CRs fall into the former/latter category for reasonable values of density and $B$ (up-to LISM-like), this will only have the effect of ``shrinking'' the effective injection zone in time and space, before the spectrum becomes ACRH-like.}

We stress that the ``injection zone'' is defined in both space and/or time. In other words, one could have intermittent injection with some duty cycle, where after a source shuts off, all CRs age together \textit{while also streaming out}. In this case we are interested in the maximum ages $\sim $\,Gyr that are reached before most of the CR energy is lost, since this will dominate the total duty cycle. Given the ratio of this age to the timescale $\sim 10^{7}$\,yr for the high-energy CRs to lose their energy, something like $90-99\%$ of systems would be in the ACRH-like state. 

Alternatively, consider continuous, steady-state injection. Then, CRs which have propagated to distance $R$ necessarily have ages $\Delta t \approx R / v_{\rm st,\,eff} \sim {\rm Gyr}\,(R/100\,{\rm kpc})\,(100\,{\rm km\,s^{-1}}/v_{\rm st,\,eff})$ as noted above. So the ``injection zone'' would refer to the zone over which the high-energy CRs can propagate before losing their energy, i.e.\ $\sim 1-10\,$kpc around the source, while the ACRH describes the volume out to $\gtrsim 100\,$kpc out to which the low-energy CRs can travel before losing their energy. In this case, the basic duty cycle and energetics arguments (now in space) implies $\gtrsim 99\%$ of the \textit{volume} occupied by a significant population of CRs, and $\gtrsim 90\%$ of the total emission from CRs, comes from the ACRH.

\subsection{Transport-Related Loss/Gain Processes}
\label{sec:cr.spectrum:transport.losses}

For completeness, we note there are additional loss/gain processes directly tied to CR transport which appear in the rigorously-derived CR transport equations \citep[e.g.][]{skilling:1971.cr.diffusion,schlickeiser:89.cr.transport.scattering.eqns,1997JGR...102.4719I,2007ApJ...662..350L,hopkins:m1.cr.closure}, including: 
differential escape/propagation to different $R$ (if $v_{\rm st,\,eff}$ depends strongly on $E_{\rm cr}$), 
magnetic focusing, diffusive shock acceleration in weak/turbulent shocks, 
the adiabatic/non-inertial-frame/CR ``work'' terms (scaling with the CR pressure and velocity shear tensor products), 
the ``streaming-loss''/asymmetric-scattering terms ($D_{p\mu}$ and $D_{\mu p}$), 
and the ``diffusive/turbulent reacceleration'' term ($D_{pp}$). 
These can variously manifest as effective loss or gain or either, and evaluating the net sign (i.e.\ if these ``net'' act like re-acceleration or losses) requires a full model for the (turbulent+mean) magnetic field, velocity, \Alf\ speed (hence density), and scattering rate structures (including their spatial-and-time cross-correlations), knowledge of the asymmetry between forward and backward-traveling \Alf\ waves in the CR-comoving \Alf\ frame, and therefore detailed micro-physics of CRs and magnetic fields on gyro-resonant scales.\footnote{For example, a common expression used for ``turbulent reacceleration'' $D_{pp} \sim p^{2} v_{A}^{2}/9 D_{xx}$ (in terms of the CR momentum $p$ and spatial diffusivity $D_{xx}$), is only valid if (1) extrinsic turbulence dominates CR scattering, with (2) symmetric forward/backward-traveling \Alf\ wave amplitudes, and (3) restricts CRs to bulk stream at speeds much slower than \Alf{ic}. This is not generally expected for the low-energy CRs of interest here, whose gyro radii are orders-of-magnitude smaller than the damping scale of the turbulence in clusters \citep{yan.lazarian.2008:cr.propagation.with.streaming,yan.lazarian.04:cr.scattering.fast.modes,yan.lazarian.02,hopkins:cr.transport.constraints.from.galaxies,hopkins:2021.sc.et.models.incompatible.obs,kempski:2021.reconciling.sc.et.models.obs}, and whose scattering is generally expected to follow from intermittent structure and/or self-excited modes \citep{hall:1981.galaxy.cr.prop.largescale.field.reversals.quasilinear.theory.weak.confinement,butsky:2023.cosmic.ray.scattering.patchy.ism.structures,kempski:2023.large.amplitude.fluctuations.and.cr.scattering,kempski.li.2024:unified.cr.scattering.plasma.scattering.from.strong.field.curvature.intermittent.ism.structure.explained.together,reichherzer:2023.micromirror.cr.confinement.hot.galaxy.cluster.icm}, with $v_{\rm st,\,eff} \gtrsim v_{A}$ (\citealt{ensslin:2011.cr.transport.clusters,wiener:cr.supersonic.streaming.deriv}; though see also \citealt{zhang:2024.cr.mirror.diffusion.in.nonlinear.turbulence,ewart:2024.cr.confinement.micromirror.cgm.of.massive.clusters}). In those cases the ``streaming losses'' ($D_{\mu p}$ and $D_{p \mu}$ terms) will be larger (and opposite in sign) from the $D_{pp}$ term, but much larger reacceleration terms can arise from the ``adiabatic'' terms and first-order Fermi acceleration through the spectrum of weak shocks \citep{ensslin:2011.cr.transport.clusters,hopkins:m1.cr.closure,hopkins:cr.multibin.mw.comparison}.} However, in more detailed calculations these generally modify the CR spectrum by at most order-unity multiplicative factors \citep{chan:2018.cosmicray.fire.gammaray,su:turb.crs.quench,hopkins:cr.mhd.fire2,hopkins:cr.multibin.mw.comparison,hopkins:2021.sc.et.models.incompatible.obs,hopkins:cr.spectra.accurate.integration,ji:fire.cr.cgm,buck:2020.cosmic.ray.low.coeff.high.Egamma,thomas:2022.self-confinement.non.eqm.dynamics}
and for the fiducial parameter choices here have rather small effects, so will not qualitatively change our conclusions here and are largely degenerate with the assumed CR injection rate/spectrum, streaming speed, and background density/loss rate. But again, for specific, quantitative comparisons to real observed cluster spectra, such order-unity effects would all be potentially important.

\section{Soft X-Ray Emission}
\label{sec:softxray}

\subsection{Continuum IC Spectra}
\label{sec:spectra}

Per \paperone\ and \citet{hopkins:2025.crs.inverse.compton.cgm.explain.erosita.soft.xray.halos}, CRs with energy $E_{\rm cr}$ (Lorentz factors $\gamma_{\rm cr} = E_{\rm cr}/m_{e}c^{2}$) will IC scatter CMB photons to energies $\gamma_{\rm cr}^{2}\,h \nu_{0}$, which in observed-frame ($z=0$) is a redshift-independent\footnote{Because of the scaling of the CMB, there is a factor $(1+z)$ in the emission/rest-frame $E_{\rm IC}$ which is redshifted back out in the observed frame, and likewise for a factor $(1+z)^{4}$ in the effective emissivity and surface brightness.}
$E_{\rm IC}^{\rm obs} \sim 3\,{\rm keV}\,E_{\rm cr,\,GeV}^{2}\,(\nu_{0}/\nu_{\rm CMB,\,peak})$ (using $h\nu_{\rm CMB,\,peak} \approx 3\,k_{B} T_{\rm CMB}$), so we expect most of the IC emission to come in soft X-rays in ACRHs. 

\begin{figure}
	\centering\includegraphics[width=0.99\columnwidth]{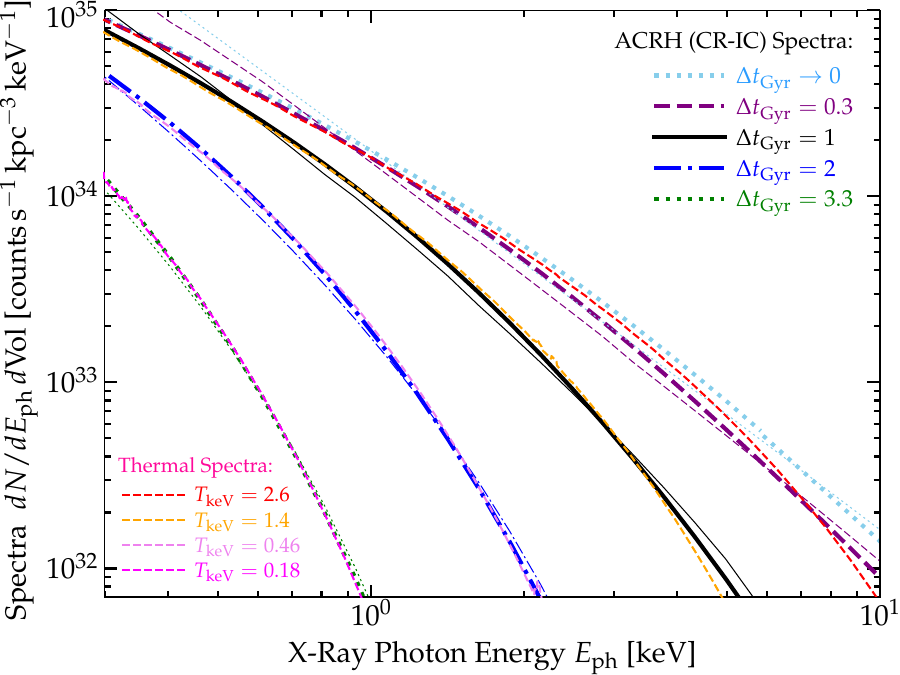} 
	\caption{Continuum X-ray spectra predicted for CR leptons in an ACRH IC scattering cosmic background light, for CRs with a near-source/injection-zone LISM-like (\textit{thick}) or pure power-law-like (\textit{thin}) spectrum (Fig.~\ref{fig:cr.spectra}). We evolve to different ages (accounting for Coulomb, bremsstrahlung, and IC losses) $\Delta t = \Delta t_{\rm Gyr}\,$Gyr, corresponding also to emission at a distance $R \sim 100\,{\rm kpc}\,\Delta t_{\rm Gyr}\,v_{100}$ from the injection site, for a mono-age population in a homogeneous background. We compare APEC thermal spectra for primordial gas with different temperatures. At $\gg 10^{7}\,$yr or equivalently $\gtrsim$\,kpc from the injection sites, continuum CR-IC spectra become almost indistinguishable from thermal X-ray spectra with $T \sim $\,keV (independent of the injection spectrum). 
	\label{fig:spectrum.xr}}
\end{figure}

Fig.~\ref{fig:spectrum.xr} shows a more detailed calculation of the continuum IC spectra from our fiducial toy models, similar to that in \paperone. We convolve over the full CMB spectrum and IC emissivity for each CR electron or positron, including Klein-Nishina corrections as well as cosmic optical and IR backgrounds \citep[e.g.][]{cooray:2016.extragalactic.background.light.compilations.review,khaire:2019.extragalactic.background.light.spectra}, and integrate over the full CR lepton spectrum. We consider the time-and-space evolved toy-model CR spectra from Fig.~\ref{fig:cr.spectra}, at different $\Delta t = {\rm Gyr}\,\Delta t_{\rm Gyr}$ or $R \sim v_{\rm st,\,eff} \Delta t \sim 100\,{\rm kpc}\,v_{100}\,\Delta t_{\rm Gyr}$.

What is notable is that for any region $R$ or $\Delta t$ outside the injection zone, or even in the injection zone if the spectrum is LISM-like, the soft X-ray IC  spectrum is \textit{not} a pure power-law (with e.g. $dN_{\rm ph}/dE_{\rm ph} \propto E_{\rm ph}^{-(\delta+1)/2}$ as expected for IC from a power-law CR population with $d N_{\rm cr}/d E_{\rm cr} \propto E_{\rm cr}^{-\delta}$), but exhibits considerable curvature -- much more closely resembling thermal spectra with effective temperatures $\sim 0.3-5$\,keV (the same as found in \paperone\ for non-mono-age CR populations at radii $R \gtrsim 10$\,kpc distances from sources with LISM-like injection spectra). The upper end of this range -- corresponding to harder CR spectra -- is more sensitive to the assumed injection-zone spectrum and loss/gain processes (e.g.\ hardening by Coulomb losses at higher densities or reacceleration processes), and for plausible assumptions could be as large as $\sim 10-15\,$keV (much harder injection spectra effectively resembling the power-law case in X-ray bands). The lower end of this range ($\ll$\,keV) can drop rapidly at very long timescales $\gg$\,Gyr but this is by definition longer than the CR-IC loss timescale so cannot contribute much to observed emission. Thus the range $\sim 1$ to $\sim$\,a few keV, at the radii where \textit{most} of the CR-IC is emitted, is qualitatively robust and depends only weakly on the form of the injection-zone spectrum -- even for a very hard pure-power-law ($\delta \sim 2$) spectrum, by $\sim 1$\,Gyr the soft X-ray spectrum closely resembles $\sim$\,keV blackbody free-free emission spectrum. 

\begin{figure}
	\centering\includegraphics[width=0.99\columnwidth]{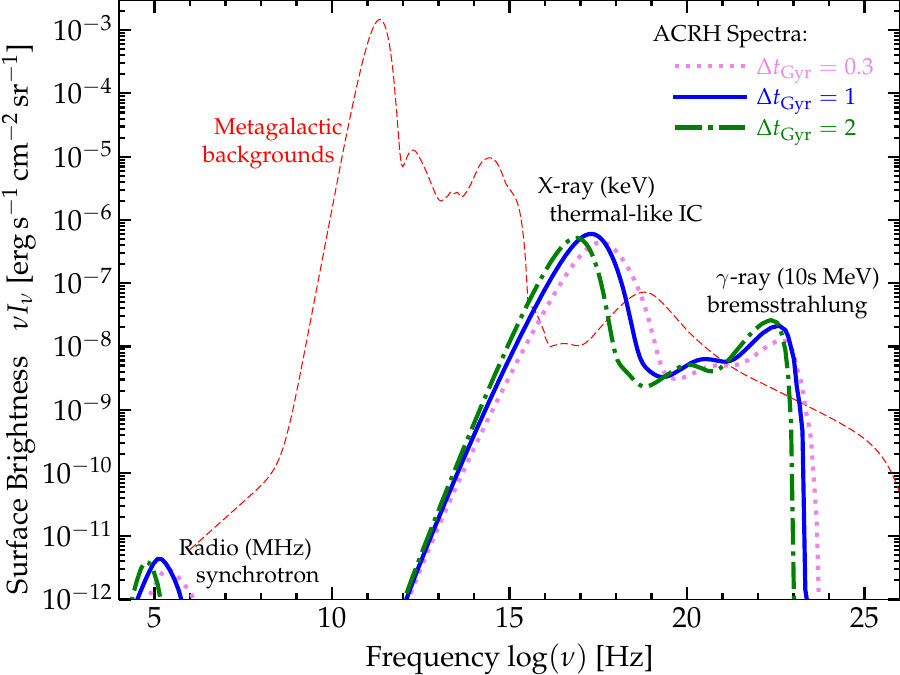} 
	\caption{All-wavelength surface brightness predicted for ACRHs leptons, for different ages where most of the radiation emerges ($\sim$\,Gyr), or equivalently integrating over all radii out to $\sim 100\,$kpc for a constant (steady-state) injection, for a representative toy-model cluster at redshift $z=0.05$ with X-ray IC luminosity $L_{\rm X,\,IC} \approx \dot{E}_{\rm cr,\,\ell} \sim 10^{43}\,{\rm erg\,s^{-1}}$. We compare the meta-galactic background light compiled in \citet{cooray:2016.extragalactic.background.light.compilations.review,khaire:2019.extragalactic.background.light.spectra,tompkins:2023.cosmic.radio.background}. The spectra are qualitatively similar over a wide range of ages/sizes and injection spectra.
	There are three peaks: most of the emission emerges in soft X-rays with a very thermal-like peak (CR-IC scattering CMB photons); there is a secondary $\sim 100\,\times$ less-luminous peak at $\sim $\,MeV soft $\gamma$-rays from relativistic bremsstrahlung, and a much less-luminous peak at low-frequency $\sim$\,MHz radio from synchrotron.
	\label{fig:spectrum.allband}}
\end{figure}

Over a much broader range of wavelengths in Fig.~\ref{fig:spectrum.allband}, plotting $\nu I_{\nu}$, we can see this more clearly. In this plot we normalize our toy model to a total CR injection rate $\dot{E}_{43} = 1$, and include synchrotron (as described below), IC, bremsstrahlung, as well as other processes such as synchrotron self-Compton and pair production which are negligible \citep[for relevant scalings, see e.g.][]{1965AnAp...28..171G,blumenthal:1970.cr.loss.processes.leptons.dilute.gases,1972Phy....60..145G,rybicki.lightman:1979.book}. Again as seen in \paperone\ for mixed CR ages at a given radius, we see the spectrum at any age here peaks clearly in the soft X-rays with a thermal ``bump'' appearance, as opposed to a simple IC power law. The secondary, much smaller, peaks in radio (synchrotron) and $\sim$\,MeV $\gamma$-rays will be discussed below.

For the purposes of comparison below, it is useful to define the effective soft X-ray emissivity of IC, around $\sim$\,keV energies, which is: 
\begin{align}
\tilde{\epsilon}_{i} &\equiv \frac{1}{10^{34}\,{\rm erg\,s^{-1}\,kpc^{-3}}} \frac{d \epsilon_{i}}{d \ln{E}_{\rm ph}} \ , \\
\label{eqn:emissivity} \tilde{\epsilon}_{\rm IC} &\approx 14\,(1+z)^{4}\,e_{\rm cr,\,0.1}\,\mathcal{F}_{\rm IC} \ , 
\end{align}
where the $\tilde{\epsilon}$ convention is simply a convenient units choice, $e_{\rm cr,\,0.1} \equiv e_{\rm cr,\,\ell}/0.1\,{\rm eV\,cm^{-3}}$, and the dimensionless function $\mathcal{F}_{\rm IC}=\mathcal{F}_{\rm IC}(E_{\rm ph},\,\Delta t) $ captures the spectral shape information from the full calculations (defined here to be $\sim 1$ at keV until ages $\gg$\,Gyr).

\subsubsection{Excited X-ray Emission from CRs}
\label{sec:excited}

If there were no gas present (only CRs and CMB), then the continuum IC spectra in Fig.~\ref{fig:spectrum.xr} would be the whole (X-ray) story. Of course, there will also be thermal and line emission from the hot gas in the absence of CRs (discussed below). But the line emission can be boosted directly and indirectly by CRs, as these (1) provide additional direct ionization and line excitation (e.g.\ direct CR-Fe ion interactions; \citealt{chung:2001.cr.electron.multiple.neon.ionization,montanari:2012.cr.ionization.rates.heavy.elements,montanari:2014.cr.ionization.electrons.heavy.elements,montanari:2017.ionization.cross.sections.e.p.beams}); (2) enhance the collisional rates by exciting free electrons via Coulomb interactions; and (3) indirectly shape the emission by heating/cooling the gas (via Coulomb, adiabatic/compressive, streaming, and thermalized hadronic interactions) and contributing pressure forces (e.g.\ driving buoyant motions, outflows, etc.). Calculating in detail what this does to the line spectra is quite challenging, as it involves integrals over the relevant CR spectrum and cross-sections (many of which are poorly-understood), accounting for multiple ionizations and excitations, secondary excitations, and the excitation spectrum induced by the high-energy free electrons from CR scattering and the revised temperature structure caused by CR heating and dynamical terms, complicated further by the fact that the salient heating/cooling times are relatively long so it is not always obvious that we can assume full ionization equilibrium \citep[see e.g.][]{segers:2017.cgm.ionization.out.of.eqm.owing.to.flickering.agn,oppenheimer:2018.cgm.eagle.nonequm.ion.important.potentially}, and the dynamical effects necessitating full CR-MHD dynamical simulations. 

We largely defer studying these effects to future work. However, we can obtain a rough estimate of some of the effects by noting that (1) we will take the ``true'' cluster gas temperature as somewhat arbitrary in our toy models anyways, and (2) we expect some fraction of the thermalized heating rate to be re-emitted by the dominant thermal cooling channel (``line emission,'' broadly speaking, as defined below in \S~\ref{sec:thermal}), with that rate given by the integral of the Coulomb rate (dominated for the spectra in Fig.~\ref{fig:spectrum.xr} by modest-energy $\sim 10-100\,$MeV electrons, since the lowest-energy electrons lose their energy too quickly), which gives an emissivity:
\begin{align}
\tilde{\epsilon}_{\rm CRl} &\sim 1.3\,e_{\rm cr,\,0.1}\,n_{-3}\,\mathcal{F}_{\rm CRl} \ , 
\end{align}
where the factor $\mathcal{F}_{\rm CRl} \sim \tilde{\epsilon}_{\rm tel} / (\tilde{\epsilon}_{\rm tel} + \tilde{\epsilon}_{\rm ff})$ combines all our ignorance about the detailed emergent re-emitted spectrum and fraction of the energy which comes out in the lines, but can be roughly approximated following \citet{mazzotta:2004.xray.temperature.measurement.modeling.and.caveats,gastaldello:2021.metallicity.groups.clusters.review.caveats.future.sensitivity} as the fraction of the true thermal emission coming out in lines ($\tilde{\epsilon}_{\rm tel}$) to free-free continuum ($\tilde{\epsilon}_{\rm ff}$) plus lines.\footnote{An absolute lower limit to $\mathcal{F}_{\rm CRl}$ in the $\sim 0.5-2\,$keV band can be obtained by calculating only the direct CR excitation rate of the salient Fe L-shell resonant lines, which (crudely following the cross-sections in \citealt{montanari:2017.ionization.cross.sections.e.p.beams}) gives $\mathcal{F}_{\rm CRl} \gtrsim 0.002\,(Z/Z_{\odot})$ -- essentially, suppressed by the ratio of the number of iron electrons to total free electrons.}

Note that the CR ``heating time'' is often still long at large radii outside cluster cores, $t_{\rm heat} \sim e_{\rm thermal}/\dot{e}_{\rm cr,\,thermal} \sim 40\,{\rm Gyr}\,T_{\rm keV}/e_{\rm cr,\,0.1}$, so this heating will not usually represent a radical change to the global energetics of the gas in clusters, though it can be non-negligible (especially for cooling flows themselves) as argued in \citet{wiener:cr.supersonic.streaming.deriv,su:2021.agn.jet.params.vs.quenching,su:2023.jet.quenching.criteria.vs.halo.mass.fire,su:2025.crs.at.shock.fronts.from.jets.injection}. There can also be interesting excitation of emission in optical/UV lines if much cooler gas is present, discussed below.

\subsubsection{Comparison to Hot Gas Emission \&\ Simple Fitting Recipes}
\label{sec:thermal}

In the absence of CRs, the emitted spectrum (at the level of accuracy we require here) is dominated by two terms: a thermal free-free continuum, plus line cooling. The effective emissivities of these around $\sim$\,keV energies are well-understood and can be approximated by:
\begin{align}
\tilde{\epsilon}_{\rm ff} &\sim 5\,n_{-3}^{2}\,\mathcal{F}_{\rm ff} \ , \\ 
\tilde{\epsilon}_{\rm tel} &\sim 5\,n_{-3}^{2} Z^{\prime} \mathcal{F}_{\rm tel} \ , 
\end{align}
where $\mathcal{F}_{\rm ff} \approx T_{\rm keV}^{-1/2}\,E_{\rm ph,\,keV}\,\exp{(1-E_{\rm ph,\,keV}/T_{\rm keV})}\,g(E_{\rm ph},\,T)$ ($g$ the Gaunt factor), and $\mathcal{F}_{\rm tel} \sim 20\,q^{2}\,(q+1)^{-1}\,\exp{(-q)}$ for $q\equiv 0.3/T_{\rm keV}$ around $E_{\rm ph} \sim $\,keV \citep{mewe:1985.xray.thermal.collisional.line.emissivities}. 

\begin{figure}
	\centering\includegraphics[width=0.99\columnwidth]{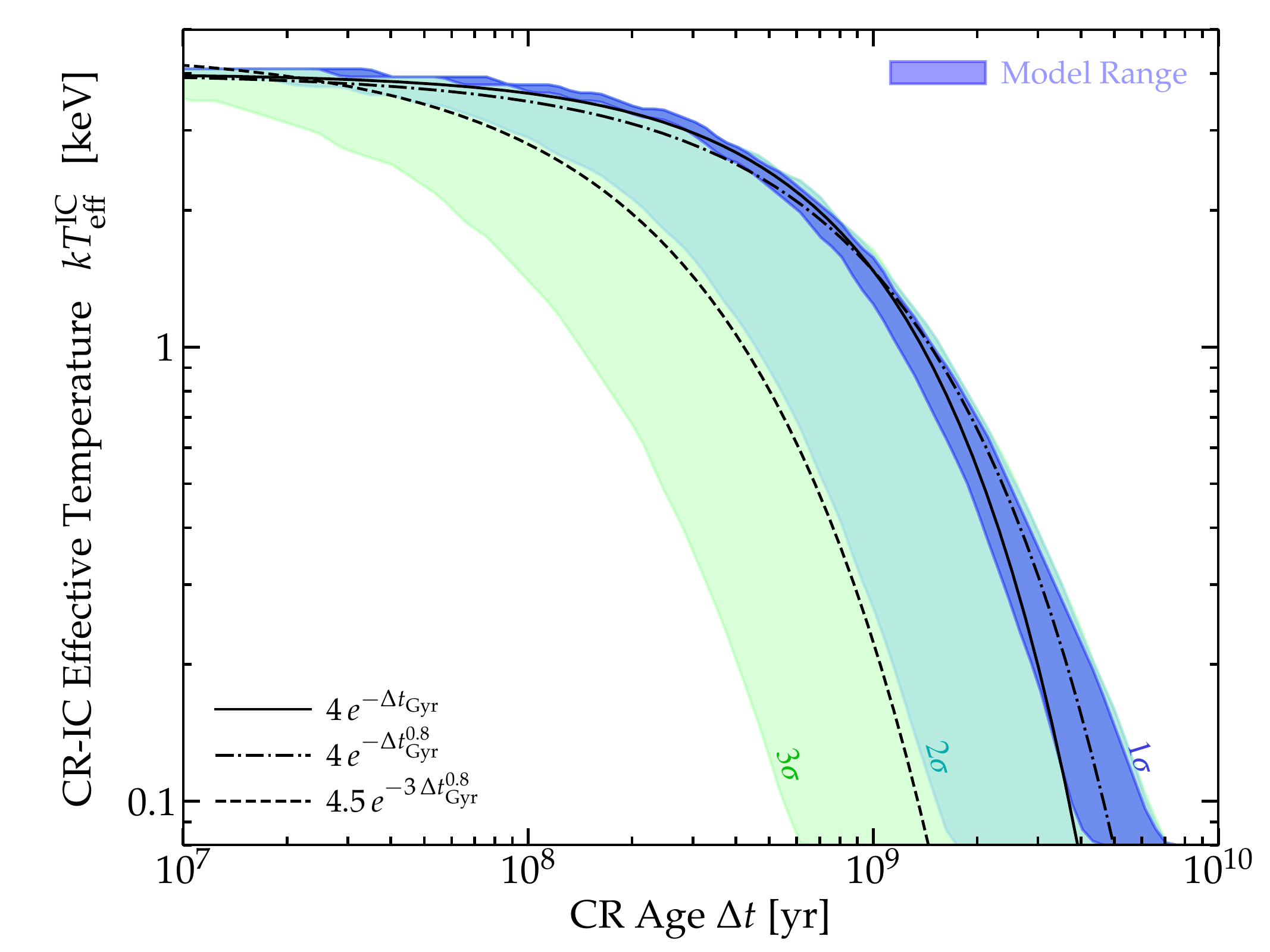} 
	\caption{Effective X-ray spectral temperature $T_{\rm X,\,eff}^{\rm IC}$ fit to pure CR-IC emission (as Fig.~\ref{fig:spectrum.xr}), as a function of the age $\Delta t$ (time since escaping injection zone) of the CRs for different mono-age CR populations (for an LISM-like injection spectrum). 
	We generate this for each age, at each radius, in several hundred mock clusters (\S~\ref{sec:thermal}) varying cluster mass, density, temperature, metallicity, and magnetic field profiles through which CRs are propagated (all at redshift $z\sim0$). Different (energy-dependent) loss rates lead to somewhat different CR spectra at a given radius and age, producing different $T_{\rm X,\,eff}^{\rm IC}$. 
	Shaded regions show the $1-3\sigma$ inclusion regions at each $\Delta t$, lines show some approximate scalings, e.g.\ $T_{\rm X,\,eff}^{\rm IC} \sim 4\,{\rm keV}\,\exp{(-\Delta t_{\rm Gyr}})$. 
	The values at young $\Delta t\rightarrow 0$ depend on the assumed injection spectrum, while those at $\gtrsim 10^{8}\,$yr are more shaped by losses, and those at $\gg 10^{9}$\,yr ($\ll$\,keV) always correspond to regimes where most CR energy has already been lost (emission is minimal).
	\label{fig:cric.temp.vs.age}}
\end{figure}

In Fig.~\ref{fig:spectrum.xr}, we compare typical thermal spectra of gas with primordial composition (primarily free-free), and note the similarity between those and ACRH-IC. Fig.~\ref{fig:cric.temp.vs.age} expands on this. The ``effective'' temperature $T_{\rm X,\,eff}^{\rm IC}$ of the CR-IC X-ray spectrum, even for a mono-age population at some $\Delta t$ since escaping the injection zone, depends on the loss history (integral over time of the loss rate, where Fig.~\ref{fig:spectrum.xr} assumed constant loss-rates in time) and CR transport history (which CRs are at a given radius at $\Delta t$, which Fig.~\ref{fig:spectrum.xr} integrates together). To approximate this, we consider the simple toy cluster model from \paperone: a spherically-symmetric, time-steady background cluster with some density+temperature+magnetic field profiles $n(r)$, $T(r)$, $B(r)$ (functional forms detailed in \S~\ref{sec:cc.profiles}), with a $\delta$-function injection of CRs at the origin at time $t=0$, evolved according to the standard time-dependent but isotropic Fokker-Planck equation for CR transport \citep{zank:2014.book,hopkins:m1.cr.closure,thomas:2021.compare.cr.closures.from.prev.papers} with some uniform (in time and space) but CR energy-dependent diffusivity $\kappa_{\rm iso}$ and energy-independent streaming speed $v_{\rm st}$. At some $\Delta t$, we then take the CR spectrum at each location $r$, and calculate the CR-IC X-ray spectrum and approximate best-fit $T_{\rm X,\,eff}^{\rm IC}$ (fitting as Fig.~\ref{fig:spectrum.xr} to primordial thermal spectra in the $0.1-10\,$keV window weighted by counts). We systematically vary the density+temperature profile shapes following the range of ``universal'' fits to NCC, WCC, and SCC clusters in \citet{mcdonald:2013.cluster.gas.profiles,ghirardini:2019.cluster.profiles.compilation.universal.fits}; $B$ assuming fixed $B_{\rm \mu G}=0.1,\,1,\,10$ or plasma $\beta=P_{\rm therm}/P_{\rm B} = 100$; diffusivity normalization ($\kappa_{\rm iso}$ at $\sim 1\,$GV between $10^{28}-10^{29.5}\,{\rm cm^{2}\,s^{-1}}$ following fits to Solar-system CR data in \citealt{korsmeier:2021.light.element.requires.halo.but.upper.limit.unconfined,dimauro:2023.cr.diff.constraints.updated.galprop.very.similar.our.models.but.lots.of.interp.re.selfconfinement.that.doesnt.mathematically.work,silver:2024.cr.propagation.low.energies.new.data,tovar:2024.inhomogeneous.diffusion.cr.spectra}); streaming speed $v_{\rm st} \sim 30-300\,{\rm km\,s^{-1}}$; cluster mass (which shifts the profiles as they scale with $R_{500}$) between $\sim 10^{13.5-15.5}\,M_{\odot}$; and injection spectra (LISM or power-law as Fig.~\ref{fig:cr.spectra}). The variations to $T_{\rm X,\,eff}^{\rm IC}(\Delta t)$ are shown in Fig.~\ref{fig:cric.temp.vs.age}, but are small, and the $1\sigma$ range can be crudely approximated by $T_{\rm X,\,eff}^{\rm IC} \sim 4\,{\rm keV}\,\exp{(-\Delta t_{\rm Gyr}})$. 

Of course, the trend as $\Delta t \rightarrow 0$ reflects the choice of input/injection-zone spectrum -- we stress as noted above that we only assume an LISM-like spectrum here and different choices could easily re-normalize this within the $\sim 1-10\,$keV range. And we can plainly see that ``effective CR-IC temperatures'' $\ll$\,keV only appear at times where almost all the CR energy has been lost, so will be largely invisible (especially because these would not have the strong line emission associated with metal-rich gas thermal emission at those temperatures). But at intermediate times, different injection spectra converge to more similar predictions in the $\sim1-{\rm few}$\,keV range as they are all sculpted by similar loss processes at both low and high energies. 
We have also plotted $T_{\rm X,\,eff}^{\rm IC}$ versus position $r$, but this depends more systematically on the CR transport parameters (the CR spectrum depends, to leading order, on $\Delta t$, while position depends on $v_{\rm st,\,eff}\,\Delta t$). There is a tail of models to shorter lifetimes: this comes from models with very large densities and $B \gtrsim 10\,{\rm \mu G}$ at $R \sim 100\,{\rm kpc}$, and are only a tiny fraction of the models considered, but even then the CR lifetime is only reduced by a factor $\sim 3$. 

Fig.~\ref{fig:cric.temp.vs.age} is an attempt to marginalize over some theoretical uncertainties at $z\sim 0$, modulo the details of the injection spectrum and possible reacceleration terms (where the uncertainties are larger). At higher redshifts (see \S~\ref{sec:redshift}), if all else is strictly fixed, we expect the CR lifetime/cutoff time to vary with the CMB energy density per Eq.~\ref{eqn:tloss}, while the effective temperature in the emission frame is boosted by higher CMB photon energies. Re-running the models at $z=0.5$, we confirm the primary effect is to shift the model track as expected giving: $T_{\rm X,\,eff}^{\rm IC} \sim 4\,(1+z)\,{\rm keV}\,\exp{[-\Delta t_{\rm Gyr}\,(1+z)^{4}]}$. We discuss this further in \S~\ref{sec:redshift}.

This suggests an extremely simple ``recipe'' to model CR-IC in observed X-ray CC spectra, as a zeroth-order approximation. To leading order, rather than modeling CR populations in detail, one can simply add CR-IC components to spectral fits, representing them as a set of metal-poor (primordial) emission components with some multi-temperature $T_{\rm X,\,eff}^{\rm IC}$, to be fit alongside the usual mix of more metal-rich thermal emission components with $T_{\rm X,\,eff}^{\rm thermal}$, and power-law like (injection-zone or re-accelerated CR) components. While imperfect, this is straightforward (and requires no significant code modification) in standard X-ray fitting codes like XSPEC. 

We discuss the interpretation of the metal line emission, and more detailed spectral fitting, in the presence of this ACRH-IC below.

\subsubsection{The Apparent Spectrum}
\label{sec:apparent}

We can crudely approximate the above emission processes as adding linearly. So for $\sim$\,keV temperatures, at a given position, the effective continuum will be given by $\sim \tilde{\epsilon}_{\rm IC}+\tilde{\epsilon}_{\rm ff}$, and effective line emission by $\sim \tilde{\epsilon}_{\rm CRl} + \tilde{\epsilon}_{\rm tel}$. 

We stress that the predicted spectra in e.g.\ Fig.~\ref{fig:spectrum.xr}, after including the excited line emission, adding a reasonable background thermal+line spectrum, and allowing for the common observational practice of fitting multi-temperature spectra to clusters, closely resemble thermal spectra of gas at wavelengths $\sim 0.1-10\,$keV (we show this explicitly wavelengths below). 

If $\tilde{\epsilon}_{\rm ff}$ and $\tilde{\epsilon}_{\rm tel}$ were negligibly small, the CR contribution will therefore ``appear,'' if it could be isolated on its own entirely, like a \textit{thermal} component with apparent gas temperature, density, metallicity, and entropy ($K\equiv T/n^{2/3}$ derived from the temperature and density) given by:
\begin{align}
\label{eqn:napp} n_{\rm app}^{\rm CR} &\sim 0.002\,{\rm cm^{-3}}\,e_{\rm cr,0.1}^{1/2}\,(1+z)^{2}\,\left( \frac{\mathcal{F}_{\rm IC}}{\mathcal{F}_{\rm ff}} \right)^{1/2} \ , \\ 
\label{eqn:Tapp} T_{\rm app}^{\rm CR} &\sim (0.5-5)\,{\rm keV}\,(1+z) \ , \\ 
\label{eqn:Zapp} Z_{\rm app}^{\rm CR} &\sim 0.1\,n_{-3}\,Z_{\odot}\,\left(\frac{\mathcal{F}_{\rm ff}\,\mathcal{F}_{\rm CRl}}{\mathcal{F}_{\rm tel}\,\mathcal{F}_{\rm IC}}\right) \ , \\
K_{\rm app}^{\rm CR} &\sim 140\,{\rm keV\,cm^{2}}\,(1+z)^{-1/3}\,e_{\rm cr,\,0.1}^{-1/3} \ .
\end{align}
These are defined as the single-valued temperature, density, and metallicity which would give a roughly similar spectrum in soft X-rays, ratio of line emission to continuum emission (equivalent width of the lines), and emissivity/surface brightness, as those induced by the CRs alone. Note that for the temperature, one could obtain slightly different results using the line ratios: to leading order, these will reflect the underlying thermal $T_{\rm keV}$ of the gas, with some second-order boost from CR excitation (\S~\ref{sec:excited}), but these are similar temperatures for the regime of interest. For metallicity $Z_{\rm app}^{\rm CR}$ should be taken with considerable uncertainty depending on the observations (see \S~\ref{sec:z.drops}).

With a thermal component included, if one fit a single-temperature spectrum, the ``effective'' quantities would be some emission-weighted average of the normal thermal component and CR-induced component, e.g. to very crude approximation, dimensionally scaling as 
\begin{align}
\label{eqn:neff} n_{\rm eff} &\sim n_{\rm gas}\,\sqrt{1 + \frac{\tilde{\epsilon}_{\rm IC}}{\tilde{\epsilon}_{\rm ff}}}  
\sim  n_{\rm gas}\,\sqrt{1 + \frac{3\, e_{\rm cr,\,0.1} (1+z)^{4}}{n_{-3}^{2}} } 
\end{align}
 (since most of the information about $n$ in spectral fits comes from the continuum normalization/brightness), 
 and some similar weight could apply to $Z_{\rm eff}$ based on emission around the lines of interest being fit (since most of the constraint on $Z_{\rm eff}$ comes from the equivalent widths of the lines), with $T_{\rm eff}$ a more complicated weighting depending on the spectral range and method used for the fit. Note we drop the $\mathcal{F}$ terms in the approximate expressions above for brevity, since they are order-of-magnitude anyways, but this does not mean they are negligible.

Essentially, if IC is a small correction to the continuum, then it will be an even smaller correction to inferred $T$ and line structure ($Z$), so will be a basically invisible component of the spectrum compared to a small component of gas with $\sim$\,keV temperatures. If IC contributes at the order-unity level to the continuum, however, the spectrum would qualitatively look like a fairly standard multi-temperature spectrum, where if it is fit to a \textit{single} temperature model, because the CRs contribute comparably more to the continuum via IC than to the line emission, the inferred metallicity would tend to be diluted. 

Because $T$ is similar to cluster temperatures in the continuum and the line ratios are weakly modified, this raising of the continuum level would primarily cause one to (1) \textit{over}-estimate the gas density ($n_{\rm app} > n_{\rm gas}$), and (2) \textit{under}-estimate the gas metallicity. The former term depends only on $e_{\rm cr}$, since it is entirely determined by CRs interacting with the CMB; the latter by the gas background density, since it is determined by CR-gas interactions. In a more standard multi-temperature fit, the apparently lower metallicity will be a much less-dramatic effect, because the line emission would (correctly) be inferred to be coming from from the distinct temperature component (the gas), given the known temperature sensitivity of the lines, as seen in e.g.\ \citet{matsushita:2002.m87.virgo.cluster.obs.metallicity.drop.temperature.fitting.challenges} and many of the observational X-ray metallicity studies reviewed in \S~\ref{sec:z.drops}. Indeed in many clusters, like Virgo and/or Perseus \citep{matsushita:2002.m87.virgo.cluster.obs.metallicity.drop.temperature.fitting.challenges,sanders:2007.perseus.profiles.claimed.hardxr.cr.vs.thermal} it is known that one obtains very different metallicity estimates fitting different X-ray lines corresponding to different energy ranges of the spectrum: this is suggestive of a non-thermal CR-IC component like that modeled here.

Related to this, we stress that Eq.~\ref{eqn:Tapp}-\ref{eqn:neff} should be taken as order-of-magnitude qualitative guidelines only, especially when it comes to the effective temperature and metallicity. The inferred density $n_{\rm gas}$ is relatively straightforward, as it depends simply on X-ray surface brightness (even more robustly, we could compare the emission measure itself). But it is well-known that $T$ and especially $Z$ in X-ray spectral fits to observed cool-cores are sensitive to (1) the precise frequency range being fit \citep{mazzotta:2004.xray.temperature.measurement.modeling.and.caveats,ghizzardi:2021.iron.cluster.profiles.sensitivity.of.fitting.different.lines.different.metallicities}, (2) the fit method (e.g.\ H/He-like line ratios, continuum shape, Fe-L shell lines, full-spectrum fits; see discussion in e.g.\ \citealt{matsushita:2002.m87.virgo.cluster.obs.metallicity.drop.temperature.fitting.challenges,ghizzardi:2021.iron.cluster.profiles.sensitivity.of.fitting.different.lines.different.metallicities,zhuhone:2023.cluster.temperature.fitting.sensitivities.simulations}), (3) priors on the temperature and metallicity structure (e.g.\ single-vs-multi-temperature, continuum lognormal vs.\ power-law temperatures, single-$Z$ versus multi-$Z$; e.g.\ \citealt{mazzotta:2004.xray.temperature.measurement.modeling.and.caveats,vijayan:2022.cluster.cgm.multitemperature.fit.challenges}), and (4) instrumental sensitivity and signal-to-noise of the spectrum \citep{ghizzardi:2021.iron.cluster.profiles.sensitivity.of.fitting.different.lines.different.metallicities,zhuhone:2023.cluster.temperature.fitting.sensitivities.simulations}. This is on top of effects which will be physically important such as (5) CR heating and ionization shifting the line excitation and ionization corrections, and (6) detailed spectral shape of the CR-IC plus pre-existing multi-phase thermal spectrum \citep{avestruz:2014.cluster.mocks.from.sims.sensitivity.temperature.measurements}. 

Indeed, empirical metallicity solutions for the same X-ray data in cool cores, just using slightly different temperature priors or different instruments or different spatial resolution binning, differ by as much as a factor of $\sim 5$ in $Z$ (see \S~\ref{sec:z.drops}), while fitted temperatures can differ by several keV \citep{zhuhone:2023.cluster.temperature.fitting.sensitivities.simulations}. So properly modeling the ``observed'' $T_{\rm eff}$ or $Z_{\rm eff}$ to compare to observations clearly requires detailed forward-modeling of the intrinsic spectrum, instrumental response, and using the identical fitting procedures and assumptions to specific individual observational studies. We expect, however, that in many cases the effect on observed temperatures will be relatively small (especially for more sophisticated observational temperature fitting methods), compared to the overall boost to the surface brightness enhancing the apparent density $n$ and cooling rate.


\subsection{Radial Profiles}
\label{sec:radial}

Now we can combine these with the simplest expected scaling for $e_{\rm cr,\,\ell}$ from a central source (Eq.~\ref{eq:ecr.r}) in \S~\ref{sec:cr.spectrum} to make some prediction for radial profiles. Since most of the $\sim 0.1-1$\,GeV CR energy escaping the injection region ($\dot{E}^{\rm inj}_{\rm cr,\,\ell,\,GeV}$) will emerge in IC, note $\dot{E}^{\rm inj}_{\rm cr,\,\ell,\,GeV} \sim L_{\rm X,\,keV,\,IC} \equiv 10^{42}\,{\rm erg\,s^{-1}}\,L_{\rm IC,\,42}$, with a cutoff radius at $R_{\rm out,\,IC}$ where $t_{\rm travel} \equiv R/v_{\rm st,\,eff} \gtrsim t_{\rm loss}$, or equivalently where integrated $L_{\rm IC} \rightarrow \dot{E}^{\rm inj}_{\rm cr,\,\ell,\,GeV}$, we can rewrite Eq.~\ref{eq:ecr.r} as: 
\begin{align}
\label{eqn:ecr.r} e_{\rm cr,\,0.1} &\sim 0.5\,\frac{f_{\rm loss}\,L_{\rm IC,\,42}}{v_{\rm st,\,100}\,R_{100}^{2}} \sim 1.7\,\left( \frac{R_{\rm out}}{ R} \right)^{2} \frac{f_{\rm loss} L_{\rm IC,\,42}}{R_{\rm out,\,100}\,(1+z)^{4}} \ .
\end{align}
Here $f_{\rm loss}$ accounts for losses (very crudely $\sim \exp{(-R/R_{\rm out})}$) and $R_{\rm out,\,IC,\,100}  \equiv R_{\rm out}/100\,{\rm kpc}  \sim 3\,v_{\rm st,\,100}\,(1+z)^{-4}$. 

With Eq.~\ref{eqn:emissivity}, this implies a surface-brightness profile from CR-IC: 
\begin{align}
\nonumber \frac{S_{\rm X,\,IC}}{{\rm erg\,s^{-1}\,kpc^{-2}}} &\sim 10^{37.4}\,L_{\rm IC,\,42} \frac{f_{\rm loss} (1+z)^{4} }{R_{100}\,v_{\rm st,\,100}} \\
& \sim 10^{38}\,L_{\rm IC,\,42}\,\frac{f_{\rm loss}(R) R_{\rm out}}{R} \ .
\end{align}
In other words, declining as $\sim 1/R\,v_{\rm st,\,eff}$ at small radii ($\propto 1/R$ for constant $v_{\rm st,\,eff}$, though this could become as shallow as $\sim$\,constant, as $R\rightarrow 0$, if the transport is diffusion-dominated with a constant diffusivity $\kappa \sim v_{\rm st,\,eff}\,R$), with a cutoff from losses ($f_{\rm loss}(R)$) at $\gtrsim R_{\rm out} \sim 100\,$kpc. 

We stress here the explicit dependence on CR transport parameters, here represented by $v_{\rm st}$, which \textit{does not} have to be constant with radius or other plasma/cluster properties. Unfortunately, as noted above and discussed in e.g. \citet{hopkins:2025.crs.review} and references therein, first-principles models for $v_{\rm st}$ and observational constraints outside of the near-Solar-system LISM remain many orders-of-magnitude uncertain. So the actual profile shapes do not have to follow the constant-$v_{\rm st}$ models assumed here -- we only take $v_{\rm st}\sim$\,constant because it is the simplest choice, lacking better models or observational constraints. If CR constraints at these large radii can be established from the multi-wavelength constraints modeled here, then this can be used to constrain or infer $v_{\rm st}$ in different environments, which would be of major importance for our understanding of CRs. 

With that caveat in mind, below in \S~\ref{sec:coolcores}, we discuss the implications of these simple constant-$v_{\rm st}$ models for the ``apparent'' density, cooling rate, temperature, entropy, pressure, potential, and metallicity profiles, for a simple model of an ACRH plus thermal emission.

\subsubsection{Local Fluctuations \&\ ``Turbulence''}
\label{sec:xr.fluct}

Briefly, because our models are simple, analytic, steady-state and spherically-symmetric, we do not model local fluctuations, turbulence, shocks, or non-axisymmetric structures in the CRs or their emission. But as shown in \paperone, numerical simulations of CR-pressure-dominated CGM/ICM conditions do show fluctuations in $e_{\rm cr,\,\ell}$ corresponding to these physics \citep{ji:fire.cr.cgm,butsky:2020.cr.fx.thermal.instab.cgm,su:turb.crs.quench,su:2021.agn.jet.params.vs.quenching,ruszkowski.pfrommer:cr.review.broad.cr.physics,weber:2025.cr.thermal.instab.cgm.fx.dept.transport.like.butsky.study} with implied similar-magnitude SB fluctuations to those seen in clusters. 
On large scales (where the CR diffusion time $\sim L^{2}/\kappa$ is larger than the turbulent eddy timescale), it is easy to understand this, as CR transport becomes effectively advective/\Alf{ic} streaming so CRs will quasi-adiabatically follow gas. Then turbulent compression will produce enhancements in $\delta E_{\rm cr} = m_{\rm cr}\,c^{2} \delta \gamma \sim (E_{\rm cr}/3) \nabla \cdot {\bf u}$ and $\delta n_{\rm cr} \sim n_{\rm cr}\,\nabla \cdot {\bf u}$ (the same as the gas $\delta n_{\rm gas} \sim n_{\rm gas}\,\nabla \cdot {\bf u}$), or if we define an rms compressive fluctuation $\langle |\nabla \cdot {\bf u}|^{2}\rangle^{1/2} \equiv \mathcal{D}_{\rm turb}$: $\langle |\delta E_{\rm cr}|\rangle /E_{\rm cr} \sim (1/3) \mathcal{D}_{\rm turb}$, $\langle |\delta n_{\rm cr}|\rangle /n_{\rm cr} \sim \mathcal{D}_{\rm turb}$, $\langle |\delta n_{\rm gas}|\rangle /n_{\rm gas} \sim \mathcal{D}_{\rm turb}$. Hence, the statistics and magnitude of fluctuations in the effective emission temperature $\delta T_{\rm IC}/T_{\rm IC} \sim \delta (\gamma^{2}) / \gamma^{2}\sim (2/3)\,\mathcal{D}_{\rm turb}$ and emissivity $\delta \epsilon_{\rm IC}/\epsilon_{\rm IC} \sim \delta (n_{\rm cr} \gamma^{2})/n_{\rm cr}\,\gamma^{2}  \sim (5/3) \mathcal{D}_{\rm turb}$ will be very similar (up to a $\sim 10\%$ normalization-constant-difference) to those for pure-thermal emission in adiabatic sub-sonic turbulence: $\delta T_{\rm gas}/T_{\rm gas} \sim (2/3)\,\mathcal{D}_{\rm turb}$, $\delta \epsilon_{\rm therm}/\epsilon_{\rm therm} \sim \delta(n_{\rm gas}^{2} T_{\rm gas}^{1/2})/n_{\rm gas}^{2} T_{\rm gas}^{1/2} \sim (7/3) \mathcal{D}_{\rm turb}$ (as often assumed modeling these fluctuations in e.g.\ \citealt{zhuravleva:2015.turbulence.estimates.in.clusters.from.surface.brightness.fluctuations.perseus,zhuravleva:2018.cluster.turb.props.from.xrays,li:2025.surface.brightness.fluctuations.nearby.clusters.claim.turb.equals.cooling.but.requires.super.extreme.model.pushing.more.like.three.dex.too.low,sanders:2020.bulk.flows.perseus.estimation.density.fluctuations.turbulence,devries:2023.chandra.gas.density.fluctuations.estimation.perseus}). 

It is also worth noting that the same simulations of CR pressure-dominated CGM/ICM conditions tend to produce weak, sub-sonic, buoyancy-driven turbulence, consistent with what is observed in CCs. In fact \citet{butsky:2022.cr.linewidth.effects.cgm} showed that more strongly CR-pressure dominated halos actually produce smaller non-thermal/kinematic/turbulent line-broadening, compared to the simulations of the same halos with negligible CR pressure. Thus the apparently weak turbulence seen in X-ray line emission in CC centers with microcalorimetric instruments (e.g.\ Hitomi and XRISM; see \citealt{hitomi:2016.perseus.weak.turb}) has a natural explanation in CR-IC models.

\section{Associated Multi-Wavelength Signatures}
\label{sec:multi.wavelength}

Next, consider the range of potential signatures of CR-IC (outside of soft X-rays), as tests of this scenario. Recall, Fig.~\ref{fig:spectrum.allband} shows a predicted pan-wavelength spectrum (from sub-MHz radio through PeV $\gamma$-rays), of the direct emission from the leptonic CRs with $L_{\rm IC,\,43}\sim 1$. The clear peak is in the soft X-rays, but we will consider each wavelength range in turn.

\subsection{Hard X-rays \&\ ``Standard'' IC Searches}
\label{sec:hardxray}

\begin{figure}
	\centering\includegraphics[width=0.99\columnwidth]{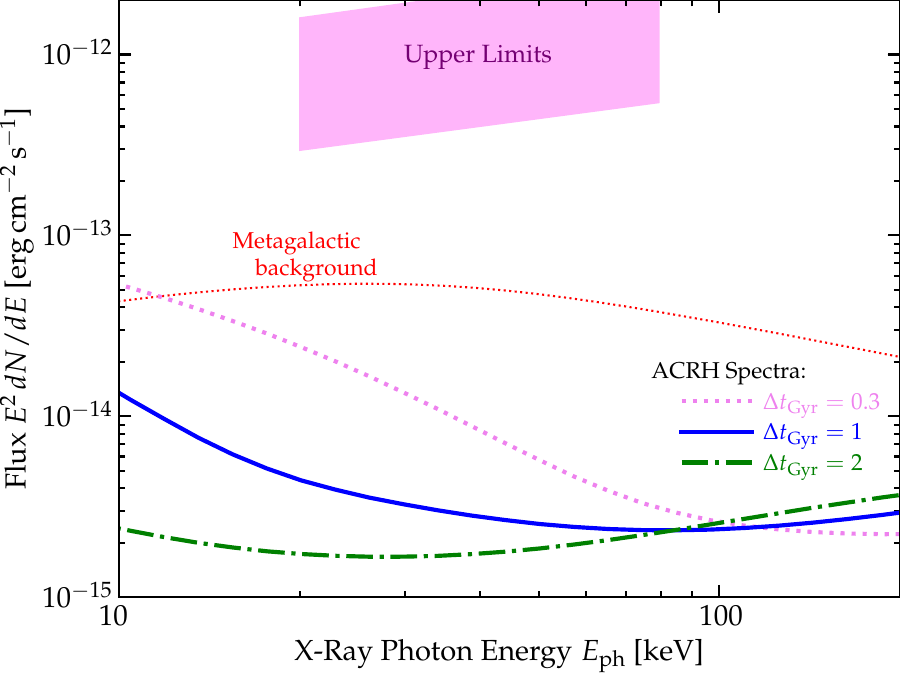} 
	\caption{Hard X-ray ACRH spectrum for a luminous cluster as Fig.~\ref{fig:spectrum.allband}, compared to the extragalactic background and to the most stringent limits possible at present from standard deep hard X-ray searches at $20-80$\,keV for IC signatures (shaded shows the range of upper limits and known detections; see \S~\ref{sec:hardxray}). The shape of the shaded region shows the ``standard'' IC spectral shape being searched for, which assumes a zero-age pure-power-law CR distribution ($E_{\rm ph}^{2} dN/dE_{\rm ph} \propto E_{\rm ph}^{+1/2}$). The ACHRs are factor $\sim 100$ less-luminous than detectable limits and much softer (owing to losses from high-energy CRs). The hard X-ray emission that is present comes primarily from $\sim 0.1-1\,$GeV leptons IC scattering cosmic infrared background photons.
	\label{fig:hardxr}}
\end{figure}

Standard IC searches and detections in clusters generally focus on looking for a very hard spectrum ($E_{\rm ph}^{2}\,dN_{\rm ph}/dE_{\rm ph} \propto E^{(3-\delta)/2} \sim E^{+0.5}$ for a power-law spectrum with $\delta \sim 2$) in the hard X-ray bands, particularly $20-80$\,keV \citep[e.g.][]{bowyer:1998.coma.ic.euv.must.be.older.electrons.than.synchrotron.more.diffuse,colafrancesco:2009.ic.ophiuchus.minihalo.possible.sources}. In Fig.~\ref{fig:hardxr}, we show the predicted ACRH hard X-ray spectra from $10-200\,$keV, compared to the meta-galactic background (from the compilations in \citealt{khaire:2019.extragalactic.background.light.spectra} and references therein), and the range of the most sensitive upper limits which have been achieved in observations of nearby clusters (e.g.\ Perseus, Ophiuchus, Coma, Virgo) with instruments like INTEGRAL, Swift-BAT, NuSTAR (compiled from \citealt{fusco:2000.cluster.hardx.ic.detection.weak.B,fusco:2003.cluster.ic.hardx.review.B.weak,fusco:2004.coma.hardx.ic.detection.B.0pt1.microG,gruber:2002.rxte.cluster.ic.B.weak,rephaeli:2003.cluster.ic.detection.B.limits.low,bonamente:2007.abell.3112.clear.xray.inverse.compton.required.luminosity.fits.models.gamma.rays.too,chen:2008.ic.searches.hardxray.B.field.lower.limits.clusters,wik:2011.swift.ic.hardx.upper.limits.clusters.b.lower.point1microg,cova:2019.cluster.ic.upper.limits.B.lower.limits,mirakhor:2022.ic.cluster.detection.B.0pt1microG}), with the lower limit taken as the sensitivity projected for hypothetical future missions like ASTRO-H \citep{bartels:2015.radio.inverse.compton.cluster.minihalo.prospects}. We immediately see three important aspects: (1) the predicted very hard emission is $\sim 100 \times$ fainter than best-case future-mission or nearest-cluster sensitivities; (2) it is also $\sim 10-30\times$ fainter than the meta-galactic background at the same energies (up to $\sim$\,MeV); and (3) the spectra are softer than standard IC searches model. 

The softer hard X-ray spectrum owes to the CR spectral curvature noted above. In fact, if we were seeing \textit{only} IC from the CMB, the predicted emission would cut off exponentially at $>10\,$keV. The predicted hard X-ray emission in Fig.~\ref{fig:hardxr} is actually dominated by IC from lower-energy CRs off the cosmic infrared background (CIB), which looks like a greybody ($T_{\rm eff} \sim 30-300\,$K) hence the IC spectrum resembling a corresponding ($\sim 10-100\,$keV) mix of temperatures, but with much lower flux owing to the orders-of-magnitude lower photon number density of the CIB compared to the CMB. 

If the injection spectrum is sufficiently hard, there could be bright IC in hard X-rays just within the injection zone (in time and space), i.e.\ for ages $\Delta t \ll 10^{7}\,$yr and spatial scales $\ll 10\,$kpc. Even for a zero-age system, our calculation (shown explicitly in \S~\ref{sec:younger}) gives a predicted IC luminosity generally below the upper limits observed except if we assume X-ray luminosities comparable to the brightest known systems ($\sim 10^{44}-10^{45}\,{\rm erg\,s^{-1}}$). Given their compact size, these would be associated with whatever ``point source like'' systems also produce hard synchrotron at $\gtrsim$\,GHz, not the ACRHs. The searches for these, in strong synchrotron sources, have indeed yielded a mix of detections and upper limits consistent with our predictions, which allows for constraints on e.g.\ the magnetic field strengths particularly {in the injection zones}, with typical $B_{\rm inj} \gtrsim 0.1\,{\rm \mu G}$ in the most centrally-concentrated injection zones and $B \sim 0.03-0.3\,{\rm \mu G}$ in the $\sim 10-100\,$kpc regions around that zone where detected \citep{bagchi:1998.inverse.compton.cluster.detection.B.0pt05microG,kaastra:1999.cluster.2199.ic.detection.B.weak,fusco:2000.cluster.hardx.ic.detection.weak.B,fusco:2003.cluster.ic.hardx.review.B.weak,fusco:2004.coma.hardx.ic.detection.B.0pt1.microG,gruber:2002.rxte.cluster.ic.B.weak,rephaeli:2003.cluster.ic.detection.B.limits.low,pereztorres:2009.ophiuchus.minihalo.xray.ic.contrib.significant.B.0pt03to0pt3microG,chen:2008.ic.searches.hardxray.B.field.lower.limits.clusters,wik:2011.swift.ic.hardx.upper.limits.clusters.b.lower.point1microg,cova:2019.cluster.ic.upper.limits.B.lower.limits,mirakhor:2022.ic.cluster.detection.B.0pt1microG}. For almost any reasonable model, the near-injection fields should be higher than the magnetic field strengths in the extended, much larger and more diffuse ACRH volume.

One potential exception is of course the brightest CC on the sky, Perseus, but even there extended diffuse hard X-rays have only just recently been detected by deep NuSTAR+Swift observations \citep{creech:2024.nustar.perseus.obs} at radii of interest for ACRHs (a $\sim 20-150\,$kpc aperture). The hard X-ray spectrum observed there is indeed quite similar to that predicted in Fig.~\ref{fig:hardxr}, and in a follow-up paper (in preparation) we show it agrees well with the prediction for a CR spectrum fit to reproduce the soft X-ray and radio data.

\subsection{Optical/UV/IR \&\ Unique CR Ionization Signatures}
\label{sec:uv.oir}

There is no significant continuum signal in infrared/optical/UV wavelengths. As shown in Fig.~\ref{fig:spectrum.allband}, at energies $\lesssim$\,keV until radio, the emission from ACRHs appears identical to the Rayleigh-Jeans tail of the thermal-like continuum soft X-ray emission, and at $\lesssim 50\,$eV, it falls well below the metagalactic background. The effective temperature of the CR-IC, even for the softest CR spectra in Fig.~\ref{fig:cr.spectra}, never falls below $\sim 0.2$\,keV.\footnote{Even when there are some even lower-energy CRs ($\lesssim 0.1\,$GeV), the CR spectral shape plus energy scaling of IC means their emission is subsumed into the Rayleigh-Jeans tail of scattered CMB light from $\sim 0.1-1\,$GeV CRs.} And somewhat higher sub-keV emission is largely restricted to the oldest, most-energetically-depleted CRs which therefore contribute little emission. Indirect signatures will also be largely undetectable in this range, with the exception of SZ effects, discussed below. 

There may, however, be telltale ionization/excitation signatures in dense, neutral (atomic or molecular) multi-phase gas within CCs. 
It is well-known that warm H$_{2}$ emission and related optical/UV/NIR lines are detected from the very small mass of cold+neutral gas in filaments at $\sim10$'s of kpc from CC centers, with extreme excitation and line ratios more akin to the Crab nebula than ISM GMCs \citep{johnstone:2007.warm.h2.emission.in.clusters.ccs}. These appear to be excited by an extremely large incident rate of high-energy particles \citep{ferland:2008.molecular.emission.in.cooling.flow.filaments.shielding.excitation.models,ferland:2009.particle.ionization.needed.for.molecular.line.emission.in.cc.perseus,salome:2011.extended.molecular.gas.around.perseus.cluster.core,vantyghem:2017.13co.detection.molecular.gas.clusters.direct.column.and.mass}, but it is unclear how much of that ionization comes from diffuse hot gas ($\sim$\,keV particles) versus CRs ($\sim$\,GeV particles). However it is generally difficult to explain these and other detailed internal properties of the molecular gas in ``hydrodynamic'' models \citep[e.g.][]{beckmann:2019.cluster.molecular.gas.sims.produce.too.clumpy.if.all.mhd.wrong.detailed.properties} without e.g.\ some large CR ``bath.'' 

We will investigate these particle-ionized lines in neutral gas in future work, as modeling them requires more detailed assumptions regarding the nature of the dense molecular gas itself, not just the bulk properties of the CRs and CC/cluster. But they can provide a strong argument in favor of CR-IC in SCC centers.

\subsection{Gamma-Rays}
\label{sec:gamma}

\begin{figure}
	\centering\includegraphics[width=0.99\columnwidth]{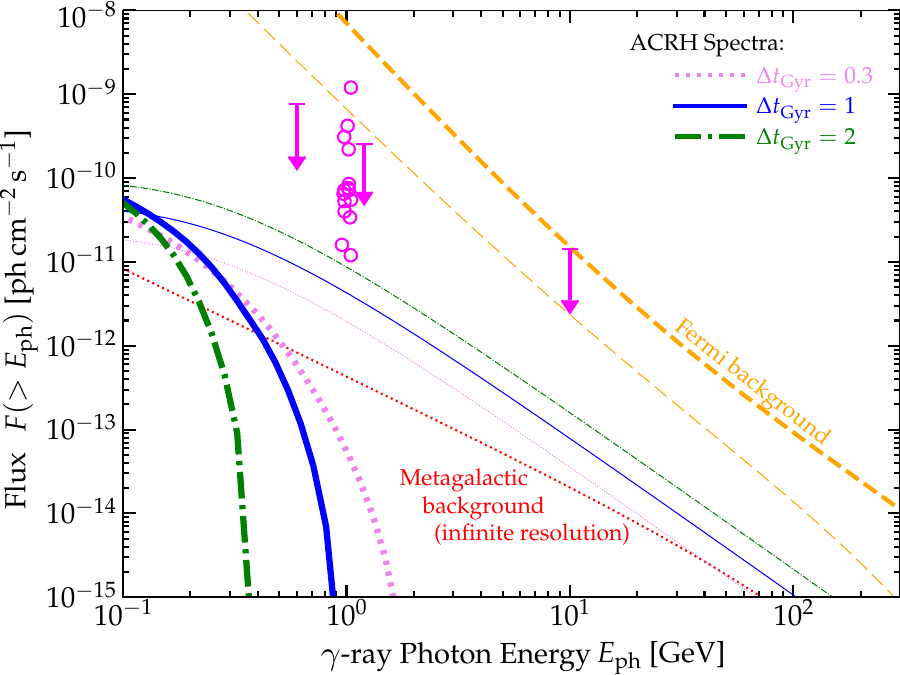} 
	\centering\includegraphics[width=0.99\columnwidth]{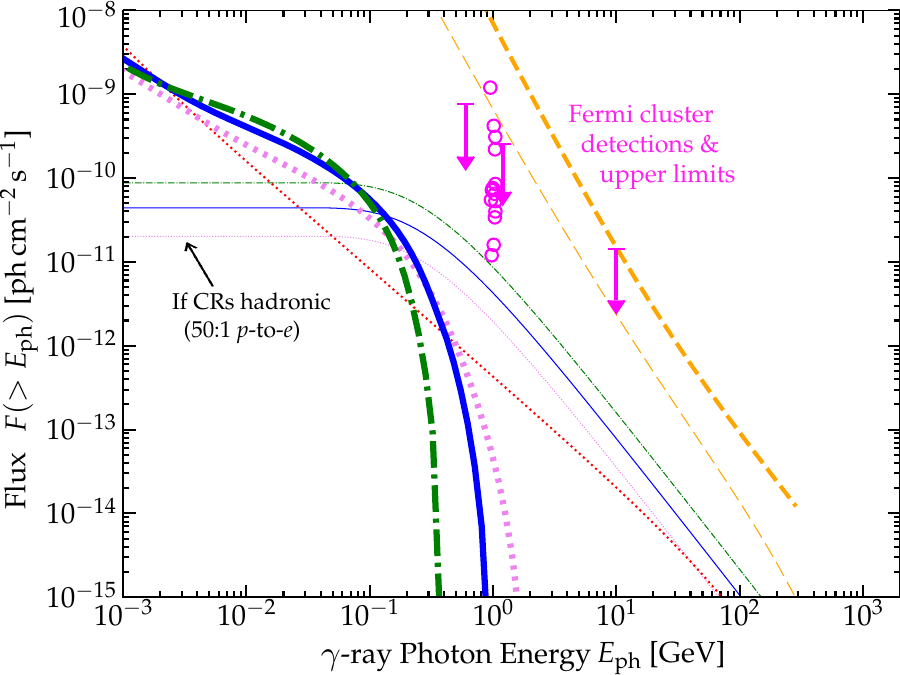} 
	\caption{Predicted ACRH $\gamma$-ray spectra as Fig.~\ref{fig:spectrum.allband}, focusing on a narrow range around \textit{Fermi} GeV bands (\textit{top}) or a broader MeV-TeV range (\textit{bottom}). 
	Thick lines show the $\gamma$-rays from the ACRH leptons which must be present (relativistic bremsstrahlung). Thin lines show emission (primarily pionic) which would be produced by hadrons if present, \textit{assuming} a 50:1 proton-to-lepton ratio (an extreme upper limit for ``hadronic'' CR injection models). 
	The meta-galactic background which would be detected with infinite $\gamma$-ray resolution and sensitivity, is shown, as is the full background in a standard \textit{Fermi} beam (\textit{thin} shows just the isotropic metagalactic backgrounds in the \textit{Fermi} beam). 
	We compare the most sensitive upper limits on clusters (\textit{arrows}) and nearby cluster-center detections (\textit{points}) compiled in \S~\ref{sec:gamma}. ACRHs are invisible in present $\gamma$-ray studies unless the leptons are associated with a $\gtrsim 100\times$ larger hadronic component.
	\label{fig:gamma}}
\end{figure}

Fig.~\ref{fig:gamma} shows the predicted $\gamma$-ray spectra from MeV through $\gtrsim$\,TeV, and focused around $\sim$\,GeV where the ratio of predicted emission to meta-galactic background is maximized.

\subsubsection{Leptonic Cosmic Rays}
\label{sec:leptonic}

First, consider the $\gamma$-rays (calculated following e.g.\ \citealt{blumenthal:1970.cr.loss.processes.leptons.dilute.gases,1972Phy....60..145G}) associated with the CR leptons known to be present already and associated with the ACRHs, and known to be produced from cluster radio galaxies and jets \citep{keenan:2021.jet.leptonic.power.1e41to1e45.easily.produced.from.modest.agn.bursts.or.steady.jets,foschini:2024.blazar.agn.jet.power.favor.leptonic.large.power.energy.much.more.than.kinetic.lobe.cavity.power}. We indeed expect, if the CRs (at least those of interest here, in the centers of SCCs and specifically associated with radio galaxy sources at cluster centers) are accelerated in relativistic jets, that they may be primarily leptonic \citep[see][and references therein]{bottcher:2013.blazar.modeling.almost.all.blazars.better.fit.by.leptonic.cr.models.not.hadronic,blandford:2019.agn.jets.review,cerruti:2020.agn.jet.leptonic.hadronic.review}. Owing to the soft CR spectrum, IC-induced $\gamma$-rays contribute relatively little, and the dominant emission mechanism is relativistic bremsstrahlung primarily at $\sim 10-100$\,MeV energies (factor $\sim 10$ below the CR spectral peak, as expected). 

This is essentially invisible, as shown in Fig.~\ref{fig:gamma}. It only just barely rises above the meta-galactic background at $\sim 10-100\,$MeV, but more importantly at energies $\gtrsim 500\,$MeV where observatories like Fermi are more sensitive, we compare to (1) the isotropic metagalactic background flux in the Fermi beam, for a cluster at typical $z\sim 0.05$ (since the size of the ACRH, at $\sim 100\,$kpc, will be un-resolved in Fermi), and (2) the total background in Fermi beam. We also compare to the compilation of upper limits and detections of clusters in Fermi (including point-source \textit{and} extended emission, though Fermi is significantly more sensitive to the former), with the range of upper limit arrows showing the minimum and maximum upper-limits derived in the compilations of \citet{huber:2013.stacked.fermi.gamma.ray.clusters.weak.upper.limits.for.hadronic.production,ackermann:2014.cosmic.ray.fermi.gamma.ray.upper.limits.galaxy.clusters.data.not.as.model.dependent,ackermann:2015.fermi.clusters.gamma.ray.upper.limits.diffuse.emission}. In either case, these are several orders of magnitude larger than the predicted leptonic $\gamma$-ray flux. Even future CTA-like missions probing down to MeV energies are not projected to reach sensitivities sufficient to detect the predicted curves here \citep{funk:2013.fermi.cta.sensitivities.mev.gev.even.cta.not.close.to.needed.for.leptonic.cluster.mw.halo.gamma.rays.detection}.

Consistent with Fig.~\ref{fig:gamma}, in future work modeling the multi-wavelength emission from specific nearby clusters in greater detail, we find that even in the brightest, closest CCs, e.g.\ Perseus and Virgo, the predicted $\gamma$-ray emission (from models fit to X-ray CR-IC and radio synchrotron observations) is more than an order-of-magnitude lower than the strongest upper limits to diffuse emission at the lowest ($\sim 100\,$MeV) energies, and a factor $\gg 100$ below the upper limits at $\gtrsim$\,GeV. And in both cases the predicted extended emission is orders-of-magnitude less-luminous than the detected compact-source spectra of NGC 1275 or M87.

\subsubsection{Hadronic Cosmic Rays}
\label{sec:hadronic}

We stress that it is very plausible, and indeed many have argued on both theoretical and observational grounds, that the CR injection in radio galaxies and blazars (the cluster-center sources of greatest interest here) are primarily leptonic (references above). Thus it is plausible that the \textit{only} $\gamma$-ray signature is the relativistic bremsstrahlung in \S~\ref{sec:leptonic}. 

But for the sake of completeness, if we imagine there were a large hadronic component injected alongside, or responsible for the leptonic CRs. Then this would boost the $\gamma$-ray emission. The upper limit in principle to the hadronic emission would come from assuming all the leptons are hadronic in origin, i.e.\ produced by proton collisions with the ICM. Given the branching ratios and relative energies (and modeling a plausible spectral slope), this gives the result that the $\gamma$-ray luminosity at $\sim 1$\,GeV would be approximately $L_{\gamma}^{\rm pure\,hadronic} \sim (1/3)\,\dot{E}_{\rm cr,\,\ell}$ -- i.e. a bit less than the X-ray luminosity. This, however, is not the scenario we are interested in, unless it occurs very close to the acceleration zone,\footnote{We are agnostic to the acceleration mechanism, so leptons could be ``initially'' produced hadronically near-source. In \S~\ref{sec:hadronic}, the hadron-to-lepton ratio we are interested in is whatever emerges on scales of the ACRH $\sim 100\,$kpc at energies $\sim 0.1-1\,$GeV.} since it would produce ``young'' electrons continuously. Moreover it is strongly disfavored by observations in many cluster cores \citep{adam:2021.gamma.ray.detection.coma.cant.make.all.electrons.as.secondaries.from.hadrons} and for plausible physical injection sources (e.g.\ blazars; \citealt{bottcher:2013.blazar.modeling.almost.all.blazars.better.fit.by.leptonic.cr.models.not.hadronic}) -- although see \citet{keshet:2025.stacked.cluster.gamma-ray.detection.claim.large.r.flat.spectrum.possible.shock.acceleration.cr.spectrum.hadronic.signature} for arguments that hadronic production may dominate in cluster outskirts. 

A slightly more plausible model would be to assume an LISM-like proton-to-electron ratio. In Fig.~\ref{fig:gamma}, we assume that in the injection zone, the protons follow an LISM-like spectrum and proton-to-electron ratio (from the same compiled observations as our LISM-like electron spectra), then propagate accounting for Coulomb and catastrophic (primarily pionic) losses \citep[from][]{Mann94,hopkins:cr.multibin.mw.comparison}. We then convolve the proton spectra with the appropriate energy-dependent cross-sections and production factors \citep[e.g.][and references therein]{yang:2018.pionic.gamma.ray.spectral.modeling.cross.sections.and.production.factors} to predict the associated $\gamma$-ray spectra. 
This does indeed boost the predicted $\gamma$-ray emission, especially at Fermi and harder (e.g.\ HESS) energies $\gtrsim 0.5\,$MeV. But even with this assumption -- that there is much more (hidden) CR hadronic energy density than leptonic -- the predicted $\gamma$-ray emission from the ACRH falls well below existing Fermi upper limits and the predicted surface brightness is only a factor of a few (at its peak $\sim 1\,$GeV) above the meta-galactic background. 

An order-of-magnitude estimate of the $\gamma$-ray flux at $\sim 1\,$GeV can be obtained by assuming LISM-like CR spectra (with a spectrally-integrated ratio of hadronic-to-leptonic energy $\xi_{\rm had} \equiv e_{\rm cr,\,had} / e_{\rm cr,\,\ell}$, with value $\xi_{\rm had,\,\odot}$ in the LISM) and some uniform density $n_{\rm gas}$, convolving as above, and using our previous estimate of the X-ray IC luminosity $L_{X,{\rm IC}} \sim \dot{E}_{\rm cr,\,\ell}$, to obtain $L_{\gamma} \sim 0.04\,(1+z)^{-4}\,\mathcal{F}_{\rm IC}^{-1}\,(n_{\rm gas}/10^{-3}\,{\rm cm^{-3}})\,(\xi_{\rm had}/\xi_{\rm had,\,\odot})\,L_{\rm X,\,IC}$, with an upper limit given by the smaller of $L_{\gamma} \lesssim L_{\rm X,\,IC}/3$ (the pure-hadronic lepton origin scenario) or $L_{\gamma} \lesssim  L_{\rm X,\,IC}/\xi_{\rm had}$ (where $\gamma$-ray losses exhaust the hadronic energy). 

Note that (as shown in Fig.~\ref{fig:gamma}), many bright individual clusters \citep{abdo:2009.fermi.detection.1275.perseus.strong.gamma.ray.source.1e45.luminosity,colafrancesco:2010.fermi.1275.perseus.gamma.rays.centrally.dominated.large.diffuse.flux.allowed,dutson:2013.fermi.gamma.ray.emission.bcgs.may.require.beaming.towards.us.so.rarer.than.expected,ahnen:2016.diffuse.perseus.gamma.rays.sensitivity.not.very.sensitive.modest.fraction.of.1275.allowed.strong.central.gamma.ray.emission.allowed} and stacked cluster samples \citep{han:2012.diffuse.cluster.gamma.rays.consistent.with.extended.cosmic.rays.but.highly.dependent.on.point.source.subtraction.dominating.signal.point.source.like.central.contrib.clearly.detected,manna:2024.stacking.clusters.sz.detected.gives.clear.gamma.ray.signal.but.could.be.dominated.by.strong.radio.sources.not.diffuse} do show some clear $\gamma$-ray emission, at the level shown. This is consistent with primarily coming from the ``injection region,'' aka the jets themselves or jet shock regions, but could easily hide a comparable level of diffuse-but-centrally-concentrated emission from an ACRH, hence the similar upper limits on pure-diffuse emission in these systems (for quantitative examples, see \S~\ref{sec:younger}, although for some claims of extended emission detections, see \citealt{kushnir:2024.cluster.halo.synchrotron.from.secondary.electrons,pshirkov:2024.claimed.gamma.ray.lstar.galaxy.halo.detection}).  As a result, we cannot rule out hadronic jets, even in the very most luminous sources ($L_{X,\,\rm cool} \sim 10^{45}\,{\rm erg\,s^{-1}}$), as these specific ultra-luminous X-ray cooling cores are precisely those which have strong centrally-concentrated observed $\gamma$-ray emission (take e.g.\ Perseus, where the apparent cooling luminosity is $L_{X,\,\rm cool} \sim 10^{45}\,{\rm erg\,s^{-1}}$, but so is the \textit{observed} $\gamma$-ray luminosity concentrated around just the source NGC 1275 within the cluster).

\subsection{Radio Luminosities \&\ Associated Synchrotron Halos}
\label{sec:radio}

\begin{figure}
	\centering\includegraphics[width=0.99\columnwidth]{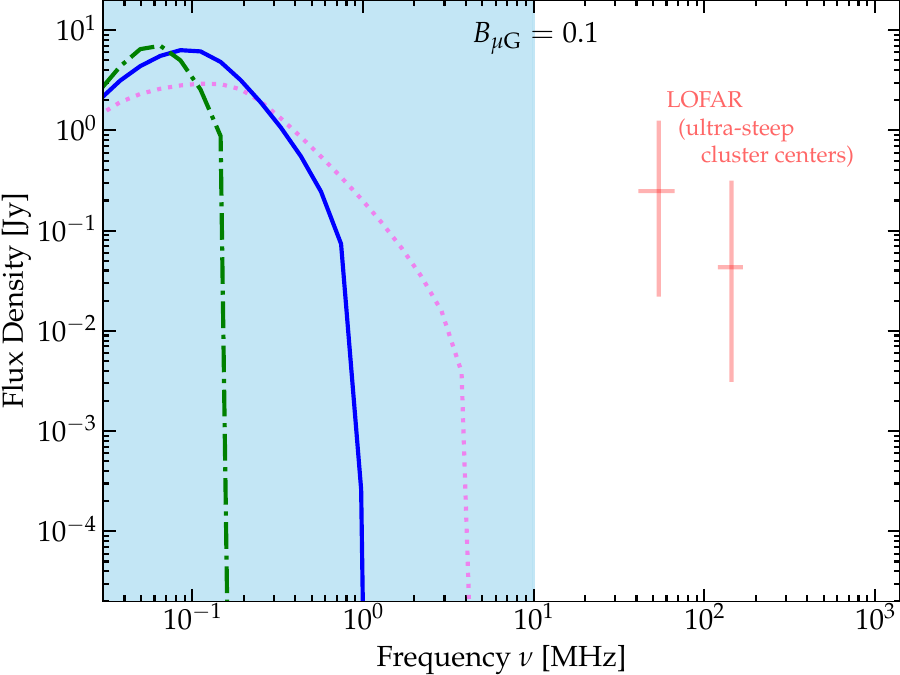} 
	\centering\includegraphics[width=0.99\columnwidth]{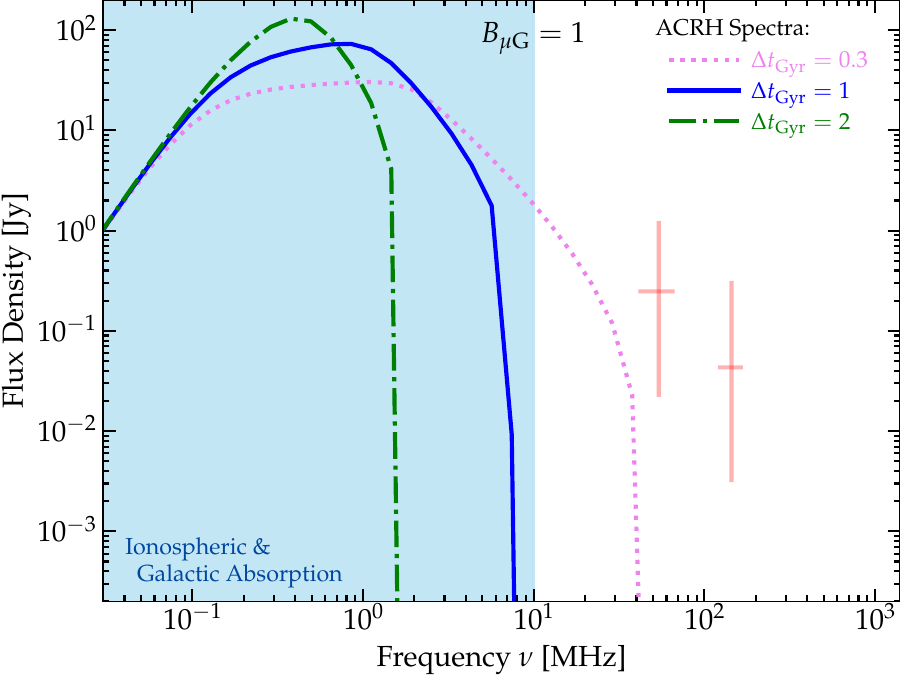} 
	\centering\includegraphics[width=0.99\columnwidth]{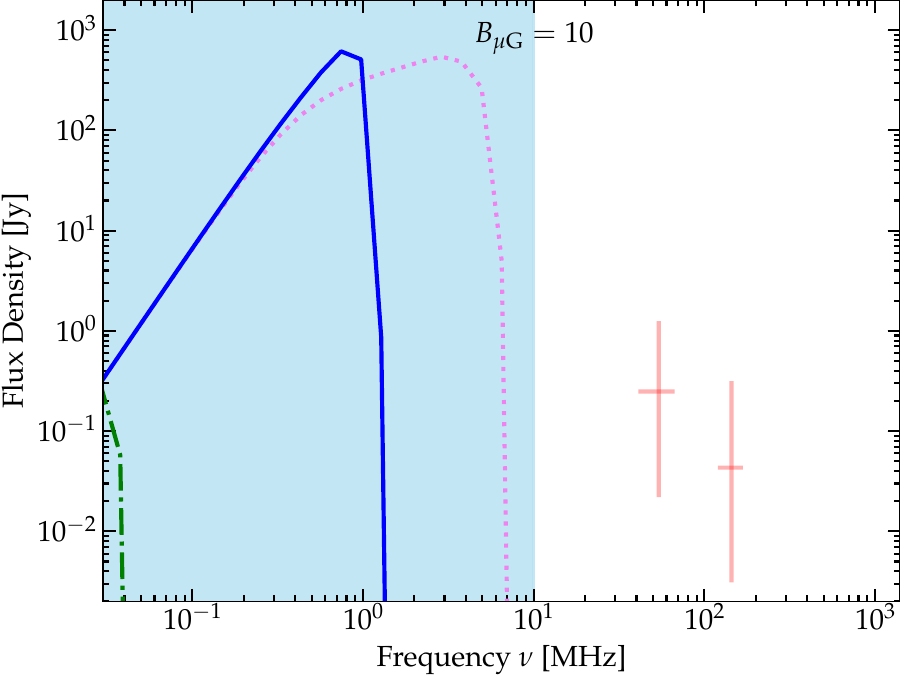} 
	\caption{Predicted ACRH synchrotron spectra as Fig.~\ref{fig:spectrum.allband} (\S~\ref{sec:radio}), for different assumed values of the diffuse, volume-filling magnetic field strength at $\sim 100\,$kpc (where most of the ACRH emission comes from), $B \equiv B_{\rm \mu G}\,{\rm \mu G}$ with $B_{\rm \mu G} \sim 0.1-1$ expected but values as large as $\sim 10$ (the most extreme known injection regions in clusters) shown. Because of losses (stronger at high-$B$) and/or weak-$B$, the synchrotron spectra are truncated at $\sim 10\,$MHz. These are orders-of-magnitude lower than the faintest detected radio sources in the lowest-frequency bands accessible to LOFAR ($\sim 53-110$\,MHz, labeled), and generally well below the frequencies (\textit{shaded}) where the ionosphere and Galaxy ISM (both neglected in the spectral calculations) are opaque. However in \S~\ref{sec:evolution} \&\ \S~\ref{sec:younger} we show that closer (in time and space) to the injection zone ($\Delta t \lesssim 10^{8}\,$yr), these resemble observed ultra-steep sources.
	\label{fig:synch}}
\end{figure}

As discussed in \S~\ref{sec:intro}, \ref{sec:cr.spectrum}, \&\ \ref{sec:hardxray} above, sufficiently close in time-and-space to injection, there is detected synchrotron emission from the sources of the leptonic ACRHs (something we also show explicitly in \S~\ref{sec:Lradio.LX} below) -- indeed, this is our motivation for considering the older ACRHs in the first place. But observed radio sources like blazars in cluster centers are, as discussed above, much more compact in volume and short-lived in time compared to the ACRHs. The observed radio emission necessarily comes from very near the effective injection zone, so the spectral index (and any inferred properties of the ICM like magnetic field strengths) represent those very near-source, young-CR quantities. Our question here is whether the older, diffuse ACRH would be radio-detectable.

Fig.~\ref{fig:synch} shows the predicted spectra and flux densities of the entire (integrated) ACRH at different ages or equivalently within different volumes,\footnote{We convolve the evolved CR spectra over the appropriate synchrotron emissivity as a function of wavelength, accounting for self-absorption \citep[][]{rybicki.lightman:1979.book}, but not foreground atmospheric or Galactic absorption.} for different values of the assumed magnetic field strength at $\sim 100\,$kpc (the extended size of the ACRH) -- a more plausible $\sim 0.1\,{\rm \mu G}$, and a ``strong field'' case of $\sim 1\,{\rm \mu G}$, with the values for the former and latter coming from the compilation of constraints on cluster magnetic field strengths in the \textit{diffuse} gas (e.g.\ far from bright observed synchrotron-emitting arcs or lobes) in \citet{garrington:1991.faraday.depolarization.density.times.magnetic.field.1eminus3.cm3.microG,carilli:2002.cluster.b.fields.obs.collection,rudnick:2003.cluster.rms.from.non.strong.synch.emitting.volume.filling.region.from.polarization.must.be.below.0pt4microG,govoni:2004.B.fields.galaxy.clusters.review,xu:2006.rm.measurement.superclusters.0pt5microG.fields.inferred.could.be.mostly.in.sources,pereztorres:2009.ophiuchus.minihalo.xray.ic.contrib.significant.B.0pt03to0pt3microG,mirakhor:2022.ic.cluster.detection.B.0pt1microG}.\footnote{Note that extrapolation of the observed ISM B-field-density relation ($B \sim 10\,{\rm \mu G}\,(n/200\,{\rm cm^{-3}})^{1/2-2/3}$; \citealt{crutcher:cloud.b.fields,ponnada:fire.magnetic.fields.vs.obs}) gives much lower $B \sim 0.003-0.02\,{\rm \mu G}$ at the gas densities of interest ($n\sim 10^{-3}\,{\rm cm^{-3}}$, observed in clusters at these radii), and upper limits in the CGM of less-massive galaxies at $\sim 100\,$kpc from their centers similarly require $B < 0.2\,{\rm \mu G}$ \citep{prochaska:2019.weak.magnetization.low.Bfield.rm.massive.gal.frb,prochaska:2019.frb.halo.constraints,lan:2020.cgm.b.fields.rm}.} We discuss likely field strengths in more detail in \S~\ref{sec:Lradio.LX} but Fig.~\ref{fig:synch} shows it is unimportant for our conclusions here.

We see that, \textit{independent of $B$}, the loss of high-energy leptons means that the radio spectrum is ultra-soft, with much of the flux at $\lesssim 1\,$MHz. In fact, $B \sim 1\,{\rm \mu G}$ actually \textit{maximizes} the flux above $\sim 1\,$MHz at these ages/radii -- if we go to larger $B \gtrsim B_{\rm cmb}$, then because increasing $B$ makes the loss time of the high-energy CRs even shorter, the $\gtrsim1\,$MHz radio emission at $\gtrsim 10^{8}$\,yr is even more strongly suppressed. This makes the very old leptons that define the ACRHs radio-unobservable, as the ionosphere and Galaxy are opaque below $\lesssim 10$ and $\lesssim 2$\,MHz, respectively. The situation is even more challenging if we recall this is spatially extended, so extremely low surface-brightness diffuse emission. 

In \S~\ref{sec:younger}, we show that we wish to produce detectable emission in the softest bands available e.g.\ LOFAR 53 and 110\,MHz (let alone GHz), then it requires (1) going to very young $\Delta t \lesssim 30\,$Myr and small $R \lesssim 3\,v_{100}\,$kpc (or $\lesssim 10\,$Myr and $\lesssim 1\,$kpc for GHz detections), \textit{and} (2) high-but-not-too-high $B \sim 1-20\,{\rm \mu G}$. This, of course, essentially defines the ``injection zones'' where strong synchrotron emission is actually seen (references in \S~\ref{sec:intro}), and the ACRH-preceding ultra-steep sources detected.  
We show this by showing the faintest detected sources in extremely deep LOFAR and/or GMRT pointings at these frequencies at known steep-spectrum cluster radio sources \citep{osinga:2021.deepest.lofar.cluster.detections.of.radio.halos,cuciti:2021.diffuse.cluster.radio.halo.fluxes.brightness.lofreq.gmrt.steeper.slopes.larger,pasini:2024.lofar.low.freq.radio.relic.detection.tend.to.extremely.steep.slopes}. The lower limit shown corresponds to the best-case models of hypothetical detectability by radio instruments projected through the next decade in \citep{bruno:2023.modeling.best.case.lofar.detectability.galaxy.clusters}. Indeed while \citet{pasini:2022.lofar.radio.survey.bcgs.often.radio.sources.all.clusters.some.bright.radio.at.low.enough.frequencies} argued essentially all BCGs are radio sources at sufficiently low frequencies and depths based on trends from GHz to LOFAR data, the extended ACRH emission predicted here would be well below their limits. 

Changing the detailed CR transport assumptions will quantitatively influence this: for example if CR transport is diffusive, there are ``tails'' of the distribution which extend to large radii even at young ages (and the effective streaming speed $v_{\rm st} \sim \kappa/r$ can become larger at small radii), if CR transport speeds depend strongly on CR energy (e.g.\ $\kappa \propto E_{\rm cr}^{0.5}$, as observed for diffusive LISM CR transport, as compared to the simple streaming/advective models here) then some higher-energy CRs will escape further out (having effectively larger $v_{\rm st}$), and re-acceleration could of course boost or retain a tail of higher-energy CRs. But the point is robust that the specific CRs important for CR-IC in ACRHs, even for the maximal CR-IC case where they contribute significantly to the total X-ray emission, will correspond to extremely low-frequency, low-surface-brightness radio.

Note that ``equipartition'' models calibrated to the local ISM and often applied to synchrotron emission (assuming an LISM-like CR spectrum at high-energies, fixed LISM-like lepton-to-hadron ratio, equal CR proton and magnetic energy densities, uniform volume-filling $B$, etc.) are clearly not applicable here. The CR spectra and lepton-to-hadron ratios in the models are qualitatively distinct. Moreover there is no mechanism enforcing equipartition in these systems and essentially all simulations of CR transport and dynamics in cluster cores and CGM/ICM conditions predict it should not hold \citep{pakmor:2012.mag.field.disk.evol.weak.fx,su:2016.weak.mhd.cond.visc.turbdiff.fx,su:2018.stellar.fb.fails.to.solve.cooling.flow,su:turb.crs.quench,hopkins:cr.mhd.fire2,ji:fire.cr.cgm,ponnada:fire.magnetic.fields.vs.obs,ponnada:2023.fire.synchrotron.profiles,werhahn:2021.electron.spectra.models,martin.alvarez:radiation.crs.galsim.similar.conclusions.fire,martin.alvarez:2023.mhd.cr.sims.synch.maps.similar.emission.regions.conclusions.to.fire.ponnada.papers.but.very.different.methods,dacunha:2024.synthetic.synchrotron.problems.with.obs.interp}, consistent with the fact that it is already known that $B$ is far below equipartition with the thermal pressure (i.e.\ $\beta \gtrsim 100$, with $\beta \sim 1$ in typical clusters already clearly ruled-out by dynamical and Faraday-rotation constraints).

\begin{figure}
	\centering\includegraphics[width=0.99\columnwidth]{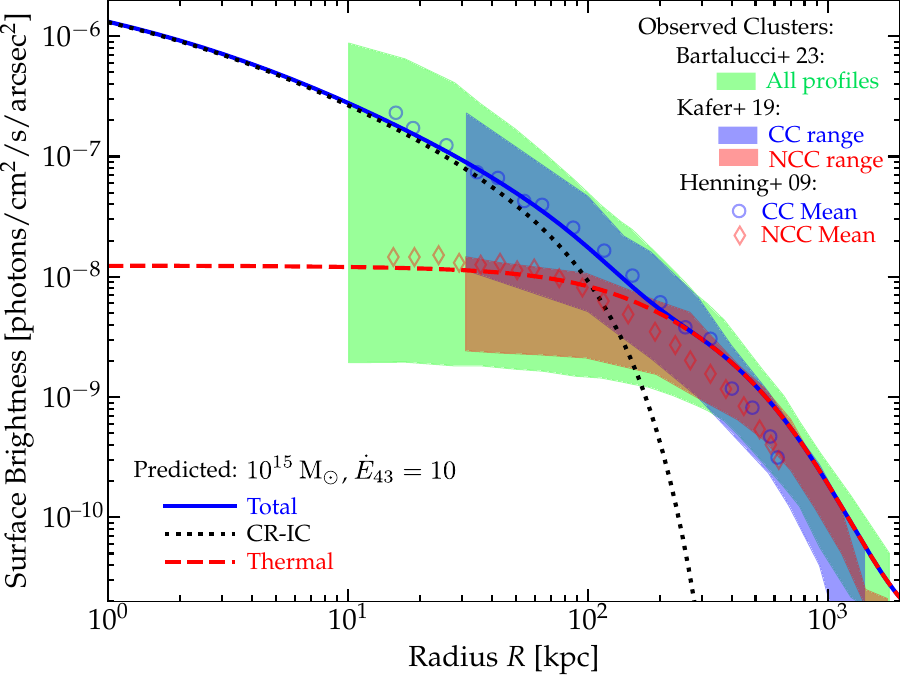} 
	\caption{Projected X-ray radial surface-brightness profile for a toy model of a cluster with a luminous ACRH (as \paperone). We assume a virial mass $M_{\rm vir} = 10^{15}\,M_{\odot}$ cluster whose true gas properties (density $n$, temperature $T$, etc.) follow standard fitted values (\S~\ref{sec:cc.profiles}) to non cool-core (NCC) clusters, with an ACRH added whose profiles assumes the simplest injection-plus-streaming model (\S~\ref{sec:cr.spectrum}, Eq.~\ref{eqn:ecr.r}) with CR injection rate $\dot{E}_{\rm cr,\,\ell} = 10^{44}\,{\rm erg\,s^{-1}}$. We then compute the effective $\sim$\,keV-band surface brightness from thermal emission and from CR-IC. 
	We overplot the distribution of surface brightness profiles from cluster samples, broken into CC and NCC cluster subsamples (labeled), and the mean CC and NCC profile from similar-mass cluster samples (\S~\ref{sec:coolcores}).
	The ``typical'' CC profile at $\lesssim 100\,$kpc is remarkably similar to the profile for a luminous ACRH, and the range from weak-to-strong CCs could be reproduced simply by changing the ACRH luminosity, rather than the ``true'' thermal cooling rate.
	\label{fig:profiles.sb}}
\end{figure}

\section{What if a ``Cooling Core'' is an ACRH?}
\label{sec:coolcores}

From \paperone, as well as \citet{hopkins:2025.crs.inverse.compton.cgm.explain.erosita.soft.xray.halos} and \S~\ref{sec:softxray} \&\ \ref{sec:multi.wavelength}, we see that the primary signature of an ACRH is extended, diffuse soft X-ray emission, with a vaguely isothermal thermal spectrum (resembling multi-temperature gas thermal emission, assuming there is some pre-existing gas present), with a surface brightness profile $S_{X} \propto (R\,v_{\rm st})^{-1}$ extending out to $\sim 100\,$kpc. But these are precisely the characteristic properties of the ``cooling cores'' of CC clusters. 

In this section, we consider in more detail the hypothetical posed in \paperone: 
What if a significant fraction of the apparent ``cooling flow'' (CF) luminosity comes, in fact, from CR-IC from an ACRH, instead of from thermal emission? 

For the sake of a quantitative model giving the most extreme plausible scenario (where essentially \textit{all} of the CF luminosity comes from an ACRH), let us  assume (for now) that the true cluster gas properties are those of a typical NCC cluster. 
Specifically, we take the quasi-universal profile fitting functions from \citet{vikhlinin:temperature.metallicity.profiles,ghirardini:2019.cluster.profiles.compilation.universal.fits}, applied to the data in \citet{mcdonald:2013.cluster.gas.profiles} for $n$ and $T$: 
$n_{\rm true,\,NCC} \sim n_{0}\,[ (r/r_{c})^{-\alpha}\,(1+(r/r_{c})^{2})^{\alpha/2-3\beta}\,(1+(r/r_{s})^{3})^{-\epsilon/3}]^{1/2}$ (with [$n_{0}/{\rm cm^{-3}},\,\alpha,\,\beta,\,\epsilon,\,r_{c}/r_{500},\,r_{s}/r_{500}] \approx [0.0035,\,0.22,\,0.35,\,2.6,\,0.1,\,0.6]$), $T_{\rm true,\,NCC} \sim T_{\rm vir}^{0}\,(1+(r/0.45\,r_{500})^{2})^{-0.3}$ (with $T_{\rm vir}^{0}$ the central virial temperature from \citealt{bryan.norman:1998.mvir.definition}). We then assume the existence of an ACRH (taking $v_{100} \sim 1$), with some leptonic injection rate $\dot{E}_{\rm cr,\,\ell} \sim L_{X,\,\rm IC}$ (i.e.\ most of the CR energy coming out in soft X-ray IC). 
Since we are modeling a radial profile, we are not mocking up mono-age CRs, but rather integrating them in space. To do so, we follow \paperone\ and solve the time-steady Fokker Planck equation for a fixed CR injection rate and (LISM-like) spectrum, CR streaming speed, and background gas density+temperature profile (giving the losses) with fixed $B_{\rm \mu G} = 1$ assumed here (varied below but not important so long as it is not $\gg 10$). But to first approximation, the predicted CR and CR-IC spectra are qualitatively similar to the mono-age spectra in \S~\ref{sec:cr.spectrum}-\ref{sec:softxray} for the equivalent $\Delta t_{\rm Gyr} \sim R/100\,{\rm kpc}$. 

We stress (as discussed below) that we are not actually arguing for this or any other specific density/temperature profile in CCs (and indeed, given the simplicity of the models here, there is no observed cluster where we would expect this to be a ``good fit'' in any rigorous or $\chi^{2}$ sense), but want to illustrate what would happen if the X-rays came primarily from an ACRH. So we specifically choose an extreme toy model case, where there is zero cooling/cold gas or true cool core. Of course, if we assume some ``real'' CC profile (e.g.\ density and temperature profiles fit to the X-rays assuming all or most of the X-ray emission comes from thermal emission), then we can (by construction) reproduce the observations with a weaker ACRH contribution. Our question here is: which CC properties can be qualitatively reproduced by ACRH emission, and what are some associated signatures?

Fig.~\ref{fig:profiles.sb} shows the resulting X-ray surface-brightness profile from such a model, for a $10^{15}\,M_{\odot}$ cluster with $\dot{E}_{\rm cr,\,\ell} \sim 10^{44}$, compared to observed profiles of NCC and strong CC clusters of similar mass \citep{henning:2009.xray.cluster.sb.profiles.obs.vs.sims,kafer:2019.cluster.xray.sb.profiles.by.cc.status,bartalucci:2023.xray.cluster.sb.profiles}.

\subsection{Cluster ``Density,'' ``Temperature,'' and ``Entropy'' Profiles}
\label{sec:cc.profiles}

\begin{figure*}
	\centering\includegraphics[width=0.32\textwidth]{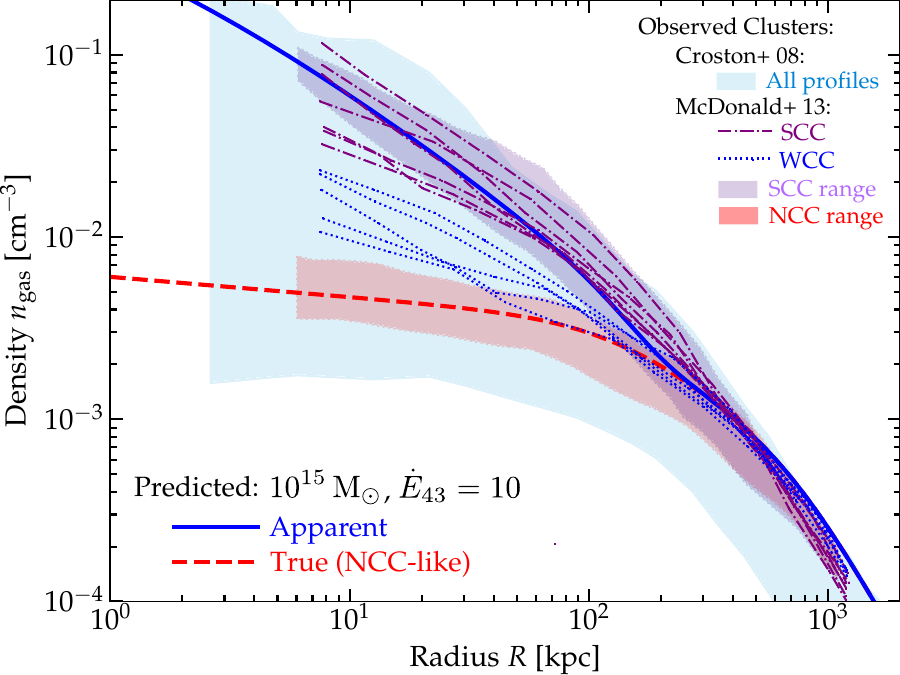} 
	\centering\includegraphics[width=0.32\textwidth]{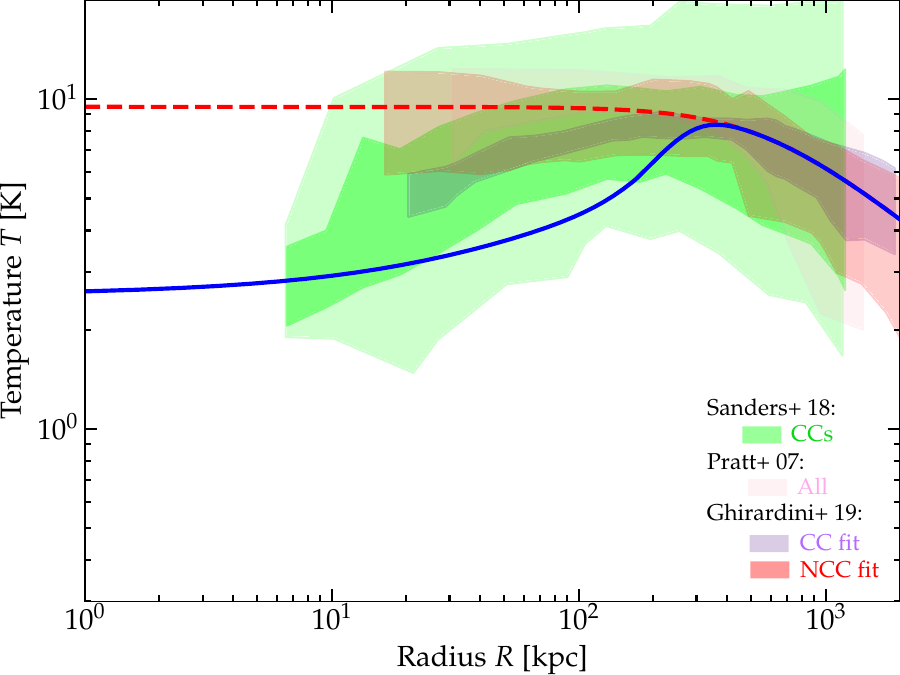} 
	\centering\includegraphics[width=0.32\textwidth]{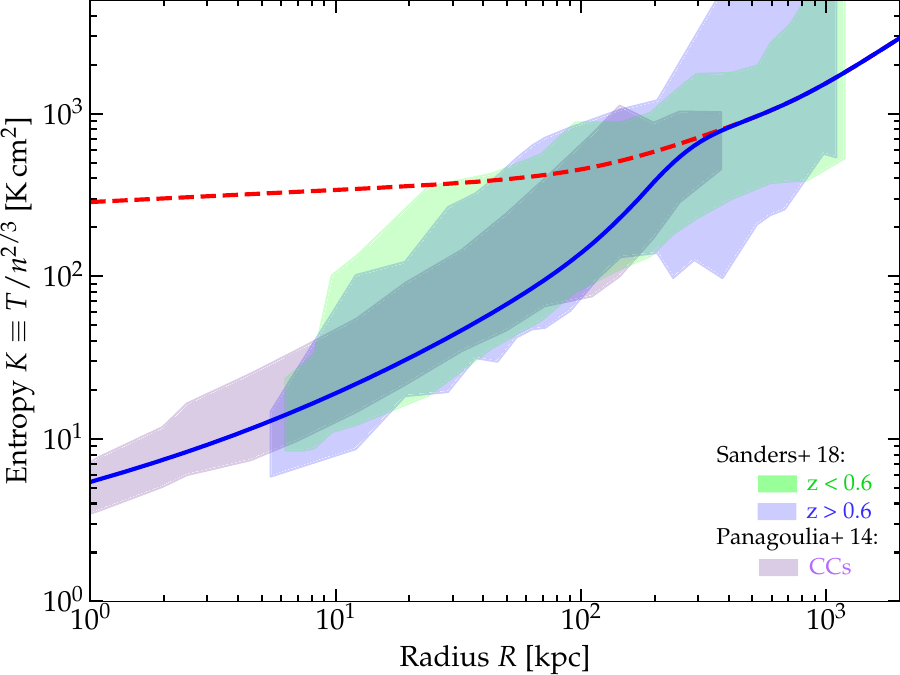} 
	\centering\includegraphics[width=0.32\textwidth]{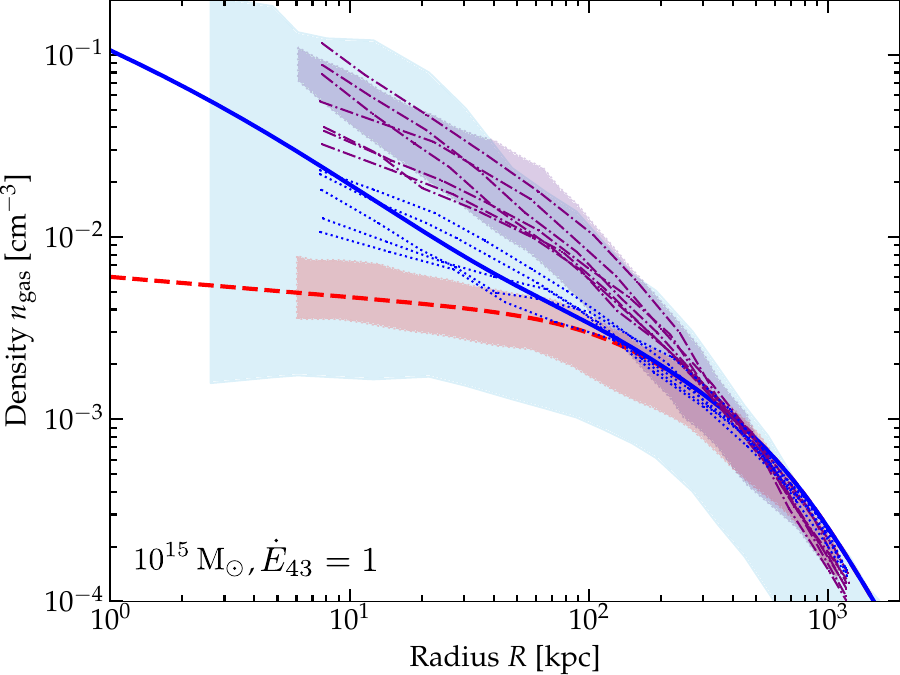} 
	\centering\includegraphics[width=0.32\textwidth]{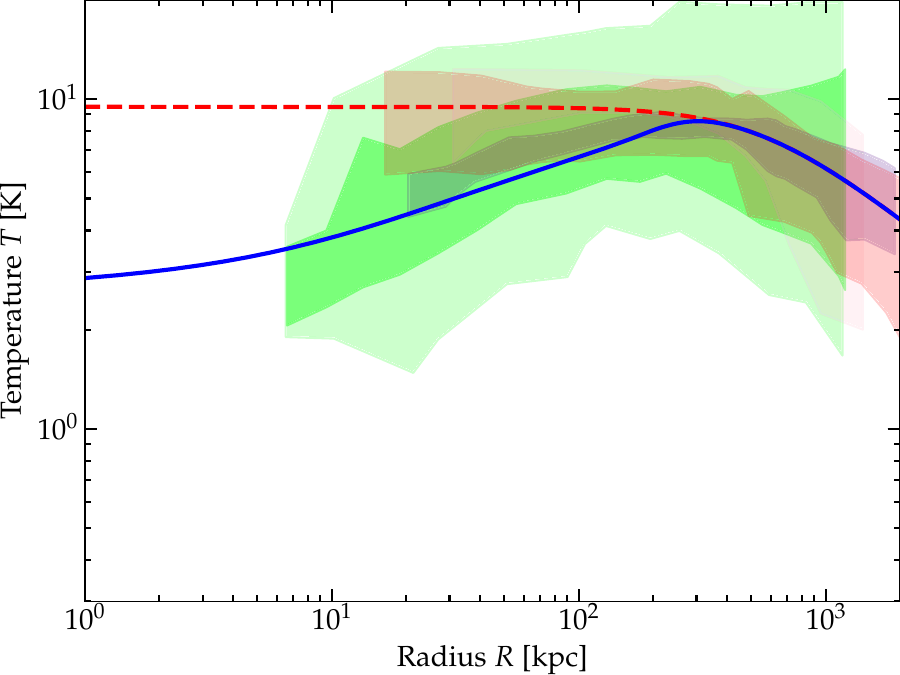} 
	\centering\includegraphics[width=0.32\textwidth]{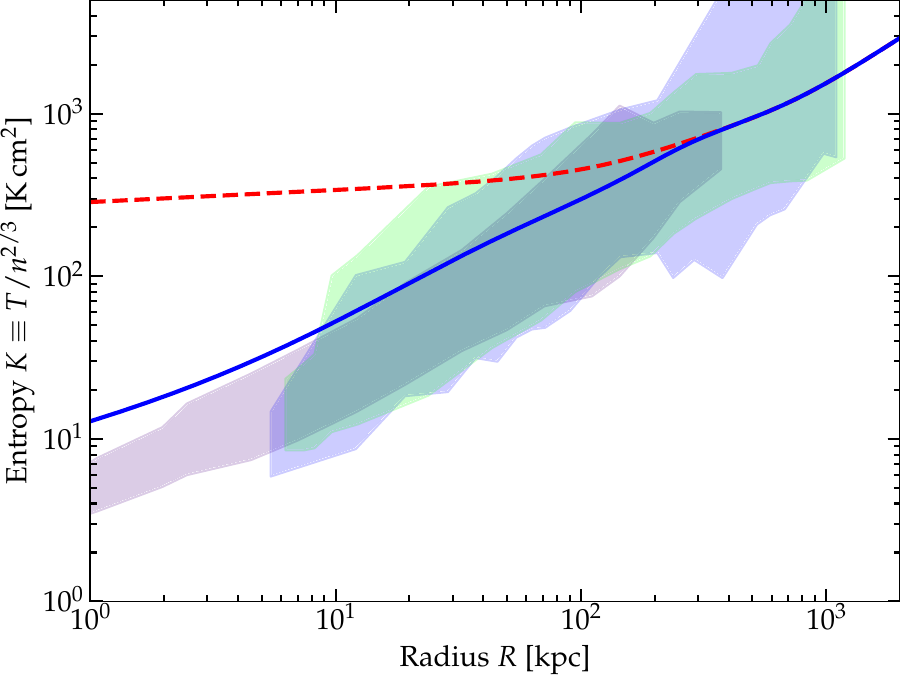} 
	\centering\includegraphics[width=0.32\textwidth]{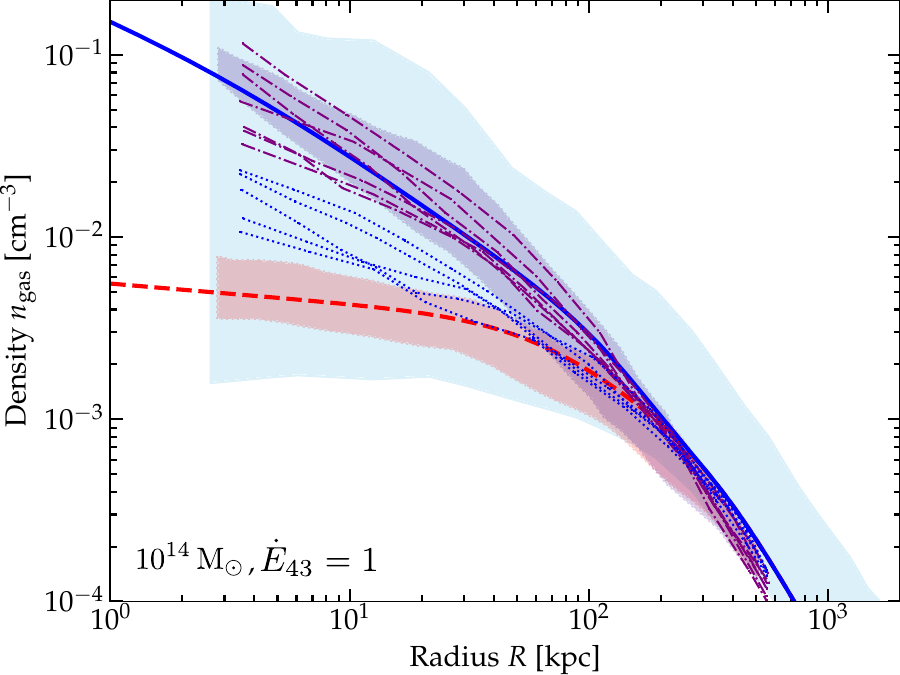} 
	\centering\includegraphics[width=0.32\textwidth]{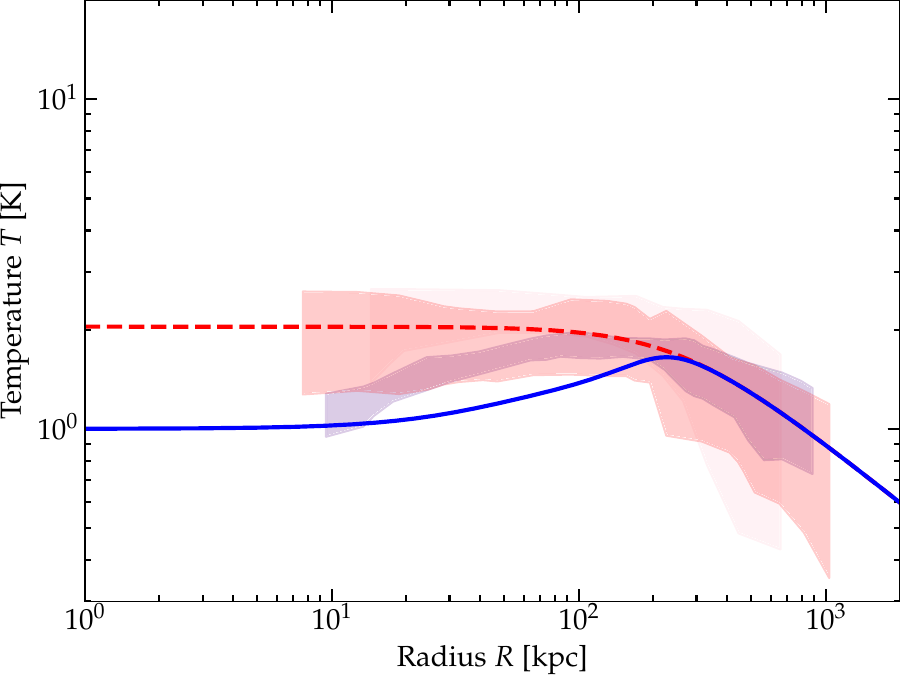} 
	\centering\includegraphics[width=0.32\textwidth]{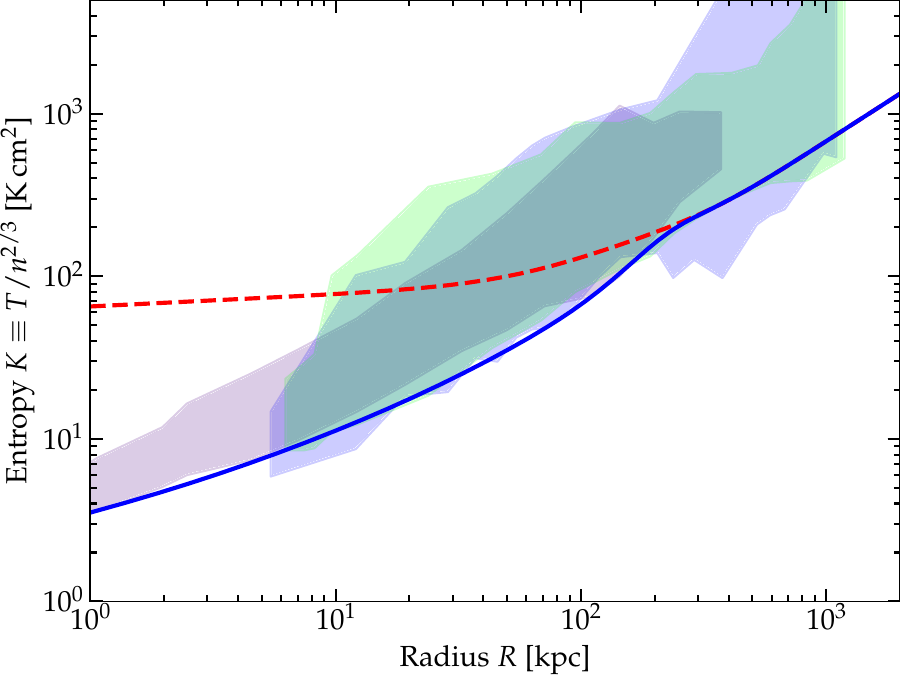} 
	\centering\includegraphics[width=0.32\textwidth]{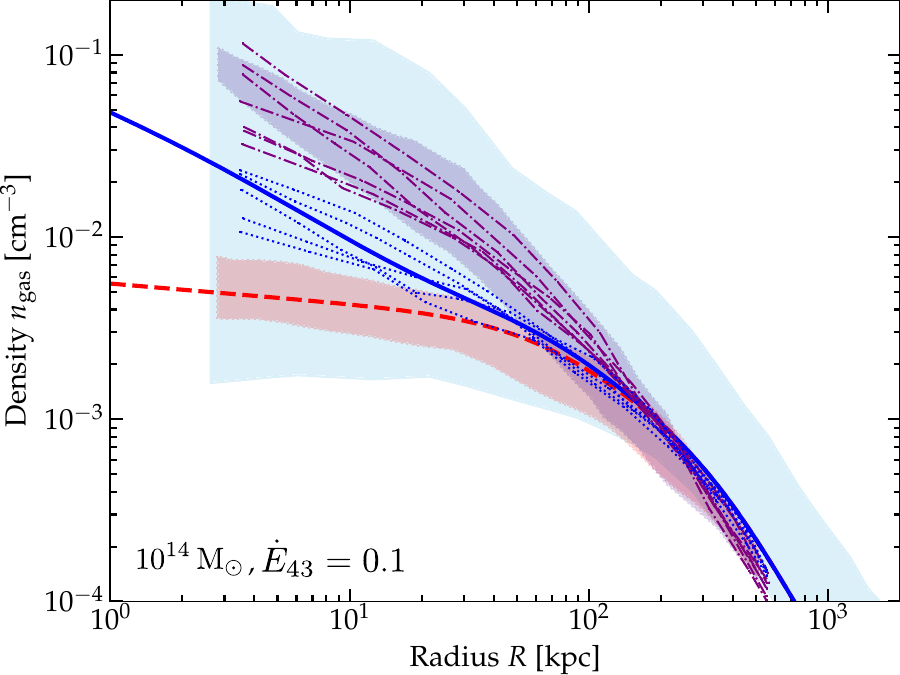} 
	\centering\includegraphics[width=0.32\textwidth]{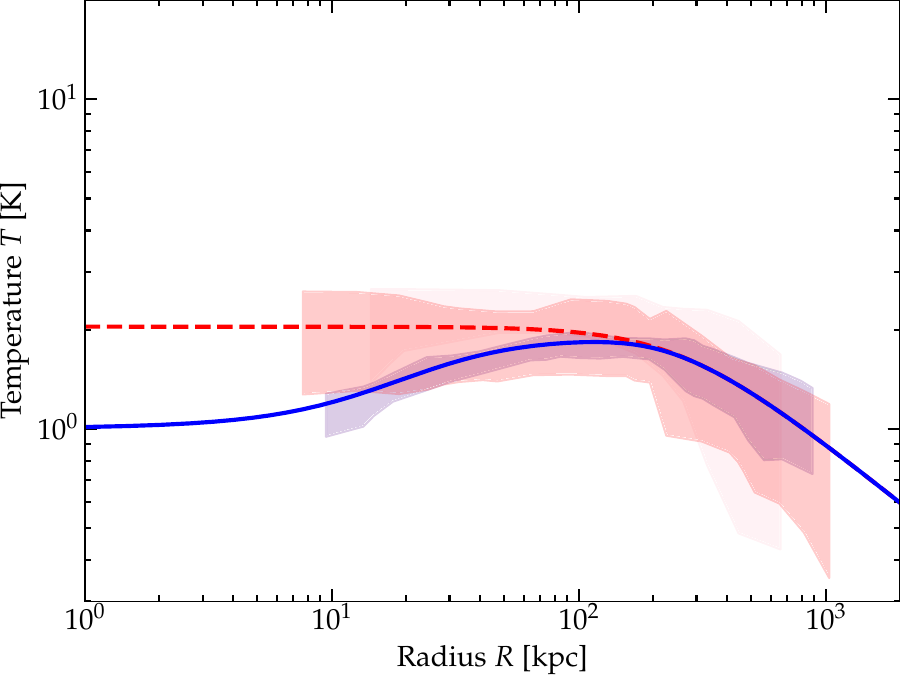} 
	\centering\includegraphics[width=0.32\textwidth]{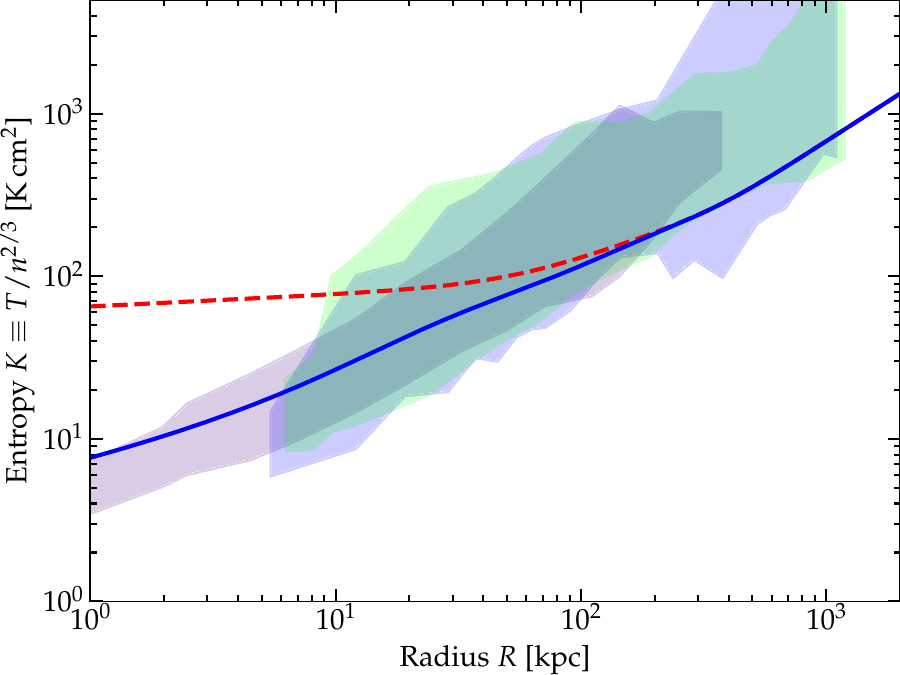} 
	\caption{Radial profiles of toy model clusters with luminous ACRHs. We assume an underlying true NCC-like density ($n$) and temperature ($T$) per \S~\ref{sec:cc.profiles}, in a cluster of the given $M_{\rm vir}$ (rows; labeled), with an ACRH as in Fig.~\ref{fig:profiles.sb} with CR injection rate $\dot{E}_{\rm cr,\,\ell}$ shown (to model the extreme case of central emission dominated by the ACRH). We then compute the ``apparent'' $n$, $T$, and entropy $K$ that would be inferred fitting the combined IC+thermal spectra to single-temperature models (\S~\ref{sec:apparent}).  We overplot observed cool-core (CC) and NCC cluster/group profiles around each mass range (labeled; \S~\ref{sec:cc.profiles}). The central ACRH emission mimics the characteristic $n$, $T$, and $K$ profiles of CC profiles for plausible parameters.
	\label{fig:profiles}}
\end{figure*}

In Fig.~\ref{fig:profiles}, we show the apparent cluster density, temperature, and entropy ($K = k_{B}T/n^{2/3}$) profiles that result from the combination of NCC+ACRH, using the simple approximate expressions for the apparent spectral temperature and density which would be inferred (\S~\ref{sec:apparent}) and expressions for $e_{\rm cr,\,\ell}(R)$ and self-consistent calculation of $f_{\rm loss}$ from \S~\ref{sec:radial}. Again we emphasize that detailed modeling of any single observation requires full spectral synthesis and fitting with the specific instrument, data reduction, and fitting package and assumptions -- these are simply heuristic estimates to provide a rough guideline of the effects. 

\paperone\ showed a single example of this: here we consider different halo masses (virial; using \citealt{bryan.norman:1998.mvir.definition} definitions) $M_{\rm vir} \sim 10^{14}-10^{15}\,M_{\odot}$, and a range of plausible $\dot{E}_{43} \sim (0.1,\,1,\,10)$ increasing with $M_{\rm vir}$. We discuss the energetics below but these are all easily within the range of $\dot{E}_{\rm cr,\,\ell}$ inferred from \textit{observed} AGN production in these halos.

We see this is sufficient for the apparent $T$, $n$, and $K$ in the central regions to be dominated by the ACRH. Within the ACRH-dominated region, the apparent values scale roughly as:
\begin{align}
\label{eqn:Tapp.cluster} T_{\rm app} &\sim (0.5-5)\,{\rm keV}\,(1+z) \propto R^{0} \, , \\
\label{eqn:napp.cluster} n_{\rm app} &\sim 0.0014\,{\rm cm^{-3}} \frac{L_{\rm IC,\,42}^{1/2}}{R_{100}\,v_{\rm st,\,100}^{1/2}} \propto R^{-1} \, ,\\ 
\label{eqn:Kapp.cluster} K_{\rm app} &\sim 170\,{\rm keV\,cm^{2}} \frac{R_{100}^{2/3}\,v_{\rm st,\,100}^{1/3}}{L_{\rm IC,\,42}^{1/3} (1+z)^{1/3}} \propto R^{2/3} \ .
\end{align}
We compare these to typical CC cluster profiles, inferred from observations \textit{assuming} all of the emission comes from thermal (the usual assumption). 
In density: we compile 
\citet{croston:2008.cluster.density.profiles} showing their full range (skyblue); and
show individual profiles from \citet{mcdonald:2013.cluster.gas.profiles}, labeled by WCC/SCC, as well as their clear CC range ($K_{0}<30$, indigo) and clear NCC ($K_{0}>150$; red) range.
In temperature: we compile profiles from 
\citet{sanders:2018.cluster.density.entropy.temperature.profiles.redshift.samples} for CC clusters (not groups), showing the full-sample and $1\sigma$ range.for clusters (lime) at $z<0.6$ (using $z>0.6$ makes little difference);
\citet{pratt:2007.cluster.temperature.profiles}, showing the full range (pink), scaled to the halo mass assumed; and 
\citet{ghirardini:2019.cluster.profiles.compilation.universal.fits} scaling their range of  ``universal fits'' to each CC ($K_{0}<30$; indigo) and NCC ($K_{0}>30$; red), scaled to the halo mass assumed.
In entropy, we compile profiles from 
\citet{panagoulia:2014.cluster.entropy.profile.compilation} for CCs (indigo); and 
\citet{sanders:2018.cluster.density.entropy.temperature.profiles.redshift.samples} at $z>0.6$ (blue) and $z<0.6$ (lime). 
Note these full-sample ranges are generally consistent with other compilations in the literature \citep[e.g.][]{vikhlinin:2006.cluster.compilation.luminosities.gas.fraction.compilation,vikhlinin:2009.cluster.profiles.compilation,pratt:2022.cluster.density.profiles.also.need.to.excise.cores.to.get.clean.lx.t.mass.relations,sayers:2023.cluster.pressure.profile.mass.redshift.dependence}, but we stress that (1) there are significant (factor $\sim 2$) differences in the inner profiles for many \textit{individual} systems, based on the fit assumptions, model priors, spatial resolution, instrument used, etc.; and (2) some samples claiming an extremely narrow range of profiles have been limited by small sample size and strict selection criteria. 

Recall, this is a toy model with a single order-of-magnitude parameter, not a fit. Given that (and the flexibility one could introduce by actually trying to fit some $\dot{E}_{\rm cr,\,\ell}$ and $v_{100}(R)$ to the observed surface brightness profiles), the agreement is rather remarkable, suggesting the radial profiles in CCs are plausibly consistent with coming from CR-IC. Their quasi-universal profiles, in the CR-IC scenario, stem simply from the fact that CR streaming and/or advection imprints an approximately $e_{\rm CR} \propto R^{-2}$ profile (for approximately-constant $v_{\rm st}$) while the background CMB is uniform. Per \paperone, we note that allowing for more cooling gas (both multi-phase cold gas in the cool core, and some decrease of the true X-ray-emitting volume-filling gas temperature $T$ in Fig.~\ref{fig:profiles}) is perfectly allowed and even expected, given that CR-dominated halos are known to promote the formation, longevity, and stability of cooler gas \citep{ji:fire.cr.cgm,hopkins:2020.cr.outflows.to.mpc.scales,ji:20.virial.shocks.suppressed.cr.dominated.halos,hopkins:2020.cr.transport.model.fx.galform,butsky:2020.cr.fx.thermal.instab.cgm,weber:2025.cr.thermal.instab.cgm.fx.dept.transport.like.butsky.study} subject to global buoyant dynamics \citep{hopkins:2020.cr.outflows.to.mpc.scales}. But it has no real effect on our conclusions or discussion here.

In contrast, in the ``standard'' cooling flow (CF) model interpretation, there is no simple physical reason why CF profiles should be quasi-universal nor should have the specific profile shapes they have. Indeed these radial profiles are famously \textit{not} what is predicted by un-regulated CFs \citep[e.g.][]{allen:2000.predicted.properties.xray.cooling.flows.not.there.missing.column.and.cold.gas,fabian:2002.classical.cooling.flow.problem.obs.definition.missing.lum.profile.of.mdot.vs.r.wrong,eckert:2021.agn.feedback.galaxy.groups.review.challenges.theory.obs.producing.realistic.coolcores,braspenning:2024.flamingo.simple.xray.modeling.sims.clusters.dont.reproduce.zdrops.other.cc.features}, so explaining them requires some AGN feedback re-heating or re-mixing the gas, but again the models (including the authors' own in this area; e.g.\ \citealt{hopkins:groups.qso,hopkins:groups.ell,2009ApJ...699..348S,sharma.2012:thermal.instability.precipitation.coolingflows,parrish:turbulence.w.conduction.regulates.cooling.flows,su:turb.crs.quench,su:2021.agn.jet.params.vs.quenching,su:2023.jet.quenching.criteria.vs.halo.mass.fire}) provide no unique shape.

\subsection{The Classical-Spectral Cooling Flow Problem}
\label{sec:cf.problem}

\begin{figure}
	\centering\includegraphics[width=0.99\columnwidth]{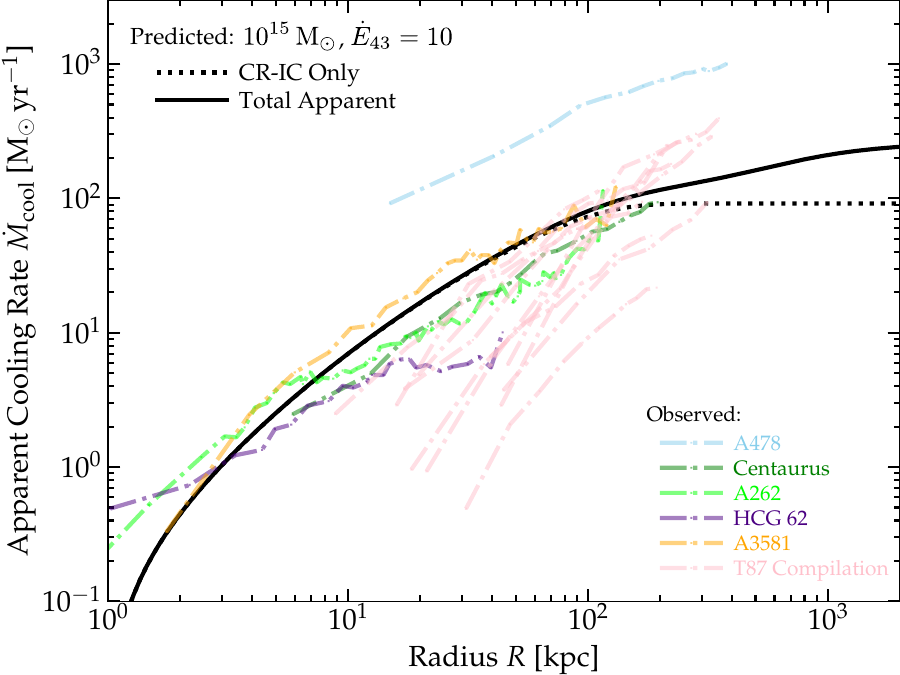} 
	\caption{Radial profile of the apparent classical cooling/mass-deposition rate $\dot{M}_{\rm cool}(<R) \equiv \int_{0}^{r} (2\mu\,m_{p}/5\,k\,T)\,{\rm d} L_{\rm X,\,cool}(r)$ which would be inferred interior to each radius in the example ACRH-boosted cluster in Figs.~\ref{fig:profiles.sb} \&\ \ref{fig:profiles}. We compare various observed profiles including the compilation in \citet[][T87]{thomas:1987.cluster.mass.deposition.rates.coolingflows.compilation}. The characteristic shape of the profile within the ``cooling radius'' arises naturally from CR-IC, and explains why $\dot{M}_{\rm cool}$ rises with $R$, while a true cooling flow should have $\dot{M}_{\rm cool}\sim$\,constant. The ``classical cooling flow problem'' is explained immediately by the fact that this apparent cooling luminosity does not actually come exclusively from cooling, but has important contributions from CR-IC (\S~\ref{sec:cf.problem}).
	\label{fig:mdot.profile}}
\end{figure}

From \S~\ref{sec:cc.profiles} and \paperone, we see that assuming the central emission comes from CR-IC immediately \textit{predicts} that the cluster should \textit{look} like it is a rapidly cooling multi-temperature medium, with inferred/apparent cooling time $t^{\rm app}_{\rm cool} \sim n\,k_{B}\,T_{\rm app} / \epsilon \ll t_{\rm Hubble}$ at $\lesssim 100\,$kpc, with a strong soft X-ray continuum emission -- i.e.\ a SCC. 
But since this primarily comes from CR-IC, and not actual thermal gas cooling, there can be a much smaller reservoir of \textit{actual} cold/cooling gas at $\ll$\,keV temperatures evident in lines or other thermal signatures (e.g.\ molecular/atomic or UV/optical), or galaxy star formation rates \citep{rafferty:2008.cluster.sfr.vs.xray.properties.no.corr.sfr.mdot.expected.coolcore.only.see.recent.sf.evidence.in.small.subsample.with.weak.radio.weak.cavity.strong.lx.central.aligned.coolcore.young.cavities}. But this is \textit{precisely} the ``classical versus spectral'' or simply ``classical'' cooling flow problem \citep{fabian:2002.classical.cooling.flow.problem.obs.definition.missing.lum.profile.of.mdot.vs.r.wrong,peterson:2003.cluster.cooling.flow.problem.missing.lowtemperature.gas.looks.multiphase.but.truncated.kev,hudson:2010.cool.core.cluster.review.central.properties.temperature.entropy.coolingtime.definition.coolcore.basic.scalings.size.luminosity.mdot}. 

This arises trivially in the CR-IC interpretation. In the ``standard'' interpretation, reconciling these is a famously challenging problem, generally involving AGN feedback to balance the cooling losses. But as we discuss below, doing this in a way that preserves the appearance of a CC (\textit{if} the emission is purely thermal) involves considerable fine-tuning and is theoretically rather challenging, because one must rather precisely offset the cooling at the observed rate, but more importantly in the same gas at the same time and place (otherwise one would simply get cold gas raining out of a hot medium; \citealt{eckert:2021.agn.feedback.galaxy.groups.review.challenges.theory.obs.producing.realistic.coolcores,altamura:2023.large.volume.sims.independent.of.free.parameters.struggle.to.make.realistic.cc.clusters}).

Quantitatively, Fig.~\ref{fig:mdot.profile} illustrates one aspect of this (similar to the argument in \paperone), plotting the apparent or inferred mass deposition rate $\dot{M}_{X,\,\rm cool}$ which would be measured, assuming the observed X-ray is all thermal/cooling emission, versus radius (for the toy models in \S~\ref{sec:cc.profiles}). Roughly speaking, in traditional cooling flow models, the inferred mass deposition rate scales with the cooling luminosity as 
\begin{align}
L_{\rm cool,\,apparent}(<R) \sim \frac{5}{2}\frac{\dot{M}_{\rm dep,\,apparent}(<R)}{\mu m_{p}}\,k T\ . 
\end{align}
So for a typical profile in the IC-dominated region, we predict $\dot{M}_{\rm dep,\,apparent} \propto L_{\rm X,\,IC}(<R) \propto R^{1}$. This is more or less exactly what is observed in classical cooling flow model fits to real clusters, which we compile and compare in Fig.~\ref{fig:mdot.profile} (compiled from \citealt{thomas:1987.cluster.mass.deposition.rates.coolingflows.compilation,white:1994.a478.cluster.cooling.flow.mass.deposition.profile.very.bright.cc,sanders:2010.cc.profiles.highres.deposition.rate.scalings}, but these are very similar in shape to other compilations with different instruments or analyses or other clusters observed later, compare \citealt{thomas:1986.elliptical.small.halo.xray.coolingflow.mass.deposition.rates,allen:1993.a478.cluster.cooling.flow.mass.deposition.rates,allen:1994.centaurus.cooling.flow.mass.deposition.rate,bohringer:2005.super.bright.chandra.coolingflow.mass.deposition.fits}). Indeed this approximate $\dot{M} \propto R$ behavior appears to be ``universal'' \citep{fabian:1994.cluster.cooling.flows.review} in clusters, and more importantly, directly contradicts the default assumption in the most standard classical cooling flow model, that $\dot{M}$ should be independent of $R$ (i.e.\ the system should be in steady-state), as well as of course directly contradicting constraints on the actual mass deposition rate (which defines the ``modern'' cooling flow problem, that said mass is not, in fact, being deposited). 

This also immediately provides a simple explanation for the apparent diversity of core profiles at fixed halo mass. For example, although there is a very tight correlation between cluster mass, effective temperature, and \textit{total} cluster X-ray luminosity integrated out to the virial radius or $R_{500}$, and moreover clusters exhibit nearly-universal profiles of $n$, $T$, $K$ down to the cooling radius, within the apparent ``cool core radius'' the scatter is much larger (see references in \S~\ref{sec:cc.profiles}, and \citealt{zhang:2008.cluster.profiles.lensing.xray.good.down.to.0pt2.r500.inner.makes.scatter.much.larger.biases.cosmological.measurements,simet:2017.weak.lensing.xray.cosmology.masses.agreement.but.large.radii,pratt:2022.cluster.density.profiles.also.need.to.excise.cores.to.get.clean.lx.t.mass.relations}). This also manifests in e.g.\ the factor of $\sim 100$ observed spread in $L_{X,\,\rm cool}$ at fixed $T_{\rm vir}$ or $L_{\rm X,\,500}$ (compared to $\sim 0.2$\,dex scatter in the cool-core-excised $L_{X,\,500}-T_{\rm vir}$ relation; \citealt{peres:no.cooling.flow.mass.corr,mittal:2011.xray.cluster.scalings.includes.rcool.lcool}). Given identical outer boundary conditions, in the standard cooling interpretation this spread is surprising \citep{ettori.brighenti:2008.cluster.core.cooling.simple.models}. But if $L_{X,\,\rm cool}$ can be boosted by CR-IC from a central AGN or outburst, which then decays, we naturally expect such a large variance.

\subsection{SZ ``Pressure Deficits''}
\label{sec:sz} 

\begin{figure}
	\centering\includegraphics[width=1.02\columnwidth]{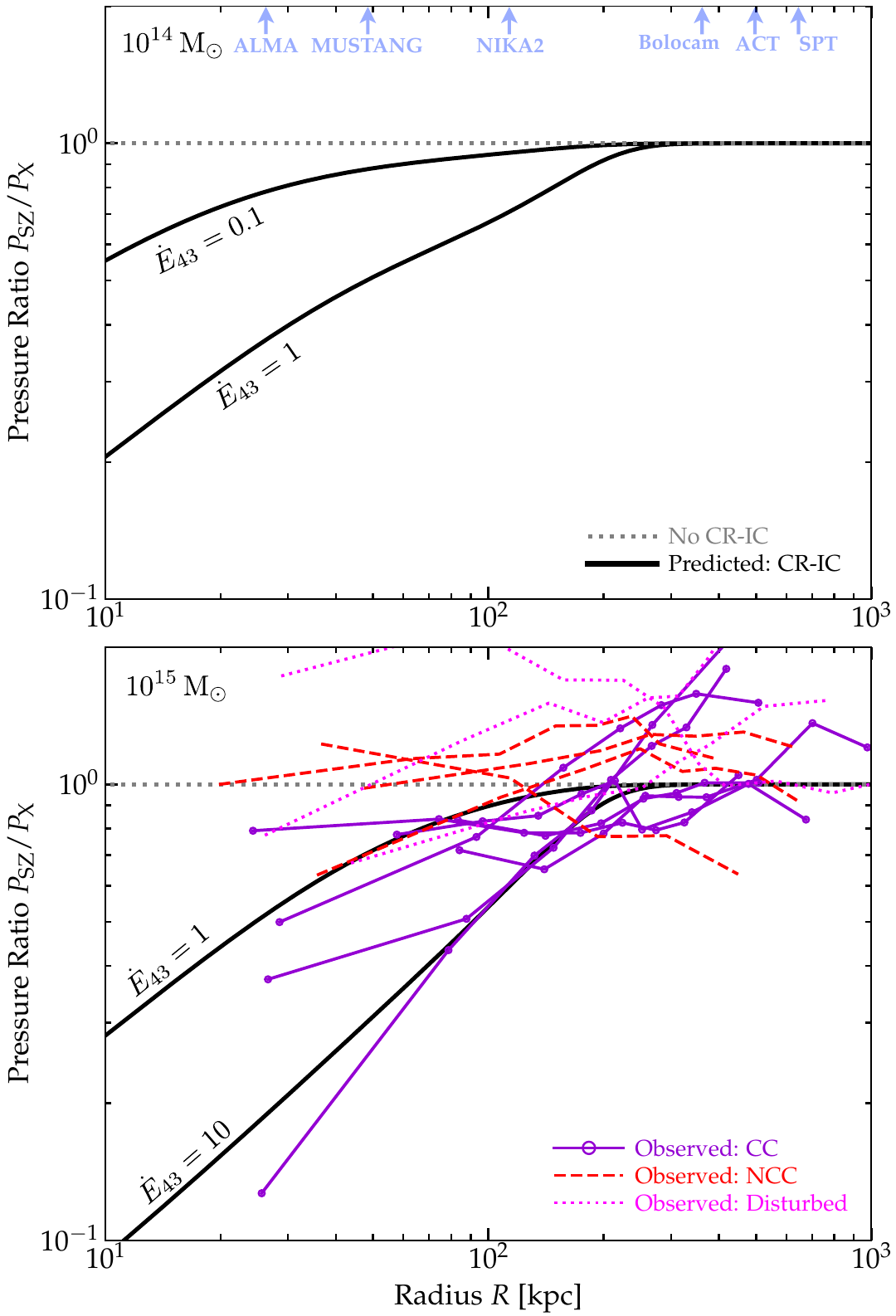} 
	\caption{Calculated ratio of the true gas pressure $P_{\rm true} = n_{\rm true}\,k_{B} T_{\rm true}$ to the \textit{apparent} (X-ray inferred) gas pressure $P_{\rm X} \approx P_{\rm eff} = n_{\rm eff}\,k_{B}\,T_{\rm eff}$ for the same toy-model profiles shown in Fig.~\ref{fig:profiles}. 
	Since SZ should trace the true pressure $P_{\rm SZ} \approx P_{\rm true}$, we plot the observed ratio $P_{\rm SZ}/P_{\rm X} \approx P_{\rm true} / P_{\rm eff}$, for CC clusters and NCC clusters with high-resolution SZ observations probing the central regions (\S~\ref{sec:sz}; \citealt{romero:2017.cluster.pressure.profiles.highres.sz.xray.cool.cores.show.central.pressure.deficit}). Arrows at \textit{top} show spatial resolution (FWHM at median sample redshift) of different SZ instruments, for reference. At large radii there is no effect of the ACRH; at smaller radii the ACRH IC emission leads to excess X-ray surface brightness which causes an over-estimate of $n$ and hence $P_{\rm X} > P_{\rm SZ}$. Remarkably, this toy-model example appears very similar to the median behavior seen in CC clusters, where $\sim 90\%$ of observed CC systems exhibit an SZ ``pressure deficit'' in their central regions, while almost no NCC clusters exhibit such a deficit.
	\label{fig:sz}}
\end{figure}

\begin{figure}
	\centering\includegraphics[width=1.02\columnwidth]{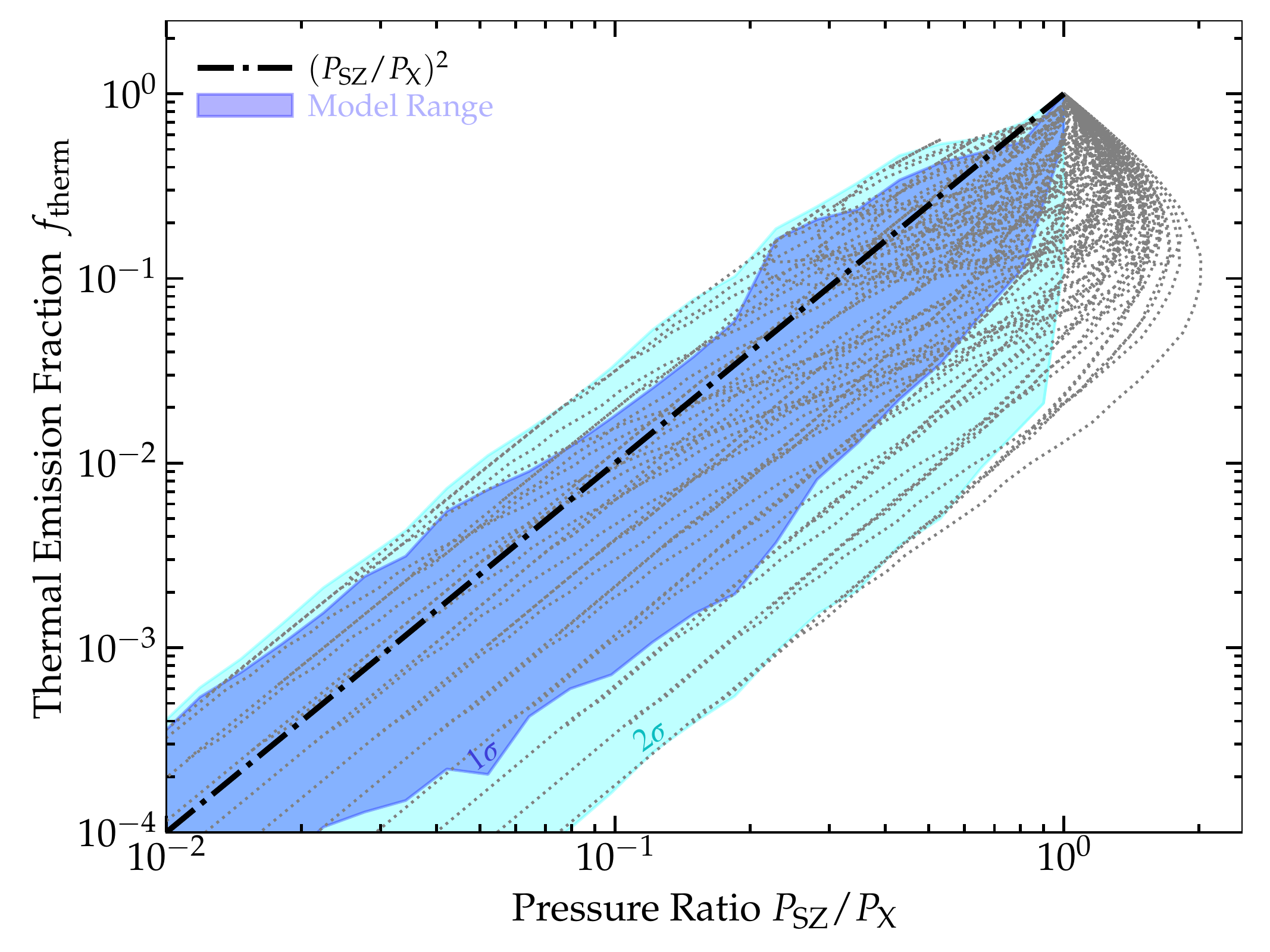} 
	\caption{Ratio of SZ-to-X-ray inferred pressure $P_{\rm SZ}/P_{\rm X}$ (as Fig.~\ref{fig:sz}), versus fraction of the X-ray emissivity (integrated $0.1-10\,$keV here) coming from true thermal emission $f_{\rm thermal} = \epsilon_{\rm thermal}/\epsilon_{\rm total}$. 
	We construct several hundred models per \S~\ref{sec:sz} (\textit{lines}), arbitrarily varying cluster mass; density, temperature, metallicity, and magnetic field profiles; CR transport speeds and injection rates; whether we assume an underlying NCC/WCC/SCC; and whether we assume all CR heating is re-radiated in thermal X-rays. Shaded regions show the $1-2\sigma$ inclusion regions at each $P_{\rm SZ}/P_{\rm X}<1$. There is a broad range of $f_{\rm therm}$, primarily driven by different temperatures in the models ($T_{\rm X}/T_{\rm true}$ varying), and temperature under-estimation can even lead to $P_{\rm SZ} \gtrsim P_{\rm X}$ by a modest amount, but $P_{\rm SZ} < P_{\rm X}$ is a robust sign of $f_{\rm therm} \ll 1$ with approximate scaling $f_{\rm therm} \propto (P_{\rm SZ}/P_{\rm X})^{2}$.
	\label{fig:sz.fullstats}}
\end{figure}

If the profiles in Fig.~\ref{fig:profiles} are correct, there is an immediate and robust test of this proposed in \paperone\ and followed-up on in \citet{silich:2025.cr.ic.tests.in.zw.3146.cluster} -- a unique and clear signature of the CR-IC CC scenario -- in the thermal Sunyaev-Zeldovich (tSZ) effect at mm wavelengths. Recall, tSZ comes from \textit{weak} Compton scattering of CMB photons by \textit{non-relativistic}, i.e.\ thermal, electrons in the cluster. The CRs do not contribute to the tSZ signal because they do not generate a small spectral shift in the CMB photons (the tSZ signature in mm) but, as we showed, scatter up to soft X-rays. Thus the tSZ signal still cleanly measures the true \textit{thermal} pressure $n\,k_{B}\,T$ of the cluster at a given radius -- i.e.\ $P_{\rm SZ} \approx P_{\rm thermal,\,true}$. 

However, if the CR-IC scenario above is true, the X-ray inferred pressure $P_{\rm X} \approx P_{\rm therm,\,apparent} = n_{\rm app}\,k_{B} T_{\rm app}$ will be biased (generally to higher values) because of the central surface brightness from CR-IC (the ``apparent'' CC). Using the profiles in Fig.~\ref{fig:profiles}, we can immediately plot the SZ-to-X-ray pressure ratio $P_{\rm SZ}/P_{\rm X} = P_{\rm therm,\,true} / P_{\rm therm,\,app}$ in Fig.~\ref{fig:sz}. As expected, we obtain $P_{\rm SZ} < P_{\rm X}$ in the central regions. By definition if the CC luminosity were dominated by thermal emission, we would have $P_{\rm SZ}=P_{\rm X}$.

\subsubsection{Models Compared to Observations}

Fig.~\ref{fig:sz} compares all available clusters for which both spatially-resolved X-ray and high-resolution (since we need to explore the central $\lesssim 100\,$kpc) tSZ data exist, as presented in \citet{romero:2017.cluster.pressure.profiles.highres.sz.xray.cool.cores.show.central.pressure.deficit}. The tSZ comes primarily from MUSTANG, as e.g.\ Bolocam/ACT/SPT/Planck can only resolve radii a large fraction of $\sim R_{500}$ where the prediction is $P_{\rm SZ} = P_{\rm X}$ independent of CR-IC. Remarkably, the tSZ data appear to exhibit a strikingly similar trend to the CR-IC prediction. The median $P_{\rm SZ}/P_{\rm X}$ for CC clusters begins to fall to $<1$ at $R \lesssim 100\,$kpc (the approximate ``cooling radii'') and falls to nearly $\sim 0.1$ (an order-of-magnitude decrement) at the smallest MUSTANG-resolved radii ($\sim 10$\,kpc). Indeed, in the sample of clusters in \citet{romero:2017.cluster.pressure.profiles.highres.sz.xray.cool.cores.show.central.pressure.deficit}, 6 of the 7 relaxed CC clusters show positive evidence for a central SZ decrement -- with the only two CC (relaxed or unrelaxed) without evidence for a decrement being either highly disturbed or having ambiguous centering (see discussion in \citealt{romero:2017.cluster.pressure.profiles.highres.sz.xray.cool.cores.show.central.pressure.deficit}). In contrast, for NCC clusters, $P_{\rm SZ} \sim P_{\rm X}$ appears to remain true in the centers, and only 1 of 4 relaxed-NCC clusters (or 1 of 7 including disturbed NCC) in \citet{romero:2017.cluster.pressure.profiles.highres.sz.xray.cool.cores.show.central.pressure.deficit} appear to show any evidence for a deficit (and this case would likely be classified, based on its other X-ray properties, as a ``hidden CC'' by the more aggressive criteria used in \citealt{fabian:2023.hidden.cooling.flows.note.includes.many.others.classified.as.cooling.flow.like.5044}).
There are hints of similar deficits in a couple other CCs with MUSTANG or ALMA data published since, but these lack the same de-projected $P_{\rm SZ}/P_{\rm X}$ comparison \citep[see][]{romero:2020.cluster.pressure.profile.sz.xr.zw3146.xr.only.large.radii.agrees.there,kitayama:2020.phoenix.cluster.alma.sz.hint.of.deficit.when.compare.but.no.comparison.in.paper}. 

In \citet{silich:2025.cr.ic.tests.in.zw.3146.cluster}, archival MUSTANG2 and Chandra data were combined to test this hypothesis in one previously-unstudied but clean (uncontaminated) CC which otherwise shows similar signatures for CR-IC in its profiles, ZwCl 3146, and the authors found that there appears to be a similar decrease in $P_{\rm SZ}/P_{\rm X}$ within the CC radius (from $P_{\rm SZ}/P_{\rm X} \sim 1$ at all $R \gtrsim 100\,$kpc to $P_{\rm SZ}/P_{\rm X} \lesssim 0.5$ at $R \lesssim 40\,$kpc), very similar to the $\dot{E}_{43}\sim1$ prediction in Fig.~\ref{fig:sz}. There, the authors survey any other possible observational systematics and/or physical effects which could give rise to a suppressed $P_{\rm SZ}/P_{\rm X}$ (e.g.\ AGN contamination, triaxial halos viewed along special sightlines, Helium sedimentation), arguing that none of these could explain more than a few percent decrease in $P_{\rm SX}/P_{\rm X}$ (and most are just as likely to bias the observations to $P_{\rm SX}>P_{\rm X}$). 

Thus the SZ data appears to provide unique positive (albeit still tentative) evidence of CR-IC contributing significantly in \textit{most} apparent CC clusters. We stress there is no explanation for these deficits in the ``standard'' thermal CC interpretation. AGN feedback, for example, can modify $P_{\rm therm}$, but if the X-ray pressure is recovered correctly, there should be no discrepancy between $P_{\rm X}$ and $P_{\rm SZ}$ \citep[see also discussion in][]{silich:2025.cr.ic.tests.in.zw.3146.cluster}. 

We emphasize that extremely high-resolution SZ data is essential here: previous studies claiming agreement between SZ and X-ray pressure profiles, from e.g.\ Planck or Bolocam or ACT or SPT data \citep{bonamente:2012.sz.xr.pressure.profiles.agree.large.radii.close.to.r500.integrated,sayers:2013.bolocam.pressure.profiles.sz,planck:2013.pressure.profiles.sz.good.agree.xray.large.radii,bulbul:2019.sz.cluster.less.selection.biased.to.cc.also.agree.outer.y.param.but.large.radii,romero:2020.cluster.pressure.profile.sz.xr.zw3146.xr.only.large.radii.agrees.there,pointecouteau:2021.planck.act.pressure.profiles.large.radii.clusters.reasonable} only resolve radii well outside the radii with predicted deviations in Fig.~\ref{fig:sz}. This is true even for the nearest clusters, like Virgo and Perseus \citep{planck:2016.sz.virgo.signal.only.large.radii.constrained.but.those.close.to.xray}. The only instruments capable of this, at present, are MUSTANG2, ALMA, and NIKA2. These generally require some synthesis of larger-beam data to construct such profiles (owing to the large-scale modes being filtered out by higher-resolution instruments): it is also important when fitting the joint profiles to allow the de-convolved SZ pressure to vary independently in bins, as in the non-parametric studies in \citet{romero:2017.cluster.pressure.profiles.highres.sz.xray.cool.cores.show.central.pressure.deficit,romero:2020.cluster.pressure.profile.sz.xr.zw3146.xr.only.large.radii.agrees.there}, as opposed to fitting parametric functional forms to the combined data (e.g.\ gNFW profiles as fit to hybrid data in \citealt{dimascolo:2019.sz.alma.cluster.xray.pressure.comparison.but.param.study.fit.to.large.r.data.anchor}, where most of the statistical constraining power for the ``small scale slope'' of $P_{\rm SZ}$ comes from data at $>100\,$kpc).

\subsubsection{How Well Can we Infer The Emission Ratio from the Pressure Ratio?}

The predictions in Fig.~\ref{fig:sz} are for one specific family of toy model, akin to the example in \paperone. More generally, Fig.~\ref{fig:sz.fullstats} shows how $P_{\rm SZ}/P_{\rm X}$, or (similarly) the observed SZ $y$ parameter ($y\equiv(\sigma_{T}/m_{e} c^{2})\int n_{e} k_{B} T_{e} {\rm d} \ell$) in projection varies with the fraction of the X-ray (here $0.1-10\,$keV integrated) emission (or intensity $I_{X}$ from a given spherical radius) coming from true thermal emission, $f_{\rm therm} \epsilon_{X,\,\rm therm} / \epsilon_{X,\,\rm tot}$. For this, we can integrate models like those in Figs.~\ref{fig:cric.temp.vs.age} \&\ \ref{fig:profiles} out to a given radius, but vary different input parameters: $\dot{E}_{43}$ (from $0.1-100$), $M_{\rm vir}$ (from $10^{14}-10^{16}\,M_{\odot}$), the density and temperature profiles (over the range shown in shaded regions in Fig.~\ref{fig:profiles}), $B_{\rm \mu G}$ (from $0.1-10$ or assuming a fixed $\beta = 100$), the assumed metallicity (from $0.1-2\,Z_{\odot}$ or following the profiles in \S~\ref{sec:z.drops} renormalized by a systematic factor of $0.5,\,1,\,2$), and the CR diffusivity (from $\sim 10^{28}-10^{30}\,{\rm cm^{2}\,s^{-1}}$) and streaming speed (from $\sim 30-300\,{\rm km\,s^{-1}}$). For each, we also vary whether or not we assume all of the CR thermalized heating rate (primarily Coulomb, here) is re-radiated in thermal emission at the local virial temperature. 
We sample a few choices along each of these model variations (spaced roughly log-uniform in each) and then measure the ratios at each $R$ for each model, and plot the range of all of these in Fig.~\ref{fig:sz.fullstats}. Note that for perfectly self-similar profiles, where one recovers $T$ exactly, and the contribution of CR-IC to emissivity is independent of radius, then the prediction is exactly 
\begin{align}
\label{eqn:ftherm.sz.xr} f_{\rm therm} \sim \frac{\epsilon_{X,\,\rm therm}}{\epsilon_{X,\,\rm tot}} \sim \left( \frac{P_{\rm SZ}}{P_{\rm X}} \right)^{2}
\end{align}
(because $I \propto n^{2}$ for thermal emission, while $P \propto n$). 

Despite the complexity and non-self-similarity of these models, this simple scaling provides a reasonably good approximation to the predictions. Most of the variable choices above actually have very weak effects on the scaling predicted in Fig.~\ref{fig:sz.fullstats}. The primary source of the $\sim 0.5\,$dex $1\sigma$ spread in $f_{\rm therm}(P_{\rm SZ}/P_{\rm X})$ is the temperature structure (since Eq.~\ref{eqn:ftherm.sz.xr} is primarily based on how the inferred \textit{density} scales with the intensity), but $P_{\rm SX}/P_{\rm X}$ also depends on the ratio $T_{\rm true}/T_{\rm X}^{\rm app}$ (i.e.\ true-to-X-ray-inferred temperature). But this is largely systematic with the cluster mass, since that scales strongly with $T_{\rm X,\,true}$. The trend in Fig.~\ref{fig:sz.fullstats} is that higher-mass clusters, with larger $T_{\rm X,\,true} \sim T_{\rm vir}$, have lower values of $f_{\rm therm}$ at a given $P_{\rm SZ}/P_{\rm X}$. 

These temperature effects also explain the tail of models which predict $P_{\rm SZ} > P_{\rm X}$. As shown in \paperone\ (and also seen, but much more weakly, in the models in Fig.~\ref{fig:sz}), in massive (hotter) clusters, at $\sim 100\,$kpc, where $P_{\rm SZ}$ first approaches $P_{\rm X}$, it can ``overshoot'', depending on how we approximate $T_{\rm X}^{\rm app}$ (the X-ray fit temperature), as this can still be biased somewhat low in the X-ray by slightly more than the density $n_{\rm gas}$ is biased high (since the inferred density scales as the square root of the emissivity), so for a small range of radii around this scale one predicts $P_{\rm SZ} \gtrsim P_{\rm X}$, albeit always by a small factor (usually a factor $\sim 1.1-1.5$ at most). As noted above the X-ray temperature modeling is highly sensitive to fitting methods at this level, so the prediction in this regime should be taken with some caution. However we note this to emphasize that while $P_{\rm SZ} < P_{\rm X}$ \textit{always}, in these models, corresponds to $f_{\rm therm} \ll 1$, it is \textit{not} true that $P_{\rm SZ} \sim P_{\rm X}$ always rules out $f_{\rm therm} < 1$ (indeed we see a range as large as $f_{\rm therm} \sim 0.01-1$ at $P_{\rm SX} \approx P_{\rm X}$). 


\subsubsection{Distinction from ``Non-Thermal Pressure'' Effects}

Briefly, we also stress that this is completely different from the \textit{indirect} modifications to SZ from CR pressure being a significant fraction of the total pressure, which have been discussed in the literature \citep[e.g.][]{colafrancesco:2004.crs.as.pressure.heating.in.cluster.centers,su:2018.stellar.fb.fails.to.solve.cooling.flow}. Generally, what has been discussed is the possibility that CR feedback from AGN could either (1) provide most of the total pressure in the cluster center, so that the thermal pressure is much less than what is required for virial equilibrium, and/or (2) heat and redistribute gas, modifying the thermal pressure directly. But these both still assume X-ray emission is primarily thermal and so predicts $P_{\rm SZ} = P_{\rm X}$, independent of the CR pressure. And as we discuss below, the CR-IC scenario we discuss does not necessarily imply either of these -- indeed it is perfectly plausible that most of the total pressure still comes from gas and CR heating/forces on gas are negligible even if CR-IC dominates the CC emission.

\subsection{Suppressed Central Metallicities (Slow-Rising Profiles, Plateaus, and Drops)}
\label{sec:z.drops}

\begin{figure*}
	\centering\includegraphics[width=0.99\textwidth]{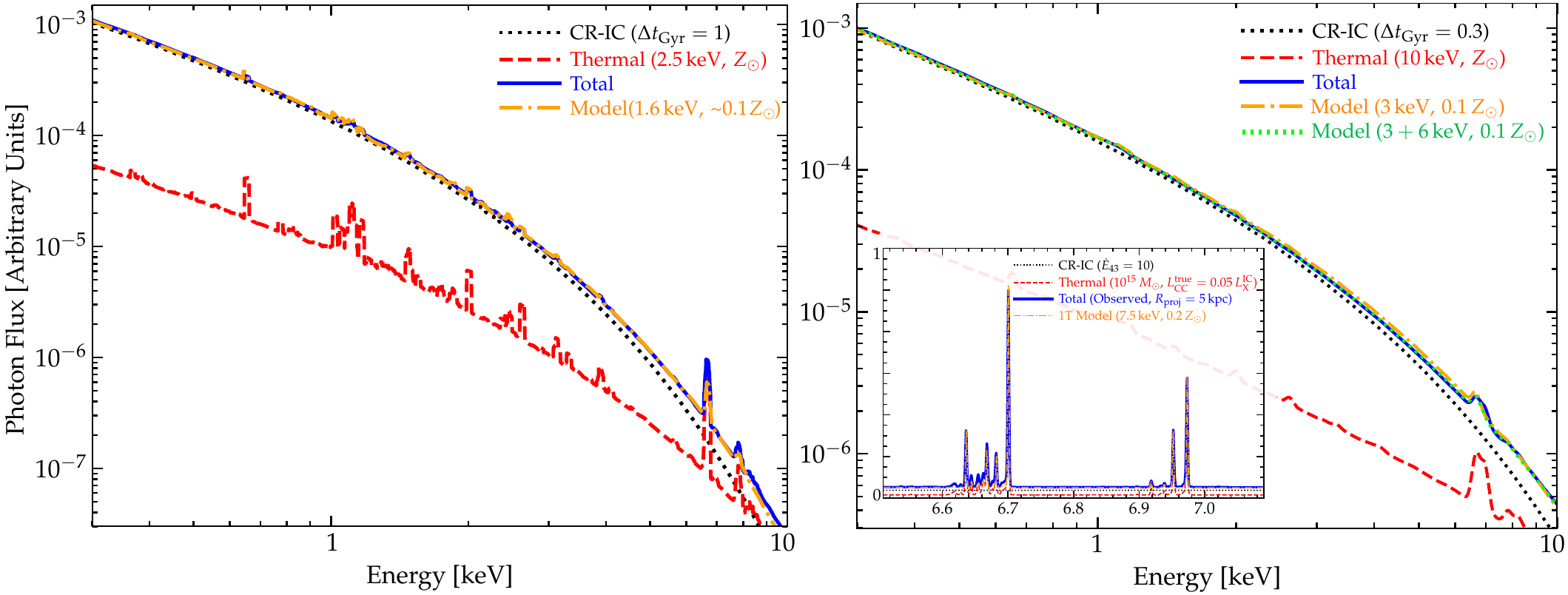} \vspace{-0.2cm}
	\caption{Heuristic illustration of how CR-IC can influence X-ray spectral estimates of metallicity $Z$ in CC centers ($R \lesssim 30\,$kpc; \S~\ref{sec:z.drops}). 
	We show CR-IC spectra at a single $\Delta t_{\rm Gyr}$ as Fig.~\ref{fig:spectrum.xr} plus APEC thermal spectra of single-phase, Solar-abundance ($Z_{\odot}$) gas with given $T$, ignoring CR heating/excitation (for group, \textit{left}, and cluster \textit{right}, like $T$), for CCD-like spectral resolution.
	These are chosen to represent the most extreme cases where the CR-IC emission is both much more luminous than the thermal emission, and has a very different effective temperature. 
	We compare a different single-phase APEC model with lower $T$ and $Z\sim0.1\,Z_{\odot}$ -- while not a fit, this illustrates how CR-IC can dilute the apparent $Z$. 
	At \textit{right} where some deviations from a single-temperature (1T) fit are evident, we also show a two-temperature (2T) fit (with a $\sim 2\%$-different absorption correction $N_{H}$), which minimizes residuals to much smaller than in observed CCs like Virgo/Centaurus/Perseus. 
	\textit{Inset:} Spectrum from one of the cluster models in \S~\ref{sec:cc.profiles}-\ref{sec:sz}: a $10^{15}\,M_{\odot}$ cluster, with $\dot{E}_{43}=10$, with a diffuse hot gas profiles as Fig.~\ref{fig:profiles}, plus very weak CF (from the hot gas $T$ down to $0.5\,$keV at each $r$) normalized so the true cooling luminosity $L_{\rm X,\,cool}^{\rm true}$ is just $\sim 5\%$ of the CR-IC luminosity. We integrate the projected spectrum from CR-IC and multi-phase APEC thermal emission in an aperture enclosing $<5\,$kpc, and zoom in to show the Fe-K line complex at $1\,$eV resolution. We compare a single-temperature fit. Separating CR-IC from mixed thermal contributions with lower $Z$ is challenging, though X-ray spectroscopy can constrain the effective CR-IC and thermal temperatures at energies and radii where the two are comparable.\vspace{-0.2cm}
	\label{fig:Z.spectral.demo}}
\end{figure*}

\begin{figure*}
	\centering\includegraphics[width=0.49\textwidth]{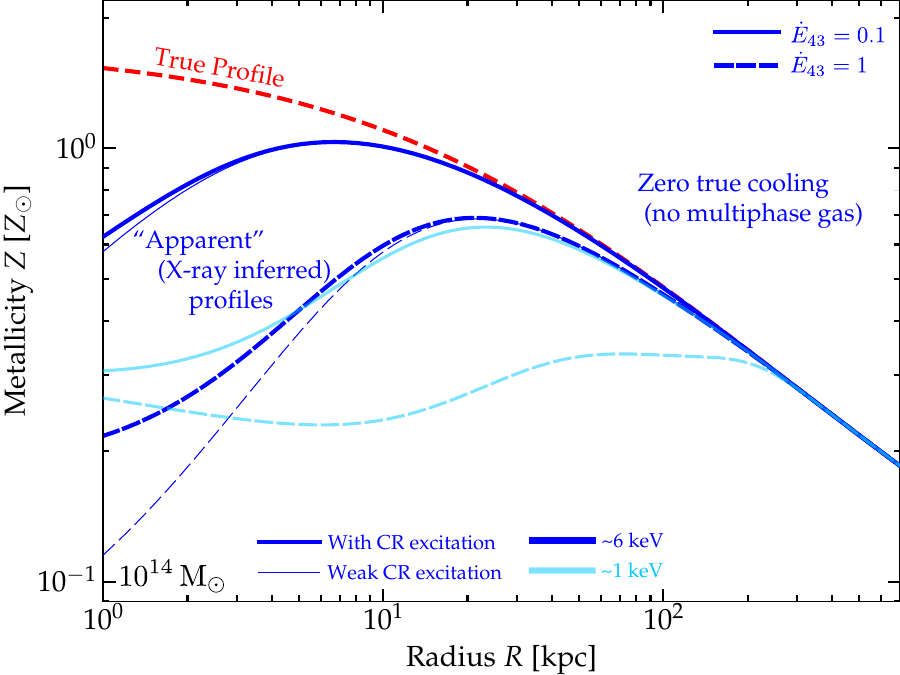} 
	\centering\includegraphics[width=0.49\textwidth]{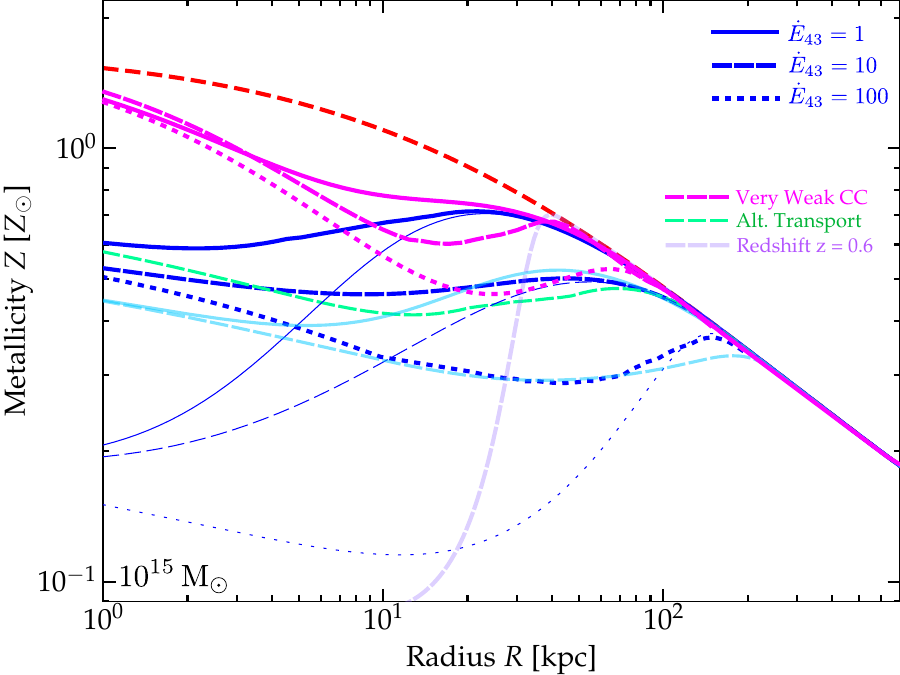} \vspace{-0.2cm}
	\caption{True and apparent \textit{X-ray observed} metallicity profiles for the cluster toy models in Fig.~\ref{fig:profiles} -- i.e.\ cluster models with \textit{no} true CC/CF (the most extreme possible case) -- assuming a simple weighted single-temperature fit following \S~\ref{sec:z.drops}. 
	Because CR-IC primarily boosts the X-ray continuum thermal-like emission (while only weakly boosting line emission) it dilutes the lines, suppressing the X-ray inferred $Z_{\rm X}$ relative to the true value.
	Assuming some true $Z(R)$ (labeled), we estimate the inferred $Z_{\rm X}$ with a toy model for dilution of line equivalent widths at a given wavelength (e.g.\ $\sim 6-7$\,keV for metallicities driven by Fe-K, or $\sim1$\,keV for Fe-L lines), including or excluding a toy model for CR excitation of lines by ionization and Coulomb heating, with different CR injection rates $\dot{E}$ (labeled). We also show one example of a different CR transport model or redshift, and for the high-mass halo, a couple models from \paperone\ which include a very weak true CC (true cooling luminosity $L_{X,\,{\rm cool}} = 10^{43}\,{\rm erg\,s^{-1}}$).  The detailed shapes are sensitive to all of these details and also to how we weight and model the $Z_{\rm X}$ ``estimator'' but all show qualitatively similar behavior, with stronger central suppression of $Z_{\rm X}$ for more luminous CR-IC contributions. This gives rise to slow-rising profiles or plateaus in the central $Z$, or even (in the most extreme cases, if we ignore CR excitation and heating) central ``drops'' in apparent abundance within a few to $\sim 30-100$\,kpc (detectable with few-arcsecond or better resolution X-ray spectroscopy).\vspace{-0.2cm}
	\label{fig:Z.profiles.model.only}}
\end{figure*}

\begin{figure*}
	\centering\includegraphics[width=0.49\textwidth]{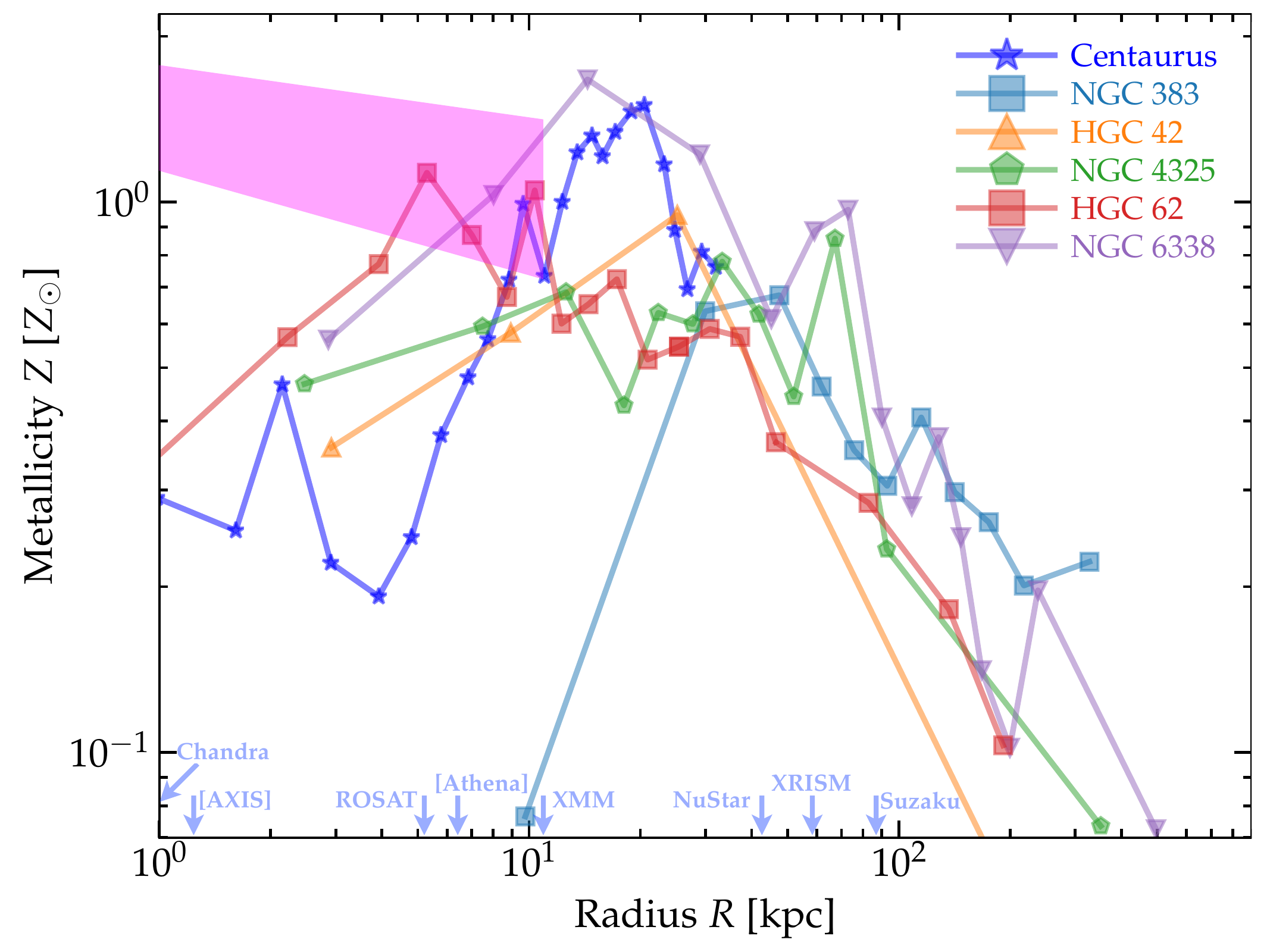} 
	\centering\includegraphics[width=0.49\textwidth]{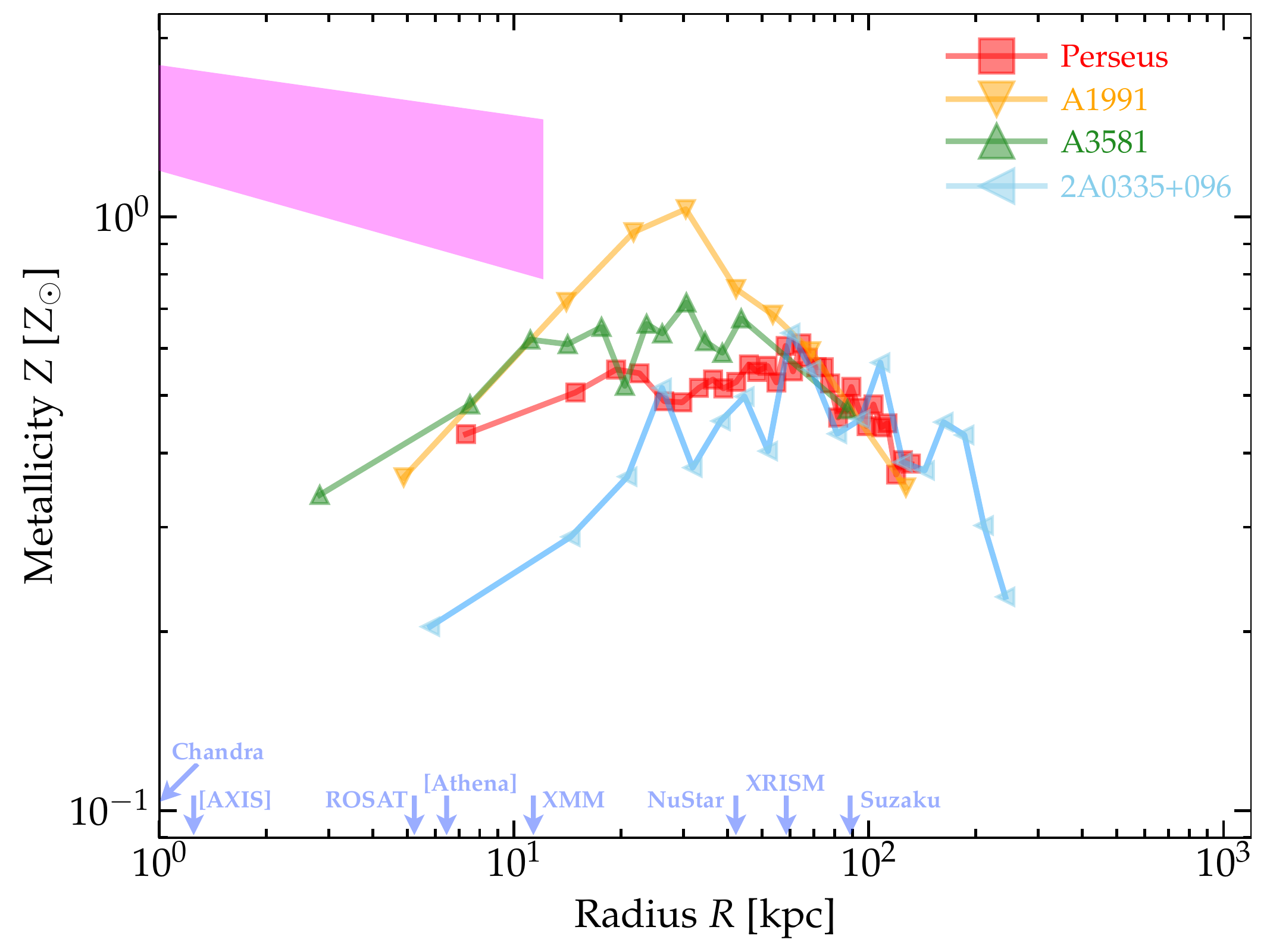} 
	\centering\includegraphics[width=0.495\textwidth]{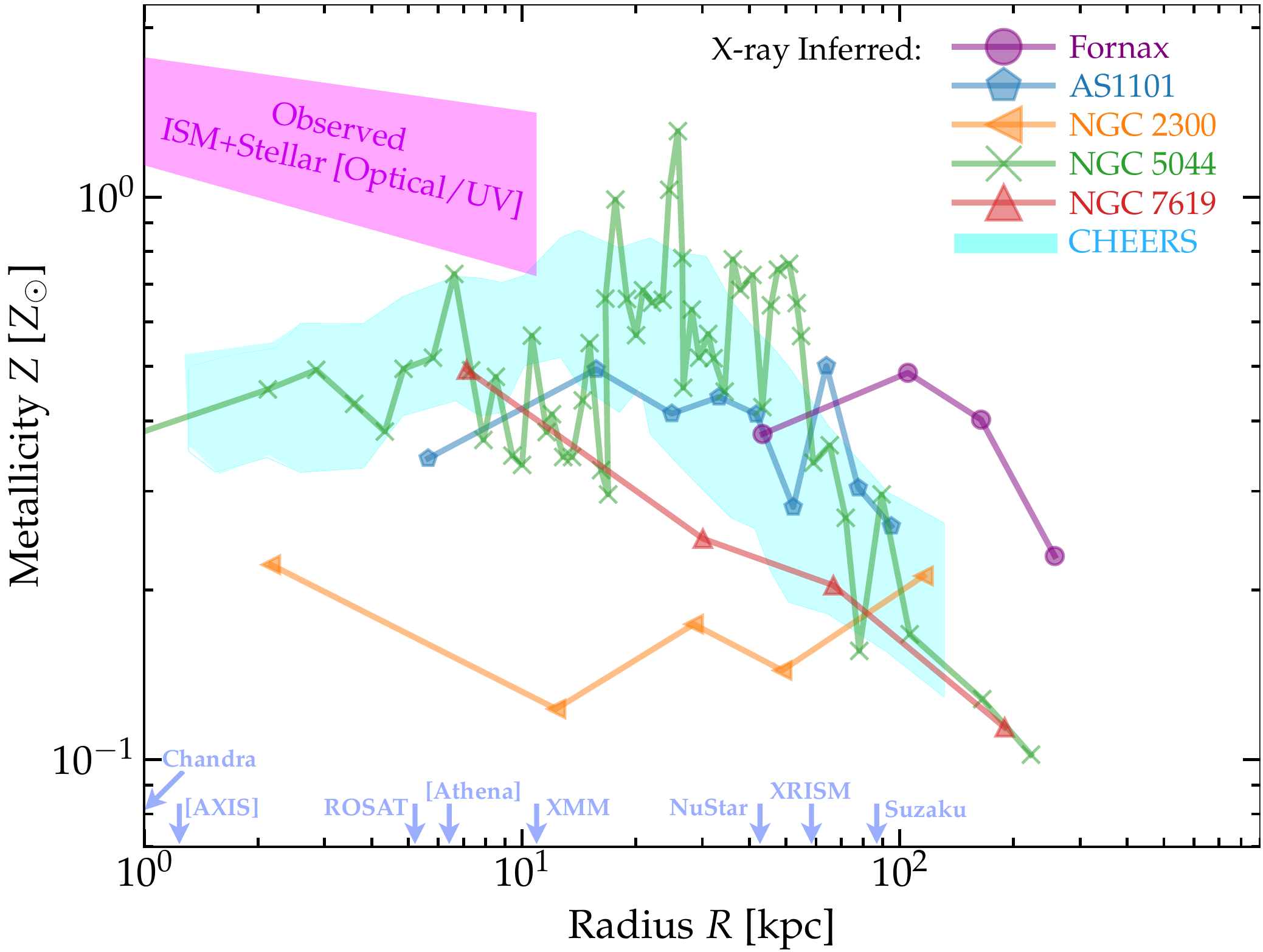} 
	\centering\includegraphics[width=0.49\textwidth]{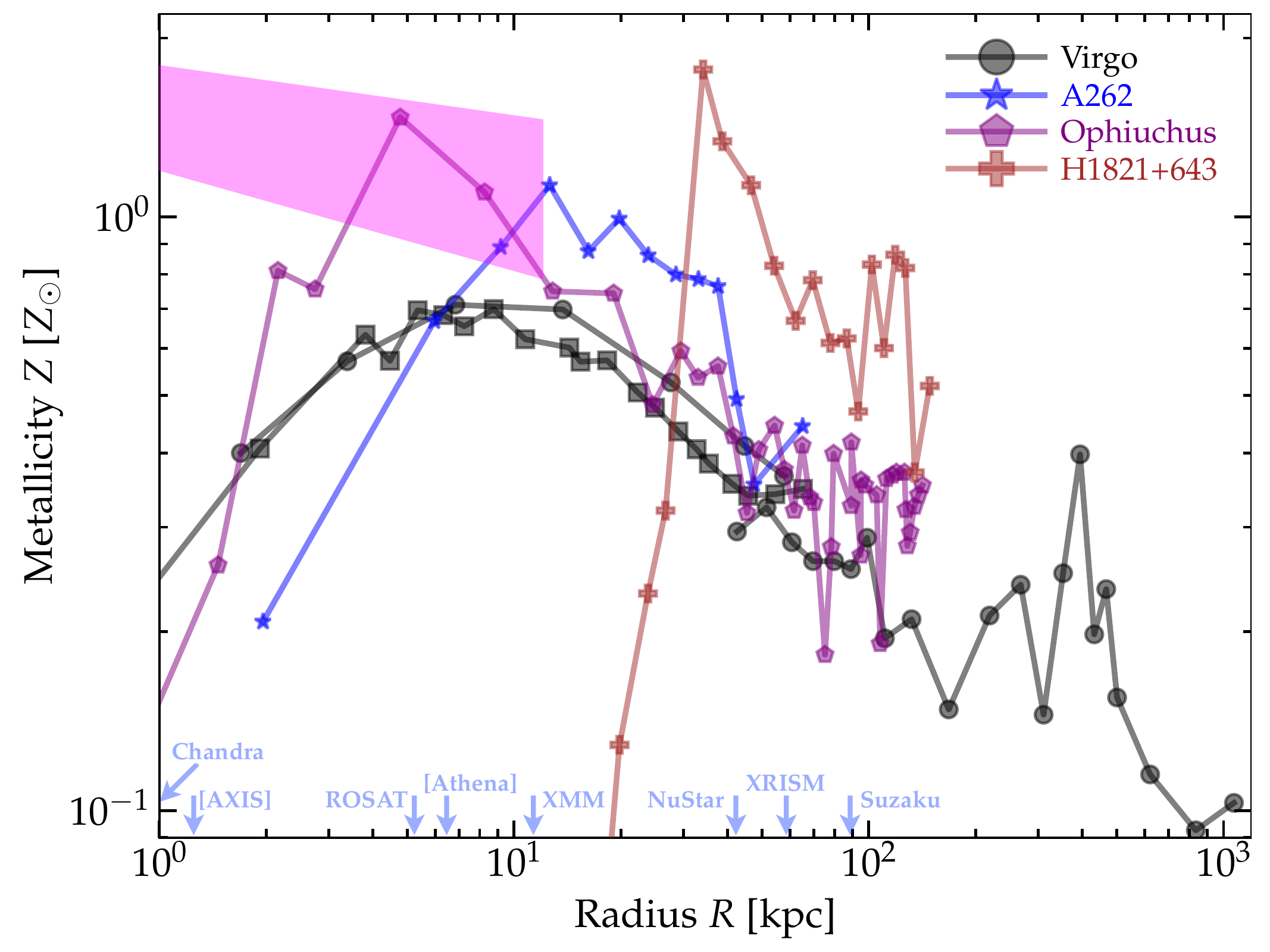} 
	\caption{Example observed cluster metallicity profiles, to compare with Fig.~\ref{fig:Z.profiles.model.only}. We plot (\textit{lines with points}) the published X-ray inferred metallicity $Z_{\rm X}$ from single-temperature fits to CC groups (\textit{left}) and clusters (\textit{right}), compiled from the references in \S~\ref{sec:z.drops}, with the highest spatial/angular resolution and signal-to-noise. 
	Angular resolution of different instruments (at the median distance $\sim 70\,$Mpc of the sample plotted) are labeled.
	We compare (\textit{shaded range}) observed ISM gas and stellar-phase metallicities measured for BCGs in halos of the same masses (including, in many cases, the \textit{same} clusters plotted) at $\lesssim 10\,$kpc from standard optical/NIR/UV metallicity diagnostics. These show, even for the same systems (e.g.\ Centaurus,Virgo, Perseus), Solar or super-Solar $Z_{\rm opt/UV} \sim 1-2\,Z_{\odot}$. Something appears to be artificially suppressing the X-ray-inferred metallicities at small radii, regardless of their shape (slow-rise, plateau, or drop).
	\label{fig:Z.profiles.obs.only}}
\end{figure*}

\begin{figure*}
	\centering\includegraphics[width=0.99\textwidth]{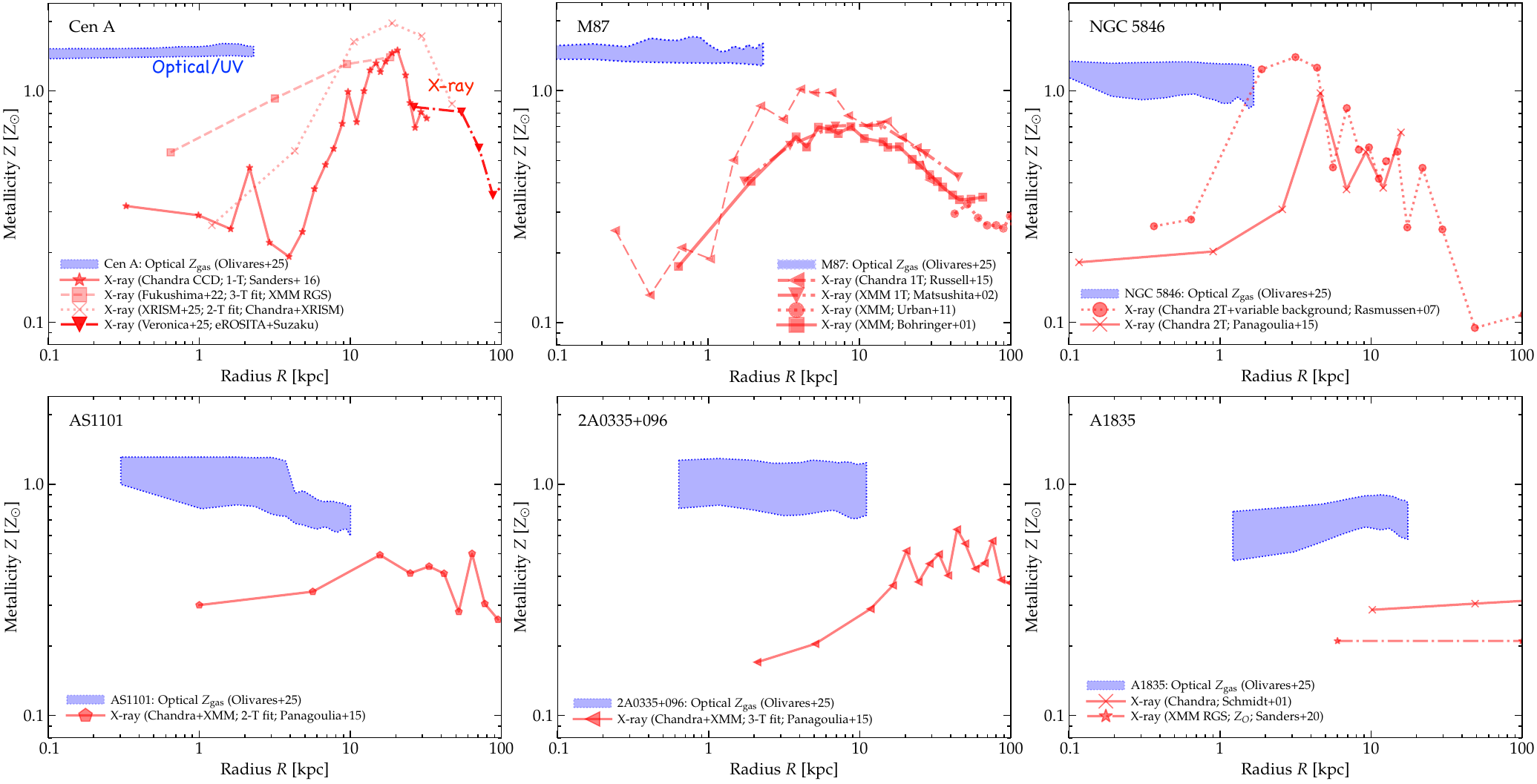} \vspace{-0.2cm}
	\caption{Examples of specific clusters with spatially-resolved, high-spectral resolution optical/UV gas-phase metallicity profiles from \citet{olivares:2025.massive.group.cluster.core.gas.metallicity.gradients.none.show.drops.like.in.xrays.even.drops.much.less.extreme.optical.UV}, compared to the X-ray-inferred metallicity profiles for the same clusters \citep{bohringer:2001.m87.cluster.profiles.no.real.cooling.flow.mass.deposition,schmidt:2001.A1835.cluster.profiles,matsushita:2002.m87.virgo.cluster.obs.metallicity.drop.temperature.fitting.challenges,rasmussen.ponman:2007.metallicity.temperature.profiles.groups,sanders:2010.A1835.cluster.properties,urban:2011.virgo.xray.profiles.to.rvir,russell:2015.m87.highres.chandra.zprofile.superstrongdrop,panagoulia:2015.cluster.metal.profiles.with.drops,sanders:2016.centaurus.metallicity.profiles.zdrop,fukushima:2022.centaurus.cluster.xray.zdrop,xrism:2025.centaurus.kinematics.metallicity,veronica:2025.erosita.centaurus.profiles}. 
	X-ray data includes CCD (Chandra/XMM) but also XMM RGS and XRISM spectroscopy. 
	Where possible (including all RGS \&\ XRISM examples) we plot the abundance of the same species (O, here) as measured in optical/UV. \vspace{-0.2cm}
	\label{fig:Z.profiles.obs.opticalvsxray}}
\end{figure*}

\begin{figure*}
	\centering\includegraphics[width=0.32\textwidth]{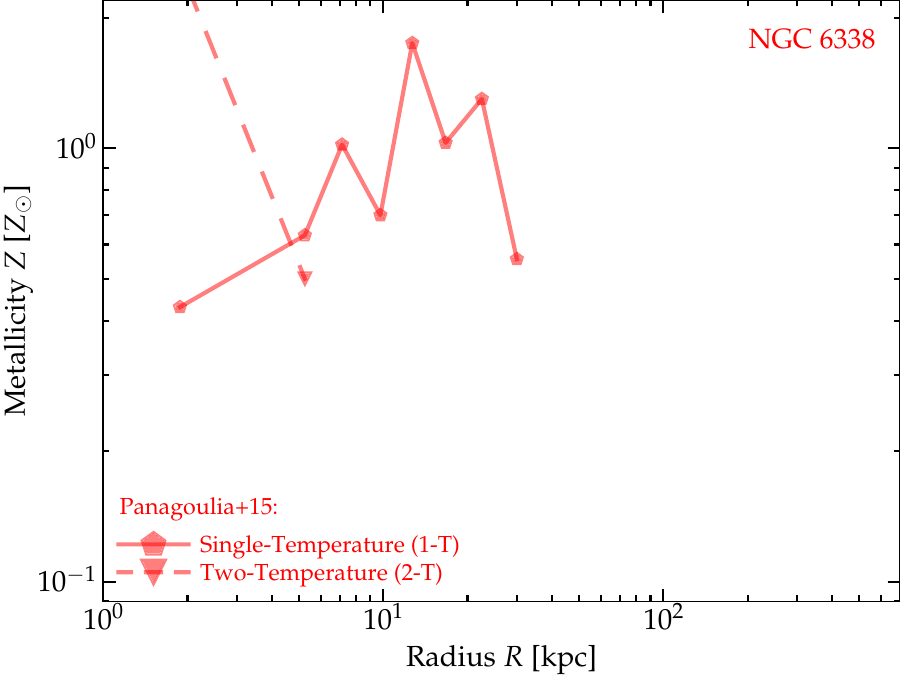} 
	\centering\includegraphics[width=0.32\textwidth]{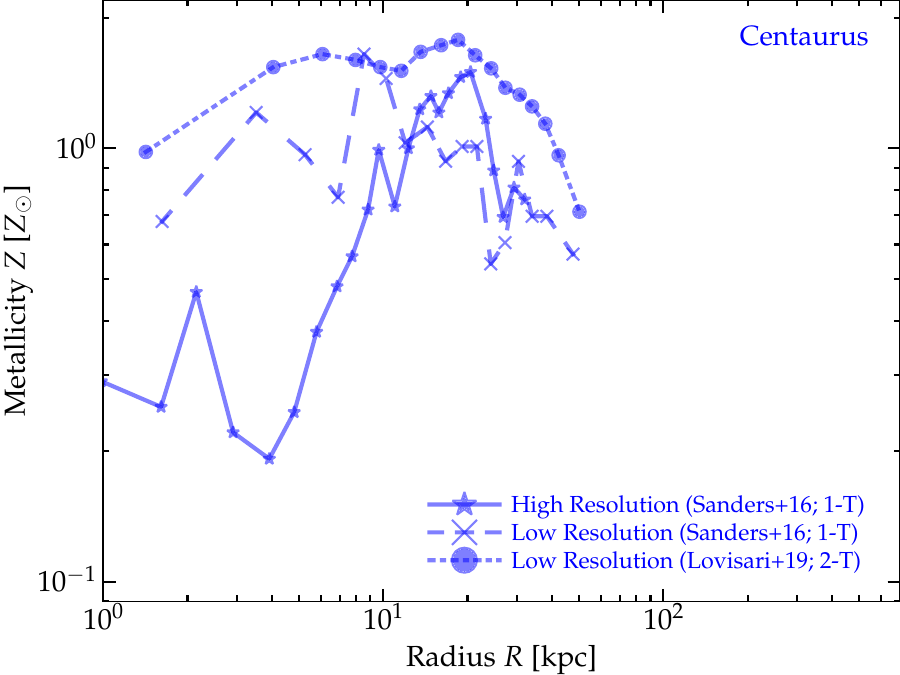} 
	\centering\includegraphics[width=0.32\textwidth]{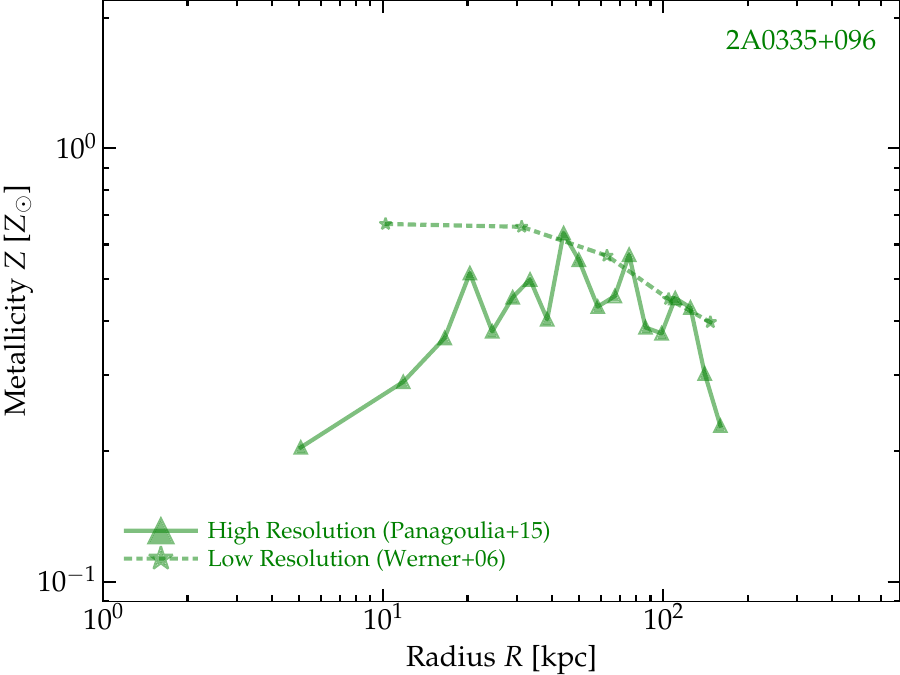} 
	\centering\includegraphics[width=0.32\textwidth]{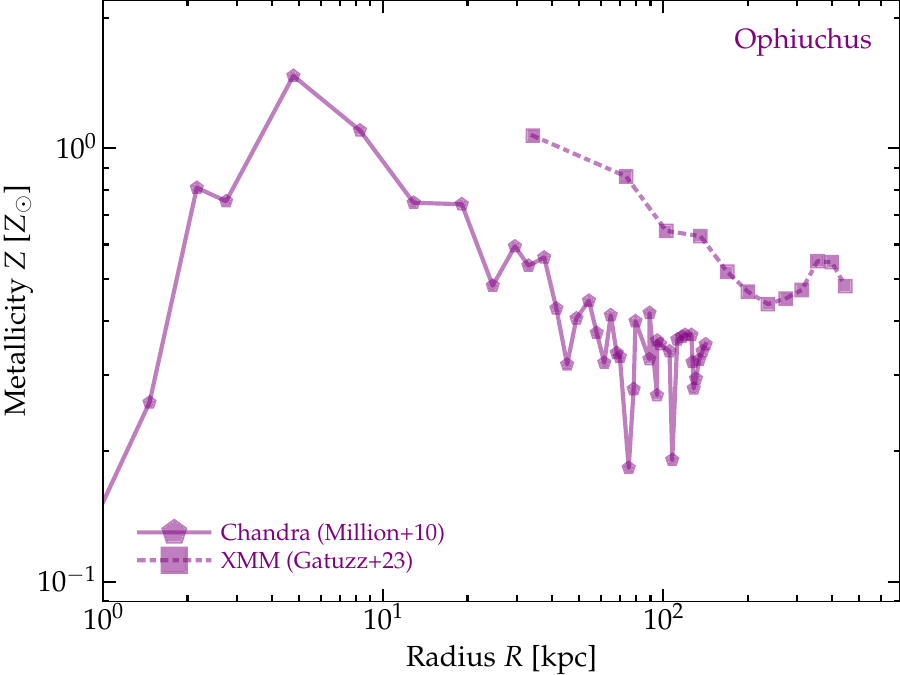} 
	\centering\includegraphics[width=0.32\textwidth]{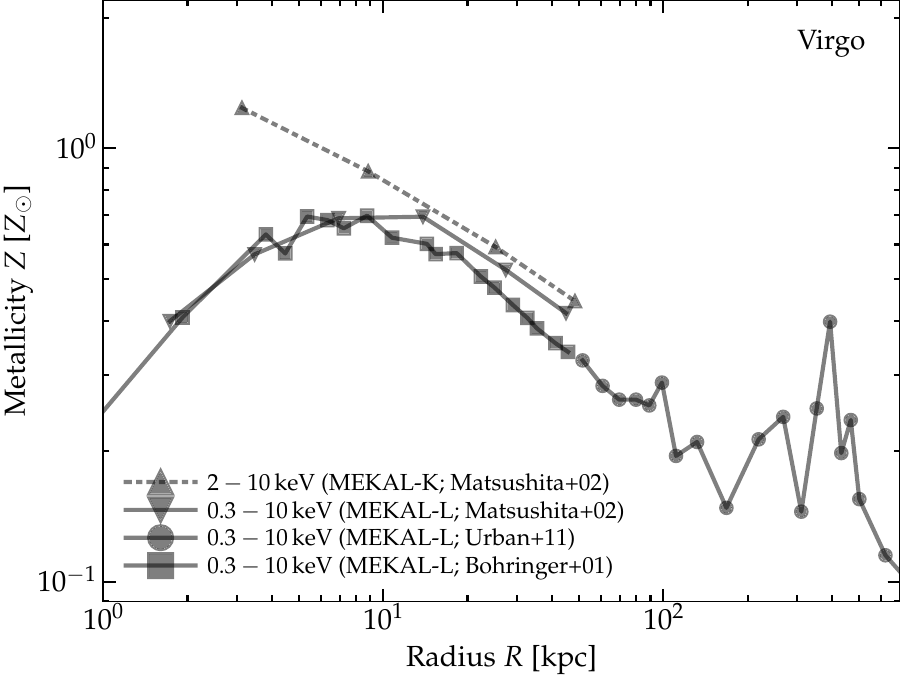} 
	\centering\includegraphics[width=0.32\textwidth]{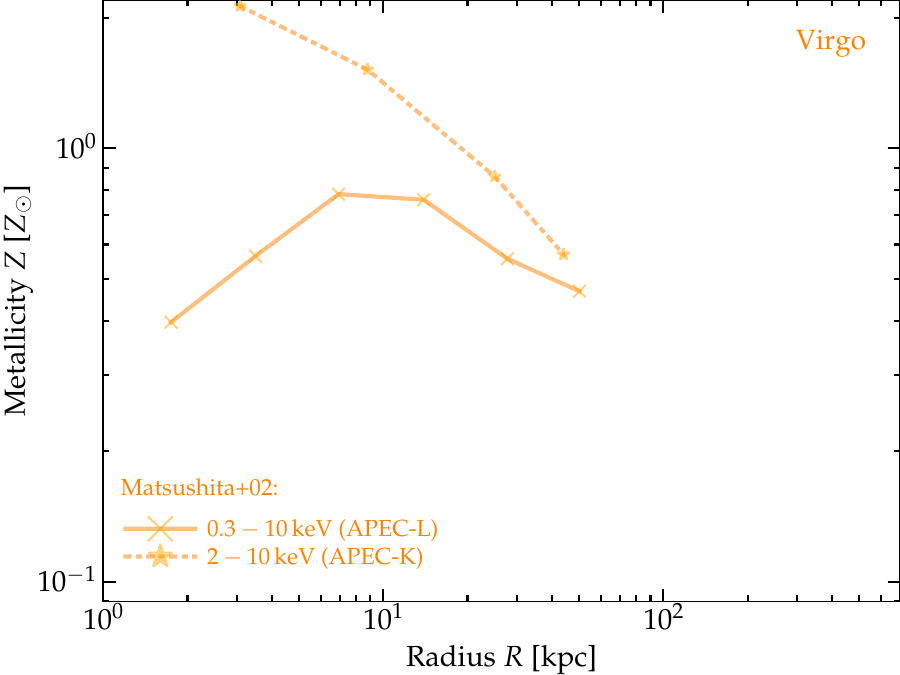} \vspace{-0.2cm}
	\caption{A few cautionary examples of observational systematics in recovery of the apparent X-ray metallicity $Z_{\rm X}$. 
	We compare literature results 
	(1) fitting the same data using single versus two-temperature models \citep{panagoulia:2015.cluster.metal.profiles.with.drops}; 
	(2) the same data re-binned to lower spatial-resolution bins and/or fit with a two-temperature model \citep{sanders:2016.centaurus.metallicity.profiles.zdrop,lovisari.reiprich:2019.cluster.metallicity.profiles.compilation.larger.radii}; 
	(3) observations with different angular resolution \citep{werner:2006.cluster.metallicity.profile,panagoulia:2015.cluster.metal.profiles.with.drops}; 
	(4) data from two different instruments \citep{million:2010.ophiuchus.central.metallicity.profile,gatuzz:2023.ophiuchus.metallicity.profile}; 
	(5) data re-fit by different authors using the same frequency range ($Z$ primarily from $\sim 1\,$keV Fe-L lines), or just $>2\,$keV (Fe-K; \citealt{bohringer:2001.m87.cluster.profiles.no.real.cooling.flow.mass.deposition,matsushita:2002.m87.virgo.cluster.obs.metallicity.drop.temperature.fitting.challenges,urban:2011.virgo.xray.profiles.to.rvir}); 
	(6) same data and model assumptions but re-fit with different codes  \citep{matsushita:2002.m87.virgo.cluster.obs.metallicity.drop.temperature.fitting.challenges}. 
	These can all strongly influence $Z_{\rm X}$, and can qualitatively change whether more dramatic features like drops appear, but this is expected for spectra which are \textit{not} actually reflective of single-temperature, single-metallicity thermal emission, as modeled here.
	\label{fig:Z.profiles.warn}}
\end{figure*}

Again as noted in \paperone, in the CR-IC interpretation, the thermal-like \textit{continuum} soft X-ray emission at small radii is boosted by CR-IC, but per \S~\ref{sec:excited}-\ref{sec:apparent}, this boosts the line emission only weakly (through e.g.\ CR Coulomb interactions, primary/secondary excitation, etc.). In the simplest models -- e.g.\ fitting the spectrum over a broad wavelength range to a single metallicity and temperature -- this will bias the observations (assuming all the emission is thermal, as usual) to a higher density but \textit{lower} metallicity $Z$, relative to the true underlying metallicity (Fig.~\ref{fig:Z.spectral.demo}). Fig.~\ref{fig:Z.spectral.demo} shows this is true for a wide range of underlying gas temperatures, and for both low-resolution CCD spectra, as well as $\sim1\,$eV spectral-resolution measurements from instruments like Hitomi/XRISM and (in the future) ATHENA.

Fig.~\ref{fig:Z.profiles.model.only} attempts to illustrate some hypothetical ``true'' and ``apparent'' metallicity profiles from our toy model in Fig.~\ref{fig:profiles}, using a broadly similar simple weight for estimating the effective inferred metallicity $Z_{\rm eff}$ and temperature $T_{\rm eff}$ from observational studies in \citet{mazzotta:2004.xray.temperature.measurement.modeling.and.caveats,vikhlinin:2006.single.multi.temperature.cluster.fits.also.includes.some.metallicity.discussion,braspenning:2024.flamingo.simple.xray.modeling.sims.clusters.dont.reproduce.zdrops.other.cc.features} (assuming their $\zeta \propto Z^{1}$ for metal line emission and $\zeta \propto Z^{0}$ for free-free, per \S~\ref{sec:apparent}). We need to assume a ``true'' metallicity profile for reference, for which we adopt the following simple function: $Z_{\rm true}^{-1} \approx Z_{\rm max}^{-1} + (Z_{\rm min} + Z_{\odot}\,\sqrt{r_{Z}/r})^{-1}$ with $(Z_{\rm min},\,Z_{\rm max},\,r_{Z})=(0.1\,Z_{\odot},\,1.7\,Z_{\odot},\,50\,{\rm kpc})$. This is motivated by the combination of observations of (1) both stellar and gas-phase metallicities and their gradients observed in UV/optical (thus immune to the IC contamination effects discussed here) from very large samples discussed below and shown in Fig.~\ref{fig:Z.profiles.obs.only}, (2) the most metal-rich observed central NCC profiles from X-ray surveys \citep[e.g.][]{mernier:2017.cluster.metallicity.profiles.mean}, and (3) the range of metallicity profiles seen at large radii (which should have minimal IC contamination) in X-ray spectroscopy as compiled in \citet{leccardi:2008.cluster.metallicity.profiles,molendi:2016.cluster.metallicity.profiles.large.radii.compilation,mernier:2017.cluster.metallicity.profiles.mean,lovisari.reiprich:2019.cluster.metallicity.profiles.compilation.larger.radii}. 

A specific example of the projected spectrum, integrating through a model cluster with a weak ``true'' CC designed to have just $\sim 5\%$ of the total ``apparent'' cooling luminosity (with $t_{\rm cool}^{\rm app} < 7\,$Gyr) is also shown in Fig.~\ref{fig:Z.spectral.demo}. But in Fig.~\ref{fig:Z.profiles.model.only} we by default neglect \textit{any} true CFs, and assume the extremal case (i.e.\ zero true multiphase gas below the virial temperature, zero true CF or CC) for the sake of illustration, akin to the profiles in Fig.~\ref{fig:profiles}. More examples with a similar idea, but somewhat different cluster models (showing different shapes as a result) are shown in \paperone. But we do note that the models in Fig.~\ref{fig:Z.profiles.model.only} can always be made to have a steeper central rise (closer to the ``true'' $Z_{\rm true}$) if we assume some non-zero ``true'' CC (even a weak CC) is present as well, as we show. 

We compare a number of observed SCC metallicity profiles in Fig.~\ref{fig:Z.profiles.obs.only} \&\ \ref{fig:Z.profiles.obs.opticalvsxray}. For the sake of straightforward comparison, we focus in Fig.~\ref{fig:Z.profiles.obs.only} on bright, SCC clusters observed at reasonably high angular resolution, with de-projected profiles of $Z$ published from {\em single-temperature}, single-phase fits. This is what our toy-model observations mimic, and where we would expect to see a $Z$ suppression, unlike the results of multi-temperature fitting which can eliminate the apparent $Z$ discrepancy in the models. We discuss these further below (\S~\ref{sec:z.drops.obs}). Fig.~\ref{fig:Z.profiles.obs.opticalvsxray} focuses on systems where specifically gas-phase optical/UV and X-ray  measurements of metallicities exist at the same radii, in the same clusters, for the same species.

Again we stress metallicity estimation is complicated and known to be sensitive to e.g.\ whether one fits single or multi-temperature models, isobaric models with varying densities, multiphase models with varying metallicities, multiple annuli with an assumed gradient, projected or de-projected models, as well as the instrument properties, effective wavelength range being fit and sensitivity of the spectrum (determining which lines contain the most information). We demonstrate this for a couple of examples in Fig.~\ref{fig:Z.profiles.warn} -- showing that changing the assumed temperature priors or instrument or spatial re-binning of the data can easily lead to factors of several change in the metallicities. However, the qualitative effects we are interested in persist in at least $\gtrsim50\%$ of the sample independent of these choices, even though the \textit{quantitative} values change dramatically. So clearly modeling a specific observed metallicity profile requires forward-modeling the full X-ray spectrum and synthetic observations and using the identical data reduction and analysis and fit routines. Our intent here is simply to illustrate the qualitative effects. 

Especially in the lower-mass cases (small and large groups), the most extreme possible cases (high CR-IC luminosity, low-mass halo, low background gas density, zero true CC/CF, weak CR heating and excitation of the gas) produces a sufficiently-rapid suppression of $Z_{\rm eff}/Z_{\rm true}$, relative to the rise in $Z_{\rm true}$, that it can produce a central ``metallicity drop,'' while in the higher-mass and/or less-extreme cases $Z_{\rm eff}$ is clearly suppressed but can continue to rise towards $R\rightarrow 0$ (a ``slow-rising'' profile). While there is always some central $Z$ suppression (relative to $Z_{\rm true}$) in the CR-IC dominated case, whether it formally produces an apparent ``drop'' (change in the sign of the gradient of $Z_{\rm eff}(R)$) depends much more sensitively on the relative rate-of-change of the fraction of the surface brightness coming from CR-IC and fraction of the emission coming from lines versus radius, versus the slope of the intrinsic underlying $Z_{\rm true}(R)$. Much of the reasonable model parameter space produces profiles more like ``plateaus'' or ``slow-risers,'' by which we mean central metallicities that rise less-rapidly than the optical/UV diagnostics imply. So high-spatial-and-spectral resolution observations of metallicity profiles, especially with corresponding optical/UV data, could provide a powerful model discriminator in the future.

We stress again that \textit{a very large and plausible parameter space for strong CR-IC produces no ``drops'' }-- just suppression below the UV/optical $Z$ -- even in lower-mass halos. But the general trend of more dramatic/pronounced suppression in lower mass halos is likely robust and occurs because: (1) their intrinsic emission is more strongly line-dominated at the energies where CR-IC emits most strongly (so adding a continuum source without lines more strongly suppresses the inferred $Z_{\rm eff}$), compared to hotter higher-mass systems where the true thermal spectrum is already more free-free dominated and the strongest lines are at $\gtrsim 6$\,keV where the fractional CR-IC emission is weaker; and (2) the scalings of the slopes/profiles of $e_{\rm cr}$, true thermal emission, and $Z(R)$ make the effect more dramatic at lower-$M_{\rm vir}$.

\subsubsection{Comparison to Observed Central Metallicity Suppression in X-rays Relative to Optical}
\label{sec:z.drops.obs}

In Figs.~\ref{fig:Z.profiles.obs.only}-\ref{fig:Z.profiles.warn}, we compare to a selection of observed X-ray systems compiled from \citet{bohringer:2001.m87.cluster.profiles.no.real.cooling.flow.mass.deposition,schmidt:2001.A1835.cluster.profiles,matsushita:2002.m87.virgo.cluster.obs.metallicity.drop.temperature.fitting.challenges,rasmussen.ponman:2007.metallicity.temperature.profiles.groups,komiyama:2009.5044.metallicity.profile,million:2010.ophiuchus.central.metallicity.profile,sanders:2010.A1835.cluster.properties,urban:2011.virgo.xray.profiles.to.rvir,murakami:2011.fornax.metallicity.profile.suzaku.xmm,walker:2013.centaurus.cluster.few.other.large.radii.entropy.metallicity.profiles,panagoulia:2015.cluster.metal.profiles.with.drops,russell:2015.m87.highres.chandra.zprofile.superstrongdrop,sanders:2016.centaurus.metallicity.profiles.zdrop,fukushima:2022.centaurus.cluster.xray.zdrop,gatuzz:2023.centaurus.metal.profile.outer,gatuzz:2023.ophiuchus.metallicity.profile,xrism:2025.centaurus.kinematics.metallicity,chatzigiannakis:2025.cheers.groups.all.show.central.zdrop.if.chosen.to.have.agn.not.along.jet.though,veronica:2025.erosita.centaurus.profiles}. The shallow profiles (plateaus and slow-risers) and rarer but even more extreme drops suggested in our simple toy-model calculations in Figs.~\ref{fig:Z.spectral.demo}-\ref{fig:Z.profiles.model.only} and \paperone\ are indeed strikingly similar to those seen in most cool-core systems. Indeed, these behaviors have been known for decades, and are much more common than the small number of high-signal-to-noise observations shown here (the cited samples include dozens of additional clusters which would show similar behaviors, albeit often with lower signal-to-noise). They are especially prominent if one fits single-temperature models (what our Figs.~\ref{fig:Z.spectral.demo}-\ref{fig:Z.profiles.model.only} assume), since as discussed at length in those papers any multi-temperature fit will allow for fitting more continuum to slightly hotter gas with weaker lines (whether this emission is \textit{actually} thermal or entirely from CR-IC) and give a higher central $Z$. 

In contrast, direct measurements of the ISM metallicities and stellar metallicities of the galaxies at $\lesssim 10\,$kpc, from traditional optical/UV high-resolution (both spatial and spectral resolution) spectroscopy, clearly indicate modestly super-Solar ($\sim 2\,Z_{\odot}$) metallicities at these masses, \textit{at the same radii probed by X-ray metallicity estimates!} We stress that these include \textit{both} stellar metallicities \citep{loubser:2009.bcg.stellar.metallicities.ssp.fits,loubser:2012.bcg.stellar.metallicity.gradients.profiles,montes:2014.m87.stellar.metallicity.age.gradients.and.central.values,zinchenko:2024.manga.stellar.and.gas.metallicities.same.galaxies.differences.overall.very.consistent,edwards:2024.manga.bcg.stellar.metallicities.gradients.lowz} \textit{and} gas-phase/ISM metallicities \citep{ellison:2009.mzr.clusters.cluster.bcgs.gas.phase.supersolar.even.moreso.if.bcg.or.cluster.center,manucci:2010.fundamental.sf.metallicity.relation.strong.sfr.dependence.low.mass.weak.high.mass,maier:2022.bcg.gas.phase.ism.abundances.supersolar.gradients.consistent.others,castignani:2022.bcg.gasphase.sfrs.metallicities.most.sf.bcgs.still.ism.supersolar,zinchenko:2024.manga.stellar.and.gas.metallicities.same.galaxies.differences.overall.very.consistent,olivares:2025.massive.group.cluster.core.gas.metallicity.gradients.none.show.drops.like.in.xrays.even.drops.much.less.extreme.optical.UV}. These optical/UV samples are orders-of-magnitude larger than X-ray spectroscopic samples at $<10\,$kpc, so have also been able to clearly show that these central metallicities extend over the entire range $\sim 1-10\,$kpc (with the median metallicity gradient from \citealt{maier:2022.bcg.gas.phase.ism.abundances.supersolar.gradients.consistent.others} shown in Fig.~\ref{fig:Z.profiles.obs.only}), that there is no strong difference between ``quenched'' or star-forming cooling-flow BCGs \citep{manucci:2010.fundamental.sf.metallicity.relation.strong.sfr.dependence.low.mass.weak.high.mass,castignani:2022.bcg.gasphase.sfrs.metallicities.most.sf.bcgs.still.ism.supersolar,zinchenko:2024.manga.stellar.and.gas.metallicities.same.galaxies.differences.overall.very.consistent}, that they specifically hold for BCGs (indeed the cluster-center galaxies seem to, if anything, have slightly \textit{higher} ISM gas-phase metallicities than galaxies of the same mass in the cluster outskirts or field; \citealt{ellison:2009.mzr.clusters.cluster.bcgs.gas.phase.supersolar.even.moreso.if.bcg.or.cluster.center}), and that the stellar and gas-phase metallicities and gradients estimates in the optical/UV are consistent with one another to $\sim 0.1\,$dex \citep[references above and][]{edwards:2024.manga.bcg.stellar.metallicities.gradients.lowz,zinchenko:2024.manga.stellar.and.gas.metallicities.same.galaxies.differences.overall.very.consistent}. Furthermore the samples in these papers \textit{specifically} include almost all of the well-studied $Z$-drop or slow-rise galaxies in Fig.~\ref{fig:Z.profiles.obs.only}, such as Virgo, Centaurus, Fornax, Hydra-A, and Perseus. 
In e.g.\ the sample of \citet{olivares:2025.massive.group.cluster.core.gas.metallicity.gradients.none.show.drops.like.in.xrays.even.drops.much.less.extreme.optical.UV} of high spatial ($\lesssim 1-10\,$kpc) and spectral resolution optical/UV gas phase metallicity profiles of BCGs in 13 clusters, \textit{every} example where there exist spatially-resolved (deprojected) X-ray maps at overlapping radii appears to show that the X-ray metallicity is systematically suppressed relative to the optical/UV (even taking measurements of the same chemical species, at the same position, in the same cluster). 
In other words, the X-ray $Z$-suppression is not simply surprising from a theoretical point of view -- the X-ray inferred metallicity in the central $\lesssim 10\,$kpc is suppressed \textit{relative to that from UV/optical observations of gas and stars at the same radii}. 

When fitting X-ray data to single-temperature profiles, \textit{most} well-resolved CC groups (where many independent radial bins can be fit to spectra from radii $\sim 1-30\,$kpc) appear to exhibit some level of $Z$ suppression (as we define it), or even drops -- including almost all of the most famous and well-studied CC groups: e.g.\ Virgo, Fornax, NGC 5044, Centaurus, Ophiuchus, HCG 62, A262. Even using multi-temperature fits (which should give metallicities closer to the ``true'' values), $\gtrsim 50\%$ of CC groups and clusters in \citet{rasmussen.ponman:2007.metallicity.temperature.profiles.groups}, and every 
CC group in the CHEERS sample (selected to have significant radio/cavity power) in \citet{chatzigiannakis:2025.cheers.groups.all.show.central.zdrop.if.chosen.to.have.agn.not.along.jet.though}, \textit{still} exhibit drops (let alone suppression). And all of the strong drops in \citet{rasmussen.ponman:2007.metallicity.temperature.profiles.groups,panagoulia:2015.cluster.metal.profiles.with.drops} are associated with apparent SCCs. As shown in \citet{panagoulia:2015.cluster.metal.profiles.with.drops}, a likely reason that abundance drops and/or strong suppression appear to be much more common in the nearest, brightest groups/clusters is simply spatial (angular) resolution: binning together or removing the central few Chandra resolution elements (rebinning to $\sim 5^{\prime\prime}$) even in Virgo leads to artificially inferring a flat central $Z$ profile (still suppressed, but less dramatically so) instead of fully-resolving the suppression (see Fig.~\ref{fig:Z.profiles.warn} or compare the lower-resolution profiles in \citealt{schmidt:2002.perseus.core.metallicity.temperature.maps} or \citealt{sanders:2002.centaurus.metallicity.profile.older.data.lower.res.still.some.drop.detected,sanders:2016.centaurus.metallicity.profiles.zdrop} to see the same for Perseus and Centaurus, respectively, if downgraded to $>10^{\prime\prime}$ resolution). Conversely, more recent higher-angular-resolution studies of SCC systems previously classified as not having drops (from older, low-resolution data) have seen ubiquitous flattening (plateaus) or even more extreme drops appear at higher angular resolution \citep{ng:2024.cluster.detailed.chandra.profiles.coolcore.zdrops.sizes.profiles.nonuniversal.norm}. In Fig.~\ref{fig:Z.profiles.warn} we attempt to show some examples of the same galaxy observed with different instruments, binned at different resolution, or fit with different assumptions, to illustrate that while the qualitative behavior of suppression can often remain, this can produce factor of $\sim 5$ systematic effects on the actual metallicities (even at large radii). In either case, almost all of the ``non-dropping'' CC profiles in these studies (e.g.\ AS1101, NGC 2300, Hydra-A, \citealt{simionescu:2009.hydra.a.cluster.profiles.not.central.drop.but.very.slow.rise.subsolar.at.kpc.scales}) are still consistent with the \textit{suppressed} but non-dropping profiles in Fig.~\ref{fig:Z.profiles.model.only} (and more in \paperone, where we allow for some ``true'' very weak CC) -- i.e.\ they exhibit extremely flat, typically sub-Solar metallicity profiles extrapolated to $R\rightarrow 0$, in tension with the UV/optical data.

The combination of these observational effects, different definitions of ``suppression'' in the first place, and lack of study means that it has been challenging to robustly quantify the demographics of said central metallicity suppression. Most of the historical study has focused exclusively on ``drops,'' defined from X-ray data alone, but we stress that not only are these sensitive to the fitting details and instrument used, they are (if present) only the most extreme cases of suppression, while many systems which show no drop are still strongly suppressed relative to the UV/optical data. For example, \citet{rasmussen.ponman:2007.metallicity.temperature.profiles.groups} attempted to construct a flux-limited sample of nearby groups (irrespective of CC/NCC status, though there is a well-known X-ray bias towards CCs), and found positive evidence for ``drops'' specifically within 13 of 15 ($87\%$), with the only 2 exceptions being the two most poorly-spatially-resolved groups in the sample (so potentially consistent with a true ``drop rate'' of $100\%$). 
They also found only two systems switch from ``drop'' to ``non-drop'' in their sample going from single to two-temperature fits. 
By our much looser definition of suppression (relative to e.g.\ $Z_{\rm true}$ in Fig.~\ref{fig:Z.profiles.model.only}), all of these, including the non-drops, would be highly ``suppressed.''
\citet{panagoulia:2015.cluster.metal.profiles.with.drops} attempted to construct a larger volume-limited ($z\le 0.071$) sample of groups and clusters with multi-temperature fits and found some positive evidence for drops in 14 of 65, however they showed explicitly that ``drop'' classification was strongly correlated with signal-to-noise and spatial resolution in their sample. Downgrading their well-resolved ``certain'' drop examples to the (much poorer) median spatial resolution of their sample, or angular resolution $\gtrsim 20^{\prime\prime}$,  led to the drop being unresolved and therefore incorrect classification of the system as a ``non-drop'' after de-projection. Restricting to their highest signal-to-noise cases ($\gtrsim 10^{5}$\,central counts) their fraction of ``certain drops'' rises to 6/11, or restricting to their highest-spatial resolution cases (central bin $<2\,$kpc; note this sample overlaps with the high S/N-sample) it rises to 9/12. And that was using multi-temperature fits: it is important to note that every example of the ``non-drops'' with high S/N in \citet{panagoulia:2015.cluster.metal.profiles.with.drops} has been classified as a ``drop'' based on \textit{single}-temperature deprojected spectral fits elsewhere in the literature. Indeed, to our knowledge, there is no example of a group-mass CC system with high-spatial resolution X-ray spectroscopy (sufficient for accurate $Z$ fits at $\sim$\,kpc resolution -- i.e.\ arcsecond angular resolution spectroscopy) which does not show evidence for a central suppression (slow-rise, plateau, or drop) from \textit{single-temperature} fits. We stress that many of the (more massive, or lower-resolution, or multi-temperature-fitted) ``non-drop'' cases give sub-Solar central $Z$, so they still provide positive evidence of strong metallicity suppression by the central ACRH, especially if these systems have any ``true'' very weak CC (per Fig.~\ref{fig:Z.profiles.model.only}). 

And again, these studies focus largely on the specific idea of ``drops'' defined as a change of sign of the slope of $d Z/d r$: there has been almost no study of the demographics of central metallicity \textit{suppression}, i.e.\ the fraction of clusters which appear to have metallicities below $\sim 1.5-2\,Z_{\odot}$ at $\lesssim 10\,$kpc inferred from single-temperature de-projected X-ray fits. Again we stress that it does require very high (arcsecond) angular resolution comparable to Chandra or future missions like AXIS, in almost all cases, to spatially resolve the X-ray suppression relative to optical/UV diagnostics (especially in de-projected 3D profiles): instruments like XRISM Resolve, with $\sim$\,arcminute resolution, cannot resolve the interesting radii in the predicted or observed profiles in Figs.~\ref{fig:Z.spectral.demo}-\ref{fig:Z.profiles.warn} in any of the examples plotted.

We therefore suggest that a better observational metric for suppression or ``dilution,'' compared to ``whether or not there is a drop'' would be something like the following: the ratio of the central $r<10\,$kpc, de-projected, single-temperature/single-phase X-ray metallicity $Z_{\rm X}$ fit to the spectrum around $\sim 1\,$keV (closer to the peak of CR-IC), over the fit from optical/UV diagnostics of similar galaxies ($Z_{\rm opt/UV} \sim 2\,Z_{\odot}$). 

Ideally, one would quantify this as a function of wavelength/X-ray energy with spectral coverage from $\sim 0.1-10\,$keV. To take the closest and therefore best-resolved CC, Virgo, as an example: many studies with XMM and Chandra \cite{bohringer:2001.m87.cluster.profiles.no.real.cooling.flow.mass.deposition,matsushita:2002.m87.virgo.cluster.obs.metallicity.drop.temperature.fitting.challenges,matsushita:2003.m87.species.by.species.abundance.profiles,urban:2011.virgo.xray.profiles.to.rvir,russell:2015.m87.highres.chandra.zprofile.superstrongdrop} have shown that the central X-ray metallicity suppression is a strong function of X-ray frequency: as shown in Fig.~\ref{fig:Z.profiles.warn}, fitting the same data with the same methods but only using the Fe-K ($\sim 6-8$\,keV) line regions shows no suppression (giving a monotonically-rising metallicity, super-Solar interior to $\lesssim 15\,$kpc, in good agreement with optical/UV metallicities), while fitting the Fe-L line complexes around $\sim 1$\,keV gives strong suppression and even drops. The central Fe abundances differ by an order of magnitude: in other words, it specifically appears as if the $\sim$\,keV X-rays are diluted in the central $\lesssim 10\,$kpc, exactly as expected for CR-IC.\footnote{As noted above, fitting multi-temperature components can modify the inferred ``drop'' behavior, but this still gives significant $Z$ suppression relative to optical/UV estimators \citep[compare][]{matsushita:2002.m87.virgo.cluster.obs.metallicity.drop.temperature.fitting.challenges,russell:2015.m87.highres.chandra.zprofile.superstrongdrop}. \citet{molendi:2001.m87.abundance.profiles.xray.fit.out.with.multi.temp.plus.single.temp.plus.power.law.plus.background} were only able to reconcile the fitted metallicities with optical/UV values by invoking several different temperature components plus an explicit non-thermal (power-law) component and free X-ray absorption in their fits.}

Similar behavior is seen in Centaurus, with Chandra plus RGS/XRISM spectroscopic fits still finding a central drop even in Fe-K, but with order-of-magnitude stronger $Z$ suppression in Fe-L ($\sim 10\times$ lower Fe at $\sim 1\,$kpc fitting the $0.7-7$\,keV spectrum versus $2-8$\,keV; \citealt{majumder:2025.phd.centaurus.profiles.chandra.and.some.xrism.data}). 

However, unlike for the SZ effect in \S~\ref{sec:sz}, even if one has both optical/UV ($Z_{\rm opt/UV}$) and some estimated X-ray ($Z_{\rm X}$) metallicities of the same system, at the same radii ($r \lesssim 10\,$kpc) and assumes $Z_{\rm opt/UV} \approx Z_{\rm true}$, there is no simple relation between $Z_{\rm opt/UV}/Z_{\rm X}$ and the fraction of the total X-ray emission coming from true thermal vs.\ CR-IC at that radius. This owes to the complex observational uncertainties in fitting $Z_{\rm X}$ but also to the fundamental physical uncertainty that the ``apparent'' $Z_{\rm X}$ from any given fitting method will depend on the true underlying multi-phase temperature/density/metallicity structure of the gas and excitation/heating by CRs, integrated along the line-of-sight. Still, for the heuristic models and observations in Figs.~\ref{fig:Z.spectral.demo}-\ref{fig:Z.profiles.warn}, the observed $Z_{\rm opt/UV}/Z_{\rm X}$ ratios are broadly consistent with fractions ranging from $\sim 1 -50\%$ of the emission coming from thermal emission at $\lesssim 10\,$kpc.

\subsubsection{Alternative Interpretations}
\label{sec:z.drops.alt.interp}

Again, in the CR-IC interpretation, these suppressed metallicity profiles, and occasional more extreme $Z$ drops (especially in single-temperature model fits to group-mass systems) are almost trivial to explain. And indeed, the possibility that the $Z$ suppression observed could be explained by some contamination from non-thermal emission has been noted before (see discussion in \citealt{mernier:2017.cluster.metallicity.profiles.mean}). However in those studies, a CR-IC type explanation was dismissed specifically because it was assumed that such non-thermal emission would have to be closely spatially associated (in e.g.\ radius and detailed morphology) with the locations of strong high-frequency synchrotron emission (e.g.\ the bright GHz jets/bubbles specifically). But of course, such a strict association is \textit{not} predicted for the ACRHs, as we have emphasized above. 

In comparison, in the standard interpretation, there is no widely-accepted explanation for such central $Z$ suppression. Indeed even cosmological models which can occasionally produce drops due to peculiar metal-mixing via feedback or mergers predict they are extremely rare (factors at least $\sim 10-100$ more rare than observed; see e.g.\ \citealt{braspenning:2024.flamingo.simple.xray.modeling.sims.clusters.dont.reproduce.zdrops.other.cc.features}), and more importantly those models necessarily predict an identical drop in optical/UV metallicities (not seen). Heuristic explanations involving sedimentation or condensation into grains have been discussed but these are not predicted naturally -- the X-ray emitting gas should be much too hot to have a large fraction of grains containing almost all of the refractory elements \citep{draine.salpeter:ism.dust.dynamics}, and indeed such strong depletion at the level needed is not observed in the ISM until one reaches temperatures $\ll 100\,$K and densities $\gtrsim 10^{3}\,{\rm cm^{-3}}$ \citep{jenkins:2009.depletion.patterns.vs.densities.in.ism.grains}. Moreover those models would predict diffuse IR luminosities much larger than observed \citep{wolfire:1995.neutral.ism.phases}. And most importantly sedimentation or depletion/condensation is clearly inconsistent with the observed abundances from the optical/UV being ``normal'' (super-Solar) in the ISM of the actual central galaxies (see extensive references above). 

Interestingly, there is a well-established association between the groups/clusters with the strongest evidence for relatively recent AGN activity (e.g.\ large, bright AGN-centered radio cavities or mini-halos, or the most luminous ongoing radio AGN) and strong $Z$-suppression or plateaus/drops (including many of the specific cases above but also see \citealt{rafferty:2013.agn.corr.with.hcg.62.z.drop.metallicity.profile,gendron:2017.4472.metal.profile.z.drop.argue.agn.mixing.but.opposite.sign.expected.from.theory,liu:2019.cluster.z.drop.compilation.model.discussion,chatzigiannakis:2025.cheers.groups.all.show.central.zdrop.if.chosen.to.have.agn.not.along.jet.though}). This is not seen in simulations (assuming the X-ray is pure thermal cooling emission) even with widely-varying AGN feedback models \citep{braspenning:2024.flamingo.simple.xray.modeling.sims.clusters.dont.reproduce.zdrops.other.cc.features,gonzalez.villalba:2024.magneticum.cluster.cc.ncc.statistics.pred,lehle:2024.simulation.cluster.profiles,nelson:2024.tng.cluster.sims.profiles.basic.properties,prunier:2024.tng.cluster.xray.cavities.properties}. And even heuristic models for the AGN-drop correlation require some complicated explanations, involving large amounts of volume-filling low-metallicity hot (X-ray emitting) gas coming into the central cluster regions (but just in the hot phase, so as not to contaminate the UV/optical metallicity diagnostics) via some convective cells while all the high-metallicity hot gas from the galaxy is vented to large radii (but not in excess of the usual observed metallicities at large radii, and without mixing with the incoming gas). However, in the CR-IC scenario, the observed association is trivial: stronger CR sources associate with $Z$-suppression because there is a larger CR-IC continuum.

Thus while less clean than the SZ pressure deficits, these central metallicity profiles again appear to provide strong positive evidence for the CR-IC scenario, in an order-unity fraction of CC systems.

\begin{figure}
	\centering\includegraphics[width=0.98\columnwidth]{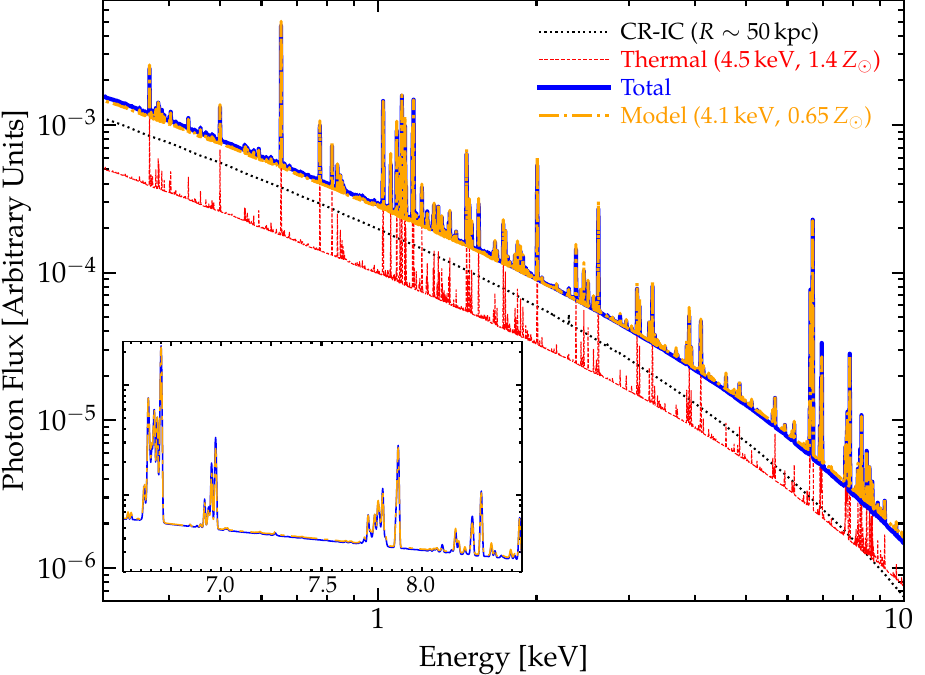} 
	\caption{Combined X-ray CR-IC plus thermal spectra as Fig.~\ref{fig:Z.spectral.demo}, but showing a more typical case for a SCC at uniformly high ($<1\,$eV) spectral resolution. Here we take a Perseus-like model in terms of CR injection rate ($\dot{E}_{43}=100$ in a $10^{15}\,M_{\odot}$ cluster) at $\sim 50-60$\,kpc from the center (the outer region of the cool core), with gas temperature $\sim 4.5\,$keV chosen to represent typical line-ratio temperature estimates of the Perseus core (\S~\ref{sec:temp.fitting}). Inset zooms in from 6.5-8.5\,keV. The overall spectral shape, line ratios, and linewidths are extremely well-fit by a single-temperature model with modestly lower metallicity and a slightly different temperature, but still well within systematic X-ray spectroscopic temperature measurement uncertainties. Two-temperature models ($T\sim 3.5-5\,$keV) can fit the spectrum almost perfectly (to much smaller deviations than line thickness) over the entire range shown. The primary bias introduced by CR-IC is to density $n_{X}$, the secondary bias to metallicity $Z_{X}$, while effects on $T_{X}$ are weak (provided the CR-IC effective temperature $T_{\rm IC}$ is not too dramatically different from the gas temperature $T_{\rm gas}$, if the two emit comparably as considered here). 
	\label{fig:Z.spectral.demo.alt}}
\end{figure}

\subsection{On Temperature Constraints and X-ray Spectral Fitting: Line-Ratios, Continuum-Shapes, and Line Widths}
\label{sec:temp.fitting}

More generally, one can ask whether there are any spectral features in high-resolution, broad-wavelength-coverage (e.g.\ $\sim 0.1-10\,$keV), high S/N X-ray spectra which would be uniquely recognizable signatures of CR-IC in CCs. For example, if CR-IC contributes significantly to the soft X-ray continuum, one might wonder whether this would lead to highly discrepant estimates of the X-ray temperature in spectral fitting, comparing different methods for fitting $T_{X}$ to X-ray spectra assuming pure-thermal emission. In principle, from sufficiently high-resolution X-ray spectra, one can estimate electron temperatures from the continuum shape, excitation and ``ionization'' temperatures (the latter assuming strict collisional ionization equilibrium) from various line ratios, and kinetic ion temperatures from thermal line-broadening, and ask how these are influenced by CR-IC. 

\subsubsection{General Considerations and Multi-Wavelength Data}

Importantly, per \S~\ref{sec:softxray}, if the \textit{only} information available is X-ray spectroscopy, then without imposing \textit{any} additional strong priors, no spectral measurement can formally rule out or distinguish significant CR-IC from thermal emission. To illustrate this trivially, note that because the CR-IC X-ray spectrum is the linear superposition of that from CRs of different energies, each of which produces a thermal continuum shape after convolution with the CMB, there always exists a CR spectrum that exactly reproduces any given thermal continuum shape, and so the ratio of continuum emission from thermal vs.\ CR-IC becomes strictly degenerate with the underlying gas metallicity (i.e.\ degree of $Z$ suppression, per \S~\ref{sec:z.drops}). This is not to say the X-rays are not constraining: they set an upper limit to the CR energy density (it cannot be larger than what would produce all the observed X-rays, obviously) and (if CR-IC is important) strongly constrain the shape of the CR spectrum where it would emit in X-rays ($\sim 0.3-2\,$GeV). Obviously, one cannot have most of the observed X-rays coming from a CR spectrum peaked at $\sim 0.4\,$GeV (which would produce CR-IC with an ``effective'' X-ray temperature $T_{\rm IC} \sim 0.5\,$keV) in a CC with negligible emission at $<1\,$keV (e.g.\ with an effective spectral temperature $\sim 10\,$keV).   

For this reason, the most powerful constraints come from systems where multi-wavelength constraints exist and can provide strong priors. For example, optical/UV gas-phase measurements can anchor the true CC center metallicity $Z$ profile, spatially-and-spectrally-resolved low-frequency radio (especially combined with Faraday rotation measurements) can strongly constrain the CR lepton spectrum at modestly higher energies than observed in X-rays (e.g.\ $\sim 100\,$MHz through $10\,$GHz radio corresponding to $\sim 4-80\,$GeV CRs depending on details of $B$), and SZ measurements can anchor the true thermal pressure $P_{\rm true} = n_{\rm gas}\,k\,T_{\rm gas}$. With all of these combined, the X-ray spectra -- fit to a multi-component CR spectrum plus multiphase gas distribution constrained by these priors -- can be used to much more meaningfully test whether or not CR-IC contributes significantly to the X-rays. To our knowledge, there are no clusters at present with all three of these constraints. Even two of three exist only in a few, nearby, very bright systems (e.g.\ Perseus, Virgo, Centaurus), which will be the subject of more detailed study in future work.

There are also of course some ``reasonable'' priors that can be imposed without direct multi-wavelength observations in every individual cluster. For example, if a given cluster already shows a super-solar X-ray de-projected, single-temperature best-fit metallicity $Z_{X} > Z_{\odot}$ in its center (at both Fe-L and Fe-K energies), then it is reasonable to infer that one cannot have a soft X-ray CR-IC surface brightness much larger than thermal emission ($S_{X,\,\rm IC} \gg S_{X,\,\rm therm}$), as that would require the underlying ``true'' metallicity $Z_{\rm true} \gg Z_{X} \gg Z_{\odot}$ be order-of-magnitude larger than Solar. But such high metallicities seem implausible and are basically never seen where optical/UV observations exist. Constraints like these involve some reasonable implicit multi-wavelength priors from what we know of the population as a whole.

\subsubsection{Example Constraints from CCD X-ray Spectroscopy}

In other cases, X-ray spectroscopy can be used to constrain theoretical priors. For example, if one \textit{assumes} that the injection spectrum+losses+transport parameters are identical to the LISM plus simple toy models here, so that the CR spectra follow those in \S~\ref{sec:cr.spectrum} and therefore $T_{\rm IC}$ follows Fig.~\ref{fig:cric.temp.vs.age}, one can ask whether or not this is allowed to contribute an $\mathcal{O}(1)$ fraction of the X-ray spectrum in a given observed cluster. The strongest constraints here will of course come from high-spectral-resolution (microcalorimeter with $\sim$\,eV resolution), high signal-to-noise, and also \textit{spatially}-resolved observations (because one needs to resolve the interior of the CC where CR-IC is most important, \textit{and} be able to de-project a radially-varying mean gas temperature along the line of sight). For a typical CC with CCD spectra (e.g.\ from Chandra or XMM) there will be more caveats/degeneracies (see \S~\ref{sec:apparent}). Fig.~\ref{fig:Z.spectral.demo} is illustrative in this respect. In the group scale example there, even a large CR-IC contribution following the toy models above combined with a single-temperature, NCC ``background,'' produces a relatively small temperature bias in the combined spectrum of $<1\,$keV (which drops to $\lesssim 10\%$ if we combine CR-IC with a ``cooling-flow-like'' multi-temperature background). These are much smaller differences than typical systematic uncertainties or instrument-to-instrument or fitting-method/code-dependent variations in high-quality CCD spectral fits for $T_{X}$ in groups \citep{matsushita:2002.m87.virgo.cluster.obs.metallicity.drop.temperature.fitting.challenges,mazzotta:2004.xray.temperature.measurement.modeling.and.caveats,avestruz:2014.cluster.mocks.from.sims.sensitivity.temperature.measurements,zhuhone:2023.cluster.temperature.fitting.sensitivities.simulations}. 

The cluster-scale example in Fig.~\ref{fig:Z.spectral.demo} is intended to illustrate a more extreme hypothetical: we assume $T_{\rm gas} \sim 10$\,keV, with a much cooler CR-IC $T_{\rm IC} \sim 2$\,keV. Since these temperatures are quite distinct, the model is no longer degenerate with a single-temperature thermal model -- as shown, with a sufficiently high-quality CCD spectrum covering the entire range from $\sim 0.5-8\,$keV (capturing curvature at $\ll 2\,$keV and the Fe-K complex at $6-8$\,keV), one would be forced to adopt at least a two-temperature model with temperatures $\sim 3-6$\,keV to match the result. So there is clearly a point where even CCD spectroscopy can rule in or out models where CR-IC is important but $T_{\rm IC}$ is ``too different'' from $T_{\rm gas}$. 

This is discussed in the specific example of Zw 3146 in \citet{silich:2025.cr.ic.tests.in.zw.3146.cluster}, where the authors obtain SZ measurements to obtain a prior on $P_{\rm true}$. There, they show that while no thermal-only model (of the range they consider) can reproduce the X-ray spectrum \textit{and} the constraint $P_{\rm X} \equiv n_{X} k T_{X} \approx P_{\rm true}$ simultaneously, anchoring the gas pressure to $P_{\rm true}$ and the true metallicity to $\lesssim 1.5\,Z_{\odot}$ does require that the CR-IC at radii $\sim 30\,$kpc have $T_{\rm IC}\sim$\,a few keV, not too different from the $\sim 5-7\,$keV inferred from single-temperature pure-thermal fits. But as the authors note, a more robust constraint on what range of $T_{\rm IC}$ and/or $T_{\rm gas}$ is allowed, even with the SZ prior, requires marginalizing over additional choices such as the absorption correction (which can be degenerate with the continuum contribution of low-energy CR-IC at $\lesssim$\,keV), background models (more degenerate with the highest-energy constraints), and allowing for multi-temperature CR and gas models (not considered therein). Complicating the matter further, assessing formal ``goodness-of-fit'' through the standard techniques used in X-ray spectral fitting (with codes like XSPEC) requires that the ``null hypothesis'' be a good statistical representation of the data (with $\chi^{2}/\nu\approx 1$), which is almost never true in well-studied clusters (e.g.\ \citealt{arnaud:2011.handbook.xray.astronomy}, and  examples in Fig.~\ref{fig:Z.profiles.obs.only}-\ref{fig:Z.spectral.demo.alt} as well as the cases in \citealt{romero:2017.cluster.pressure.profiles.highres.sz.xray.cool.cores.show.central.pressure.deficit,silich:2025.cr.ic.tests.in.zw.3146.cluster}, and the Hitomi Perseus spectra discussed below). 

A more extreme example might be Virgo, as shown in Fig.~\ref{fig:Z.profiles.warn}. Per \S~\ref{sec:z.drops.obs}, there is a stark difference in the CCD spectral-fit metallicity profile from both Chandra and XMM within the CC ($\lesssim 10\,$kpc), depending on whether one fits the $\sim$\,keV Fe-L or $\sim 6-8$\,keV Fe-K region of the spectrum, implying ``preferential'' dilution of the low-energy spectrum, and this is exactly what spectral models invoking a combination of multi-temperature and additional non-thermal components have done in the past to reconcile the X-ray spectra with a single metallicity profile consistent with optical/UV observations \citep{molendi:2001.m87.abundance.profiles.xray.fit.out.with.multi.temp.plus.single.temp.plus.power.law.plus.background,russell:2018.m87.inner.kpc.multi.temp.and.strong.agn.maps.with.zdrop.many.signatures.of.cr.ic}. Those spectral fits allow for $T_{\rm IC}$ as low as $\sim 0.2-1$\,keV (though there can be CR-IC at all $T_{\rm IC} \lesssim 3\,$keV) with non-negligible thermal gas emission at $T_{\rm gas} \sim 2-4\,$keV. Similar behaviors are seen in Centaurus as well \citep{majumder:2025.phd.centaurus.profiles.chandra.and.some.xrism.data}. 

For a massive case like Perseus, from CCD spectroscopy which can spatially-resolve the core, the best-fit $T_{X}$ in the core at a given $R$ spans a factor $\sim 3$ range ($2-6$\,keV) along different azimuthal angles \citep{sanders:2004.perseus.profiles}, and the mean (azimuthally-averaged) $T_{X}$ differs between XMM measurements in \citet{churazov:2003.perseus.profiles} and Chandra measurements in \citet{sanders:2004.perseus.profiles} by a factor $>2$ at $<10\,$kpc (the inner core, where XMM recovers hotter $T_{X}$ up to $\sim 12\,$keV) and at $60-100\,$kpc (the outer core, where Chandra recovers hotter $T_{X}$), allowing for central temperature components containing an $\mathcal{O}(1)$ fraction of the luminosity between $\sim 2-10\,$keV (in part degenerate with how backgrounds and AGN scattered light contributions are modeled/subtracted).

\subsubsection{Example Constraints from High-Resolution X-ray Spectroscopy}

By far the most detailed high spectral-resolution X-ray temperature modeling of a (angularly-resolved, high signal-to-noise) CC comes from the Hitomi studies of the Perseus core \citep{hitomi:2018.perseus.temperature.structure.hard.emission}, integrated within a central $\sim 80\,$kpc aperture. So in Fig.~\ref{fig:Z.spectral.demo.alt}, we consider a simple spectral comparison akin to Fig.~\ref{fig:Z.spectral.demo} but for Perseus-core-like parameters, taking a model from Fig.~\ref{fig:profiles} with Perseus (NGC 1275)-like CR lepton jet injection rate \citep[e.g.][]{tavecchio:2014.jet.leptonic.luminosity.ngc.1275.2e45,hodgson:2021.perseus.jet.minimum.kinetic.luminosity.gamma.rays} and surface-brightness at $\lesssim 80\,$kpc (the strong CC), with a simple single-temperature thermal background of $\sim 4.5\,$keV taken to be similar to the Hitomi line-ratio observations. We see that, akin to the less-extreme cases in Fig.~\ref{fig:Z.spectral.demo}, the CR-IC contribution, while dominant in $L_{X}$ in the core, biases the best-fit temperature by only $\sim 10\%$ to $\approx 4.1$\,keV, and reproduces the line ratios and linewidths at $1$\,eV resolution. 
For comparison, fitting the core region \citet{hitomi:2018.perseus.temperature.structure.hard.emission} found: 
(a) different lines and line-ratio combinations give different best-fit temperatures ranging from $2.7-5\,$keV; 
(b) additional temperature components ranging from $0.5-8$\,keV are required to simultaneously match the Hitomi, XMM, and Chandra spectra; 
(c) modifying the instrumental background model (ARF) can vary the range of $T$ in two-temperature models to as large as $\sim 2.5-20\,$keV; 
and 
(d) minor changes in the assumed atomic line models (e.g.\ adopting versions 3.0.8 vs.\ 3.0.9 of AtomDB) can shift the favored temperature range from $\sim 3-5$\,keV to $\sim 1.6-4\,$keV. 
Also as noted above all of the models considered have $\chi^{2}/\nu \gg 1$, consistent with their being prior+systematics dominated.
And even with these spectra the ion temperature measured from line broadening is only constrained to be in the range $\lesssim 17\,$keV \citep{hitomi:2018.perseus.core.turbulence.temperatures}. 
This is all to illustrate that the more realistic $\sim 10\%$ temperature bias introduced by CR-IC is very small compared to both (1) systematic observational uncertainties, and (2) real physical uncertainties and/or multi-temperature effects. 

So, even in Perseus with Hitomi-quality spectra, significant CR-IC components with ``effective temperature'' $T_{\rm IC}$ in the range $\sim 0.5-20\,$keV are allowed, though a \textit{dominant} CR-IC component averaged within the central $\sim 80\,$kpc aperture would require a aperture-emission-averaged effective $T_{\rm IC}$ closer to the range $\sim 2.5-5$\,keV. 

Interestingly, in Fig.~\ref{fig:Z.spectral.demo.alt}, there is a very weak residual trend where the best-fit single-temperature thermal model fit to different frequency ranges (including the lines) gives slightly different $T_{X}$, increasing from $T_{X} \sim 3.9\,$keV at $E_{X} \sim 3\,$keV to $T_{X} \sim 4.1-4.2\,$keV at $E_{X} \gtrsim 7\,$keV. This is strikingly similar to (although somewhat weaker in magnitude than) the pattern seen in Hitomi observations \citep{hitomi:2018.perseus.temperature.structure.hard.emission} of the Perseus core (see e.g.\ their Fig.~5, ``entire core'' region) as one goes from lines at $\sim 2.5-3$\,keV (S \&\ Si Ly$\alpha$) to those at $\gtrsim 6.5\,$keV (Fe complex). So it even appears that the sense of residuals/deviation from single-temperature thermal-only spectra is the same as observed, in this example.
All of this could, of course, be the result of genuinely multi-temperature gas in these systems (and some of it must be). 
Our point here is simply that current X-ray spectroscopy from $0.1-10$\,keV, even in the best-studied clusters, is plausibly consistent with an $\mathcal{O}(1)$ contribution of CR-IC to $L_{X,\,\rm cool}$. 

    

\subsubsection{On Coupling Between CR Spectra and Cluster Thermodynamic Properties}

For simplicity, in our toy models we assume that the CR spectra and cluster thermodynamic properties are decoupled -- e.g.\ in Fig.~\ref{fig:cric.temp.vs.age}, there is no explicit dependence of CR spectra (and therefore $T_{\rm IC}$) on gas temperature $T_{\rm gas}$. But there are many mechanisms which could couple the two in principle. (1) Since our ``injection zone'' refers to the collective central/young radio core/galaxy/jet region at $\ll 10\,$kpc, even if the true CR acceleration spectrum is universal, the effective injection-zone spectrum could be shaped by losses and/or reacceleration (diffusive/turbulent, shock, or adiabatic/convective) which will depend on gas densities, stellar+AGN radiation and magnetic fields, and gas velocity fields, all of which could depend directly or indirectly on cluster thermodynamic properties like $T_{\rm 500}$, $R_{\rm 500}$, etc. (2) In propagation, CRs can ``feel'' thermodynamic properties: not only are CR losses sensitive to cluster properties ($B$, $u_{\rm rad}$, $n_{\rm gas}$), but in almost any first-principles model for CR transport, the transport properties/speeds are also sensitive to these (through e.g.\ different scattering wave-growth and damping processes having different dependencies on properties like magnetic fields, turbulence, temperature, CR flux, density, spatial scale, metallicity, etc.). The challenge, as discussed in e.g.\ \cite{hopkins:2020.cr.transport.model.fx.galform,hopkins:cr.transport.constraints.from.galaxies,hopkins:2021.sc.et.models.incompatible.obs,hopkins:2025.crs.review}, is that different assumptions produce qualitatively different scalings with these parameters, and there are almost no direct observational constraints outside the local ISM. And (3) gas in the ICM can ``feel'' CR properties, through both direct heating/ionization (\S~\ref{sec:excited}), but also through CR pressure and dynamical effects (\S~\ref{sec:discussion}) which will modify thermal instability, drive buoyant motions, and differentially accelerate different gas phases. 

Thus while our toy models here effectively assume $T_{\rm IC}$ and $T_{\rm gas}$ are independent, it is completely plausible to construct models where the two are strongly coupled in various ways. 
Given the large uncertainties in all of (1)-(3), the question of how low-energy CR spectra (hence $T_{\rm IC}$) do or do not systematically depend on cluster properties must be an empirical one. A wide range of potential scalings may be plausible, but if on the other hand the only way to reproduce X-ray spectra were to require $T_{\rm IC}= T_{\rm gas}$ \textit{exactly}, this would probably require an implausible degree of fine-tuning. 
So if CR-IC is indeed important in some clusters, X-ray spectral fitting presents a unique and powerful tool to address these questions.

\subsection{Apparent ``Jet Power'' Versus ``Cooling Luminosity''}
\label{sec:pjet}

\begin{figure}
	\centering\includegraphics[width=0.99\columnwidth]{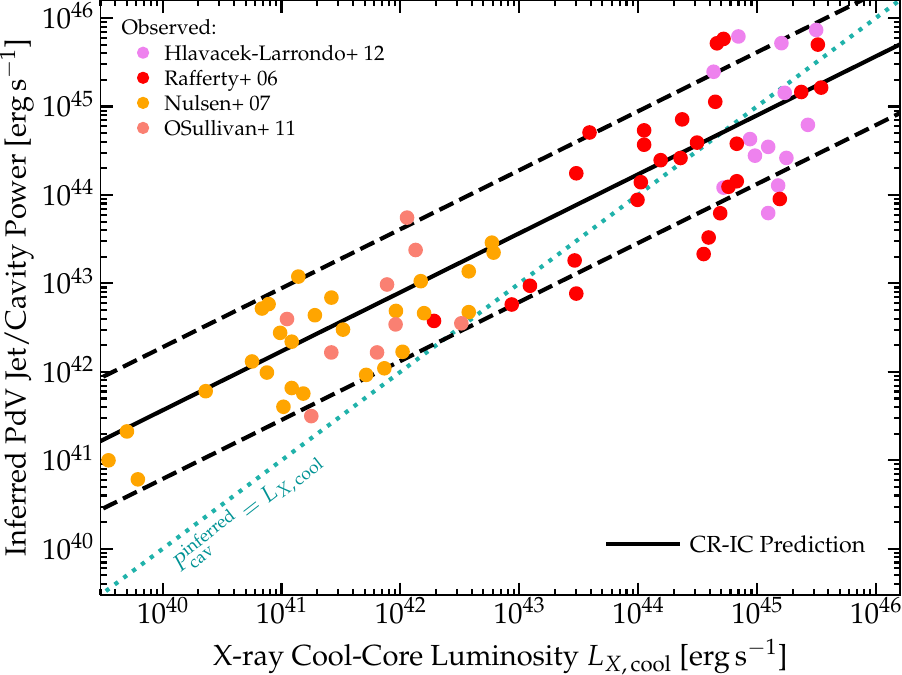} 
	\caption{Apparent ``cavity'' or ``jet'' power $P_{\rm cav} \sim 4\,P_{\rm eff}\,V/t_{\rm buoy}$ (in terms of ambient apparent X-ray inferred pressure $P_{\rm eff}$ and buoyancy time $t_{\rm buoy}$ plus cavity volume $V$) versus apparent X-ray cool-core ``cooling luminosity'' $L_{X,\,{\rm cool}}$. We show various observational compilations (\textit{points}).
	Dotted green line shows the standard interpretation that $P_{\rm cav}$ from AGN feedback represents work balancing X-ray cooling.
	Thick black lines show the \textit{prediction} from \S~\ref{sec:pjet} if we assume that the apparent cooling luminosity is dominated by ACRH IC emission ($L_{\rm X,\,{\rm cool}} \sim L_{\rm IC}$, so $P_{\rm eff}$ and $t_{\rm buoy}$ follow the apparent CR-IC-dominated scalings in \S~\ref{sec:apparent}). Solid/dashed lines adopt the median/$90\%$ range of the observed volume filling-factor of cavities (the only other input needed). Remarkably, this appears to fit the data and its scatter better than the standard model (notably the non-linear slope), without any tunable/fitted parameters. In the ACRH interpretation, this correlation arises because $P_{\rm cav}$ and $L_{\rm X,\,cool}$ actually measure \textit{the same thing}, namely the IC emission from the ACRH (as this determines $P_{\rm eff}$ and $t_{\rm buoy}$ within a given emitting volume). 
	\label{fig:pjet}}
\end{figure}

One of the arguments most often cited for the ``standard'' thermal interpretation of the CF problem is the apparent correlation between ``jet'' or ``cavity power'' ($P_{\rm jet}$, $P_{\rm cav}$) and ``cooling luminosity'' ($L_{X,\,{\rm cool}}$). The latter is essentially the soft X-ray luminosity interior to the cooling radius. The former is defined by the apparent ``PdV work'' associated with X-ray cavities or bubbles. Specifically, the most common observational method to define $P_{\rm cav}$ is given by $P_{\rm jet} \equiv P_{\rm cav} \equiv 4\,p_{\rm ext}\,V_{\rm cav}/t_{\rm buoy}$, where the $4$ comes from assuming an ultra-relativistic fluid filling the cavity, $p_{\rm ext} \equiv P_{X,\,{\rm ext}} \equiv n_{\rm app,\,ext}\,k_{B}\,T_{\rm ext}$ is the apparent X-ray inferred (as \S~\ref{sec:sz}) thermal pressure from the X-ray surface brightness profile and spectra outside the cavity radius (at the same $R$), $t_{\rm buoy} \equiv R/v_{\rm buoy} \equiv R \sqrt{S\,C/2\,g\,V_{\rm cav}}$ is an approximate simple analytic estimate of the buoyancy time assumed for such a cavity to inflate/rise to distance $R$ (defined usually as the distance from the cavity center to the BCG center at $R=0$), taking $S\equiv \pi R_{\rm cav}^{2}$ to be the cross-sectional area (in terms of the effective cavity radius $R_{\rm cav}$), $V_{\rm cav} \equiv (4\pi/3)\,R_{\rm cav}^{3}$ the volume, $C\equiv 0.75$ a constant, $g\equiv 2\,\sigma^{2}_{\ast,\,\rm gal}/R$ the acceleration in terms of an assumed-constant isothermal velocity $\sigma_{\ast,\,\rm gal} \approx \langle \sigma\rangle_{\ast,\,\rm gal} \sim 280\,{\rm km\,s^{-1}}$ (taken as a constant in most studies), $R_{\rm cav} \sim  ({a_{\rm cav}\,b_{\rm cav}})^{1/2}$ in terms of the semiminor/major cavity axis, and $R\sim R_{\rm cav}/\alpha$ the projected distance from cavity center to radio core. The observed correlation commonly cited, with all compiled data points from \citet{birzan:2004.radio.cavity.jet.power.vs.xray.power.correlation,rafferty:2006.cluster.cavity.jet.power.vs.xray.luminosity.agn.feedback.arguments.but.accretion.rates.dont.match.obvious,rafferty:2007.xray.radio.cluster.data.compilation,nulsen:2007.cluster.jet.power.versus.cooling.luminosity.correlation.compilation,mcnamara:2007.agn.cooling.flow.review.cavity.jet.power.vs.xray.luminosity.scalings.emph.compilation,osullivan:2011.radio.jet.power.radio.emission.and.xray.cooling.luminosities,hlavacek.larrondo:2011.cluster.radio.cavity.jet.power.vs.cooling.xray.luminosity.extension,hlavacek.larrondi:2012.expanding.sample.luminous.clusters.xray.cooling.luminosity.vs.radio.cavity.jet.power}, is shown in Fig.~\ref{fig:pjet}.

What is the prediction of the CR-IC scenario? Per \S~\ref{sec:sz} in this case, $p_{\rm ext}$ that would be measured is not the true gas pressure but is dominated by the ``apparent'' $P_{\rm app} = n_{\rm app} \,k_{B}\,T_{\rm app}$, giving $P_{\rm cav} \sim 10^{44}\,{\rm erg\,s^{-1}}\,\alpha_{1/3}^{7/2}\,n_{\rm app,-3}\,R_{100}^{2}\,T_{\rm app,\,keV}$ (with $\alpha_{1/3} \equiv \alpha/(1/3)$, $n_{\rm app,\,-3} \equiv n_{\rm app}/10^{-3}\,{\rm cm^{-3}}$, $R_{100} \equiv R/100\,{\rm kpc}$, $T_{\rm app,\,keV}\equiv T_{\rm app}/{\rm keV}$). But also in this limit $L_{X,\,\rm cool} \sim L_{\rm IC}$ scales as above, and we can insert our expressions for $n_{\rm app}$ and $T_{\rm app,\,keV}$ above (with the size of the X-ray cooling region $R_{X} \sim \alpha_{X}\,R$). This gives $P_{\rm cav} \sim 1.4 \times 10^{44}\,e_{\rm cr,\,eV}^{-1/6}\,\mathcal{F}_{\rm IC}^{-1/6}\,\mathcal{F}_{\rm ff}^{-1/2}\,(1+z)^{1/3}\,T_{\rm app,\,keV}\,(\alpha_{1/3}/^{7/2}/\alpha_{X}^{2})\,L_{X,\,44}^{2/3} $, i.e.\ $P_{\rm cav,\,44} \sim L_{X,\,44}^{2/3}\,(\alpha_{1/3}^{7/2}/\alpha_{X}^{2})$ (with $L_{X,\,44} \equiv L_{X}[R<R_{\rm cool}]/10^{44}\,{\rm erg\,s^{-1}}$, and $e_{\rm cr,\,eV} \equiv e_{\rm cr}(R=R_{\rm cool})/{\rm eV\,cm^{-3}}$ ). Although the dependence on $e_{\rm cr}$ is already extremely weak (the $1/6$ power), we can eliminate it entirely using the definition of $R_{\rm cool}$ (see \S~\ref{sec:rcool}),  
Neglecting the very weak dependence on order-unity numbers like $\mathcal{F}_{\rm IC}$, this gives the \textit{predicted}: 
\begin{align}
\frac{P_{\rm cav}}{10^{44}\,{\rm erg\,s^{-1}}} \sim \frac{1.5}{(1+z)}\,\left( \frac{L_{X,\,\rm cool}}{10^{44}\,{\rm erg\,s^{-1}}} \right)^{2/3}  \frac{\alpha_{1/3}^{7/2}}{\alpha_{X}}\ .
\end{align} 
This is shown in Fig.~\ref{fig:pjet}.

Note that this predicts a $P_{\rm cav}-L_{X,\,\rm cool}$ correlation which depends appreciably only on the geometric parameter $\alpha^{7/2}/\alpha_{X}^{2}$, which is not a ``free'' or ``adjustable'' parameter of the model, but is directly measured in the observations. It is approximately the observed volume filling factor $f_{V,\,{\rm cav}}^{X}$ -- i.e.\ the fraction of the volume of the apparent X-ray cooling region (the region in which IC dominates the emission, in this toy model) which is filled by the cavities. In the observed samples compiled, median values are $\alpha \sim 0.5$ and $\alpha_{X} \sim 2$, equivalent to $\alpha \sim 1/3$, $\alpha_{X} \sim 1$, but the values range from $\alpha \sim 0.2-0.7$ and $\alpha_{X} \sim 1-4$, with the effective $f_{V,\,{\rm cav}}^{X}$ having a factor $\sim 30$ spread. We use these values to illustrate upper/lower intervals in Fig.~\ref{fig:pjet}.

While the derivation above is somewhat opaque, the physical reason this correlation emerges in the CR-IC scenario is trivial: \textit{both $P_{\rm cav}$ and $L_{X,\,\rm cool}$ are measuring different manifestations of the same thing}. $P_{\rm cav}$ does not (entirely) trace the true ``PdV work'' done by the cavities/jets because $p_{\rm ext}$ and $t_{\rm buoy}$ are not tracing the true pressure and buoyancy time of the gas, but just different apparent functions of the lepton energy density $e_{\rm cr,\,\ell}$. Likewise, $L_{X,\,\rm cool}$ does not trace the true thermal cooling rate of the gas, but rather the CR-IC emission, which is just proportional to $e_{\rm cr,\,\ell}$. So the two necessarily trace one another up to the volume filling factor term above. Remarkably, the CR-IC predicted correlation not only agrees with observed CC clusters in Fig.~\ref{fig:pjet}, but it predicts the normalization, slope, and scatter of the relation, \textit{and} correctly predicts that it is, in fact, sub-linear. Indeed it appears to predict this relation significantly more accurately than the standard (thermal) interpretation.

\subsection{Apparent ``Cooling Radii''}
\label{sec:rcool}

\begin{figure}
	\centering\includegraphics[width=0.99\columnwidth]{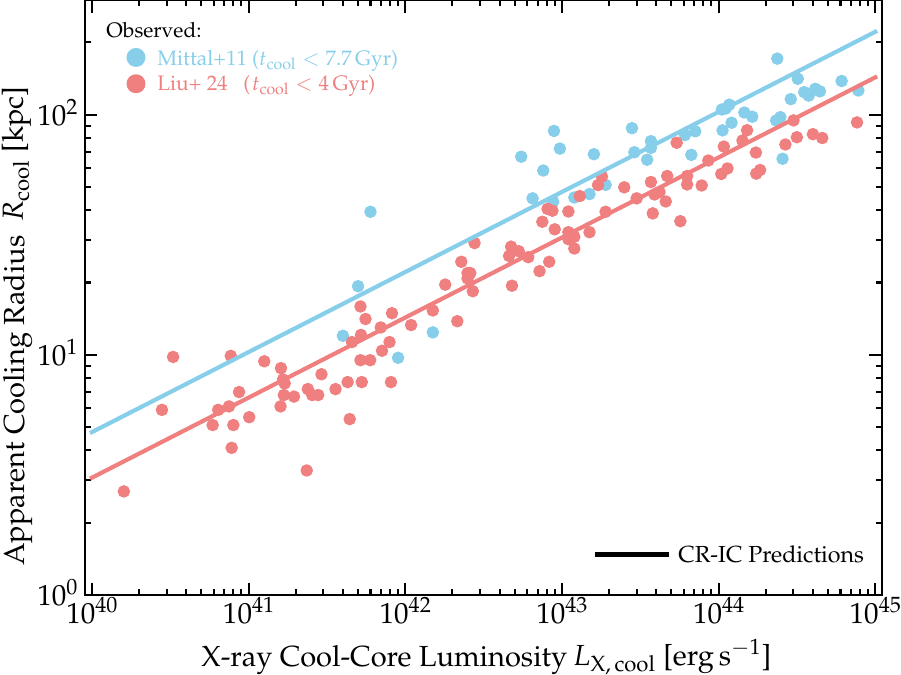} 
	\caption{Apparent ``cooling radius'' (defined as the radius where the X-ray inferred gas cooling time would drop below a critical value of $4$ or $7.7$\,Gyr, labeled), versus apparent cool-core luminosity. We compare observed clusters compiled in \S~\ref{sec:rcool} to the predicted relation if the ``cooling'' luminosity is dominated by CR-IC (Eq.~\ref{eqn:rcool}).This relation emerges automatically because of how the apparent temperature and density scale with $e_{\rm cr,\,\ell}$ (which determines $L_{\rm cool}$) in the CR-IC regime.
	\label{fig:rcool}}
\end{figure}

The common definition of CC ``size'' $R_{\rm cool}$ is the radius where the cooling time $t_{\rm cool} \approx t_{c,\,7}\,7\,$Gyr (we use the parameter $t_{c,\,7}$ since some studies adopt $4-10\,$Gyr definitions). But in the CR-IC scenario $t_{\rm cool} \equiv e_{\rm therm}/\epsilon \sim (3/2)\,n_{\rm app}\,k_{B}\,T_{\rm app} / \epsilon_{\rm IC}$. Inserting the definitions of these from \S~\ref{sec:apparent} and solving for $R_{\rm cool}$ allows us to immediately note that if, at $R_{\rm cool}$, the surface brightness is CR-IC dominated, then the leptonic energy density must be $e_{\rm CR,\,eV}(R\sim R_{\rm cool}) \sim 2\,(1+z)^{-4}\,T_{\rm app,\,keV}^{2}/(\mathcal{F}_{\rm IC}\,\mathcal{F}_{\rm ff} \, t_{c,\,7}^{2}) \sim 2\,(1+z)^{-2}$. Using $L_{\rm X,\,cool} \sim (4\pi/3)\,R_{\rm cool}^{3}\,\epsilon_{\rm IC}(R\sim R_{\rm cool})$, this then allows us solve for $R_{\rm cool}$ in terms of $L_{\rm X\,cool}$, obtaining $R_{\rm cool,\,100} \sim 0.93\,(t_{c,7}/T_{\rm app,0,\,keV})^{2/3} (\mathcal{F}_{\rm ff}\,L_{X,43})^{1/3}(R<R_{c})$ or more simply
\begin{align}
\label{eqn:rcool} R_{\rm cool} \sim 45\,{\rm kpc}\,(3\,t_{c,7}/T_{\rm app,0,\,keV})^{2/3}\left( \frac{L_{X,\,\rm cool}}{10^{43}\,{\rm erg\,s^{-1}}} \right)^{1/3}\ .
\end{align}

Fig.~\ref{fig:rcool} plots this predicted correlation against observed CC clusters compiled from \citet{liu:2024.strong.lradio.lxray.connection.luminous.cool.cores.strong.central.xray.only.when.strong.radio.source} and \citet{mittal:2011.xray.cluster.scalings.includes.rcool.lcool} (themselves with X-ray datasets including \citealt{cavagnolo:cluster.entropy.profiles} and \citealt{hudson:2010.cool.core.cluster.review.central.properties.temperature.entropy.coolingtime.definition.coolcore.basic.scalings.size.luminosity.mdot}, respectively). We see this appears to explain the correlations observed, with modest scatter owing to e.g.\ the details of profile shapes, variation in redshift (all else held fixed; see \S~\ref{sec:redshift}), difference between $L_{X,\,\rm cool}$ estimated via different techniques and calculated this way as compared to the bolometric $L_{\rm IC}$, etc. It also correctly predicts the dependence of $R_{\rm cool}$ on $t_{\rm c}$. Again, we stress this is a prediction with no free or adjustable parameter, as opposed to a fit. So the sizes of CC clusters appear to be naturally explained in the CR-IC scenario, as well. Again, in contrast, there is no particular obvious reason for a specific value of $R_{\rm cool}$ or $R_{\rm cool}-L_{X,\,\rm cool}$ relation, in the standard (thermal) interpretation.

For comparison the radius where the CR-IC will begin to be strongly truncated by losses, and so CR-IC surface brightness will fall below any background thermal profile, is 
\begin{align}
R_{\rm cool}^{\rm max} \sim (100-300)\,{\rm kpc}\,(1+z)^{-4}\,v_{\rm st,\,100}
\end{align}
(with the exact pre-factor depending on $\dot{E}_{\rm cr}$ and the background profile brightness). This is quite similar to the largest $R_{\rm cool}$ over the range of plausible redshifts, and suggests that even for the most extremely luminous sources $L_{X,\,\rm cool} \sim 10^{45}\,{\rm erg\,s^{-1}}$, $v_{\rm st,\,100}$ only needs to be as large as $\sim 1-5$ (a plausible value in such extreme systems) for the CR-IC halo to extend over the entire cooling radius. But if the CR-IC loss radius is much smaller than $R_{\rm cool}$ as estimated above, the ``apparent'' CC will simply have a smaller radius, consistent with some of the scatter and tentative behavior of the very highest-luminosity systems at $L_{X,\,{\rm cool}} \sim 10^{45}\,{\rm erg\,s^{-1}}$  in Fig.~\ref{fig:rcool}.

\subsection{Radio Power Versus ``Cooling Luminosity'' and the SCC-Radio Connection}
\label{sec:Lradio.LX}

\begin{figure}
	\centering\includegraphics[width=0.99\columnwidth]{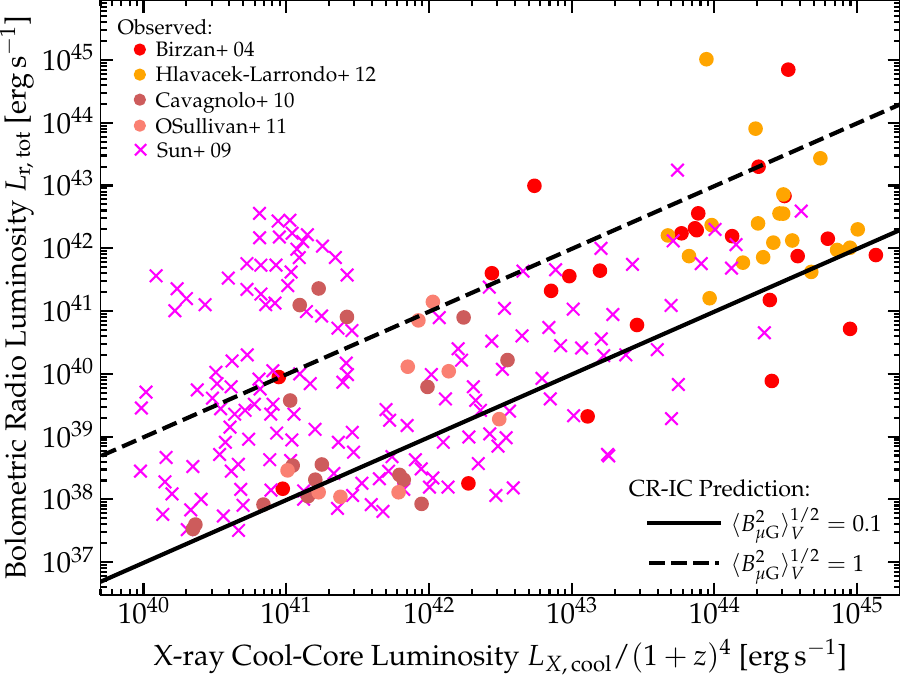} 
	\caption{Bolometric radio luminosity $L_{\rm r,\,tot}$ (integrating over all wavelengths observed from the compilations shown; \textit{circles}, or using a simple spectral index correction from single-wavelength GHz observations, $\times$'s; \textit{labeled}) versus apparent X-ray cool-core ``cooling luminosity'' $L_{X,\,{\rm cool}}$. 
	Lines show the predicted correlation for different values of the volume-filling diffuse-gas magnetic field strengths at core radii $\sim 100\,$kpc (where $L_{X,\,{\rm cool}}$ arises), assuming that the $L_{X,\,{\rm cool}}$ comes primarily from ACRH IC emission, for typical observationally-inferred diffuse $B$ strengths (\S~\ref{sec:Lradio.LX}). 
	The radio arises from much smaller spatial/timescales as it depends on high-energy leptons which lose energy in $\sim 10^{7}\,$yr (Fig.~\ref{fig:synch}), while the X-ray comes from $0.1-1\,$GeV leptons with lifetimes $\gtrsim$\,Gyr, so there should be scatter owing to time-variability of the injection sources over $\sim$\,Gyr as well. 
	\label{fig:lradio}}
\end{figure}

\subsubsection{Radio Luminosity versus X-ray Cooling Luminosity: Global Correlations and Local Variations}
\label{sec:Lradio.LX.cool}

It is also well-established that there is a correlation between bolometric radio galaxy/AGN/core luminosity of clusters $L_{r,\,\rm tot}$ and apparent cooling luminosity $L_{\rm X,\,\rm cool}$, for X-ray luminous clusters. We show this in Fig.~\ref{fig:lradio}, with the data compiled in Fig.~\ref{fig:pjet} \citep{birzan:2004.radio.cavity.jet.power.vs.xray.power.correlation,rafferty:2006.cluster.cavity.jet.power.vs.xray.luminosity.agn.feedback.arguments.but.accretion.rates.dont.match.obvious,rafferty:2007.xray.radio.cluster.data.compilation,nulsen:2007.cluster.jet.power.versus.cooling.luminosity.correlation.compilation,mcnamara:2007.agn.cooling.flow.review.cavity.jet.power.vs.xray.luminosity.scalings.emph.compilation,osullivan:2011.radio.jet.power.radio.emission.and.xray.cooling.luminosities,hlavacek.larrondo:2011.cluster.radio.cavity.jet.power.vs.cooling.xray.luminosity.extension,hlavacek.larrondi:2012.expanding.sample.luminous.clusters.xray.cooling.luminosity.vs.radio.cavity.jet.power}, plus the addition of data not in that compilation from \citet{sun:2009.every.radio.agn.has.xray.cc.strong.lradio.lxray.connection.clusters.all.xray.agns.also.radio.agns}. Note we emphasize datasets where multi-wavelength observations allow $L_{r,\,\rm tot}$ to be calculated, but also show the (more approximate) set of clusters for which only $\sim$\,GHz radio data exist, corrected to bolometric using the GHz spectral slope. We plot $L_{r,\,\rm tot}(L_{\rm X,\,\rm cool})$, but similar studies have shown that almost every strong radio AGN appears to also exhibit an X-ray cool-core \citep[e.g.][and references therein]{sun:2009.every.radio.agn.has.xray.cc.strong.lradio.lxray.connection.clusters.all.xray.agns.also.radio.agns,mittal:2009.radio.cluster.properties.vs.xray.strong.radio.fraction.increases.with.coolcore.strength.all.scc.strong.radio.correlation.with.mdot.lx,liu:2024.strong.lradio.lxray.connection.luminous.cool.cores.strong.central.xray.only.when.strong.radio.source}. 

In the CR-IC scenario, such a correlation is trivial, because $L_{\rm r,\,\rm tot}$ is just driven by the young CR leptons being injected, so proportional to the ``instantaneous'' or $\lesssim 10^{7}$\,yr time-averaged lepton injection rate $\dot{E}_{\rm cr,\,\ell}$, while $L_{\rm X,\,\rm cool}$ is given by the old CR leptons in the ACRH, so proportional to the $\sim 10^{9}$\,yr time-averaged lepton injection rate $\langle \dot{E}_{\rm cr,\,\ell} \rangle_{\rm Gyr}$. In other words, Fig.~\ref{fig:lradio} basically plots a quantity against its time-average. So there should be scatter, corresponding to both variation in time and magnetic field strength (which determines the steady-state proportionality factor), but we would expect a roughly linear correlation, as observed, even though (as we emphasized above) the radio comes from \textit{different} regions in space and time, compared to the X-rays. More specifically, if both the IC (soft X-ray) and synchrotron (radio) come from the same parent distribution  of relativistic electrons, then the ratio of their \textit{bolometric} luminosities (what we plot in Fig.~\ref{fig:lradio} since both are spectrally-integrated) is simply given by $L_{X,\,\rm cool} \sim ( B_{\rm cmb}^{2} / \langle |{\bf B}|^{2} \rangle_{\rm Vol})\,\langle L_{r,\,\rm tot} \rangle_{\rm Gyr}$, where $B_{\rm cmb} \equiv 3.2\,{\rm \mu G}\,(1+z)^{2}$, $\langle |{\bf B}|^{2} \rangle^{1/2}_{\rm Vol}$ represents the mean magnetic field strength or energy density \textit{averaged over the entire volume of the X-ray emitting ACRH}, and $\langle L_{r,\,\rm tot} \rangle_{\rm Gyr}$ represents the average radio luminosity \textit{averaged over the entire lifetime ($\sim$\,Gyr) of the X-ray emitting low-energy electrons}. More simply:
\begin{align}
\frac{L_{\rm X,\,cool}}{(1+z)^{4}} \sim \frac{10}{\langle B_{\rm \mu G}^{2} \rangle_{\rm Vol}} \langle L_{r,\,\rm tot} \rangle_{\rm Gyr}\ .
\end{align}

These predicted correlations are shown in Fig.~\ref{fig:lradio} for reasonable estimates of the $\sim 100\,$kpc diffuse volume-filling magnetic field strength $\langle B_{\rm \mu G}^{2} \rangle_{\rm Vol}^{1/2} \sim 0.1-1$ motivated by Faraday rotation from background lines of sight and other probes that actually sample the diffuse gas (as compared to sychrotron-sensitive probes that sample only the injection regions), in \citet[][]{garrington:1991.faraday.depolarization.density.times.magnetic.field.1eminus3.cm3.microG,carilli:2002.cluster.b.fields.obs.collection,rudnick:2003.cluster.rms.from.non.strong.synch.emitting.volume.filling.region.from.polarization.must.be.below.0pt4microG,govoni:2004.B.fields.galaxy.clusters.review,xu:2006.rm.measurement.superclusters.0pt5microG.fields.inferred.could.be.mostly.in.sources,pereztorres:2009.ophiuchus.minihalo.xray.ic.contrib.significant.B.0pt03to0pt3microG,mirakhor:2022.ic.cluster.detection.B.0pt1microG}.\footnote{We stress again we are interested in the diffuse, volume-filling $\langle B_{\rm \mu G}^{2} \rangle_{\rm Vol}^{1/2}$ at $\sim 100\,$kpc, where these measurements argue generally for $\langle B_{\rm \mu G}^{2} \rangle_{\rm Vol}^{1/2} \sim 0.03-1\,{\rm \mu G}$. Larger values of a few ${\rm \mu G}$ from some synchrotron rotation measure studies, are shown from multi-wavelength and/or spatially-resolved radio, X-ray and depolarization studies to primarily come from a very small region ($\sim 1\,$kpc) around the acceleration/injection regions \citep{bicknell:1990.cluster.jet.rotation.measures.modeling.rm.with.high.b.comes.from.very.small.region.around.acceleration.region,feretti:1995.strong.B.from.cluster.RMs.but.1kpc.from.synchrotron.outer.more.0pt1.micron,taylor:2002.strong.B.in.clusters.from.synchrotron.rotation.measures.self,vogt:2003.cluster.B.faraday.high.values.in.emitting.regions.not.cluster.as.a.whole,osinga:2022.synchrotron.depolarization.clusters.strongB.but.perhaps.in.sources}. This is predicted given the strong sensitivity of such measurements (in an integrated sense) to $B$, which means they are necessarily dominated by the highest-$B$ sub-volumes of the cluster \citep{ponnada:2023.fire.synchrotron.profiles,martin.alvarez:2023.mhd.cr.sims.synch.maps.similar.emission.regions.conclusions.to.fire.ponnada.papers.but.very.different.methods,whittingham:2024.cluster.Bfield.synch.measurements.biased.to.strongest.B.subregions}.} 
We see that the observed clusters lie along this relation for all $L_{\rm X,\,\rm cool} \gtrsim 3\times 10^{41}\,{\rm erg\,s^{-1}}$. For the very weak CCs with $L_{X,\,\rm cool} \lesssim 2\times 10^{41}\,{\rm erg\,s^{-1}}$, there is much larger scatter, if one only uses the $1.4\,$GHz-inferred $L_{r,\,\rm tot}$, but interestingly (though the sample size is small) the correlation remains reasonable if we only consider the clusters with multiwavelength $L_{r,\,\rm tot}$ coverage. So it may be this correlation persists over the whole range, but GHz radio becomes an increasingly un-reliable proxy for $L_{r,\,\rm tot}$ at low-$L_{X,\,\rm cool}$. Alternatively, if the systems with apparently large GHz radio luminosity relative to $L_{X,\,\rm cool}$ are ``rejuvenating'' there may not have been time yet for the ACRH to develop -- i.e.\ these may be caught in the ``un-contaminated cooling flow'' stage (see \S~\ref{sec:evolution}). But another plausible explanation is simply that at low $L_{X,\,\rm cool}$, the identification of the systems as CC at all, let alone robust calculation of $L_{X,\,\rm cool}$, is challenging.

For individual well-studied clusters where there is also time-and-spatially-resolved Faraday rotation, even stronger constraints are possible, and these will be studied in future work with detailed modeling of individual systems. But for example, in M87/Virgo, time and spatial variability places a strong upper limit on the rotation measure from diffuse gas at $\gtrsim$\,kpc of ${\rm RM} \lesssim 130\,{\rm rad\,m^{-2}}$ \citep[][and references therein]{park:2019.m87.rms.detailed.modeling.huge.spatial.temporal.datasets.strong.upper.limits.outside.bondi.radius,tseunetoe:2022.m87.detailed.rm.modeling.jet.sheaths.and.bases.not.larger.icm,peng:2024.m87.jet.structure.tiny.scale.structure.rms.inside.jet.and.very.close.to.smbh}, or $B\lesssim {\rm \mu G}\,(n_{e}/0.01\,{\rm cm^{-3}})^{-1}\,(R/{\rm 10\,kpc})^{-1}$. Given the large central densities $n_{X}$ inferred if one assumes the X-rays are entirely thermal in origin (which therefore lower $B$), this makes it hard to reconcile any pure-thermal model with the observed radio and X-ray luminosities without predicting some significant CR-IC contribution. 

We also stress that many of the studies cited above, as well as e.g.\ \citet{giacintucci:2011.cluster.minihalo.fills.cool.core.morphologically.similar,giancintucci:2019.expanding.radio.cluster.minihalo.sample.no.good.corr.total.cluster.mass.or.total.cluster.xray.but.very.strong.corr.cooling.radius.xray.luminosity.consistent.with.linear.standard.correlation,bravi:2016.minihalo.luminosity.strong.corr.xray.luminosity.clusters,2023MNRAS.524.4939M,balboni:2024.lofar.xray.surface.brightess.corr.indiv.halos.well.corr.positive.as.expected.for.ic.same.particles.but.sublinear.because.spread.by.B.scatter,riseley:2023.cluster.radio.minihalo.correlated.xray.brightness.disturbed.re.energized.example,riseley:2024.large.minihalo.reenergized.with.merger.but.center.shows.strong.radio.xray.with.steeper.radio.hotter.gas.and.local.ir.ix.corr.as.expected.for.ic.nearly.linear}, have shown that the $L_{R,\,\rm tot}-L_{X,\,\rm cool}$ relation applies not just in an integrated sense, but locally (looking at the surface brightness in different pixels/beams, or at different annular projected radii) \textit{within} the mutually detected diffuse gas areas (generally e.g.\ mini-halos within the cool-core). 
This is of course expected for SB fluctuations arising from CR-IC (as predicted in CR-dominated halos; see \S~\ref{sec:xr.fluct}). 
For such a local relation the slope predicted by CR-IC can be sub-linear, because if a given region has, say, a fixed $e_{\rm cr,\,\ell}$ but varying $B$, it will ``smear out'' the correlation, but the qualitative behavior should remain. As we discuss below, for the brightest ultra-steep low-frequency radio sources, the radio and X-ray morphology and profiles also closely trace one another as predicted if they have a common leptonic origin. 

A potential exception to that observed scaling are the most central, X-ray-dim cavities, which we discussed above and explore further below. The cavities themselves are already believed to be dominated by a CR-like fluid with relatively low gas density, as we predict (in modeling both their power and size). However it is worth noting that in the most prominent central cavities -- associated with the termination of the ``active'' jet in SCCs -- the X-rays locally \textit{anti}-correlate with the high-frequency radio. This is generally interpreted as the cavities being ``inflated'' by CRs from the jet, and that interpretation is perfectly compatible with the models here. Given our simplistic, spherically-symmetric, time-steady, point-source injection models for clusters, we obviously cannot model the structure of cavities in detail, but they are, in this interpretation, effectively just part of the ``injection zone'' of CRs, as we discuss in \S~\ref{sec:cr.spectrum:injection}. Viewed in that light, there are many plausible reasons one would expect the cavities to be relatively dim in X-ray CR-IC compared to radio: the CRs could be primarily accelerated at the termination shocks/edges (via diffusive shock acceleration); or be initially more hadronic (converting to leptonic when they pass through high gas densities at the shocks); the low-energy CRs could be weakly confined within the jets (as in some CR-dominated regions like pulsar wind nebulae in the LISM); the CRs could simply be too high-energy (as the radio spectra in these regions in systems like Virgo and Perseus is much harder than that of diffuse gas just outside, and implies CR-IC from the CMB would primarily be emitted at $\sim$\,MeV energies, not X-rays) to produce X-rays until they lose energy escaping the injection zone. Or as we discuss below in \S~\ref{sec:pressure}, if these reflect regions of very high CR-to-gas pressure, the CRs could efficiently drive the cavity expansion and, in the process, advectively sweep the low-energy (more confined) CRs relevant for CR-IC out of the cavities rapidly (as seen in some CR-MHD simulations; \citep{su:2025.crs.at.shock.fronts.from.jets.injection,goyal_effects_2025}.

However, once CRs stream out into the more diffuse, volume-filling medium, radio and X-ray surface brightness become positively, rather than negatively correlated, as observed (see references above and e.g.\ \citealt{vanweeren:2024.perseus.giant.radio.halo.filled.electrons.just.tiny.fraction.high.energy} for an example of this in Perseus, outside of the central $\sim 10\,$kpc-sized cavities around NGC 1275). And it is worth recalling (as shown by direct comparison of the cavity and cool-core sizes in \S~\ref{sec:rcool} \&\ \ref{sec:pressure}) that those cavities typically occupy only $\sim 1\%$ of the volume of the cool core.

\subsubsection{Jet/Cavity Power versus Radio Luminosity}
\label{sec:pjet.lradio}

\begin{figure}
	\centering\includegraphics[width=0.99\columnwidth]{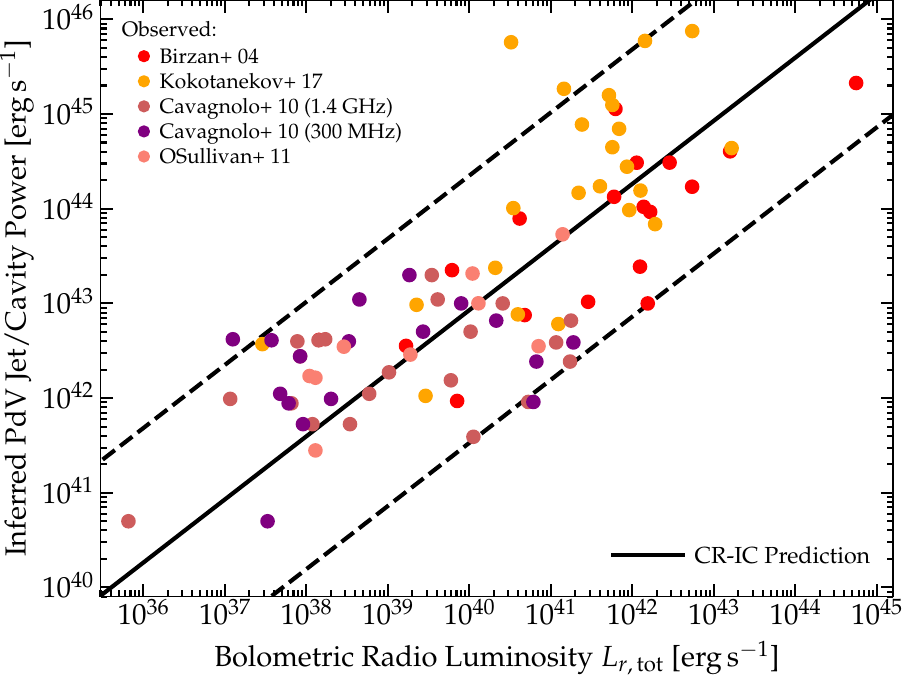} 
	\caption{Apparent ``jet/cavity'' power $P_{\rm cav}$ (as Fig.~\ref{fig:pjet}) versus radio luminosity $L_{r,\,\rm tot}$ (as Fig.~\ref{fig:lradio}). We show various observational compilations (\textit{points}).
	Thick black lines show the \textit{prediction} from \S~\ref{sec:pjet.lradio} if we assume that the apparent cooling luminosity is dominated by ACRH IC emission (with the range as Fig.~\ref{fig:pjet}), while the radio follows the median $L_{X,\,\rm IC}-L_{r,\rm tot}$ relation (equivalent to $\langle B_{\rm \mu G}^{2} \rangle_{\rm Vol}^{1/2} \sim 0.3$, allowing for scatter in the range $\sim 0.1-3\,{\rm \mu G}$). In the ACRH interpretation, this correlation arises because $P_{\rm cav}$ measures (through X-ray IC) properties of the old leptonic halo, while $L_{r,\,\rm tot}$ measures the current (injection-zone) lepton injection/acceleration rate.
	\label{fig:pjet.lradio}}
\end{figure}

From Figs.~\ref{fig:pjet} \&\ \ref{fig:lradio} showing the jet/cavity power $P_{\rm cav}$ versus (apparent) X-ray cooling luminosity $L_{X,\,\rm cool}$ and $L_{X,\,\rm cool}-L_{r,\,\rm tot}$ relations, it trivially follows that CR-IC halos predict a corresponding $P_{\rm cav}-L_{r,\,\rm tot}$ relation, just with additional scatter from the variance introduced by different $B$ in $L_{X,\,\rm cool}-L_{r,\,\rm tot}$. This is shown in Fig.~\ref{fig:pjet.lradio}, using the compilations above where possible as well as \citet{birzan:2008.cluster.radio.cavity.jet.power.correlation.radio.luminosity.sublinear,cavagnolo:2010.agn.cavity.jet.power.radio.luminosity.sublinear.relation.wide.dynamic.range,osullivan:2011.radio.jet.power.radio.emission.and.xray.cooling.luminosities}. Note that all the key behaviors here appear to remain if we use exclusively low-frequency observations to map the radio power (see \citealt{kokotanekov:2017.radio.vs.cavity.jet.power.lofar.low.frequencies.similar.relation.highly.sublinear}). 

The predicted CR-IC correlation of course follows, as it must, from the equations above. Combining the previous predicted scalings, we obtain:
\begin{align}
\frac{P_{\rm cav}}{10^{44}\,{\rm erg\,s^{-1}}} \sim \frac{1.5\,(1+z)^{5/3} \alpha_{1/3}^{7/2}}{(B_{\rm \mu G,\, Vol}/0.3)^{4/3} \alpha_{X}} \,\left( \frac{\langle L_{r,\,\rm tot} \rangle_{\rm Gyr}}{10^{42}\,{\rm erg\,s^{-1}}} \right)^{2/3}  \ .
\end{align} 
The interesting thing which is more obvious in this space, however, is the highly sub-linear correlation between jet/cavity power $P_{\rm cav}$ and bolometric radio luminosity $L_{r,\,\rm tot}$, which is closer to $P_{\rm cav} \propto L_{r,\,\rm tot}^{0.6-0.7}$ in all the studies above. In the CR-IC scenario this trivially emerges as $L_{\rm X,\,\rm cool} \propto L_{r,\,\rm tot}$ from the same leptons and $P_{\rm cav} \propto L_{\rm X,\,\rm cool}^{2/3}$ from the built-in correlations between $n$, $T$, and $e_{\rm cr}$ (hence $L_{\rm X,\,\rm cool}$) for ACRHs. In contrast, in the ``standard interpretation'' where $P_{\rm cav}$ traces the true jet injection energy, most models would predict a very different correlation, e.g.\ $P_{\rm cav} \sim L_{r,\,\rm tot}$ in a simple calorimetric+equipartition model (like what is assumed for synchrotron in most ISM and CGM studies), which predicts both the wrong slope and normalization (by a factor $\sim 100$). While it is possible in principle to come up with models for how jets are energized, introducing free parameters to represent their lepton-loading, escape efficiency, values of ISM/central and CGM/halo magnetic field strength, fraction of energy which goes into inflating bubbles, and allowing these to vary as a function of accretion rate \citep[e.g.][]{willott:1999.model.for.radio.power.jet.vs.agn.power.lots.of.assumptions.and.ad.hoc.parameters.to.explain.sublinear.scaling}, the CR-IC model here again makes a much simpler, effectively single (order-unity)-parameter prediction for the scaling seen.

\subsubsection{Implications for Energetics and Comparison with Standard Models}
\label{sec:lradio.lx.energetics}

One more important note from this is that Fig.~\ref{fig:lradio} and other studies even of \textit{purely} synchrotron emission in luminous CCs \citep{ignesti:2020.cluster.minihalo.leptonic.edot.1e44.1e46.signatures.radio.and.xrays,duchesne:2021.diffuse.cluster.radio.lowfreq}, make it clear that the leptonic injection rate $\dot{E}_{\rm cr,\,\ell}$ must exceed $\sim 10^{44}-10^{46}\,{\rm erg\,s^{-1}}$ in the strongest radio systems -- comparable to or more luminous than the brightest known CCs apparent X-ray cooling luminosity $L_{X,\,\rm cool}$. In other words, regardless of the (still poorly-understood) question of how these leptons are accelerated, it is clear that there is no intrinsic ``energy problem'' accounting for $L_{X,\,\rm cool}$ via CR-IC.

In any case, CR-IC trivially predicts the $L_{r,\,\rm tot}-L_{X,\,\rm cool}$ and $P_{\rm cav}-L_{r,\,\rm tot}$ correlations and scatter. In the standard interpretation, these correlations have a much less natural interpretation: the most common argument is that $L_{X,\,\rm cool}$ represents a cooling luminosity, which leads to some cool gas episodically accreting from the CF to the galaxy (though this gas remains undetected in the X-rays in the classical cooling-flow problem sense), then from the galaxy to the BH, which in turn produces some jet/cavity power $P_{\rm cav}$ which self-regulates via feedback regulation cycles to cancel out the cooling luminosity on large scales, which is in turn correlated in a non-linear fashion with $L_{r,\,\rm tot}$ via various arguments about how jets are energized and escape from galaxies and inflate bubbles \citep[e.g.][]{willott:1999.model.for.radio.power.jet.vs.agn.power.lots.of.assumptions.and.ad.hoc.parameters.to.explain.sublinear.scaling}, which (combined with the non-linear $P_{\rm cav}-L_{X,\,\rm cool}$ correlation) just happens to produce a linear $L_{X,\,\rm cool} - L_{r,\,\rm tot}$ correlation in SCCs.

\section{Discussion: Implications for AGN Feedback \&\ Cooling Flows}
\label{sec:discussion}

\subsection{Prevalence of True Strong Cool Cores}
\label{sec:cfs}

If indeed many SCCs are only ``apparently strong'' owing to CR-IC from ACRHs mimicking thermal gas cooling and so boosting the apparent cooling luminosities/rates, this has dramatic implications for the CF problem. It immediately explains many otherwise puzzling aspects of the classical and spectral CF problems, the dependence of $\dot{M}$, $n$, $T$, $K$, and $Z$ on radius, SZ pressure deficits, central metallicity suppression, the $P_{\rm cav}-L_{\rm X,\,cool}$ and $L_{r,\,\rm tot}-L_{X,\,\rm cool}$ correlations, and the sizes of cool cores. It could even mean that in many apparent CC cases, there may actually be no ``observational'' classical or spectral CF \textit{problem} per se -- by which we mean a large discrepancy between the steady-state temperature distribution which should be present given the observed apparent soft X-ray cooling rate, and the actual temperature distribution. One could simply be observing a weaker CC system (or even a NCC system, in the most extreme cases), with a bright ACRH mimicking larger cooling rates.

We stress that we are not arguing wholesale against the existence of CFs/CCs. Many SCCs clearly exhibit signs of multi-phase gas and some inflowing material in their cores, with famous examples like Perseus showing atomic and molecular gas that appears to be cooling out of the inner $\lesssim 30\,$kpc (though others like Virgo and Centaurus show essentially no cold gas at all). But in all these cases, the ``CF problem'' still appears: the observed cold gas and ``real'' cooling rate appears to be much smaller than that implied from the X-ray luminosity.\footnote{For example $\sim 2-3\,{\rm M_{\odot}\,yr^{-1}}$ star formation and atomic/molecular gas cooling/deposition/inflow rates inferred in Perseus \citep[][and references therein]{canning:2014.small.molecular.formation.deposition.rate.inferred.perseus}, versus $\sim 200-500\,{\rm M_{\odot}\,yr^{-1}}$ nominal ``mass deposition rate'' inferred from the usual cooling flow analysis (\S~\ref{sec:cf.problem}).} The alternative explanation here is that these are simply weak CCs, with real cooling rates more consistent with what is actually observed in the dense/cold phases, and the \textit{apparent} cooling rate (as well as the inner apparent density) is boosted by an ACRH.

It is worth noting that some clusters do not exhibit all of the telltale signatures of an ACRH mimicking a CC. For example, a small fraction ($\sim 10\%$) of the CC clusters observed with high-resolution tSZ do not appear to show a central SZ deficit (i.e.\ they maintain $P_{\rm SZ}/P_{\rm X} \approx 1$ to the center). 
Likewise some few percent of CC clusters show steep central metallicity profiles in single-temperature fits reaching $\sim 2\,Z_{\odot}$ at galaxy ($\lesssim 10\,$kpc) scales (consistent with stellar+ISM metallicities), and show distinct profile shapes in $n$ and $T$. These may also be outliers in the $L_{r,\,\rm tot}-L_{X,\,\rm cool}$ and $P_{\rm cav}-L_{X,\,\rm cool}$ relations. A possible explanation is that this subset represents the systems where CR-IC makes a negligible contribution to the central X-ray surface brightness. But from our comparisons above, what is surprising is that such CC systems appear to be the exception, while the majority exhibit all the properties expected of clusters with strong ACRHs in their centers. So this suggests that ``uncontaminated CC'' systems do exist, but their duty cycle could be much lower: while ``apparent'' (ACRH-boosted) SCCs constitute an $\mathcal{O}(1)$ fraction of groups/clusters \citep[e.g.\ $\sim30\%$ in SZ-selected, or $\sim 60\%$ in bright X-ray selected samples;][]{rossetti:2017.planck.vs.xray.selected.clusters.xray.coolcore.selection.bias.well.known.also.fraction.coolcore.depends.strongly.on.coolcore.definition.resolution.likely.as.noted.other.refs.in.paper}, ``uncontaminated SCCs'' (SCCs with very weak ACRH contributions) could represent as little as a few percent of said population.

\subsection{Implications for AGN FB Models}
\label{sec:feedback}

Regardless of the origin of apparent or boosted CCs, we stress that some version of the ``theoretical CF problem'' or ``overcooling problem'' remains. By this, we specifically mean the theory challenge often referred to as the ``CF problem'' -- namely that in cosmological calculations absent something like strong AGN feedback, groups and clusters will ``over cool'' predicting (1) far too many star-forming (blue instead of red) cluster galaxies, with (2) far too-large stellar and gas masses and star formation rates, and (3) incorrect morphologies, internal structure and kinematics, density profiles, velocity dispersions, and (4) central cluster gas properties (e.g.\ $n$, $T$, $K$) which resemble neither observed CC nor NCC clusters \citep[see e.g.][for reviews]{silk:2012.galaxy.formation.review,fabian:2012.agn.fb.obs.review,morganti:2017.agn.feedback.review,naab.ostriker:2017.galaxy.formation.theory.review,eckert:2021.agn.feedback.galaxy.groups.review.challenges.theory.obs.producing.realistic.coolcores,harrison:2024.agn.feedback.review}. This is a well-known and universal problem, and the need for AGN feedback in clusters persists \textit{regardless} of the origins of CC soft X-rays.

Indeed, if CCs are actually boosted by ACRHs, then we are directly observing AGN feedback in action, in the form of CR leptons. Ironically, if the CR-IC interpretation is correct, then in the $P_{\rm cav}-L_{X,\,\rm cool}$ relation, even though $P_{\rm cav}$ is not really tracing the AGN feedback power (as traditionally assumed), $L_{X,\,\rm cool}$ \textit{is} tracing this power. Rather than a cooling luminosity, in the strong-CR-IC case $L_{X,\,\rm cool} \sim L_{\rm IC} \sim \dot{E}_{\rm cr,\,\ell}$ directly traces the $\sim$\,Gyr time-averaged CR (leptonic) injection rate into the cluster. Likewise instead of $L_{r,\,\rm tot}-L_{X,\,\rm cool}$ tracing a response of radio power to cooling triggering AGN, both variables trace the same AGN and their correlation tells us about the ICM magnetic fields and AGN duty cycle. The apparent CF profile in e.g.\ $n$ and $K$, rather than telling us about inflows to the galaxy, directly reflect $e_{\rm cr,\,\ell}$, allowing one to uniquely determine the CR profile and properties like injection rates and streaming speeds as a function of cluster position. 

So our interpretation of certain quantities like the ``apparent'' Bondi accretion rates, jet/cavity power, cooling luminosities, etc., may change dramatically, but AGN feedback is very much still part of the story. There is still a strong association between apparent SCC and strong AGN, for example, but the causality can be reversed: \textit{systems resemble SCCs because they have had strong AGN feedback episodes (generating an ACRH)}, rather than the traditional interpretation that the SCCs have triggered the AGN. 

We discuss energetics below, but in the broadest sense, there is no change in the implied energetics for AGN feedback models. In the standard interpretation, AGN feedback needed to produce some effectively-coupled thermal energy (via direct heating, shocks, mixing, etc.) to offset the apparent $L_{X,\,\rm cool}$. In the ACRH interpretation, the same energy is required, it is simply prescribed that it be in the form of CR leptons. But it has been well-established for decades that producing or balancing $L_{X,\,\rm cool}$ requires only a tiny fraction of the BH accretion energy available \citep[e.g.][]{silkrees:msigma,croton:sam,mcnamara:2007.agn.cooling.flow.review.cavity.jet.power.vs.xray.luminosity.scalings.emph.compilation}. Moreover we know this energy is plausibly present in the form of leptons, from the radio synchrotron constraints. 

In fact, from the theory point of view, the CR-IC interpretation makes AGN feedback models ``easier,'' in two important ways. First, it means that the actual AGN energy which must be coupled to gas and converted into heat can be much smaller than the apparent $L_{X,\,\rm cool}$, because the real duty cycle of SCC halos may be lower and/or fraction of energy actually lost in the CF much lower, which is especially important for models where momentum-driven outflows and circulation or mixing (rather than pure heating) play a major role \citep[e.g.][]{su:2021.agn.jet.params.vs.quenching}. Second, for decades the most challenging aspect of AGN feedback ``solutions'' to the CF problem has been reproducing a reasonable population of steady-state SCC CF clusters. While not impossible, the vast majority of models (without careful fine-tuning or introduction of additional sub-grid model fitting/tuning parameters) will feature some short duty cycle of cooling, which triggers AGN, which then ``blow out'' or heat up the central baryonic mass (often to temperatures so hot it cannot re-cool in a Hubble time; \citealt{scannapieco:2004.qso.feedback.heating.clusters.blowout}). This produces a NCC or WCC cluster, or often clusters with some hybrid CC-NCC features, or in some cases clusters with features of neither -- but it is quite challenging to reproduce an $\mathcal{O}(1)$ fraction of clusters which look like ``ongoing'' CFs, as observed. From a thermal balance point of view this is obvious: if the heating rate is slightly smaller than $L_{X,\,\rm cool}$, cooling continues rapidly, while if it is slightly larger, runaway heating will lead to a Sedov-Taylor like explosion. This has been discussed in recent examples from a wide variety of different simulations with different AGN feedback physics, see e.g.\ \citet{altamura:2023.large.volume.sims.independent.of.free.parameters.struggle.to.make.realistic.cc.clusters,su:2023.jet.quenching.criteria.vs.halo.mass.fire,su:2024.fire.jet.sim.using.acc.jet.prescriptions.from.cho.multiscale.experiments,gonzalez.villalba:2024.magneticum.cluster.cc.ncc.statistics.pred,lehle:2024.simulation.cluster.profiles,nelson:2024.tng.cluster.sims.profiles.basic.properties,prunier:2024.tng.cluster.xray.cavities.properties}. The CR-IC interpretation relieves this tension: even if AGN feedback does eject or heat up the central cluster gas, but \textit{also} injects leptons, then the cluster will not immediately stop looking like a CC -- it will instead have a $\gtrsim$\,Gyr duty cycle as an ACRH whose CR-IC emission makes it look just like an observed ``apparent'' CC system (\paperone), without any need to fine-tune the amount of feedback or where it is distributed in the cluster.

\subsection{Energetics \&\ Demographics}
\label{sec:energetics}

As we have noted throughout, the energetics of the CR leptons invoked in the ACRHs are plausible and known to exist in the radio and (where detected) $\gamma$-ray sources observed -- even in very luminous sources like Perseus with $L_{X,\,\rm cool} \sim 10^{45}\,{\rm erg\,s^{-1}}$, the implied leptonic luminosity needed to explain the X-ray emission in the CC is less than implied by modeling of the radio and $\gamma$-rays from NGC 1275 alone (let alone other radio sources like NGC 1270). Moreover in Perseus, NGC 1270 is believed to host a BH mass in excess of $10^{10}\,M_{\odot}$, and at least 10 other galaxies in the cluster center (within the center of the CC) appear to host BHs well above $10^{9}\,M_{\odot}$ \citep{meusinger:2020.galaxy.bh.population.in.perseus,liepold:2020.massive.survey.some.outlier.BHs,liepold:2024.massive.bhs.lots.of.scatter.lots.of.big.bhs.for.ptas.but.also.tension.with.soltan.argument}. This means the integrated accretion energy available is $\gtrsim 5\times 10^{64}\,{\rm erg}$ -- sufficient to power the observed extremely bright CC for several Gyr if just a fraction $\epsilon_{\rm cr,\,\ell} \sim 10^{-3}$ of the accretion energy goes into accelerating leptons. 

Of course, most clusters do \textit{not} have cooling flows nearly so luminous/extreme. Rather than thinking of demographics in terms of CC/NCC (or weak/strong CC) where the boundary is somewhat arbitrary and depends sensitively on spatial resolution, X-ray selection effects, and various definitions, it is better to think of the demographics in terms of $L_{X,\,\rm cool}$, and how common different values are. We have estimated this in three different ways: (1) taking the ``direct'' cumulative number density of systems above a given mass deposition rate (or equivalently, $L_{X,\,\rm cool}$, after correcting for temperature) from \citet{peres:no.cooling.flow.mass.corr}; (2) taking the much better-studied luminosity function of \textit{total} (integrated out to $R_{500}$) X-ray luminosity of groups plus clusters ($L_{X,\,500}$) in volume and flux-limited samples \citep{koens:2013.warps.cluster.xray.luminosity.functions,bohringer:2014.cluster.xray.luminosity.functions,pacaud:2016.cluster.xray.luminosity.functions}, convolved with the observed relation between $L_{X,\,\rm cool}$ and $L_{X,\,500}$ \citep{mittal:2011.xray.cluster.scalings.includes.rcool.lcool}; or (3) taking the even-better studied radio luminosity functions \citep[e.g.][and references therein]{smolcic:2017.cosmos.radio.luminosity.functions.source.counts,tucci:2021.radio.agn.luminosity.function.compilation,vardoulaki:2023.cosmos.radio.group.luminosity.functions} and convolving with the radio luminosity $L_{r,\,\rm tot}-L_{X,\,\rm cool}$ relation we studied above. All of these give similar results to within a factor of a couple. Specifically, one obtains a Schechter or double power-law-like function with a slope $d N/dL_{\rm X,\,\rm cool} \propto L_{\rm X,\, cool}^{-1.5}$ at $L_{X,\,\rm cool} \ll 10^{43}\,{\rm erg\,s^{-1}}$, with a steeper cutoff/slope at higher luminosities (exponential or at least as steep as $\propto L_{\rm X,\,cool}^{-2.5}$). 

Regardless of the details, if we then ask from these fits what fraction of the Universe-integrated energy $\int L_{X,\,\rm cool} \, d n(L_{X,\,\rm cool})$ comes from clusters with a cooling luminosity greater than some threshold $L_{X,\,\rm cool}^{\rm min}$, we obtain the result that most of the cooling luminosity comes from CCs with $L_{X,\,\rm cool} \lesssim 10^{43}\,{\rm erg\,s^{-1}}$, with only $\sim 20\%$ coming from $L_{X,\,\rm cool} \gg L_{X,\,\rm min} \sim 10^{43}\,{\rm erg\,s^{-1}}$, a couple to a few percent coming from clusters with $L_{X,\,\rm cool} \gtrsim 10^{44}\,{\rm erg\,s^{-1}}$, and only $\sim 0.1\%$ of the total energy comes from clusters with $L_{X,\,\rm cool}\gtrsim 10^{45}\,{\rm erg\,s^{-1}}$. In other words, the vast majority of ACRHs in the Universe, even if they account for much of the apparent CC emission and CF problem, should be quite modest luminosity.\footnote{Recall we are estimating Hubble-volume-integrated demographics. As such this is not a statement about the relative proportion of non/weak/strong CCs within massive clusters, but instead reflects the well-known result that there are exponentially more lower-mass, lower-$L_{X}$ halos (e.g.\ many more small than large groups, many more large groups than massive clusters).} Integrating the \textit{total} $L_{\rm X,\,cool}$ over all systems (e.g.\ obtaining the total CC luminosity density of the local Universe), and comparing this to the BH mass density in groups and clusters inferred from BH-host scalings in e.g.\ Virgo and more massive systems \citep{kormendy:2013.review.smbh.host.correlations,mcconnell:mbh.host.revisions,liepold:2020.massive.survey.some.outlier.BHs} implies a leptonic ``efficiency'' of $\langle \epsilon_{\rm cr,\,\ell} \rangle  \sim 10^{-4}$ if this BH mass was accreted over $\sim 10\,$Gyr. We obtain a similar result if we consider a system much closer to the median of the scalings above (though still luminous), like Virgo, with $L_{X,\,\rm cool} \approx 10^{43}\,{\rm erg\,s^{-1}}$ and central-only (M87) BH mass $6\times10^{9}\,M_{\odot}$, implying that the system could be powered continuously at this level since $z\sim 1$ for an efficiency of $\sim 2\times10^{-4}$. Of course, AGN growth and feedback is intermittent and time-variable, so these are only extremely crude estimates, but they make the point that typical ACRHs require only a very small fraction of BH accretion energy, and that \textit{most} of the energy comes out at quite modest luminosities $L_{X,\,\rm cool} \lesssim 10^{42}-10^{43}\,{\rm erg\,s^{-1}}$, similar to the power in fairly weak optical/UV/X-ray AGN and/or typical group/cluster radio sources. 

The microphysical origin of this leptonic acceleration remains an open question, but such an efficiency is well within the range of plausible theoretical models for blazar acceleration efficiencies \citep{bottcher:2013.blazar.modeling.almost.all.blazars.better.fit.by.leptonic.cr.models.not.hadronic,blandford:2019.agn.jets.review,keenan:2021.jet.leptonic.power.1e41to1e45.easily.produced.from.modest.agn.bursts.or.steady.jets,foschini:2024.blazar.agn.jet.power.favor.leptonic.large.power.energy.much.more.than.kinetic.lobe.cavity.power}, and it is similar to the values argued for in e.g.\ \citet{wellons:2022.smbh.growth,byrne:2023.fire.elliptical.galaxies.with.agn.feedback} for AGN feedback injection in CRs from AGN, used there to explain the quenching and star formation histories of lower-mass (halo mass $M_{\rm vir} \lesssim 10^{13.5}\,M_{\odot}$) systems. It is also similar to or lower than the more well-known efficiency of supernovae remnant lepton acceleration \citep{2018AdSpR..62.2731A,kronecki:2022.cosmic.ray.gamma.ray.review.not.calorimeters}. Physically, CRs could be accelerated ``directly'' within the accretion or jet-launching region around the SMBH, at jet termination shocks, potentially via some turbulent or merger-shock induced re-acceleration as they propagate, or even be secondary CRs produced by primary hadrons, though the latter does run into more potential challenges with over-producing CR pressures and conceivable tension with $\gamma$-ray constraints (see above). Our arguments in this paper are largely agnostic to the acceleration mechanisms -- the primary difference between these for our purposes would manifest in the exact central profile of $e_{\rm cr,\,\ell}$ (e.g.\ if they are accelerated at jet termination shocks, we might expect their injection to be distributed spatially over the jet, rather than being in a point-source like fashion, which would flatten the profiles at $\lesssim$\,tens of kpc), but these are degenerate with propagation parameters in our simple toy model methodology.

\subsection{Implied Cosmic Ray Pressures and Their Dynamical Role}
\label{sec:pressure}

\begin{figure}
	\centering\includegraphics[width=0.99\columnwidth]{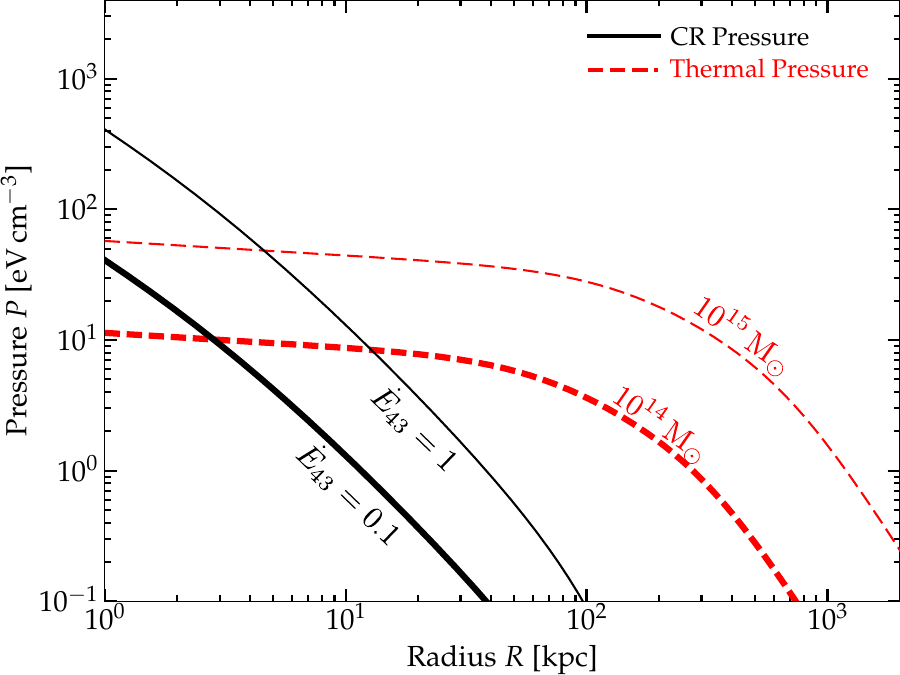} 
	\centering\includegraphics[width=0.99\columnwidth]{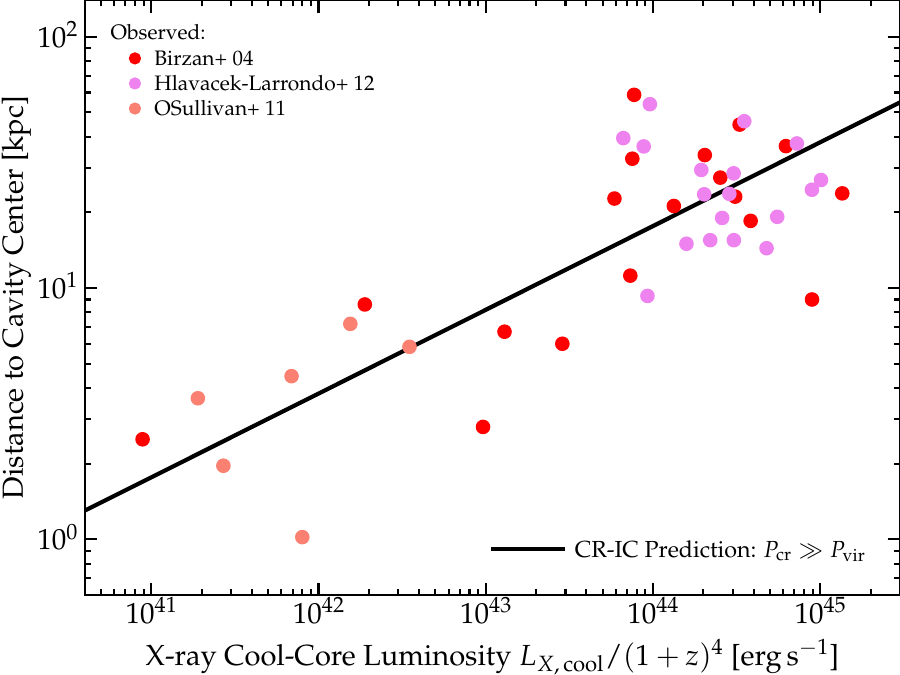} 
	\caption{\textit{Top:} Profiles of the true thermal pressure and CR pressure, for a couple example toy-models from Fig.~\ref{fig:profiles}.  Despite CR-IC dominating the X-ray surface brightness out to $\sim 100\,$kpc in these models, the CR pressure would only dominate over the thermal pressure at $\lesssim 3-10\,$kpc, well interior to the central galaxy and at radii where our simple constant-streaming-velocity and point-like CR injection model likely break down (more realistic streaming+diffusion models with spatially-extended diffusion will flatten this central kpc-scale peak). 
	\textit{Bottom:} Distance from cluster center to the center of the observed X-ray cavities, versus apparent X-ray cooling luminosity. Solid line shows $1/2$ the predicted radius where $P_{\rm cr} \approx P_{\rm vir}$ (i.e.\ where CR pressure can be significant relative to gravity) assuming the entire $L_{\rm X,\,\rm cool}$ owes to CR-IC (\S~\ref{sec:pressure}) and that the true central density is a constant $\sim 5\times 10^{-3}\,{\rm cm^{-3}}$. The scales where CR pressure should become dominant correspond broadly to the extent of cavities/bubbles, as expected if they are CR-pressure driven.
	\label{fig:Pcr}}
\end{figure}

It is important to note that the soft X-ray surface brightness being dominated by CR-IC \textit{does not} necessarily mean that CRs dominate the pressure at the same radii. Likewise, even if CRs dominate the X-ray surface brightness and pressure, this does not necessarily mean CR heating of the gas is significant.

If CR-IC dominates the X-ray surface brightness, then that $S_{X}$ (or equivalently the emissivity $\epsilon_{\rm X,\,keV} \sim S_{\rm X,\,keV}/3\,R$) is directly proportional to the leptonic CR pressure, since for an ultra-relativistic CR population (assuming scattering mean-free-paths smaller than the CC size, easily satisfied in any model where CRs do not immediately free-steam out of the cluster at speeds $\sim c$) $P_{\rm cr,\,\ell} \approx e_{\rm cr,\,\ell}/3 \approx 0.2\,{\rm eV\,cm^{-3}}\,(S_{\rm X,\,keV}/10^{38}\,{\rm erg\,s^{-1}\,kpc^{-2}})\,(100\,{\rm kpc}/R)$. Thus we have a direct constraint on the leptonic CR pressure; the total CR pressure depends on the hadronic contribution (\S~\ref{sec:gamma}), as $P_{\rm cr} \approx P_{\rm cr,\,\ell}\,(1+\xi_{\rm had})$ (where $\xi_{\rm had} \equiv e_{\rm cr,\,hadrons}/e_{\rm cr,\,\ell}$)

The challenge in knowing the dynamical importance of this pressure is that, in the CR-IC scenario, the X-ray observations in CC centers no longer directly measure the gas thermal pressure $P_{\rm th,\,true}$ or density $n_{\rm true}$ (only the apparent $P_{X}$, \S~\ref{sec:sz}, or $n_{\rm app}$, \S~\ref{sec:cc.profiles}). CR pressure is unimportant if $P_{\rm cr} \ll P_{\rm th,\,true}$, and/or $|\nabla P_{\rm cr}| \ll |\rho_{\rm true} \nabla \Phi | \sim \rho_{\rm true} V_{c}^{2}/R$ (the pressure required to support some gas against gravity), but $P_{\rm true}$ and $\rho_{\rm true}$ are no longer straightforward to measure.  It is perhaps easiest to write the CR-thermal pressure ratio in terms of the ``apparent''  versus ``true'' properties in the CR-IC dominated-limit: $P_{\rm cr}/P_{\rm th} \sim 0.008\,(1+z)^{-4}\,n_{\rm app,\,-3}^{2} / (n_{\rm true,\,-3}\,T_{\rm true,\,keV})$. So $P_{\rm cr} \gtrsim P_{\rm th}$ when $n_{\rm app} \gtrsim 0.01\,(1+z)^{2}\,{\rm cm^{-3}}\,(n_{\rm true,-3}\,T_{\rm true,\,keV}\,(1+\xi_{\rm had}))^{1/2}$. 
Of course $P_{\rm th}$ could be small in e.g.\ the centers of CCs if gas is quite cold and/or low-density -- what matters more (and what is most often actually of interest for either feedback or cosmology studies) is the ratio of CR pressure to the ``virial pressure'' -- i.e.\ how important CR pressure support is to virial equilibrium (or outflows/inflows). Assuming the potential follows a \citet{nfw:profile} profile with a typical concentration-mass scaling from \citet{bullock:concentrations}, then the CR pressure-induced acceleration $|\nabla P_{\rm cr}|/\rho$ becomes comparable to or larger than halo gravity $|\nabla \Phi |$ when 
\begin{align}
n_{\rm app} \gtrsim 0.03\,{\rm cm^{-3}}\,(1+z)^{2}\,\left( \frac{n_{\rm true,-3}\,R_{100}} {1+\xi_{\rm had}} \right)^{1/2}\ . 
\end{align}

Fig.~\ref{fig:Pcr} illustrates this (for the models, where we know $n_{\rm true}$), assuming leptonic injection ($\xi_{\rm had} \ll 1$). If the CRs are leptonic, then even for conservative (low $n_{\rm true}$) background group/cluster profiles, for realistic $L_{\rm X,\,cool}$ at observable radii this means that if $L_{X,\,\rm cool} \ll 10^{43}\,{\rm erg\,s^{-1}}$, the profile is essentially never CR-dominated at observable radii. In order to have $P_{\rm cr} \gtrsim P_{\rm th}$ at $\sim 10\,$kpc at any reasonable group/halo mass or redshift this requires $L_{X,\,\rm cool} \gtrsim 10^{44}\,{\rm erg\,s^{-1}}$. To have $P_{\rm cr} \gtrsim P_{\rm th}$ at $\sim 100$\,kpc is non-linearly more challenging, because one must account for additional suppression by IC losses -- we find in the set of toy models we consider above that this only occurs if we reach $L_{X,\,\rm cool} \gtrsim 5 \times 10^{45}\,{\rm erg\,s^{-1}}$, comparable to the couple most extreme known CC systems in the Universe. So in short, the \textit{majority} of CC clusters and groups (\S~\ref{sec:energetics}) will essentially never have $P_{\rm cr} \gtrsim P_{\rm th}$ at $\gg$\,kpc scales (note at these or smaller scales, our toy model assumes point-like CR injection, which is clearly not valid, so the profiles would in reality be much more smoothed out on the smallest scales), while the very most extreme CCs would have CR pressure comparable to gravity in the central tens of kpc. But this is fine -- those systems, as discussed above and widely in the literature, all show large cavities/jets on these scales, so clearly something \textit{is} exerting pressure in excess of gravity on the same scale. 
Indeed, Fig.~\ref{fig:Pcr} shows this predicted scale closely matches the actual size scale of the large cavities observed.
And quantitative models of AGN-CR with CR-pressure-dominated profiles on these scales show no tension with observations \citep{su:2021.agn.jet.params.vs.quenching,su:2023.jet.quenching.criteria.vs.halo.mass.fire,su:2024.imbh.jet.selfreg.supereddington.accretion.challenging.idealized.flow,su:2024.fire.jet.sim.using.acc.jet.prescriptions.from.cho.multiscale.experiments}. But the CRs are only a tiny fraction of the pressure in a cluster-averaged sense or at $\sim R_{500}$ -- where more traditional observational constraints on the CR-thermal pressure ratio apply \citep[e.g.][]{ackermann:2014.cosmic.ray.fermi.gamma.ray.upper.limits.galaxy.clusters.data.not.as.model.dependent}.

On the other hand, if CR acceleration is hadronic and LISM-like, so $\xi_{\rm had} \sim 50$ or some other very large number, then CR pressure could be important out to $\gtrsim 100\,$kpc. This is still not most of the cluster, and so most constraints on ``non-thermal pressure'' in clusters would not apply. And recall from Fig.~\ref{fig:gamma} that even in this case, the system remains basically undetectable in $\gamma$-rays. This at first appears to contradict some quoted $\gamma$-ray limits on CR pressure in the literature, but we stress (1) those limits assume pure hadronic CRs with a specific, very hard power-law spectrum; (2) they are generally integrated over the cluster volume out to $\sim R_{500}$ or $\sim R_{\rm vir}$, where most of the total energy resides, so still unaffected by anything like this within $\lesssim 100\,$kpc; and (3) used a strongly-model dependent method \citep[e.g.\ the scalings in][]{pinzke.pfrommer:2010.cluster.gamma.ray.emission.simple.scalings.for.specific.advection.acceleration.models} to infer $P_{\rm cr}/P_{\rm th}$ which makes specific (strong) assumptions about cluster density profiles, CR acceleration physics and sources, $\xi_{\rm had}$, streaming and diffusion speeds (essentially strictly adiabatic transport, so only valid for very small diffusion/streaming speeds), and co-variance of CR and gas densities. Moreover, (4) since the pionic $\gamma$-ray emissivity scales $\propto e_{\rm cr,\,had}\,n_{\rm gas,\,true}$, if the halo is CR-IC dominated, $n_{\rm gas,\,true}$ is much smaller than it appears, meaning the predicted $\gamma$-ray luminosity $L_{\gamma}$ for a given $e_{\rm cr,\,had}$ is also much smaller, which in turn means that upper limits to $L_{\gamma}$ are much less constraining than one might otherwise assume.

\subsection{Implications for Cluster Masses, Lensing, Cluster Cosmology, and Dynamical ``Non-Thermal Pressure'' Constraints}
\label{sec:cosmology}

\begin{figure}
	\centering\includegraphics[width=0.99\columnwidth]{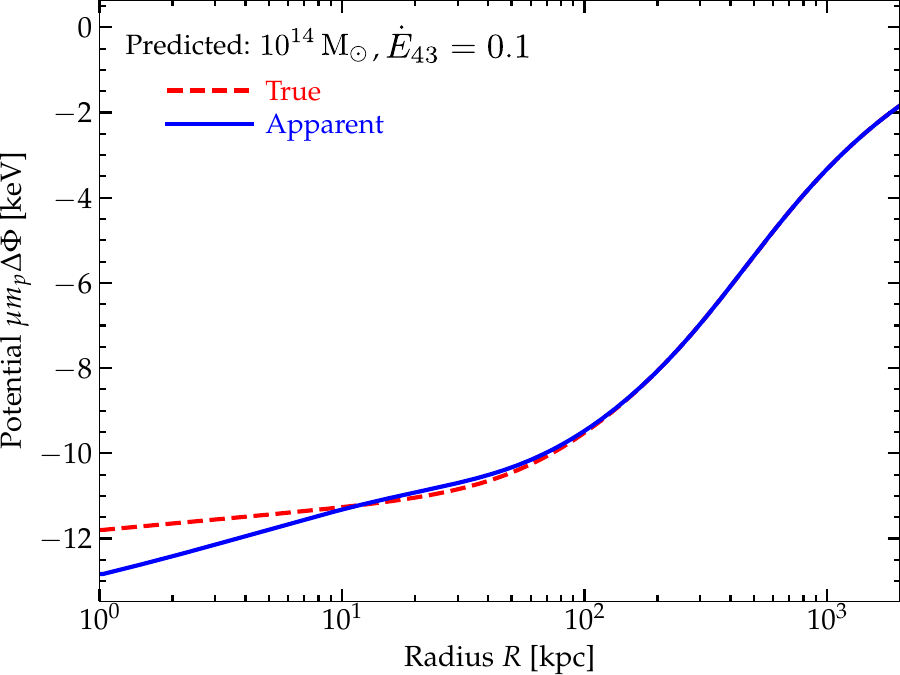} 
	\centering\includegraphics[width=0.99\columnwidth]{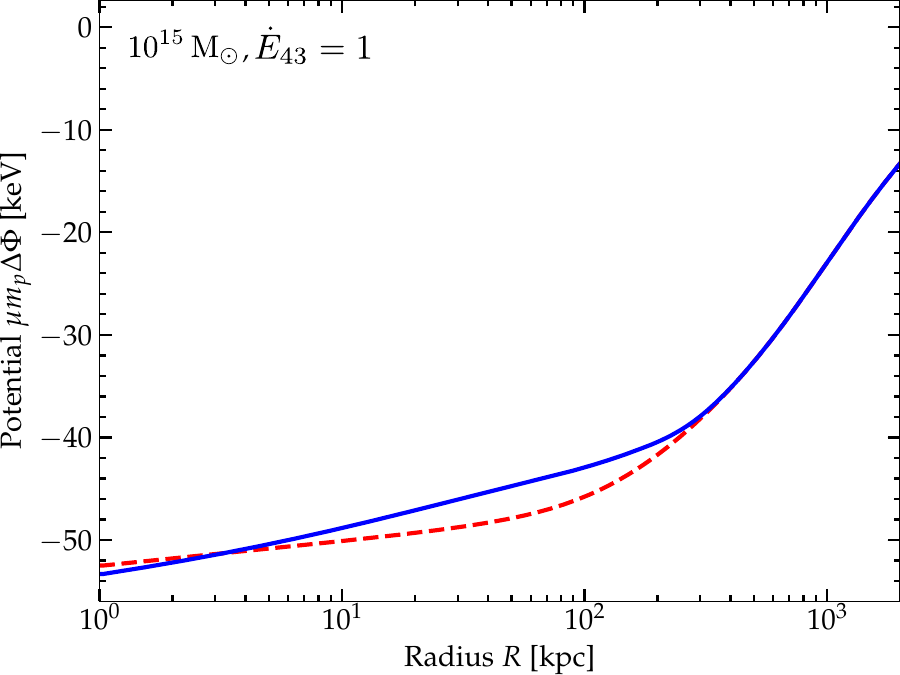} 
	\centering\includegraphics[width=0.99\columnwidth]{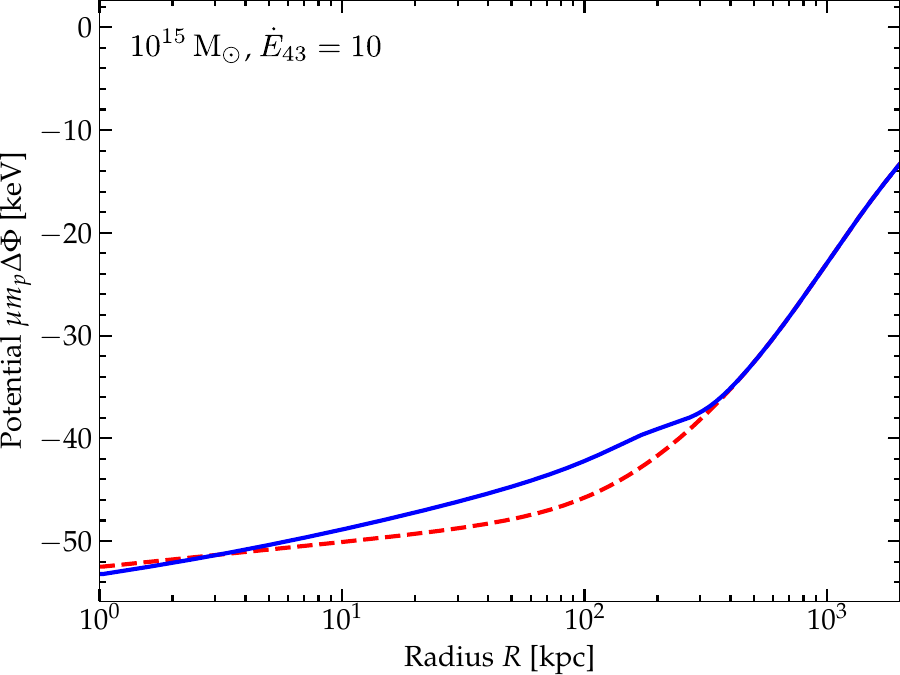} 
	\caption{Profiles of the reconstructed potential from X-ray observations (\S~\ref{sec:cosmology}) for the toy-model profiles in Fig.~\ref{fig:profiles}, using the ``true'' potential with the true density $n$ and temperature $T$, or using the ``apparent'' profile ($n_{\rm eff}$, $T_{\rm eff}$) including a ACRH with CR-IC dominating the central X-ray luminosity. At large radii relevant for cosmology (outside $\gtrsim 0.1\,R_{200}$) the differences are sub-percent. Even at the smallest $R$, there is never more than a $\sim 5-10\%$ bias in the inferred potential even at the very center (similar to typical observed offsets in lensing versus X-ray inferred masses/potentials; \S~\ref{sec:cosmology}). This is because potential reconstruction is only weakly sensitive to the apparent density, and the temperature-density biases from CR-IC offset one another in potential/mass reconstruction. 
	\label{fig:potential}}
\end{figure}

Even in the strongest possible interpretation, where most apparent cool-cores are ACRHs, we stress that this has very little effect on the use of and interpretation of clusters for large-scale structure or many other types of cosmological studies. There are several reasons for this. 

First, CR-IC would only modify certain X-ray properties (like density profiles) of clusters in the central region around central radio galaxies. Studies reliant on gravitational lensing (weak or strong), or the thermal or kinetic Sunyaev-Zeldovich effect, or satellite kinematics/dynamics, or peculiar velocities of clusters and/or their constituent members, or imaging/richness, would all be unaffected. Moreover X-ray studies reliant on bulk dynamics or geometry or positions of gas -- e.g.\ constraints on cosmology or particle dark matter from the Bullet cluster -- would be unaffected as well.

Second, even for classical X-ray studies reliant on X-ray luminosities, temperature, or (hydrostatic) masses, only the very center of the cluster is affected. Radii where cluster masses like $M_{500}$ (i.e.\ $R_{500}$) are estimated are well outside the ACRH radii. This also means only a small fraction of the luminosity is affected. While we emphasize $L_{X,\,{\rm cool}}$ could arise from CR-IC, in almost all clusters -- and every toy example plotted in e.g.\ Fig.~\ref{fig:profiles} -- the majority of the \textit{total} X-ray luminosity (what matters for standard cosmological scaling relations) is largely unaffected, so likewise for related quantities like the cluster (luminosity-weighted) mean temperature, baryon fraction (calculated out to $\sim R_{200}$), and related observables \citep{simet:2017.weak.lensing.xray.cosmology.masses.agreement.but.large.radii}. 

Third, most modern X-ray cosmology studies for e.g.\ large-scale structure or cosmological parameter estimation of $\Omega_{m}$ or $\sigma_{8}$ already \textit{exclude} CC radii from consideration. This is because even in the ``standard'' interpretation, such cores often have multi-phase gas, and it is well-known that different fitting methods, weighting schemes, different priors, and different fitted frequency ranges produce factor $\sim 2$ or larger differences in quantities like the core temperature (see extensive references in \S~\ref{sec:apparent}). It is also well-established that \textit{if} the cores are included, the scatter in cluster scaling relations and the scaling of apparent enclosed mass (from X-rays versus dynamics or lensing), and even basic cluster scalings like $L_{X}-M_{500}$ or $L_{X}-T$, increases significantly, rendering them less reliable by the standards of modern cosmological precision \citep{zhang:2008.cluster.profiles.lensing.xray.good.down.to.0pt2.r500.inner.makes.scatter.much.larger.biases.cosmological.measurements,simet:2017.weak.lensing.xray.cosmology.masses.agreement.but.large.radii,pratt:2022.cluster.density.profiles.also.need.to.excise.cores.to.get.clean.lx.t.mass.relations}. 

Fourth, even in a ``worst-case'' scenario, if one focused only on small radii in a cluster where CR-IC dominated the central emission -- for example, if one wished to use hydrostatic X-ray mass profiles to constrain the nature of ``cores'' vs.\ ``cusps'' in the centers of cluster dark matter halos \citep{mantz:2016.cluster.cosmology.testing.nfw.profiles,eckert:sidm.profiles.xcop.clusters} -- the direct CR-IC effect on cosmological observables like the hydrostatic mass profile or potential inferred from X-ray observations is quite small. For example, assuming spherical symmetry, the usual expression for the potential inferred from X-ray profiles is given by: 
\begin{align}
\Phi(r) - \Phi(r_{0}) &\approx -\frac{k}{\mu m_{p}}\left[T(r)-T(r_{0}) + \int_{r_{0}}^{r} T\,\frac{{\rm d} \ln n}{{\rm d} r} {\rm d} r \right] \ .
\end{align}
For an isothermal profile this is just $\mu m_{p} \Delta \Phi \sim -kT \ln{[n/n(r_{0})]}$. Thus the primary sensitivity in the mass or potential reconstruction is to the apparent temperature, $kT$ (with only a logarithmic dependence on density $n$). But recall, CR-IC biases the X-ray inferred $n$ significantly, but only biases the inferred $kT$ weakly (logarithmically). Moreover, these biases have opposite signs here: $T$ is biased lower, while $\ln{n}$ is biased higher, so they should partially cancel. Fig.~\ref{fig:potential} shows explicit examples of this, comparing the true potential one would obtain if one knew the ``true'' values of $T_{\rm true}$, $n_{\rm true}$ versus those using the X-ray observed $T_{\rm eff}$, $n_{\rm eff}$ from the toy models in Fig.~\ref{fig:profiles}. As expected, there is no difference at larger radii, but even at $\sim 1\,$kpc, the differences between the ``true'' $\Phi_{\rm true}$ and ``apparent'' ($\Phi_{\rm eff}$) potential reconstruction are not larger than $\sim 5-10\%$. But $\sim 10\%$ deviations between e.g.\ lensing, stellar/galaxy/star cluster dynamics, SZ, and X-ray inferred stellar masses are commonly known already, throughout cluster cores, with observations allowing  larger deviations at smaller radii \citep{sayers:2021.cluster.core.constraints.nonthermal.motions}. So caution is needed for some of these applications of hydrostatic mass estimation, but it is not obvious that these caveats are much larger than the already-known caveats of these methods at small radii around the central radio galaxies in SCCs. 

This also means that apparent consistency at the tens-of-percent level between e.g.\ observed mass profiles from lensing and/or stellar dynamics and X-ray observations does \textit{not} rule out an ACRH dominating the X-ray emission. For example, even in an exceptionally well-studied and high signal-to-noise, very weak cool-core case like Virgo and Fornax, \citet{churazov:2008.virgo.mass.profile.kinematics.vs.xrays.as.measurement.of.nonthermal.pressure} showed (1) different X-ray measurements (using different frequency ranges or priors on temperature) give different different $\Delta \Phi(r)$ with the deviations increasing from $\sim 5\%$ to $\sim 20\%$ to $\sim 60\%$ at $R\sim 100,\,10,\,1$\,kpc, respectively; and (2) in the central $\sim 20\,$kpc, the best-fit $\Delta \Phi$ from kinematics and X-rays only agree at the $\sim 30\%$ level -- significantly larger deviations than we predict in Fig.~\ref{fig:potential}. 
Moreover new IFU data and improved Jeans modeling (allowing for e.g.\ anisotropy) have changed the M87 (Virgo) stellar dynamical modeling significantly such that the two methods now disagree by a factor up to $\sim 2$ in $\Phi(r)$ at $\lesssim 10$\,kpc \citep{gebhardt:2009.bh.mass.revision.new.models.m87,murphy:2011.new.m87.models,liepold:2023.m87.dynamical.masses}, and in SCC centers the deviations seen are much larger \citep{zhang:2008.cluster.profiles.lensing.xray.good.down.to.0pt2.r500.inner.makes.scatter.much.larger.biases.cosmological.measurements,newman:2013.cluster.mass.profiles.multi.method,simet:2017.weak.lensing.xray.cosmology.masses.agreement.but.large.radii,pratt:2022.cluster.density.profiles.also.need.to.excise.cores.to.get.clean.lx.t.mass.relations,allingham:2023.clusters.kinematic.lensing.vs.xray.mass.profiles.large.disagreement.qualitative.similar.nfw.profiles}
So the discrepancy CR-IC ACRHs would predict fits comfortably within already-measured systematic deviations.

\begin{figure}
	\centering\includegraphics[width=0.99\columnwidth]{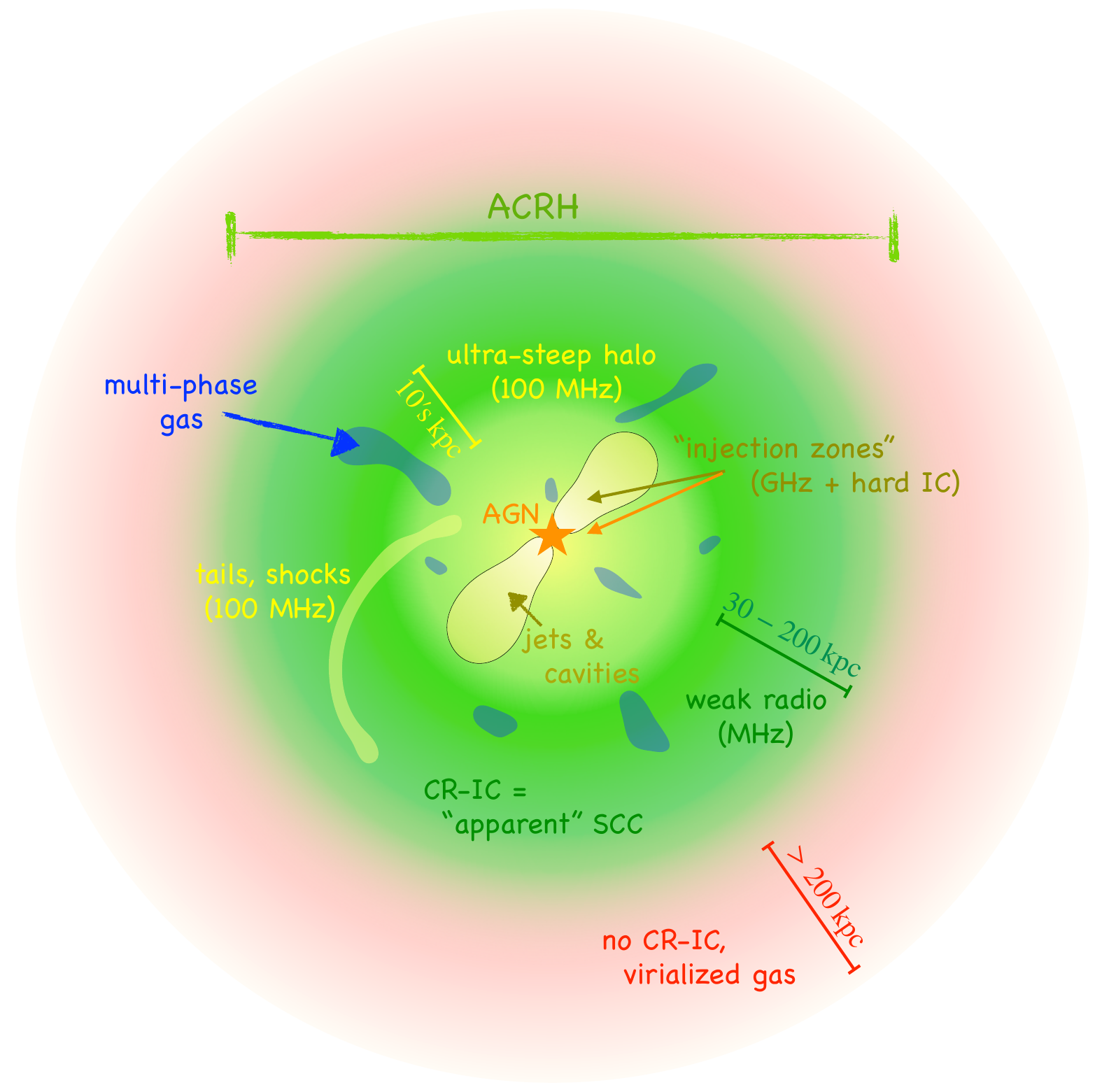} 
	\caption{Illustration of an ACRH and apparent SCC, linking different cluster and radio source categories (\S~\ref{sec:evolution}). A cluster with some weak cooling can fuel an AGN (radio galaxy), which in turn launches jets/bubbles which all inject CRs and form a central $\sim$\,GHz radio source. These diffuse out, producing an intermediate ultra-steep spectrum halo (and potential ``minihalo'') which emits at $\sim 100\,$MHz. At larger radii losses drop the volume-filling radio emission to $\sim$\,MHz and its surface brightness declines rapidly, so the halo is only detectable via thermal-like CR-IC in soft X-rays (plus non-volume-filling shocks/tails/``hot spots'' that appear as ultra-steep spectrum radio sources). CR-IC boosts the apparent X-ray cooling luminosity to appear as a SCC, while explaining the ``universal'' CC X-ray profiles and correlations between X-ray and AGN/radio properties. Eventually even $\sim 0.1-1\,$GeV CRs lose all their energy and drop out at $\gg 100\,$kpc, and the outer halo emission comes from hot virialized, diffuse gas. We label some spatial scales, but these are just heuristic (and not to scale).
	\label{fig:cartoon}}
\end{figure}

\begin{figure}
	\centering\includegraphics[width=0.98\columnwidth]{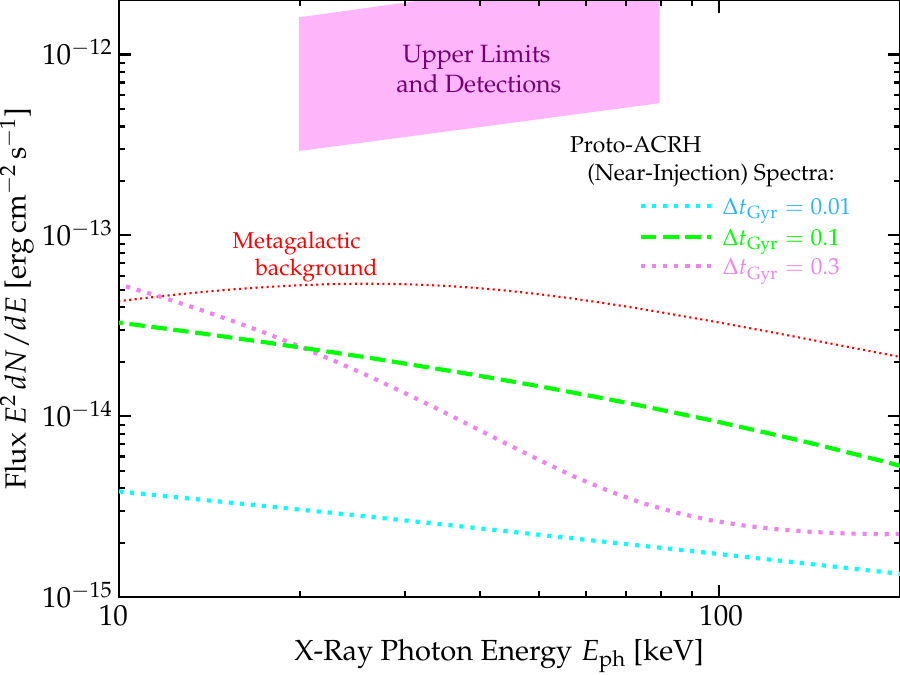} 
	\caption{Hard X-ray spectra from CRs predicted by the simple model from the text (\S~\ref{sec:hardxray}), as Fig.~\ref{fig:hardxr}, but considering leptons at much younger pre-ACRH ages $\Delta t_{\rm Gyr} \sim 0.01-0.3$, or equivalently at radii $R \sim 100\,v_{\rm st,\,100} \Delta t_{\rm Gyr}\,{\rm kpc} \sim 1-30\,$kpc from the ``injection zone'' around a persistent source. This model is scaled to a lepton injection rate/total IC luminosity $\dot{E}_{\rm cr,\,\ell} \sim 10^{43}\,{\rm erg\,s^{-1}}$. We predict therefore that the ``mini-halo'' core with leptons $\sim 10^{8}$\,yr old becomes marginally detectable in traditional hard X-ray IC searches only for injection luminosities $\dot{E}_{\rm cr,\,\ell} \sim 10^{45}\,{\rm erg\,s^{-1}}$ comparable to the most extremely-luminous CC clusters. But note that the spatial scale of the emission predicted is $R \lesssim 10\,$kpc, difficult to distinguish the bright central point source AGN for those sources (requiring $\sim$\,arcsecond resolution). 
	\label{fig:hardxr.young}}
\end{figure}

\begin{figure}
	\centering\includegraphics[width=0.98\columnwidth]{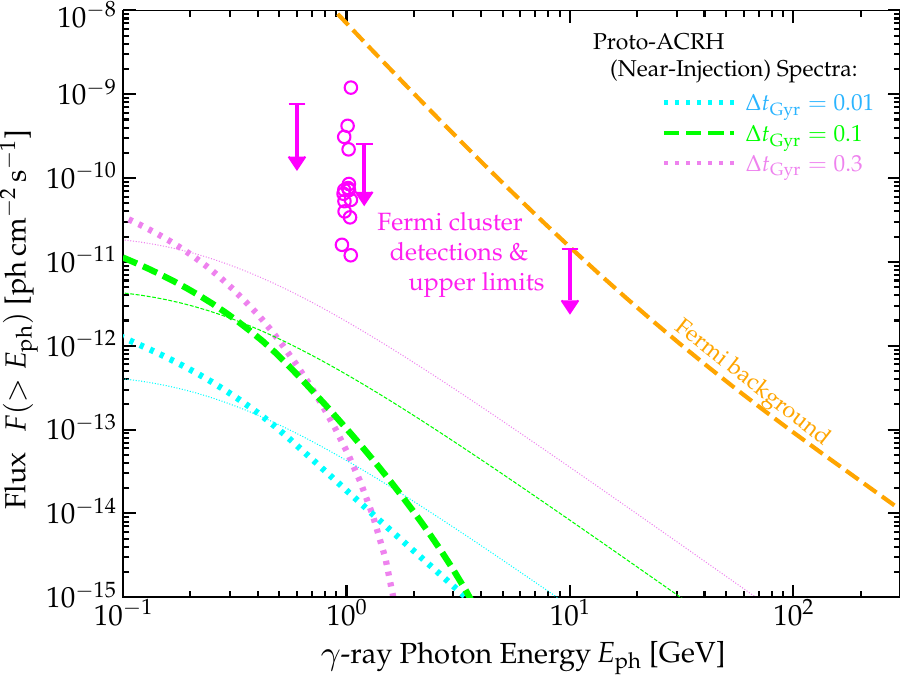} 
	\caption{$\gamma$-ray spectra from CRs predicted by the model from the text (\S~\ref{sec:gamma}), as Fig.~\ref{fig:gamma}, for a source with $\dot{E}_{\rm 43}=1$, but leptons at younger pre-ACRH ages and smaller radii $1-30\,$kpc as Fig.~\ref{fig:hardxr.young}. 
	For leptonic injection (\textit{thick}), this only becomes comparable to the faintest compact Fermi radio galaxies/cluster centers at $\dot{E}_{\rm cr,\,\ell} \gtrsim 10^{45}\,{\rm erg\,s^{-1}}$, for $\sim 1-3\times10^{8}$yr-old populations at $R \sim 10-30\,$kpc (angular scales $10-30''$ at $z\sim 0.05$ or $1.6-5''$ at $z\sim 0.5$). Distinguishing this from the bright central point source is impossible given typical $\sim$\,degree Fermi resolution.
	\label{fig:gamma.young}}
\end{figure}

\begin{figure}
	\centering\includegraphics[width=0.98\columnwidth]{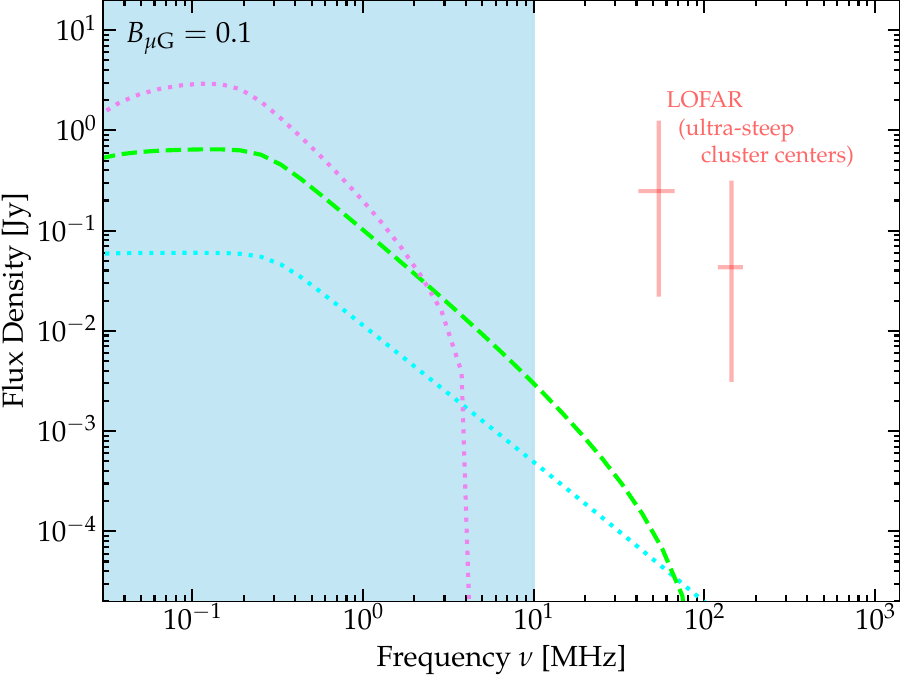} 
	\centering\includegraphics[width=0.98\columnwidth]{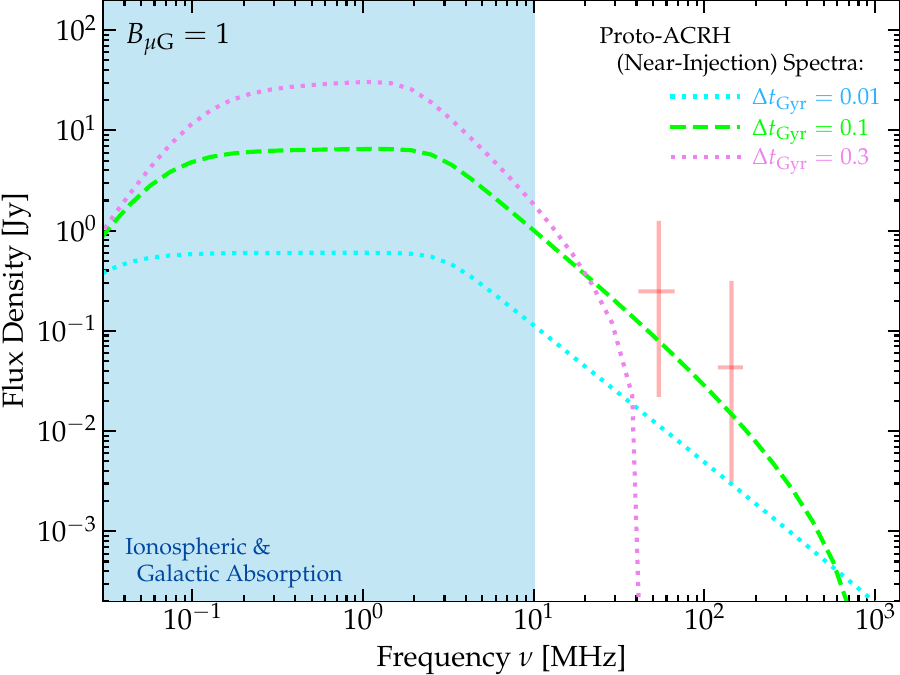} 
	\centering\includegraphics[width=0.98\columnwidth]{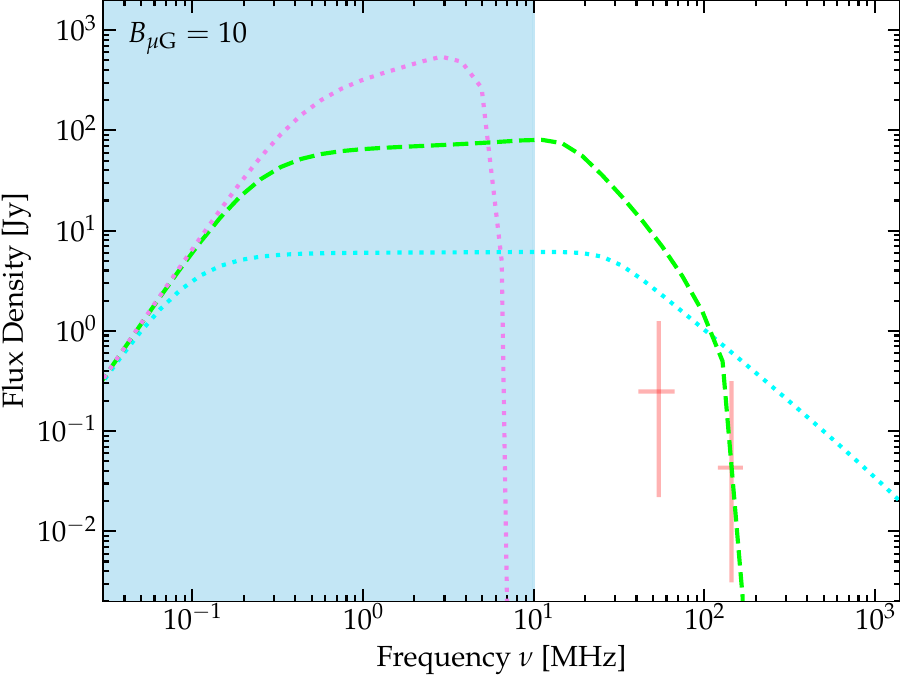} 
	\caption{Synchrotron spectra from CRs predicted by the model from the text (\S~\ref{sec:radio}), as Fig.~\ref{fig:synch}, for a source with $\dot{E}_{\rm 43}=1$, but leptons at younger pre-ACRH ages and smaller radii $1-30\,$kpc as Fig.~\ref{fig:hardxr.young}.
	  At $\sim 10^{8}$\,yr, the spectra are very similar to spectra of observed LOFAR ultra-steep sources in e.g.\ minihalos and decaying relics (assuming $B\sim 1\,{\rm \mu G}$ in the central $\lesssim 10\,$kpc, a reasonable value for these smaller scales), before vanishing completely at $\gtrsim 3\times10^{8}$\,yr (the ACRH phase, even ignoring that $B$ should decrease with increasing age/distance $R$). Without some reacceleration/rejuvenation, only the ``injection zone'' ($t \lesssim 10^{7}$\,yr, or within a few kpc from the acceleration regions) would be detectable as a traditional GHz source.
	\label{fig:synch.young}}
\end{figure}

\section{A Possible Evolutionary Scenario Unifying CCs, ACRHs, and Radio Sources}
\label{sec:evolution}

Motivated by arguments like those in \S~\ref{sec:discussion}, in \paperone, we outlined a (speculative) potential evolutionary scenario for CCs and ACRHs. Here we expand on this and discuss how it could unify many of the different objects seen. We stress this leans heavily on similar previous models and conventional wisdom -- our goal is to put the ACRHs, in particular, in context.

{\bf (a)} Imagine some ``initial'' NCC cluster which develops cooling (e.g.\ from new accretion, mergers, dense gas buildup), or a cluster/group first becoming massive, where rapid cooling still remains. This will form some ``un-contaminated'' (i.e.\ ``pure thermal'') CC and CF which leads to accretion onto the central galaxy, star formation, and, ultimately, can trigger AGN activity (much like the ``standard'' picture). Depending on where gas accretes from, the duration of this phase would be a few central dynamical times, $\sim 0.1-1$\,Gyr. Briefly, this could appear as a CC without significant CR-IC contributions to the emission, but likely with relatively weak cooling luminosity, and with a relatively low radio luminosity (e.g.\ consistent with the rare radio-undetected sources in Fig.~\ref{fig:lradio} and \citet{sun:2009.every.radio.agn.has.xray.cc.strong.lradio.lxray.connection.clusters.all.xray.agns.also.radio.agns} with $L_{\rm r,\,tot} \lesssim 10^{37-38}\,{\rm erg\,s^{-1}}$ in the weakest CCs). 

{\bf (b)} The AGN produces jets and feedback, which begin to heat and push gas out. This causes the system to (1) ``light up'' in radio, (2) inject CR leptons which diffuse outwards, leading to a rapidly-growing ACRH, and (3) eject/heat/mix/displace some of the cooling gas, regulating/suppressing the true CC/CF from the inside-out. This phase will occur on of order the jet expansion and CR diffusion/streaming timescale, $\lesssim$\,Gyr, so could be relatively short-lived. Unlike in the standard picture, there is no need for the AGN power to be fine-tuned to match the true original cooling luminosity, so it is possible that sources caught in this phase will appear as the (also rare) radio over-luminous sources in the $L_{\rm r,\,tot}-L_{\rm X,\,cool}$ relation (e.g.\ Cygnus A, which is radio-overluminous by a factor of $\sim 100$ for its $L_{\rm X,\,cool}$). CR pressure could be itself important in inflating bubbles and/or ejecting/mixing gas from the central regions, at the observed cavity sizes/radii (Fig.~\ref{fig:Pcr}; \citealt{su:turb.crs.quench,su:2021.agn.jet.params.vs.quenching}). 

{\bf (c)} The original CC is now ``boosted'' by CR-IC. The ACRH persists and has expanded to $\sim 100\,$kpc so CR-IC generates an apparently much brighter/stronger CC, obeying all of the correlations and properties in Table~\ref{tbl:properties}. The apparent CC properties in this limit become effectively independent of the true or original CC properties. For a time, if the central source is either persistent or intermittent (so long as it does not remain ``shut off'' for $\sim$\,Gyr), the ACRH luminosity will be maintained at a roughly constant rate $L_{\rm X,\,\rm cool} \propto \langle \dot{E}_{\rm cr} \rangle_{\rm Gyr} \propto \langle L_{\rm r,\,tot} \rangle$, with the older CRs in the outskirts (at $\sim 100\,$kpc). While the central source persists, there is likely to be some radio ``mini-halo'' associated with the inner radii of the ACRH (where the higher-energy CRs can propagate before losing all of their energy), with an intermediate range of radii being closer to ultra-steep halos. We discuss this below. The duration of this phase is set by the duration of intermittent activity of the central source, which could range from as short as $\sim 0.1$\,Gyr to as long as a few Gyr. 

{\bf (d)} If/when the central source is exhausted and remains ``off'' for $\sim$\,Gyr, the higher-frequency radio sources will fade and steepen, passing through a phase globally (even in the center) as ultra-steep extended radio halos. The ACRH and CR-IC will persist and become even more thermal-like, and the cluster will remain on  the correlations in Table~\ref{tbl:properties}, until the low-energy CRs in the ACRH lose all their energy, $\sim$\,a few Gyr. For IC-dominated losses, the expressions in \S~\ref{sec:cr.spectrum} give that the ACRH X-ray IC luminosity will decay with time \textit{after} turnoff of the central source as 
\begin{align}
L_{X}(t > t_{\rm turnoff}) \approx \frac{L_{X,\,0}}{[1+ ( t- t_{\rm turnoff}) / t_{\rm loss,\, IC}^{0}]^{2}} \ , 
\end{align}
 where $t_{\rm loss,\, IC}^{0} \sim $\,Gyr. So this decays relatively gradually (as a power-law in time after Gyr) and each halo will spend a differential time 
 \begin{align}
 \left| \frac{ dt }{ d\ln{L} } \right|  \propto L_{X}^{-1/2}
 \end{align}
 at each $L_{X} < L_{X,\,0}$ after this tur off. Eventually per \S~\ref{sec:softxray} it will drop out of soft X-rays, which is also where bremsstrahlung and Coulomb losses can become important (though these only modify the decay lightcurve to give $|dt / d\ln{L} | \propto 1/\sqrt{L_{X}/L_{X,\,0} + \epsilon}$ where $\epsilon$ depends on the initial ratio of Coulomb+bremsstrahlung to IC loss times), but in the meantime, this can span a non-trivial fraction of the X-ray cooling luminosity distribution observed.

{\bf (e)} If the ACRH decays completely, an observed NCC cluster is left behind. This can persist for several Gyr, until some accretion/central cooling/mergers/etc.\ rejuvenate it or lead to new cooling, potentially re-starting the cycle.

While heuristic, this would naturally explain the $\mathcal{O}(1)$ duty cycle of CCs/NCCs, the CC properties in Table~\ref{tbl:properties}, and many of the observed connections with other types of radio systems, which we discuss below. It immediately and trivially explains why essentially every known CC X-ray system has a radio galaxy located at the global X-ray surface-brightness peak: because that radio galaxy helps \textit{power} the X-ray peak.

\subsection{Connection To Younger Radio Populations}
\label{sec:younger}

In Figs.~\ref{fig:hardxr.young}, \ref{fig:gamma.young}, \&\ \ref{fig:synch.young}, we revisit the basic spectral models used in \S~\ref{sec:multi.wavelength} to predict the multi-wavelength signatures of strong CR injection, but extend these to much younger CR ages and/or much smaller ($\sim$\,kpc) distances from the ``injection zone,'' in order to more clearly illustrate the link between older ACRHs and observed radio galaxies, mini-halos, and ultra-steep-spectrum cluster cores (as discussed in \S~\ref{sec:evolution} and Fig.~\ref{fig:cartoon}). 

As shown already above in Fig.~\ref{fig:spectrum.xr}, in the soft X-rays ($\lesssim 10\,$keV) even as the CR age/radius approach the injection zone ($\Delta t\rightarrow 0$, $R\rightarrow 0$), the emergent X-ray spectrum is closer to thermal than pure power-law-like (for reasonable injection-zone CR spectra). This owes to the fact that plausible injection-zone spectra should (like the ISM) balance losses from synchrotron+IC, bremsstrahlung, Compton, and diffusive/streaming CR escape with injection, giving a spectrum more qualitatively akin to the LISM which shows non-negligible curvature (indeed, peaking) around $\sim 0.1-1$\,GeV, rather than being a pure power-law through these frequencies. It is also easy to verify that there remains negligible emission at any age at infrared/optical/UV wavelengths, per Fig.~\ref{fig:spectrum.allband}.

In Figs.~\ref{fig:hardxr.young} \&\ \ref{fig:gamma.young}, we see that in harder bands like hard X-rays ($\gtrsim 20\,$keV) for IC, and $\gamma$-rays, the predicted spectra only become comparable to existing detected sources if we restrict to young ages $\lesssim 10^{8}\,$yr (radii $\lesssim 10\,$kpc) and consider the brightest known sources with CR injection rates $\dot{E}_{\rm cr,\,\ell} \gtrsim 10^{45}\,{\rm erg\,s^{-1}}$. This is an important consistency check, since these sources \textit{are those detected} in these bands (discussed below). However, at these observed wavelengths, discriminating these compact hard X-ray or $\gamma$-ray halos (with sizes $\lesssim 10\,$kpc around the injection zones) from the ``central'' point source (e.g.\ AGN or some acceleration region around, say, jet termination shocks) is effectively impossible at present, as it would require arcsecond angular resolution (compared to e.g.\ state-of-the-art $\sim$\,degree resolution with Fermi). 

So all we have done is verify that indeed, the ``injection zone'' in our toy models resembles injection zones observed, which motivated our comparisons in the first place.

The situation is more interesting in the radio, illustrated in Fig.~\ref{fig:synch.young}. If the fields are sufficiently weak even in the injection zone, then no detectable emission appears. This could explain rare (but observed; \citealt{sun:2009.every.radio.agn.has.xray.cc.strong.lradio.lxray.connection.clusters.all.xray.agns.also.radio.agns}) systems which are highly radio-under-luminous in the $L_{\rm radio}-L_{\rm X}$ relation. But it is probably uncommon. More plausibly, we expect (given higher gas densities, stronger turbulence, and larger CR pressure, plus CR injection/acceleration/streaming processes themselves) stronger $B$ in these regions. For $B \sim 1-10\,{\rm \mu G}$, the systems would appear as GHz sources at the youngest ages ($\lesssim 10^{7}\,$yr, corresponding to $R \sim$\,kpc), i.e.\ \textit{within} the injection zone. Between $\sim 10^{7}-10^{8}\,$yr -- i.e.\ within $\sim10\,$kpc of the injection zones (though perhaps somewhat larger spatial regions for the higher-energy CRs dominating the radio, if they have larger diffusivities/streaming speeds), both the luminosities and spectral slopes predicted are very similar to detected ultra-steep-spectrum cluster cores in LOFAR \citep{osinga:2021.deepest.lofar.cluster.detections.of.radio.halos,cuciti:2021.diffuse.cluster.radio.halo.fluxes.brightness.lofreq.gmrt.steeper.slopes.larger,pasini:2024.lofar.low.freq.radio.relic.detection.tend.to.extremely.steep.slopes}. 
We discuss these connections further below.

\subsubsection{Connection to Radio Mini-Halos}
\label{sec:minihalos}

A growing number of nearby CC clusters (e.g.\ Ophiuchus and Perseus being well-studied examples) are believed to host radio ``mini-halos'' in the central CC region \citep{bartels:2015.radio.inverse.compton.cluster.minihalo.prospects,gitti:2002.perseus.minihalo.xray.inverse.compton,gitti:2004.minihalo.abell.2626.reaccel,sanders:2005.perseus.minihalo,gitti:2016.radio.minihalos.coolcore.clusters.candidates.review}. In mini-halos, it is well-known that there is a strong correlation between the morphology and spatial extend/structure \citep{giacintucci:2011.cluster.minihalo.fills.cool.core.morphologically.similar,riseley:2023.cluster.radio.minihalo.correlated.xray.brightness.disturbed.re.energized.example}, total luminosity \citep{bravi:2016.minihalo.luminosity.strong.corr.xray.luminosity.clusters}, and local surface brightness \citep{balboni:2024.lofar.xray.surface.brightess.corr.indiv.halos.well.corr.positive.as.expected.for.ic.same.particles.but.sublinear.because.spread.by.B.scatter,riseley:2024.large.minihalo.reenergized.with.merger.but.center.shows.strong.radio.xray.with.steeper.radio.hotter.gas.and.local.ir.ix.corr.as.expected.for.ic.nearly.linear} of the (especially ultra-steep) radio emission and cool-core soft X-ray emission. Indeed, \citet{giancintucci:2019.expanding.radio.cluster.minihalo.sample.no.good.corr.total.cluster.mass.or.total.cluster.xray.but.very.strong.corr.cooling.radius.xray.luminosity.consistent.with.linear.standard.correlation} showed that the \textit{best} predictor of minihalo radio luminosity was the cool-core X-ray cooling luminosity $L_{X,\,{\rm cool}}$ (estimated in the usual fashion as above), while minihalo properties correlate relatively poorly with other properties like cluster mass, BCG mass, or total cluster X-ray luminosity (out to $R_{500}$ or $R_{200}$). This is consistent with the inner regions ($\Delta t \lesssim 10^{8}\,{\rm yr}$, or $R \lesssim 10\,$kpc) of steady-state, extremely-luminous sources, predicted in Fig.~\ref{fig:synch.young}, and is predicted almost trivially in a CR-IC scenario.

And minihalo leptonic energy injection rates easily reach levels comparable to the brightest-known CC X-ray cooling luminosities, $\sim 10^{44}-10^{46}\,{\rm erg\,s^{-1}}$ \citep{ignesti:2020.cluster.minihalo.leptonic.edot.1e44.1e46.signatures.radio.and.xrays}. Indeed it has already been argued that minihalos must be ``energized'' by such bursts of lepton injection from AGN \citep{richard.laferriere:2020.minihalos.agn.closely.correlated.agn.energize.minihalos.produce.them.but.shortlived}. 

Further, minihalos are one place where more traditional hard-IC searches have seen significant detections of clear IC emission \citep{bowyer:1998.coma.ic.euv.must.be.older.electrons.than.synchrotron.more.diffuse,bonamente:2007.abell.3112.clear.xray.inverse.compton.required.luminosity.fits.models.gamma.rays.too,colafrancesco:2009.ic.ophiuchus.minihalo.possible.sources,pereztorres:2009.ophiuchus.minihalo.xray.ic.contrib.significant.B.0pt03to0pt3microG,murgia:2010.ophiuchus.cluster.minihalo.xray.inverse.compton}, albeit only in sources approaching $\dot{E}_{\rm cr,\,\ell} \gg 10^{44}\,{\rm erg\,s^{-1}}$, which also tend to show $B \sim 0.1\,{\rm \mu G}$ fields in the more volume-filling ICM. This is also consistent with the prediction for IC, for similarly luminous sources in Fig.~\ref{fig:hardxr.young}. These same, very bright sources are also the ones which have some central $\gamma$-ray detections shown and predicted in Fig.~\ref{fig:gamma.young}, but for those the predicted $\gamma$-ray emission at $\ll 10\,$kpc cannot be spatially resolved out from the central source.

Again we stress that this IC and the minihalo emission itself still come from relatively high-energy CRs, with $E_{\rm cr} \gtrsim 10\,$GeV. So as these age we expect the sources to steepen -- which appears to be observed in some ``transition'' minihalos with ultra-steep low-frequency radio spectra in the more diffuse, outer regions \citep{raja:2020.diffuse.radio.clusters.steep.spectrum.transitioning.minihalo.feedback.coolcore.to.noncoolcore.event,lushet:2024.lofar.minihalo.extended.ultrasteep.surrounding.different.particle.energization.in.different.locations.primary.vs.reenergized}, leaving behind an ACRH with $\sim 0.1-1\,$GeV CRs (which vanishes from traditional radio synchrotron and hard X-ray IC searches with a more thermal soft X-ray spectrum) surviving for $\sim$\,Gyr. There could be effects like reacceleration producing and sustaining harder emission within the cool core, but in that case the CRs invoked here are required by those models, in order for said reacceleration to operate. And with some recent mini-halo observations probing to much lower frequencies $\sim 10-100\,$MHz in the cluster cool core, it becomes possible to constrain these models and check for consistency with CR-IC in more detail in these individual systems.

\subsubsection{Connection to Ultra-Steep Spectrum (and ``Dying'' or ``AGN-Relic'') Radio Sources}
\label{sec:ultrasteep}

More broadly in the last few years, observations at ultra-low radio frequencies ($\sim 100\,$MHz) with e.g.\ GMRT and LOFAR, have shown the expected trend: strong cluster-center radio sources at e.g.\ 1.4\,GHz tend to have more extended, ultra-steep radio extended halos around them \citep{cuciti:2021.diffuse.cluster.radio.halo.fluxes.brightness.lofreq.gmrt.steeper.slopes.larger,biave:2021.lofar.ultrasteep.low.freq.radio.cluster.halo.rejuvenated.recently,pasini:2024.lofar.low.freq.radio.relic.detection.tend.to.extremely.steep.slopes}, with the radio spectral index steepening with distance from the central source in the diffuse regions (excepting specific acceleration regions like shocks; \citealt{savini:2018.lofar.ultrasteep.radio.emission.surrounding.minihalo.larger.radii.steepening.as.expected.in.coolcore,ignesti:2022.lofar.zdrop.cluster.central.ultrasteep.radio.relic.lofar.losing.energy.outside.of.center,edler:2022.abell1033.cluster.case.study.lofar.decaying.crs.super.steep.radio.rejuvenated.at.special.location.older.at.larger.r.as.expected}). These extend and generalize older categories of ``dying'' or ``AGN relic'' central radio sources, where those with more traditional radio galaxies and bright minihalos at their center would be persistent central sources, while those with a fully shut-down inner source would appear only in the ultra-steep phase as they fade \citep[e.g.][]{kempner:2004.cluster.radio.taxonomy.classifications,parma:2007.fading.agn.radio.small.halos.relics.in.clusters}. The apparent duty cycle of these ultra-steep, low-frequency sources also appears to be longer than typical GHz radio sources \citep{osinga:2021.deepest.lofar.cluster.detections.of.radio.halos}. 

But even at these frequencies, the \textit{volume-filling} detectability drops rapidly outside of tens of kpc -- most of the regions emitting even at $\sim 100\,$MHz are visibly non-volume-filling structures like shocks or ``tails'' of ram-pressure-stripped galaxies or other ``hot spots'' (references above). As such these larger-scale radio ultra-steep sources can still serve as tracers, but generally only lower-limits can be placed on volume-filling CR ages larger than a few hundred Myr, with the implication from these studies that most of the volume at $\sim 100\,$kpc is filled by even steeper-spectrum ($\sim$\,MHz), older leptons \citep[see also][]{salunke:2022.radio.relic.ultra.steep.spectrum.gmrt.100mhz.most.of.volume.steeper.than.observable,vanweeren:2024.perseus.giant.radio.halo.filled.electrons.just.tiny.fraction.high.energy}. These general considerations, as well as the implied ages, and the observed radio spectral shapes and fluxes for ultra-steep spectrum halos, agree well with the predictions for the closer-in, younger CR populations in Fig.~\ref{fig:synch.young}.

This is all what is expected for the transition between ``traditional'' cluster GHz radio sources and ACRHs, either in time (for an aging, intermittent source) or space (for a continuously-active source, where age and propagation distance are correlated). And others have pointed out that these different halos, including minihalos, can be unified by a simple model for central injection with effective streaming velocities in the hundreds of km/s \citep{keshet:2024.radio.minihalos.relics.unified.with.steepening.spectra.models.for.vstream.300.to.3000.kms.model}. This would inevitably predict ACRHs with $\sim 1-5\,$Gyr lifetimes at $\sim 100\,$kpc, similar to those we model here.

\subsection{Typical vs.\ Extreme CCs \&\ Their Radio-ACRH Connection}
\label{sec:typical.extreme}

We also stress that \textit{almost all} of the nearby, best-studied and best-resolved CC systems (e.g.\ Centaurus, Virgo, Perseus, etc.) show the telltale signatures of an ACRH: they have compact radio mini-halos or radio galaxies, which (already-known from the radio) easily provide sufficient leptonic injection rates to account for most of the cooling flow-luminosity via CR-IC (they are consistent with the $L_{X,\,\rm cool}-L_{R,\,\rm tot}$ relation predicted); their central profiles show the characteristic behavior in $n$, $T$, and $K$ expected for CR-IC-dominated profiles; they show central metallicity suppression in single-temperature model fits well below their gas-phase metallicities measured by traditional optical/UV indicators at the same radii; they lie on the CR-IC-predicted apparent cavity-cooling luminosity and cooling radius-luminosity and mass deposition-radius correlations predicted; and they appear similar to the systems which show measured SZ deficits in their centers. 

There are many other case studies we could consider as well, including the most extreme systems. Take for example cluster 1821+643: this is one of the most extreme cooling flow systems known ($L_{X,\,\rm cool} \sim 5\times10^{45}\,{\rm erg\,s^{-1}}$), in a $\sim 10^{15}\,M_{\odot}$ halo at $z\approx 0.3$ with an extremely luminous quasar (bolometric UV/optical/IR luminosity of a few $10^{47}\,{\rm erg\,s^{-1}}$) at its core. The apparent central entropy and temperature drop is one of the sharpest and most extreme at $\sim 100\,$kpc known \citep{walker:2014.case.study.cluster.with.extremely.luminous.2e47.qso.super.low.central.entropy.looks.like.very.strong.coolingflow.central.100kpc.but.modeling.quasar.as.causing.cooling.with.thermal.leads.to.factor.30.discrepancy}, and the apparent central density would be extraordinarily high, but these all agree remarkably well with our predicted toy-model profiles taking the same mass and $L_{\rm X,\,IC} \sim L_{X,\,\rm cool} \sim 5 \times 10^{45}\,{\rm erg\,s^{-1}}$ (\S~\ref{sec:cc.profiles}). It shows a strong central metallicity drop \citep{russell:2024.highres.qso.xray.coolingflow.zdrop.strong.entropydrop.verysmall.but.superluminous.coolingflow.much.larger.than.can.be.explained.physical.cooling.mechanisms} (which those authors attribute to photo-ionization but note that the models in \citealt{reynolds:2014.qso.coolingflow.spectrum.unusual.low.metallicity.and.soft.excess.central.spectra} for the central X-ray spectrum explicitly included ionization but still fit an apparent $Z<0.4\,Z_{\odot}$ from X-rays even in the quasar accretion region), while traditional optical/IR metallicity diagnostics of the galaxy and quasar spectrum give the usual super-Solar metallicity \citep{fukuchi:2022.qso.host.extreme.coolingflow.shows.sf.but.also.solar.metallicity.contracting.xray.zdrop}. The cluster shows a strong radio mini-halo and ultra-steep-spectrum giant halo whose spatial extent, morphology, and surface brightness correlate with the soft X-ray and the cluster shows no optical/UV/IR evidence for any external disturbance to rejuvenate the source \citep{bonafede:2014.giant.radio.halo.radio.quiet.qso.coolingflow,boschin:2018.qso.coolingflow.relaxed.halo.despite.radio.contour.arguments.fully.dynamically.relaxed.radio.cannot.be.from.merger.xray.radio.morph.extent.trace.each.other,duchesne:2021.diffuse.cluster.radio.lowfreq}. And despite being technically ``radio quiet'' (primarily because the optical is so luminous) the low-frequency ($\sim 100\,$MHz) radio is extremely luminous  with an FR I jet \citep{blundell:2001.quasar.large.radio.structure.coolingflow}, implying all of the X-ray ``cooling luminosity'' $L_{X,\,\rm cool}$ could come from CR-IC for diffuse magnetic fields of $\sim 1\,\sqrt{1+u_{\rm qso}/u_{\rm cmb}}\,{\rm \mu G}$ (where $u_{\rm qso}/u_{\rm cmb}$ is the local quasar radiation energy density relative to the CMB) at $\sim 100\,$kpc. This would require only a fraction $\sim 10^{-3}$ of the accretion energy go into leptons in the jet. 
Furthermore several of the papers above point out that the physical cooling flow properties do not make sense in the context of the ``standard'' interpretation: to actually cool so much gas in the center, standard mechanisms would not work, and Compton cooling of the gas from the central AGN would need to be more than $100$ times stronger than observed \citep{walker:2014.case.study.cluster.with.extremely.luminous.2e47.qso.super.low.central.entropy.looks.like.very.strong.coolingflow.central.100kpc.but.modeling.quasar.as.causing.cooling.with.thermal.leads.to.factor.30.discrepancy,russell:2024.highres.qso.xray.coolingflow.zdrop.strong.entropydrop.verysmall.but.superluminous.coolingflow.much.larger.than.can.be.explained.physical.cooling.mechanisms}.

\subsection{Connection to ``Rejuvenation'' and Deviations from the Above}
\label{sec:rejuvenation}

Of course, Fig.~\ref{fig:cartoon} is just a cartoon, and should not be taken as complete nor rigid. Additional ``bursts'' of real cooling or AGN activity or mergers could ``reset'' or ``reverse'' the spatial/chronological of events temporarily in Fig.~\ref{fig:cartoon}, meaning we should not expect it to be always be a monotonic process in either, and the exact scales labeled are purely heuristic. Moreover Fig.~\ref{fig:cartoon} is not intended to include all possible radio clusters. 

There is not, for example, an obvious direct connection between ACRHs and ``giant'' radio halos/relics/gischt covering $\sim$\,Mpc scales. While rare, it is well-known that the GHz synchrotron-radiating leptons in such halos ($\gg 10-100\,$GeV), have lifetimes $\lesssim 10^{7}-10^{8}\,$yr, making it unlikely they propagated to these radii from a central source (it would require effective CR streaming speeds an appreciably fraction of lightspeed). More likely, in our estimation, is the standard interpretation that those leptons are accelerated in situ, via e.g.\ cluster merger shocks or compressible turbulence at Mpc scales. 

There is, however, a natural connection between ACRHs and ``rejuvenation,'' as invoked for some mini-halos \citep{zuhone:2013.turbulent.reaccel.secondary.electrons.as.potential.minihalo.explanation.radio.emission,timmerman:2021.phoenix.cluster.minihalo.rejuvenated.recent.merger.still.centrally.concentrated.strong.radio} and/or centers of some giant radio halos \citep{kale:2019.minihalo.to.giant.radio.halo.in.transition.caught.after.outburst.during.merger.disturbance,giancintucci:2024.example.large.radio.minihalo.related.to.icm.sloshing.potential.disturbance.reenergizing}, and most notably radio ``phoenix'' systems \citep[which can include minihalos and centers of giant halos at $\lesssim 100\,$kpc;][]{burnetti:2004.cluster.particle.rejuvenation.radio.theory.cr.loose.review,ferrari:2008.radio.cluster.review.obs}. 
The $\gtrsim$\,Gyr-old, low-energy ($\lesssim$\,GeV) CRs present in the ACRH are precisely the ``seed'' population invoked in these ``rejuvenation'' or ``reacceleration'' models in order to explain the re-population of synchrotron-emitting (higher-energy) CRs from some tail of CRs pushed back up to higher energies, as a consequence of strong shocks and/or compressible turbulence driven in the cluster core. This could occur at any time in the long-lived ACRH stage (d) above.


\section{Redshift Evolution}
\label{sec:redshift}

Surveys have discovered clusters at ``high'' redshifts $z\sim 1-2$, so it is useful to discuss expected redshift effects in a CR-IC scenario. 
Given that very hard X-rays ($\gg10\,$keV) and $\gamma$-rays are already at best marginally-resolved and largely undetectable at $z=0$, and there is no significant UV/optical/IR signature (\S~\ref{sec:hardxray}-\ref{sec:gamma}), these wavelengths are uninteresting at high-$z$ (they become even more undetectable and/or degenerate with point source-emission from the AGN). Radio is challenging: redshifting pushes the emission to even lower frequencies, and high-resolution radio imaging is generally not possible at such low surface brightness (SB). Integrated radio detections ($L_{\rm radio}$) are possible but restricted to high emission-frame frequencies and usually degenerate with compact radio galaxies. Thus we focus on soft X-rays, where observations often focus on quantities like $n_{X}$, $T_{X}$, $P_{X}$, $K_{X}$ (Fig.~\ref{fig:profiles}). 
But these are \textit{model-dependent} in several ways: at high-$z$ one must  
(1) assume thermal or CR-IC models (giving very different results); 
(2) adopt single or multi-temperature (or cooling-flow, or multi-metallicity) fits (which can differ by factors of several at $z=0$, \S~\ref{sec:apparent}); 
(3) de-project observed profiles to 3D; 
(4) assume simple functional forms for $n(r)$ (usually \citealt{vikhlinin:2006.cluster.compilation.luminosities.gas.fraction.compilation}) owing to poorer S/N and resolution (but this means $n(r)$ at $r \lesssim 10\,$kpc is often determined by extrapolation of parameters whose fits are controlled in a $\chi^{2}$ sense by data at $\sim 100\,$kpc); 
(5) adopt a center, which can be ambiguous (and noisy maps bias $n(r)$ to flatter slopes; \citealt{mcdonald:2017.similarity.clusters.highz.lowz}); 
(6) $k$-correct observations (from fixed observed wavelengths); 
(7) assume global values or fits to $T_{X}$ and $Z_{X}$ as X-ray spectra often lack S/N to measure these directly at small $R$ \citep{girardini:2021.evolution.cluster.profiles.highz.with.uncertainties}.
So we will largely focus on a direct observable instead: the SB profile $I(R)$ at fixed X-ray energy.

\subsection{CR-IC: Weak Redshift Evolution, Regular AGN}
\label{sec:redshift:cr}

\subsubsection{Weak Apparent Thermodynamic Evolution}
\label{sec:redshift:cr:weak}

Briefly, note that even though (as discussed above and in \paperone\ and \citet{hopkins:2025.crs.inverse.compton.cgm.explain.erosita.soft.xray.halos}) at higher redshifts the redshifting of the CMB would naively (all else fixed, which it is not) lead to brighter and more-compact ACRHs (at a given, fixed $\dot{E}_{\rm cr}$ and $v_{\rm st}$), Eqs.~\ref{eqn:Tapp.cluster}, \ref{eqn:napp.cluster}, \ref{eqn:Kapp.cluster} give very weak redshift dependence in the apparent densities and entropies of SCC clusters selected at a given luminosity, at a given radius. Indeed the predicted apparent minimum entropy of clusters does not much evolve, changing for fixed luminosity by just $20\%$ at redshift $z=1$. These trends are completely consistent with observations of the (lack of) strong redshift dependence in these quantities observed \citep{sanders:2018.cluster.density.entropy.temperature.profiles.redshift.samples,sayers:2023.cluster.pressure.profile.mass.redshift.dependence}.

\begin{figure}
	\centering\includegraphics[width=0.98\columnwidth]{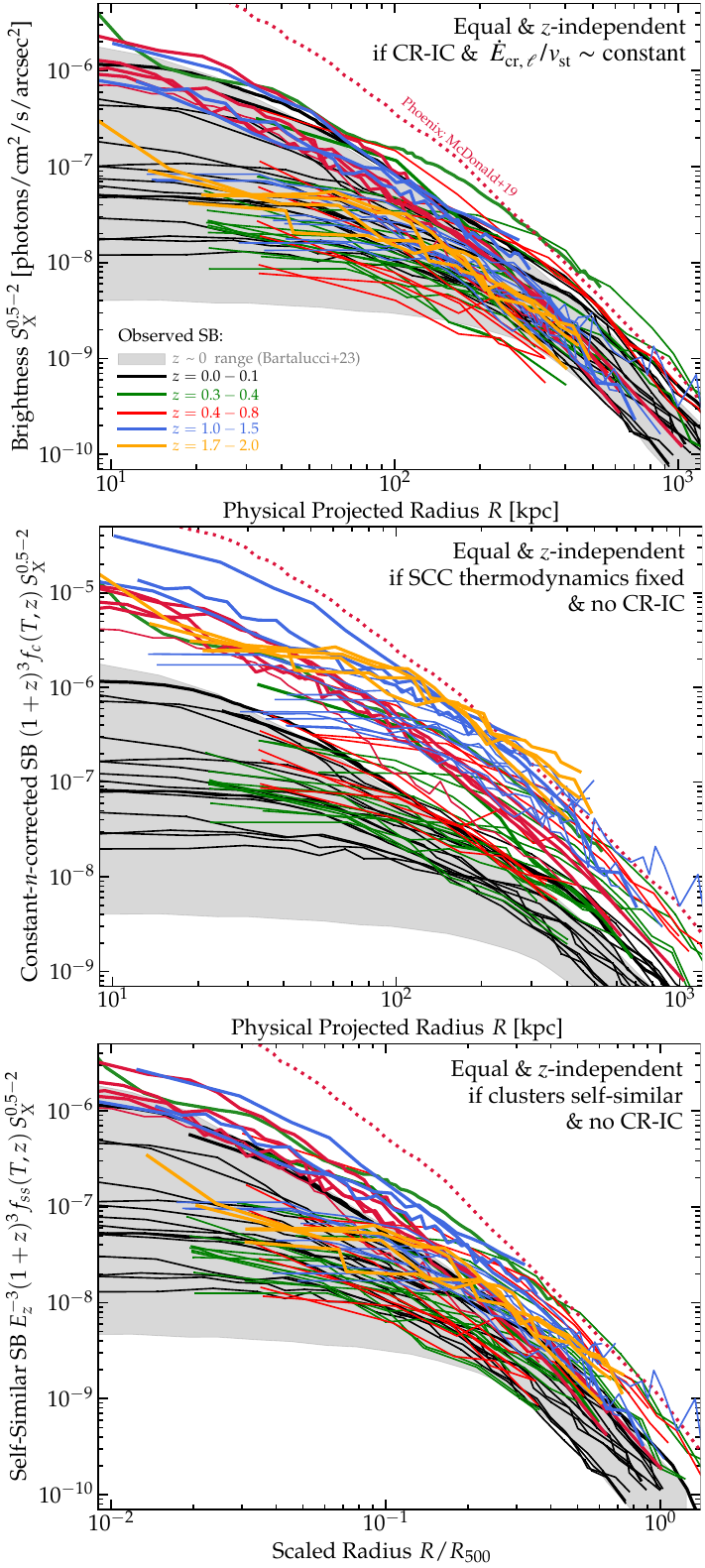} 
	\caption{\textit{Top:} Observed soft X-ray ($0.5-2$\,keV) surface brightness (SB) of clusters (\S~\ref{sec:redshift:cr:sb}), colored by redshift (Phoenix cluster \textit{dotted}), versus projected radius $R$. If CR-IC is important, observed SB in a fixed band evolves per Eq.~\ref{eqn:sb.evol.cric} ($\propto \gamma\,e_{\rm cr,\,\ell}(\gamma)$; $z$ effects on emission/dimming exactly cancel), so (if $\dot{E}_{\rm cr,\,\ell}/v_{\rm st}$ is $\sim$\,constant) would be constant in the steep, rapidly-centrally-rising SCCs where CR-IC is important. This appears consistent with the observations.
	\textit{Middle:} SB corrected for expected evolution if thermal emission dominates and central CC properties ($n$ at fixed $T$) do not evolve (Eq.~\ref{eqn:sb.evol.fixedphys}; \S~\ref{sec:redshift:thermal}), as predicted in e.g.\ precipitation-limited models. Because of the fixed band, there is a SB-dimming correction plus $k$-correction $SB_{\rm obs} \propto (1+z)^{-3}\,f^{-1}_{c}(T,\,z)$ (rather than $(1+z)^{-4}$). The larger spread and systematic $z$-dependence this introduces compared to \textit{top} means this is inconsistent with the observed SCC profiles.
	\textit{Bottom:} SB corrected for expected evolution if emission is thermal and clusters are self-similar (Eq.~\ref{eqn:sb.evol.selfsim}). This predicts scalings closer to \textit{top}, probably explaining evolution at large-$R$, though it appears less-able to explain the central (non) evolution.
	\label{fig:sb.evol.z}}
\end{figure}

\begin{figure}
	\centering\includegraphics[width=0.99\columnwidth]{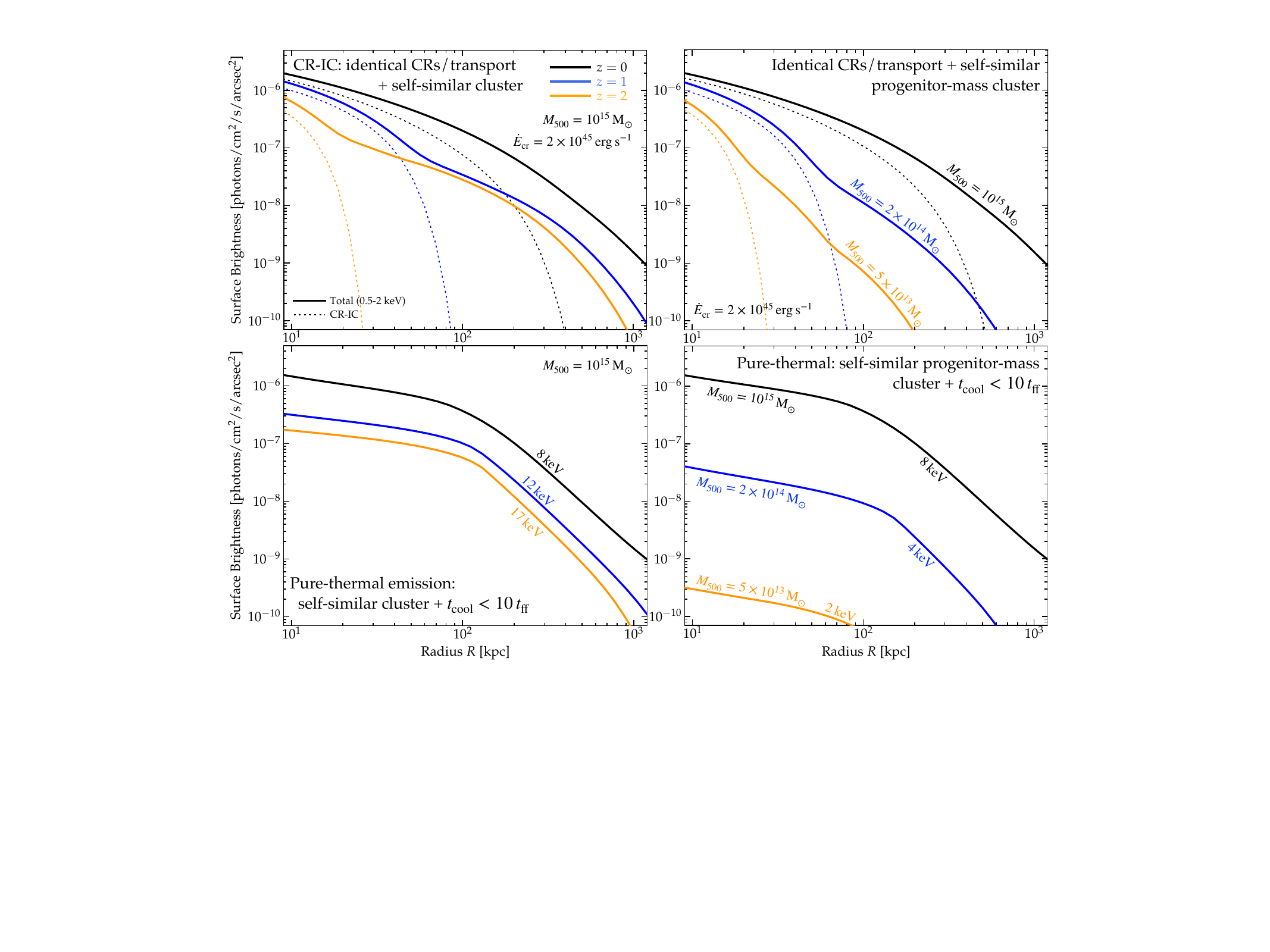} 
	\caption{Illustrative X-ray SB profiles (as Fig.~\ref{fig:sb.evol.z}) for toy models of CR-IC and pure-thermal emission at $z=0-2$. 
	\textit{Top:} CR-IC models with CR injection/transport ($\dot{E}_{\rm cr}$, $v_{\rm st}$,\,$\kappa$) properties identical, on top of a simple fit $n$, $T$ in a massive cluster (as Fig.~\ref{fig:profiles.sb}) designed to reproduce the brightest low-$z$ profiles. 
	We assume $n = n_{0}(R/R_{500})\,E^{2}(z)$ and $T=T_{0}(R/R_{500})\,(M_{500}[z]/10^{15}\,{\rm M_{\odot}})^{2/3}\,E^{2/3}(z)$ evolve strictly self-similarly, and show total SB+CR-IC. 
	We compare models with the same mass $M_{500}=10^{15}\,{\rm M_{\odot}}$ at each redshift (\textit{left}), and models taking the median progenitor mass of a present-day $10^{15}\,M_{\odot}$ cluster at $z=1$ \&\ $2$ (\textit{right}; $M_{500}$ \&\ virial $T$ labeled).  
	\textit{Bottom:} Same for a toy model of pure-thermal emission ($n \sim n_{X}$, $T\sim T_{X}$ for a SCC at $z=0$), assuming strictly self-similar evolution of $T(z)$ and $n(z)$ but with a maximum $n \le n^{\rm cond}_{\rm max}$ at the precipitation-limited value above which $t_{\rm cool}(n,R,T,...) < 10\,t_{\rm ff}$. 
	CR-IC predicts similar central SB for the same CR properties, but if nothing else (e.g.\ $v_{\rm st}$,\,$\kappa$) evolves, central cores are more compact at $z>1$. 
	The thermal model here predicts decreasing SB with $z$. At fixed $M_{500}$, the effect is modest (though still central SB decreases $\propto (1+z)^{-3}$ from $z=0-1$) if clusters are self-similar because central densities increase as $\sim E^{2}$ (allowed because $T_{\rm vir}$ and $V_{\rm ff}$ increase), but it is a strong function of mass (central ${\rm SB} \propto  (1+z)^{-8}$ at \textit{right}). 
	\label{fig:toymodel.cric.vs.thermal.zevol}}
\end{figure}

\begin{figure}
	\centering\includegraphics[width=1.02\columnwidth]{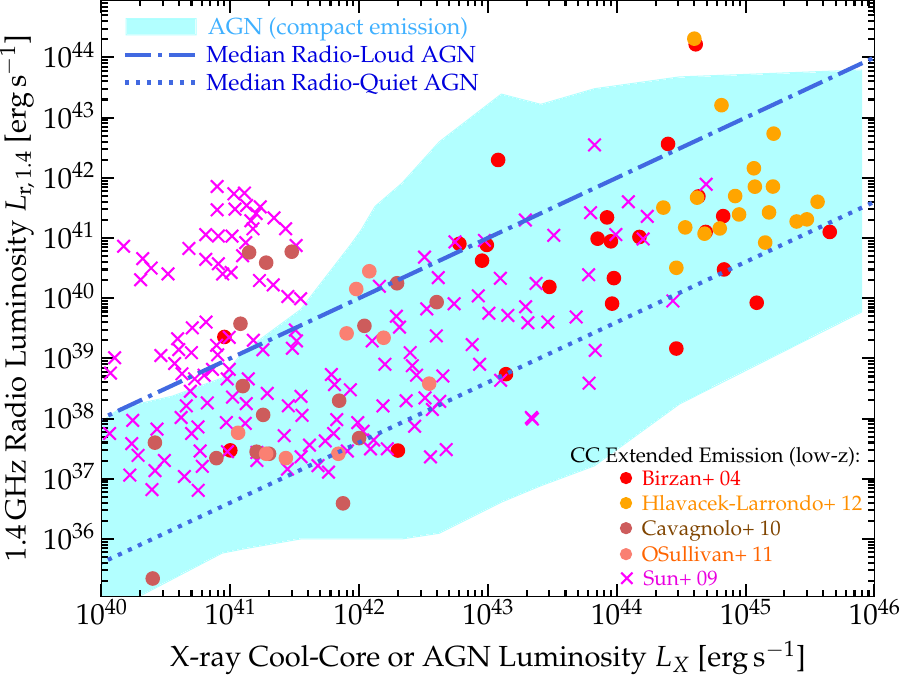} 
	\caption{Demonstration of the challenge distinguishing un-resolved CR-IC from a central AGN from ``regular'' AGN emission (\S~\ref{sec:redshift:cr:agn}). 
	We plot extended X-ray CC luminosity vs.\ $1.4$\,GHz radio jet luminosity in low-$z$ resolved sources (as Fig.~\ref{fig:lradio}, same datasets), 
	and compare the \textit{compact} luminosities at the same soft X-ray and radio wavelengths from large AGN samples \citep{panessa:2014.lradio.lx.compilation.agn.lowz.seyferts,pennock:2025.agn.xray.radio.correlations}, as well as the relation expected from the median radio-loud and radio-quiet AGN spectral templates in \citet{richards:seds}.
	The overlap between these means that if, at high-$z$, increasing CR-IC losses shrink the CR loss radius (maximum extent of CR-IC; $R_{\rm out,\,IC}$) to un-resolvable radii ($\lesssim 10\,$kpc, in Chandra), the CR-IC region would become completely indistinguishable from the ``normal'' AGN emission already present (since an AGN must power the CR-IC). 
	\label{fig:lradio.vs.agn}}
\end{figure}

\subsubsection{Surface Brightness as a Test}
\label{sec:redshift:cr:sb}

But for the reasons above, consider just SB, at fixed projected physical radius $R$ and X-ray frequency $h \nu_{\rm obs}\sim {\rm keV}$, e.g.\ the commonly-quoted SB in a relatively narrow band like $0.5-2\,$keV.  
Per \S~\ref{sec:spectra}, $h \nu_{\rm obs}$ corresponds to emission-frame $h \nu_{\rm em} = h \nu_{\rm obs}\,(1+z)$, which for CR-IC from CRs with Lorentz factor $\gamma$ (energy $E_{\rm cr} \gamma\,m_{e}\,c^{2}$) and background photons with frequency $\nu_{\rm bg}$ corresponds to $h \nu_{\rm em} = h \nu_{\rm bg} \gamma^{2}$. For the CMB $\nu_{\rm bg} = \nu_{0}\,(1+z)$, so $h \nu_{\rm obs} = \gamma^{2} h \nu_{0}$. The SB from this will be given by $\nu S_{\nu} |^{\rm obs} = (1+z)^{-4} \nu S_{\nu}^{\rm em} = (1+z)^{-4} \int (4\pi)^{-1}\,(4/3 \sigma_{T}/c)\,h \nu_{\rm em} n_{\rm photons} n_{\rm cr}(\gamma) \propto (1+z)^{-4} (1+z)^{4} h \nu_{\rm obs} n_{\rm CMB}(z=0)\,\gamma\,e_{\rm cr,\,\ell}(\gamma) \propto \gamma e_{\rm cr,\,\ell}(\gamma)$. In other words, as noted in \S~\ref{sec:spectra} and in \paperone, for the same CR spectrum $e_{\rm cr,\,\ell}$, at fixed \textit{observed} energy, one sees CR-IC from the same lepton energy $\gamma$, with the same SB $\nu S_{\nu} |^{\rm obs} \propto \gamma e_{\rm cr,\,\ell}(\gamma)$, i.e.\ 
\begin{align}
\label{eqn:sb.evol.cric} \frac{\nu S_{\nu}(z)}{\nu S_{\nu}(0)} {\Bigr|}^{\rm CR-IC}_{\nu_{\rm obs},\,R} &\approx \frac{\gamma e_{\rm cr,\,\ell}(R,\,\gamma \, | \, z )}{\gamma e_{\rm cr,\,\ell}(R,\,\gamma \, | \, z=0 )}\ .
\end{align}
The up-scaling of the CMB with redshift and subsequent redshifting/SB dimming after emission exactly offset one another, as they must. 

Thus, \textit{if} $e_{\rm cr}(r) \sim f_{\rm loss}\,\dot{E}_{\rm cr,\,\ell} / v_{\rm st,\,eff} r^{2}$ is the same, i.e.\ losses, injection rates, and streaming speeds do not evolve systematically with redshift, we have an almost-trivial prediction of redshift-independent SB \textit{at the radii where CR-IC dominates}. Of course, it is completely reasonable that those quantities would evolve with redshift, and they are almost totally unconstrained. However, it is instructive to consider the model with constant $e_{\rm cr,\,\ell}(\gamma)$, as a sort of ``fixed-CR-IC'' reference case. 
We test this in Fig.~\ref{fig:sb.evol.z}. Here we compile observed X-ray SB profiles, at $0.5-2\,$keV,\footnote{We correct profiles quoted in larger/smaller bands to the same bandpass using their quoted model-fit spectra from XSPEC, and including the correction for instrument effective area and quoted exposure time where SB is quoted in e.g.\ ${\rm counts}/{\rm arcsec^{2}}$ instead of ${\rm counts}/{\rm cm^{2}\,s\,arcsec^{2}}$, verifying these with APEC forward-modeling of the true SB from their best-fit models at the same radii. We also convert all angular radii to physical using a fixed flat $\Lambda$CDM cosmology with $\Omega_{m}=0.3$, $h=0.7$.} from \citet{arnoud:2002.rosat.xray.cluster.sb.profiles.evol}, who assemble samples with the exact same instrument and reduction pipeline (so they are truly ``apples-to-apples'') supplemented by the samples at $z\sim0$ in \citet{bartalucci:2023.xray.cluster.sb.profiles} and the Perseus observations of \citet{churazov:2003.perseus.profiles,urban:2014.azimuthally.resolved.perseus.profiles}, plus a number of observations at higher redshifts \citep{siemiginowska:2010.3c186.radio.loud.z1.cluster.xrays.profiles,osullivan:2012.gmrt.chandra.type2.radio.loud.qso.in.cooling.flow.intermediate.z.cric.candidate,santos:2012.z1.cluster.xray.coolcore.profiles,brodwin:2016.massive.highz.cluster.profile.xray,mantz:2018.xxl.clusters.highz.xray.profiles.z2.cluster,mantz:2020.highz.spt.cluster.xray.profiles,sanders:2018.cluster.density.entropy.temperature.profiles.redshift.samples,mcdonald:2019.most.relaxed.spt.clusters.xray.profiles.properties,donahue:2020.clusters.mistaken.for.xray.quasars.examples,andreon:2021.highz.cluster.xray.profiles,girardini:2021.evolution.cluster.profiles.highz.with.uncertainties,duffy:2022.xxl.cluster.xray.profiles.highz,calzadilla:2023.highz.relaxed.scc.cluster.starburst.radio.galaxy.xray.profiles,russell:2024.highres.qso.xray.coolingflow.zdrop.strong.entropydrop.verysmall.but.superluminous.coolingflow.much.larger.than.can.be.explained.physical.cooling.mechanisms}. The latter are particularly important because they extend our baseline, and many are chosen to be around radio-luminous AGN, which are excluded from many SZ surveys (and were in some cases excluded from X-ray surveys; \citealt{donahue:2020.clusters.mistaken.for.xray.quasars.examples}).

We immediately see that (1) there is a broad distribution of SB (as discussed above, Fig.~\ref{fig:profiles.sb}), and (2) there is no evidence for any systematic evolution in the \textit{observed} SB at a given physical $r$, especially the upper envelope of SB in clusters with a steeply-rising ($S_{X} \propto R^{-1}$) ``SCC-like'' profiles. We stress this is a heterogenous literature compilation, not a sample selected at some uniform mass/flux/temperature, so conclusions should be taken with some caution. But nonetheless this is exactly what is predicted for the simplest possible model here, with no evolution in $e_{\rm cr,\,\ell}(\gamma)$ in SCCs. In addition, many of the studies above note that the most steeply-rising central profiles in Fig.~\ref{fig:sb.evol.z} are almost all associated with radio-loud quasars or radio galaxies/AGN, and several exhibit radio mini-halos and/or hints of a central ``SZ deficit'' compared to X-ray pressure at $\ll 100\,$kpc.

\subsubsection{High-$z$ CR-IC Can Appear as Normal AGN in Non-CC Clusters}
\label{sec:redshift:cr:agn}

Per \S~\ref{sec:radial}, what \textit{does} evolve in a ``fixed-CR-IC'' scenario (no evolution in $v_{\rm st,\,eff}$) is the outer radius out to which CR-IC can be significant in soft X-rays, as losses limit  this to a maximum $R_{\rm out,\,IC} \sim 100-300\,{\rm kpc}\,v_{\rm st,\,100}\,(1+z)^{-4}$ if transport is strictly advective/constant-velocity-streaming or $\sim 100\,{\rm kpc}\,(\kappa_{\rm eff}/3\times 10^{29}\,{\rm cm^{2}\,s^{-1}})\,(1+z)^{-2}$ if transport is diffusive (depending on X-ray frequency and details of the transport and losses, with weaker scaling with $z$ for e.g.\ sub-diffusion). With fixed $v_{\rm st,\,100} \sim 1-2$ and $\kappa$ at LISM-like values (or the values calibrated to match local clusters like Perseus, a couple times larger), this implies radii $\sim 30-70\,$kpc ($\sim 1^{\prime\prime}-5^{\prime\prime}$) at $z\sim 1$ and $\sim 3-30\,$kpc ($\sim 0.1^{\prime\prime}-0.8^{\prime\prime}$) at $z\sim2$. 
In our toy models from \S~\ref{sec:coolcores}, this would imply a SB profile dominated by thermal emission at large radii -- evolving according to whatever physics (e.g.\ self-similar evolution, \S~\ref{sec:redshift:thermal}) is important (not modeled here) -- with the more steeply-rising CR-IC central SB steep-slope portion of the profile (${\rm SB} \propto R^{-1}$ at small $R$, or $n \propto r^{-1}$, or the $\alpha \gtrsim 2$ inner region in the standard \citealt{vikhlinin:2006.cluster.compilation.luminosities.gas.fraction.compilation} density fitting function) moving to smaller radii at high-$z$, with the shift becoming noticeable at redshifts $\sim 1$ and the steeply-rising region becoming unresolveable even in Chandra (let alone XMM) at redshifts significantly greater than $z \gg 1$. We show a heuristic example of what this might look like in Fig.~\ref{fig:toymodel.cric.vs.thermal.zevol}: taking Perseus-like NCC/WCC fits for the outer $n$, $T$ and a plausible $v_{\rm st}$, $\kappa$ from a case study of the brightest low-$z$ clusters (in prep., as Fig.~\ref{fig:profiles.sb}), assuming CR injection and transport are fixed with redshift while $n$ and $T$ evolve strictly self-similarly (\S~\ref{sec:redshift:thermal} below).\footnote{In Fig.~\ref{fig:toymodel.cric.vs.thermal.zevol} we consider both a fixed mass at each $z$ of $10^{15}\,{\rm M_{\odot}}$ for simplicity (though we caution that these are exponentially more rare at $z \sim 1-2$, and as shown would have $T \gg 10\,$keV much hotter than observed systems), and more realistically evolving masses at different redshifts (with $T$ shown). By construction (since the CR properties are fixed) this primarily influences the outer, thermal-dominated profile and its extent.}

The suggested trend of more-compact steep central regions at high-$z$ appears to be observed. We see it directly in Fig.~\ref{fig:sb.evol.z}: the steep central-rise profiles which extend to $\sim 100$\,kpc at $z\lesssim 0.3$ become slightly more rare at $z\sim 0.5-1$, but at $z > 1$ the steep-rise is only seen in a couple of sources around very luminous, radio-loud AGN at Chandra resolution interior to $\lesssim 50\,$kpc, and essentially none of the sources show such a steep rise resolved outside of the PSF at higher-$z$.\footnote{There is a hint, from \citealt{brodwin:2016.massive.highz.cluster.profile.xray}, of a steep rise in IDCS J1426.5+3508 ($z=1.75$) at $\lesssim 15\,$kpc, if and only if they center the profile on the brightness peak with Chandra, which happens to be a compact point source in X-rays as expected and to show tentative evidence for a central SZ deficit \citep{andreon:2021.highz.cluster.xray.profiles}, but \citealt{andreon:2021.highz.cluster.xray.profiles} notes this could simply be a different AGN in the cluster, or could be PSF contamination. 
The same is true of MQN01 \citep{travascio:2025.hyperluminous.z3.qso.cr.ic.evidence} at $z\approx 3.25$, which shows a hint (albeit PSF subtraction-dependent) of a very similar SB profile to IDCS J1426.5+3508 at $<30\,$kpc (with a steep rise at $\lesssim 7-15\,$kpc), around a hyper-luminous and X-ray bright QSO.} Indeed this trend is well-established: \citet{mcdonald:2013.cluster.gas.profiles} showed that cores with steeply-rising central SB out to $\sim 100\,$kpc ($\sim 0.1\,R_{500}$), common in SCCs in their low-$z$ ($z<0.75$) sample, essentially vanish in their $z>0.75$ SZ-selected sample (noting ``we find an apparent lack of classical, cuspy, cool-core clusters at $z >0.75$''). And as those authors note, this had been seen earlier in \citet{vikhlinin:low.cooling.flows.at.highz}, who noted the ``lack of cooling flow clusters at $z > 0.5$ (defined specifically by such steep central-rise profiles, in X-ray selected samples), and \citet{santos:2008.cc.rare.highz,santos:2010.cc.rare.highz.less.cuspy.clusters}, who found the same at $z>0.7$ using the surface-brightness concentration $c_{\rm SB}$ as their criterion. This is further reinforced in more recent studies like \citet{mcdonald:2017.similarity.clusters.highz.lowz,sanders:2018.cluster.density.entropy.temperature.profiles.redshift.samples,bulbul:2019.sz.cluster.less.selection.biased.to.cc.also.agree.outer.y.param.but.large.radii,girardini:2021.evolution.cluster.profiles.highz.with.uncertainties,sayers:2023.cluster.pressure.profile.mass.redshift.dependence} and those in Fig.~\ref{fig:sb.evol.z} at $z\sim 2$. 

What about the un-resolved compact CR-IC emission? If CR-IC is significant but $v_{\rm st}$ does not evolve and the ``injection zone'' is compact, so the emission is concentrated on arcsecond or smaller scales, it will appear as an X-ray point source. And as expected, a large fraction (comparable to the fraction at $z\sim0$) of the luminous X-ray clusters at high-$z$ do indeed show some bright point source emission near the cluster center. Indeed, quite often \citep[see][]{donahue:2020.clusters.mistaken.for.xray.quasars.examples} the limitation to determining whether a high-$z$ system is an X-ray cluster is resolving the extended emission out from any un-resolved/PSF-contaminated central source (which can dominate the diffuse emission to several PSF sizes, e.g.\ up to tens of arcseconds in deep XMM observations). We stress that we do not expect all clusters to show such a source, no more than we expect all clusters at $z=0$ to show an extremely luminous radio AGN. Just like at $z=0$, only those clusters which happen to host a strong, radio-luminous source (strong lepton accelerator) should exhibit CR-IC. So there would \textit{already be an AGN present, if CR-IC is important}. So can the compact/unresolved CR-IC be distinguished from the ``expected'' X-ray emission from an AGN, when one is present? This seems extremely difficult. We show this by comparing in Fig.~\ref{fig:lradio.vs.agn} the predicted relation between X-ray and radio emission (same datasets as Fig.~\ref{fig:lradio}), with the well-studied compact and bolometric X-ray-radio luminosities of AGN from \citet{panessa:2014.lradio.lx.compilation.agn.lowz.seyferts,pennock:2025.agn.xray.radio.correlations}, with illustrative lines showing standard AGN template spectra for radio-quiet and radio-loud AGN \citep{richards:seds}. The overlap between these relations is striking, and means that if CR-IC emission is unresolved, it contributes only a modest fraction, at most, to the observed AGN X-ray emission, and the system is effectively indistinguishable from any other AGN. 

In other words, in a fixed-CR-IC model (nothing else evolving), the ``false CCs''  from CR-IC begin to {disappear} at high-$z$ ($z \sim 2$), and the extended diffuse emission at $\sim 10-100\,$kpc becomes more reliably attributable to thermal emission. This again appears consistent with the observed SB profiles.

We stress though that all this assumes (1) point-like injection, and (2) un-evolving $v_{\rm st}$ and $\kappa$. If the injection is extended, of course, so too will CR-IC be extended. This is likely to be at play in at least some observed systems: for example, 4C 41.17 \citep{scharf:2003.z4.cluster} at $z\sim 4$ appears to show extended X-ray emission out to $\sim 100$\,kpc but this is clearly associated in morphology, position, and spectral shape with observed extended radio jets and lobes that reach the same radii at GHz frequencies. Indeed, the authors argue the emission is almost certainly CR-IC associated with those jets. Similarly clusters like the Spiderweb ($z\sim 2.2$) show a low surface brightness as expected at $\gtrsim 10\,$kpc ($\lesssim  10^{-7}\,{\rm photons\,cm^{-2}\,s^{-1}\,arcsec^{-2}}$ at $>10\,$kpc) except within a $\sim 10$\,kpc region around each of the radio-bright jet termination regions, where the emission is consistent with being entirely dominated by compact CR-IC  \citep{lepore:2024.spiderweb.cluster.properties.highz.emission.from.jets.cr.ic}.

\subsection{Thermal-Only Expectations}
\label{sec:redshift:thermal}

To compare Fig.~\ref{fig:sb.evol.z} to the expectations of a pure-thermal-emission model, just like how for CR-IC we must assume something about how $\gamma e_{\rm cr,\,\ell}(\gamma)$ evolves, for thermal emission we must assume something about how $n(r)$ and $T(r)$ evolve. There are two simple cases widely discussed in the X-ray literature, (1) self-similar evolution, and (2) non-evolution. In either case, for pure thermal emission in a hot cluster (the case of interest, given the bright systems observed at high-$z$ and their inferred temperatures; references in Fig.~\ref{fig:sb.evol.z}), we have approximately $\nu S_{\nu} |^{\rm em} \propto R\,n^{2} T^{-1/2} h \nu_{\rm em} \exp{[ -h \nu_{\rm em} / k T ]}$, or at a fixed observed band (accounting for redshifting and surface-brightness dimming) $\nu S_{\nu} |^{\rm obs} \propto (1+z)^{-3} R n^{2} T^{-1/2} h \nu_{\rm obs} \exp{[ -h \nu_{\rm obs}\,(1+z) / k T ]}$. 

First take the self-similar model (1), where $R = x\,R_{500}$ and all quantities scale self-similarly so at fixed $x$, $R \propto R_{500} \propto M_{500}^{1/3} E(z)^{-2/3}$, $n = f_{n}(x) \rho_{\rm crit} \propto E(z)^{2}$, $T = f_{T}(x) T_{500} \propto M_{500}^{2/3} E(z)^{2/3}$. This gives 
\begin{align}
\label{eqn:sb.evol.selfsim} \frac{\nu S_{\nu}(z)}{\nu S_{\nu}(0)} {\Bigr|}^{\rm therm,\,SS}_{\nu_{\rm obs},\,R/R_{500}} &\approx \frac{E(z)^{3}}{(1+z)^{3}}\,e^{\frac{h \nu_{\rm obs}}{k T_{0}} \left[ 1 - \frac{(1+z)}{(M_{z}/M_{0})^{2/3} E(z)^{2/3}} \right] }\ .
\end{align}
As shown in Fig.~\ref{fig:sb.evol.z}, for reasonable $T_{0}=T(z=0)$, $M_{0}=M_{500}(z=0)$, $M_{z}=M_{500}(z)$, this gives fairly weak evolution (decreasing from $\sim 1$ to $\sim 0.5$ from $z=0-0.5$, then roughly fixed at $\sim 0.5$ from $z=0.5-2$), which combined with the evolution of $R_{500}$ shifting the profiles, gives a similar effect to the ``fixed-CR-IC'' model. 
Thus the fact that there is weak evolution in the SB at large radii $\sim $\,Mpc should not be taken to imply that CR-IC is dominant at those radii, but rather that a model where the outer parts of clusters evolve self-similarly, while their inner regions are dominated by CR-IC, is completely consistent with the X-ray data. 

Second, consider the fixed-physical model, $n_{z} = n_{0}$, which gives: 
\begin{align}
\label{eqn:sb.evol.fixedphys} \frac{\nu S_{\nu}(z)}{\nu S_{\nu}(0)} {\Bigr|}^{\rm therm,\,fixed}_{\nu_{\rm obs},\,R} &\approx \frac{1}{(1+z)^{3}} e^{-\frac{h \nu_{0}}{k T_{0}}\,z} \ ,
\end{align}
which differs slightly from the bolometric $I \propto (1+z)^{-4}$ expectation because of the effective K-correction to the fixed observed band. 
This appears to be in stark contradiction to the observations in Fig.~\ref{fig:sb.evol.z} -- even at $\sim 10\,$kpc, this systematically predicts the SB of the $z\sim 0.5-2$ bright clusters would be a factor $\sim 10-100$ lower than actually observed. 
This might appear to contradict \citet{mcdonald:2017.similarity.clusters.highz.lowz}, who argued that the mean $n_{X}(r)$ at $\sim 10-100\,$kpc evolves weakly with redshift. However, in addition to the model-dependencies above, and the fact that $n_{X}(r)$ even in a CR-IC model will appear to evolve more weakly $n_{X}(r) \propto (1+z)^{3/2}\,f(T,\,z)$ (equating Eqs.~\ref{eqn:sb.evol.cric} \&\ \ref{eqn:sb.evol.fixedphys}; factor $\sim 2-3$ from $z=0-1$), this is not a fair comparison. The \citet{mcdonald:2017.similarity.clusters.highz.lowz} result is a statement about the median/mean in large SZ-selected cluster samples, including clusters which would be classified as NCC or very weak CC and do not exhibit steep $S_{X}(R)$. It very explicitly is \textit{not} a statement about the upper envelope of SCCs: clusters like Phoenix ($z=0.6$; \citealt{mcdonald:2019.most.relaxed.spt.clusters.xray.profiles.properties}) or 3C 186 ($z=1$; \citealt{siemiginowska:2010.3c186.radio.loud.z1.cluster.xrays.profiles}) or SPT-CLJ2215-3537 ($z=1.2$; \citealt{calzadilla:2023.highz.relaxed.scc.cluster.starburst.radio.galaxy.xray.profiles}) and others near the ``upper envelope'' in Fig.~\ref{fig:sb.evol.z} have central inferred $n_{X}(r)$ much larger than the mean/median in \citet{mcdonald:2017.similarity.clusters.highz.lowz}. 

Models for condensation/precipitation-limited cooling predict something close to Eq.~\ref{eqn:sb.evol.fixedphys} \citep{parrish:turbulence.w.conduction.regulates.cooling.flows,sharma.2012:thermal.instability.precipitation.coolingflows,Voit_2017}. In those models, an upper limit $n_{\rm max}^{\rm cond}$ appears because denser gas condenses into cooler phases when cooling times drop below $\sim 10$ freefall times: $t_{\rm cool}^{\rm min} \sim (3/2)\,n^{\rm cond}_{\rm max}\,k_{B} T / ((n^{\rm cond}_{\rm max})^{2} \Lambda(T,\,Z) )\sim 10\,t_{\rm ff} \sim 10\,r/V_{c}$. For hot clusters, this gives $n_{\rm max}^{\rm cond} \sim 0.003\,{\rm cm^{-3}}\,T_{\rm keV}^{1/2}\,(V_{c}[r]/400\,{\rm km\,s^{-1}})\,(r/100\,{\rm kpc})^{-1} \sim 0.0016\,{\rm cm^{-3}}\,T_{\rm keV}/R_{100}$, or a SB (in ${\rm photons\,cm^{-2}\,s^{-1}\,arcsec^{-2}}$) $\nu\,S_{\nu}^{\rm obs} \lesssim 10^{-6}\,(1+z)^{-3.1}\,\exp{[-(1+z)/T_{\rm keV}]}\,  (T_{\rm keV}/4)^{3/2}\,(10\,{\rm kpc}/R)$. This is consistent with the upper envelope at $z\sim0$, but predicts factor $\sim 10$ ($\gtrsim 100$) lower SB at $z\sim1$ ($z\sim2$), given central $T_{\rm keV}$ observed. We illustrate this in Fig.~\ref{fig:toymodel.cric.vs.thermal.zevol}, by taking the same self-similar mock cluster models at $z=0-2$ we considered with CR-IC, and now modeling them with pure thermal emission from a self-similar cluster capped at a maximum density $n^{\rm cond}_{\rm max}$. If high-$z$ systems were extremely hot ($\sim 20\,$keV) and massive, this would offset some of the predicted evolution by having the central density $n$ \textit{increase} rapidly with redshift, as $(1+z)^{1 \rightarrow 3}$ or so (though even this model still predicts order-of-magnitude decrease in SB at $z\sim1$), but given the cooler temperatures/lower masses seen and expected in $z\gtrsim 1$ clusters, this predicts dramatic high-$z$ evolution in SB.

\section{Conclusions}
\label{sec:conclusions}

Strong cool-core (CC) clusters identified in X-rays ubiquitously contain central strong radio galaxies located near the peak of the global X-ray surface brightness, observed in synchrotron (and sometimes $\gamma$-rays) to inject CR leptons at prodigious rates. Such systems must necessarily produce/contain ``ancient'' cosmic ray halos (ACRHs) populated by the $\sim 0.1-1\,$GeV leptons with lifetimes $\gtrsim$\,Gyr much longer than the higher-energy leptons generating higher-frequency synchrotron within the central radio source. For plausible streaming/diffusion speeds, these ACRHs will extend to $\sim 100\,$kpc, and their primarily loss mechanism should be inverse-Compton (IC) scattering of CMB photons. We show that these ACRHs produce a remarkably thermal soft X-ray ($\sim\,$keV) spectrum nearly independent of their injection spectrum, as losses suppress much lower and higher-energy CRs, and the convolution of a perfect blackbody (the CMB) with any peaked IC scatterer produces a blackbody X-ray spectrum. This also means these extended halos are almost undetectable in ``traditional'' hard X-ray IC diagnostics, $\gamma$-ray emission, and synchrotron (emitting at $\sim$\,MHz), but could easily account for a significant fraction of the apparently-thermal X-ray surface brightness in the central $\lesssim 100\,$kpc.

We show, as argued in \paperone, that the naively predicted apparent X-ray spectra and their radial profiles (including the apparent X-ray-inferred surface brightness, gas density, temperature, entropy, and pressure) from this CR-IC emission from ACRHs reproduce the qualitative properties of CCs. In other words, a cluster without a true strong CC (i.e.\ whose true central gas properties are more akin to weak CCs or even non-CCs, with lower densities and/or higher entropies) but with a luminous ACRH, could \textit{appear} observationally to be a prototypical strong CC in essentially every X-ray diagnostic. For fixed CR properties (injection spectrum and streaming/diffusive speeds), the predicted profiles form a one-parameter family in the apparent X-ray ``cooling luminosity'' from the CC, really the CR-IC luminosity. 

If true, this could provide an almost-trivial interpretation for the classical cooling-flow (CF) problem: X-ray ``cooling'' in CCs is not producing all of the expected massive flows of cold gas and star formation because it is not entirely thermal cooling radiation being observed. In short, the apparent cooling luminosity is boosted by the CRs. We show here that this provides immediate predictions -- with no fitted or fine-tuned parameters -- for the various correlations observed in CCs between the apparent cooling luminosity/cooling rates, cavity/jet/AGN feedback power, central AGN radio/X-ray activity, cooling-core radii, cavity sizes/radial extents/distances, inferred X-ray ``mass deposition rates,'' and more. We compare all of these to observations and find remarkably good agreement, given the simplicity of our models and lack of fitted parameters. These arise again almost trivially because \textit{the AGN powers the apparent CC} by providing the leptons that are actually giving the CC luminosity (and of course the radio and X-ray come from the same source, ultimately). This essentially reverses the traditional interpretation of said correlations.

We emphasize that there is no diagnostic from X-ray spectra alone (no matter the energy range and continuum/lines recovered, signal-to-noise, or spectral resolution) which can uniquely rule out CR-IC as a significant contribution to the X-ray continuum if one allows for the full freedom of the problem (a mix of variable multi-phase thermal emission, abundances, absorption, plus a free lepton spectrum and incident radiation spectrum at every point in space) without imposing strong priors on the form of the CR lepton spectrum (what is usually done, e.g.\ assuming that CR-IC would emit a pure-power-law spectrum). The X-rays are still constraining, and one can argue for plausibility of different models (e.g.\ how ``easy'' or ``difficult'' it would be to produce the observed spectral shape in different models), but to definitively rule out CR-IC or pure-thermal explanations, multi-wavelength constraints are crucial. 

We show that measurements from hard X-rays, radio, and $\gamma$-rays are, at present, not strongly constraining of CR-IC models for the cluster population in general, because of the lack of high-energy CRs, the very low surface brightness nature of this emission, and the small/compact nature of the emission region. As such a wide range of CR ``injection spectra,'' loss rates, and hadron-to-lepton ratios are allowed in CR-IC models (and for many reasons previous claims of upper limits to CR pressure from $\gamma$-ray or kinematic/dynamical constraints do not apply to the models here), though these can be constrained further with future observations. We stress however that integrated radio luminosities and radio X-ray correlations, especially at low frequencies, as well as jet/cavity power and cavity size measurements, and constraints on weak turbulence in cluster cores, all agree well with the predictions of models here where CR-IC contributes significantly to the X-rays. Moreover in the nearest, brightest clusters, like Virgo and Perseus, more detailed and constraining spatially-resolved multi-wavelength tests are possible. Case studies of these clusters comparing CR-IC and pure-thermal models therefore represents an important subject for future work. Further multi-wavelength connections to ultra-low-frequency radio halos being probed by new radio instruments, and IR/optical/UV line emission diagnostics of particle excitation in dense neutral and molecular gas, can represent powerful tests of these models and will also be discussed in future work.

We discuss two specific multi-wavelength tests of this scenario from \paperone\ in detail. First and most unambiguously, if CR-IC mimics bright thermal emission in the center of an apparent CC, then the X-ray inferred central gas pressure $P_{X}$ will often be over-estimated compared to the true thermal pressure $P_{\rm true}$. But the latter should still be recovered correctly in SZ measurements ($P_{\rm true} \approx P_{\rm SZ}$). So the ratio $P_{\rm SZ}/P_{\rm X}$ should drop in CC centers (at $\ll 100\,$kpc), with stronger drops in more extreme CCs. Strikingly, of the small sample of clusters with sufficiently high-resolution SZ pressure profiles for meaningful comparison \citep{romero:2017.cluster.pressure.profiles.highres.sz.xray.cool.cores.show.central.pressure.deficit}, 6 of 7 CC clusters show such a deficit (with a median factor $\sim 3$ deficit at the MUSTANG resolution limits), while none of the observed NCC clusters show a clear deficit. Follow-up tests in \citet{silich:2025.cr.ic.tests.in.zw.3146.cluster} on a different candidate CC also found such a deficit.

Second, since CR-IC produces a thermal \textit{continuum} spectrum, but CRs less-strongly excite line emission, the apparent equivalent width of X-ray line emission in CC centers should be diluted, leading to the X-ray inferred metallicity from simple weighted single-temperature soft X-ray spectral fits being under-estimated. This will lead to artificially slowly-rising, sub-Solar central X-ray-inferred metallicities in direct contradiction to super-Solar central metallicities measured from optical/UV diagnostics, and in the most extreme cases would even explain X-ray inferred central ``metallicity drops'' (where the central X-ray-inferred metallicity appears to \textit{decrease} at $\lesssim10\,$kpc). But this is actually a well-established phenomenon, with such slow-rises and even drops seen in X-ray observations in almost every one of the most nearby, most (spatially) well-resolved bright CC clusters (e.g.\ Centaurus, Virgo, Perseus, HGC 62, Fornax, NGC 5044, Hydra-A, Ophiuchus, and many more). However, here caution is needed as it has been clearly demonstrated observationally that these apparent central X-ray metallicity profiles are sensitive to spatial resolution/binning, spectral resolution, signal-to-noise, fitted wavelength range and instrumental properties, actual fitting methods/analysis, and priors on the temperature distribution. Moreover even given perfect observations, quantitative predictions of the line structure and effect of CRs depend on the true multi-phase temperature/density/metallicity/ionization structure and CR heating/ionization/excitation integrated through the cluster, as well as the exact CR spectrum (itself dependent on the injection spectrum, transport physics, and possible reacceleration mechanisms) which are not a priori known. Related to this, we show that effects on X-ray temperature estimation, line ratios, and line widths, from CR-IC are quite small, and often degenerate with multi-temperature modeling, though future work quantitatively modeling the effects on X-ray spectra in cases where SZ and metallicity and synchrotron observations from other wavelengths provide stronger priors for CR-IC models is clearly warranted.

We also discuss different scenarios and predictions for redshift evolution and their consequences for high-redshift clusters. Here the predictions depend on (highly uncertain) questions regarding both how cluster thermodynamic properties do (or do not) evolve, and how CR injection rates and transport physics may (or may not) evolve. Nonetheless we show that the simplest possible model, in which CR injection rates and transport physics are un-evolving and CR-IC dominates the central surface brightness, makes a simple prediction that the observed central X-ray surface brightness of cool cores should evolve weakly or not at all, while (similarly) inferred thermodynamic properties of cores like their central entropy would appear to evolve weakly as well. This appears to be consistent with the behavior of observed bright cool cores at redshifts $z \sim 0-2$, and differs significantly from some thermal model predictions, but expanding high-redshift ($z\gtrsim1$) high-angular-resolution CC-cluster samples would greatly strengthen these constraints.

Motivated by all of this, we expand on the (speculative) scenario in \paperone\ wherein weak ``true'' cooling flows or other sources of accretion lead to AGN activity, which produces an ACRH that persists for $\gtrsim$\,Gyr (even after the AGN begins to fade), producing a long-lived ``apparent'' much stronger CC from CR-IC (even if the AGN disrupts most of the central cool gas by expelling it or raising its entropy). 
This would naturally explain the apparent order-unity fraction of CC clusters with the specific central profiles predicted by CR-IC. 
It would also potentially bring models of AGN feedback into much better agreement with observations, as historically the most difficult aspect of the ``theoretical cooling flow problem'' in simulations has been reproducing a large population of clusters with standard CC central surface brightness/density/entropy/temperature profiles (but these historical comparisons neglected CR-IC shaping those profiles). In this context, ACRH halos naturally extend the known mini-halo and ultra-steep spectrum halo categories, with those representing the younger CRs at smaller radii (closer to the source) within more extended ACRHs whose primary radio emission emerges at $\sim$\,MHz wavelengths.

In the near future, expanding the sample of clusters with extremely high-resolution non-parametric SZ pressure profiles (reaching $\ll 10-100\,$kpc, ideally) would enable the most direct, clean, and powerful tests of this scenario. Similar tests would be possible with independent estimates of column densities from outside the X-rays: for example, future samples of fast radio burst (FRB) dispersion measures whose foreground lines-of-sight pass through clusters could be used for a version of this test, but this requires background FRBs with impact parameters $\ll 100\,$kpc from the cluster BCG (which, given the areas, will be much more rare than FRBs passing through cluster outskirts; see \citealt{connor:2023.frb.foreground.cluster.outskirts}). Looking further ahead, it may be possible to obtain more detailed constraints, especially for the line effects above, with future high-spatial-and-spectral resolution X-ray microcalorimeter instruments. But given the caveats above, such comparisons clearly require detailed forward-modeling of actual synthetic X-ray spectra from simulations including the thermal emission, CR-IC off all backgrounds (allowing for a variety of CR spectra and transport assumptions), and additional effects like CR Coulomb heating and excited line ionization and emission. 
On the theory side, as more simulations of AGN feedback include explicit integration of CR populations \citep{chan:2018.cosmicray.fire.gammaray,su:2018.stellar.fb.fails.to.solve.cooling.flow,su:turb.crs.quench,su:2021.agn.jet.params.vs.quenching,su:2025.crs.at.shock.fronts.from.jets.injection,wellons:2022.smbh.growth,byrne:2023.fire.elliptical.galaxies.with.agn.feedback} and even the full lepton spectrum \citep{girichidis:cr.spectral.scheme,girichidis:2021.cr.transport.w.spectral.reconnection.hack,ogrodnik:2021.spectral.cr.electron.code,hopkins:cr.multibin.mw.comparison,hopkins:2021.sc.et.models.incompatible.obs,hopkins:cr.spectra.accurate.integration,ponnada:2023.fire.synchrotron.profiles,ponnada:2023.synch.signatures.of.cr.transport.models.fire,ponnada:2024.fire.fir.radio.from.crs.constraints.on.outliers.and.transport}, direct predictions for ACRHs can be made, and future comparisons with observed X-ray cluster properties should not ignore CR-IC. In particular, this re-emphasizes the value of theorists actually forward-modeling (and observers publishing) actual observed quantities like cluster surface brightness profiles, hardness, line ratios, and equivalent widths, rather than comparing to the usual inferred quantities like density, entropy, temperature, pressure, and metallicity, which are strongly-model dependent quantities indirectly inferred, but not directly measured, by X-ray observations.

\begin{acknowledgements}
We thank Michael McDonald, Mark Voit, John ZuHone, Kung-Yi Su, Daisuke Nagai, Andy Fabian, Peng Oh, Tim Heckman, Irina Zhuravleva, Paola Dominguez Fernandez, Joe Lazio, Ian Smail, and Eliot Quataert for helpful conversations and suggestions. Support for PFH was provided by a Simons Investigator Grant. Numerical calculations were run on allocation AST21010 supported by the NSF and TACC.
\end{acknowledgements}

\bibliographystyle{mn2e}
\bibliography{ms_extracted}

\end{document}